\documentclass[11pt]{report}
\usepackage[toc,page]{appendix}
\usepackage{amsmath, amssymb,hyperref,enumitem,graphicx,feynmp, color, physics, simpler-wick, multicol}
\usepackage{bm}
\usepackage{mathtools}
\usepackage{fancyhdr}
\usepackage[normalem]{ulem}
\usepackage{simpler-wick}
\usepackage[compat=1.1.0]{tikz-feynman}
\usepackage[utf8]{inputenc}
\usepackage{slashed}
\usepackage[paperwidth=24cm,paperheight=28cm]{geometry}
\usepackage{lipsum}
\usetikzlibrary{calc}

\newenvironment{lyxlist}[1]
{\begin{list}{}
		{\settowidth{\labelwidth}{#1}
			\setlength{\leftmargin}{\labelwidth}
			\addtolength{\leftmargin}{\labelsep}
			}}
	{\end{list}}

\newcommand{\stkout}[1]{\ifmmode\text{\sout{\ensuremath{#1}}}\else\sout{#1}\fi}

\tikzset{
	photon/.style={decorate, decoration={snake}, draw=black},
	fermion/.style={draw=black, postaction={decorate},decoration={markings,mark=at position .55 with {\arrow{>}}}},
	vertex/.style={draw,shape=circle,fill=black,minimum size=3pt,inner sep=0pt},
}

\NewDocumentCommand\semiloop{O{black}mmmO{}O{above}}
{%
	\draw[#1] let \p1 = ($(#3)-(#2)$) in (#3) arc (#4:({#4+180}):({0.5*veclen(\x1,\y1)})node[midway, #6] {#5};)
}

\DeclareMathOperator*{\SumInt}{%
	\mathchoice%
	{\ooalign{$\displaystyle\sum$\cr\hidewidth$\displaystyle\int$\hidewidth\cr}}
	{\ooalign{\raisebox{.14\height}{\scalebox{.7}{$\textstyle\sum$}}\cr\hidewidth$\textstyle\int$\hidewidth\cr}}
	{\ooalign{\raisebox{.2\height}{\scalebox{.6}{$\scriptstyle\sum$}}\cr$\scriptstyle\int$\cr}}
	{\ooalign{\raisebox{.2\height}{\scalebox{.6}{$\scriptstyle\sum$}}\cr$\scriptstyle\int$\cr}}
}

\makeatletter
\newcommand{\mathleft}{\@fleqntrue\@mathmargin0pt}
\newcommand{\mathcenter}{\@fleqnfalse}
\makeatother

\title{ 
	Finite Temperature Corrections in an Infinite Derivative Theory of Gravity}
\author{Casper Dijkstra \\
Supervisor: prof. dr. Anupam Mazumdar}

\newcommand{\mathz}{ \mathcal{Z} }
\newcommand{\bal}{ \begin{aligned} }
\newcommand{\eal}{ \end{aligned} }

\newcommand{\bse}{\begin{subequations}}
\newcommand{\ese}{\end{subequations}}

\newcommand{\map}{\mathbf{p}}

\newcommand{\nnn}{\newline \newline \noindent }
\newcommand{\tb}{\textbf}
\newcommand{\ti}{\textit}

\begin{document}
 
\begin{titlepage}

\begin{center}
\textbf{	Candidate: Casper Dani\"{e}l Dijkstra. }\\
\textbf{	Student Number: s2026104.}
\end{center}
 	
 	\begin{center}
\large{Master's thesis for Quantum Universe.} 

\large{Specialization: Theoretical Physics. }
	\end{center}

	\begin{center}
		\Huge{Local and Nonlocal Thermal Field Theory}
	\end{center}

	\begin{figure}[h!]
	\centering
	\includegraphics[scale=0.3]{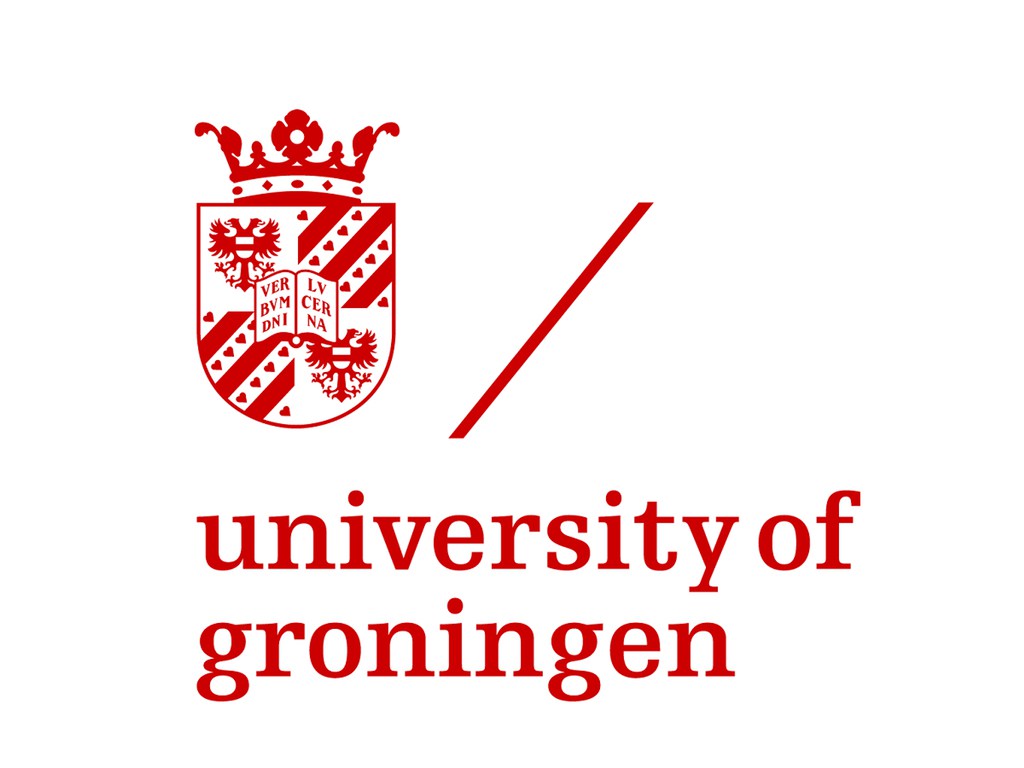}
\end{figure}

\vspace{5cm}

\begin{minipage}{.9\linewidth}
	\begin{flushleft}                           
\textbf{Supervisor: Prof. Dr. Anupam Mazumdar.} \\
\textbf{Second examiner: Prof. Dr. Diederik Roest.} \\
\textbf{Academic year: 2017/2018.} 
	\end{flushleft} 
\end{minipage}

\end{titlepage}

\newpage

\subsubsection*{Abstract}
Firstly, nonlocal field theories will be motivated, primarily in the gravity sector. We discuss how nonlocal theories of gravity can circumvent typical problems of finitely-many higher derivative theories and can, among other things, be \emph{either ghostfree and (potentially) renormalizable} and yield a non-singular Newtonian potential.
Afterwards we motivate finite temperature field theory, also known as thermal field theory, and study the thermodynamic behavior of both a local and a nonlocal scalar field theory. 
We compute (primarily leading-order) thermal corrections to the partition function for  
in low- and high- temperature expansions
and calculate the thermal mass which is acquired through continuous interactions with the heat bath. 
\nnn
We prove that the nonlocality does not contribute to the partition function of the free nonlocal theory. The presence of the nonlocality can only be noticed through interactions and we conclude that the partition function of the local and the nonlocal scalar theory are fundamentally different when interactions are included. 
We explain from both a mathematical and physical point of view why the nonlocality can only be noticed at the level of interactions and conclude that this cannot be generalized to arbitrary nonlocal modifications - only those which do not introduce new poles in the propagator. 
We study whether our results can be reconciled with the stringy thermal duality and the conjectured Hagedorn phase and conclude that the former is violated, while Hagedorn behavior emerges in the high temperature expansion of the nonlocal scalar theory. We will interpret all these results. 
\subsubsection*{Acknowledgements}
I would like to thank my supervisor prof. dr. Anupam Mazumdar for supervising my master research project and for the useful feedback which I received whenever we discussed new results. Moreover, I'm grateful for fruitful discussions with 
Folkert Nobels, Stan Lambregts, Luca Buoninfante and Shubham Maheshwari. Additionally, I would like to thank prof. dr. Joseph Kapusta and prof. dr. Tirthabir Biswas for discussing questions related to nonlocal thermal field theory with me via email correspondence.

\newpage
\thispagestyle{empty}
\listoffigures

\thispagestyle{empty}
\tableofcontents
\thispagestyle{empty}

\newpage

\pagenumbering{roman}

\section*{Conventions and notation
\hypertarget{conventions}{}
}
\addcontentsline{toc}{chapter}{Conventions and notation}

\subsubsection*{Natural units}
Natural units can simplify particle physics considerably in relativistic quantum mechanics, since quantum mechanics introduces factors of $\hbar$ and special relativity introduces factors of $c$ which obfuscate equations. We can bypass this by using natural units where $\hbar = h/2\pi =  1$ (turns Joule into inverse seconds) and $c = 1$ (turns meters into seconds). This makes all quantities have dimensions of energy (or mass, using $E = mc^2$) to some power. \newline \newline
\noindent
Quantities with positive mass dimension, (e.g. momentum $p$ and $\partial_\mu$) can be
thought of as energies and quantities with negative mass dimension (e.g. position $x$ and time $t$) can
be thought of as lengths. Denoting the dimensionality of $`\cdots'$ by $[\cdots]$, some examples are:
\begin{subequations}
	\begin{align}
\left[ \partial_{\mu} \right] = [p_\mu] = [k_\mu] = [m] &= E, \label{M1} \\
[dx] = [x] = [t] &= E^{-1}, \\
[\text{velocity}] = [x] / [t] &= E^0, \\
\left[d^4x \right] &= E^4.
\end{align}
\end{subequations}
The fact that the action $\left[ S \right] = \int \left[d^4x \ \mathcal{L} \right]$ in 4D field theories is a dimensionless quantity implies that the Lagrangian density has dimensionality energy to the power four, i.e.
$\left[ \mathcal{L} \right] = E^4$.\footnote{In $\hbar \neq 1$ units, $[S] = \hbar$.} Additionally, it is important to realize that the often recurring terms $[T] = [M] = E$, where $T$ is temperature and $M$ the so-called scale of nonlocality (the meaning of which will be discussed in length in chapter \ref{chapterIDG}).
\nnn
Additionally we set the Boltzmann constant equal to unity, i.e. $k_B = 1$. This entails that the Lagrange multiplier for the mean energy of a system $\beta$ is given by $\beta = \frac{1}{k_B T} = T^{-1}$.
\newline \newline
\noindent
It is worth mentioning the exact expressions of several important constants in natural units. The Einstein equation in general relativity is given by
\begin{equation}\label{key}
\mathcal{R}_{\mu \nu} - \frac{1}{2}\mathcal{R}g_{\mu \nu} = \kappa \tau_{\mu \nu},
\end{equation}
where $\tau_{\mu \nu}$ is the energy-momentum tensor  and $\kappa$ is the constant which guarantees that Newtonian mechanics is recovered for low energies. Its expression is well-known in SI units, namely: $\kappa = {8 \pi G \over c^4}$ and this thus reduces to $8\pi G$ in natural units.
It is often desirable to display the Planck mass (which sets the scale where gravitational quantum effects become important) instead of $\kappa$ in gravitational equations, let us therefore mathematically relate these quantities. The Planck mass is defined as
\bse
\begin{equation}\label{key}
m_p := \sqrt{\frac{\hbar c}{G}} \simeq 1.2 \cdot 10^{19} \text{ GeV}
\end{equation}
and since $\hbar = c = 1$ in natural units we can relate $\kappa$ and $m_p$ ($\kappa = 8\pi m_p^{-2}$), which can be written more conveniently after introducing the \textit{reduced Planck mass} which already contains the requisite factor of $\sqrt{8\pi}$
\begin{equation}\label{key}
M_p := \sqrt{\frac{\hbar c}{8 \pi G}} \simeq 2.4 \cdot 10^{18} \text{ GeV},
\end{equation}
which entails that
\begin{equation}\label{key}
\kappa = \frac{1}{M_p^2}.
\end{equation}
\ese
Gravitational field equations can now be expressed in terms of the reduced Planck mass.

\subsubsection*{Dimensional analysis}
Free massive scalar particles are described by the Klein-Gordon equation, whose Lagrangian density is\footnote{We will often refer to the Lagrangian density as ``Lagrangian'' in this thesis.}
\begin{equation}\label{key}
\mathcal{L}_{\text{free}} = \frac{1}{2} \left[ \partial_\mu \phi  \partial^\mu \phi - m^2 \phi^2 \right]
\end{equation}
so the dimensionality of the boson field can immediately be deduced using Eq. \ref{M1}, namely: $[\phi] = E$. Free spin 1/2 particles are described by the Dirac equation
\begin{equation}\label{key}
\mathcal{L}_{\text{free}} = \bar{\psi} (i \slashed{\partial} - m) \psi,
\end{equation}
where $\slashed{\cdots} = \gamma^{i} (\cdots)$. The fact that only one partial derivative and one mass term occurs in this Lagrangian density implies that $[\psi] = E^{3/2}$ and this can be generalized to other types of bosonic fields. \nnn
The dimensionality of coupling parameters can be straightforwardly obtained in a similar fashion. In scalar $\lambda \phi^4$ theory for instance, the quartic couplic is dimensionless whereas $\lambda = E$ in a cubic ($\lambda \phi^3$) theory.

\subsubsection*{Indices and the Einstein summation convention}
Latin indices $i,j,k$ and so on are used for spatial components only and thus take on values 1,2,3 in a (3+1)-space-time and values $1,2,\cdots,D-1$ in a $D$-dimensional space-time. Greek indices $\mu, \nu, \rho$ et cetera run over both spatial components and the temporal component. \nnn
Indices can be lowered and raised by letting either the covariant or contravariant metric tensor act upon the corresponding tensor.
We will show in the subsubsection \emph{tensors} how Ricci tensor and curvature scalar can be obtained by taking contractions (using the metric tensor) of the 4-rank Riemann tensor. \nnn
For notational brevity one can choose to suppress indices, this will be done in this thesis when introducing the spin projector operators in Appendix \ref{spinproj}, for instance, $P_{\mu \nu \rho \sigma}^2$ will be written as $P^2$.\nnn
Moreover, the Einstein summation convention, which states that repeated indices are summed over, is used in order to achieve notational brevity. It implies that
\bse
\begin{equation}\label{key}
\sum_{\mu=0}^{4} x_{\mu}x^{\mu} \equiv x_{\mu}x^{\mu} \ (=x^2).
\end{equation}
and the Einstein summation convention likewise allows for the following short-hand notation 
\begin{equation}\label{key}
\mathbf{x} \cdot \mathbf{y} = x^{i}y_{i}.
\end{equation}
\ese

\subsubsection*{Metric signature}
The signature of the metric which is used throughout this thesis is $\eta_{\mu \nu} = \text{diag}(-+++)$ when in Minkowski space. 
Euclidean space however plays a significantly more imporant role in thermal field theory than in ordinary quantum field theory, = we will work exclusively in Euclidean space in this thesis.
In the imaginary time formalism, to be introduced in Chapter 1, a \textit{Wick rotation} $t \rightarrow -i\tau$ is performed which entails that Minkowski space is replaced by Euclidean space with $\eta_{\mu \nu} = \text{diag}(++++)$ in thermal field theory computations. \nnn
The d'Alembertian operator will frequently recur in this thesis. Since its form is dependent on the sign convention of the metric we will explicitly specify its definition here.
\bse
\begin{align}
\square &= -\partial_t^2 + \nabla^2	\  \ \ \text{in Minkowski space-time,} \\
\square &= \ \ \partial_\tau^2 + \nabla^2	\ \ \ \text{in Euclidean space-time.}
\end{align}
\ese
\subsubsection*{Fourier transforms}
Different conventions are employed in the literature as to where the factors of $2\pi$ should be mentioned in Fourier transforms. The origin of these factors lies in
\begin{subequations}
\begin{equation}\label{key}
\delta(x) = \int_{-\infty}^{\infty} dp \ e^{2\pi ikx}
\end{equation}
which holds true for either sign.
Either $p$ or $x$ can be rescaled in order to remove the factor $2\pi$ from the exponent. We choose to rescale $p$, since position is generally not an angular coordinate. The result is
\begin{equation}\label{key}
2\pi \delta(x) = \int_{-\infty}^{\infty} dp \ e^{ ikx}.
\end{equation}
The convention where \textit{momentum-space integrals contain $(2\pi)^{-D}$ factors (where $D$ is the dimensionality of space-time) and position space integrals have no $2\pi$ factors} is employed in this thesis. This implies in $4D$ that
\begin{equation}\label{key}
f(x) = \int {d^4p \over (2\pi)^4} \tilde{f}(p) e^{\pm ikx}
\end{equation}
and
\begin{equation}\label{key}
\tilde{f}(x) = \int d^4x \ f(x) e^{\pm ikx}.
\end{equation}
\end{subequations}
The sign is determined by our convention that the Fourier transform of the partial derivative $\partial_\mu$ is $-ik_\mu$, it thus satisfies
\begin{equation}\label{key}
\partial_{\mu} f(x) = -ik_\mu \tilde{f}(x).
\end{equation}
The sign for $f(x)$ is consequently ``$-$'' and the sign for $\tilde{f}(p)$ is ``$+$''.

\subsubsection*{Tensors}
Let us introduce a convenient notation
for symmetric and antisymmetric tensors.
When indices are enclosed in curly braces or brackets, these denote
properties of either symmetry or antisymmetry respectively, defined by the following rules
\begin{equation}\label{key}
T_{\{\mu \nu \}} = \frac{1}{2}\left[T_{\mu \nu} + T_{\nu \mu} \right], \ \
T_{[\mu \nu]} = \frac{1}{2}\left[T_{\mu \nu} - T_{\nu \mu} \right]. 
\end{equation}
The (anti-)symmetry only applies to the \textit{first and last index} enclosed by the curly braces or brackets, so for instance $\mathcal{R}_{\{\mu \nu \rho \sigma\}}$ should be interpreted as
\begin{equation}\label{key}
\mathcal{R}_{\{\mu \nu \rho \sigma\}} =
\frac{1}{2}
\left[
\mathcal{R}_{\mu \nu \rho \sigma}
+
\mathcal{R}_{\sigma \nu \rho \mu}
\right].
\end{equation}
The linearized forms of the curvature tensors are
\bse
\begin{align}
\mathcal{R}_{\mu\nu\lambda\sigma} &= 
\frac{1}{2}\left[\partial_{\nu}\partial_{\lambda}h_{\mu\sigma}
+\mathcal{\partial_{\sigma} \partial_{\mu}}h_{\nu\lambda}
-\mathcal{\partial_{\sigma}\partial_{\nu}}h_{\mu\lambda}
-\partial_{\mu}\partial_{\lambda}h_{\nu\sigma}\right] \\
&=\big[\partial_{[ \nu}\partial_{\lambda}h_{\mu\sigma ]}
+\mathcal{\partial_{\sigma} \partial_{[\mu}}h_{\nu]\lambda}
 \big]
, \\
\mathcal{R_{\mu\nu}}&=g^{\alpha\rho}\mathcal{R}_{\alpha\mu\rho\nu}=  \frac{1}{2}\left[\partial_{\rho}\partial_{\nu}h_{\mu}^{\rho}+\partial_{\rho}\partial_{\mu}h_{\nu}^{\rho}-\partial_{\mu}\partial_{\nu}h-\square h_{\mu\nu}\right], \\
\mathcal{R} &=  \partial_{\mu}\partial_{\nu}h^{\mu\nu}-\square h.
\end{align}
\ese
\subsubsection*{Abbreviations}
The magnitude of three-vectors is abbreviated by small variables, so $x \equiv \abs{ \mathbf{x} }$ and $k \equiv \abs{\mathbf{k}}$. Both Euclidean and Minkowski spacetime play an important role in thermal field theory and we would like to introduce short-hand notations for calculations in both cases. 
Capital letters  are used for Euclidean cases, $X$ and $Y$ denote four-vectors whose first entry is the imaginary time variable and whose second entry is the spatial three-vector, i.e.
\begin{equation}\label{key}
X^{\mu} \equiv  (\tau, x^{i}).
\end{equation}
The Euclidean action is defined as the space-time integral over the Euclidean Lagrangian
\begin{subequations}
\begin{equation}\label{key}
 S_{E} = \int_{X} L_{E},
\end{equation}
where the following short-hand notations are used
\begin{equation}\label{key}
\int_{X} \equiv 
\int_{0}^{\beta} d \tau \int d^{3}x
= \int_{0}^{\beta} d \tau \int_{\mathbf{x}},
\end{equation}
where
\begin{equation}\label{key}
\int_{\mathbf{x}} = \int d^{3}x
\end{equation}
so $\int_{X}$ denotes the integral over Euclidean space and time and $\beta = T^{-1}$ is the Lagrange multiplier of the canonical ensemble which determines the average energy of the system.
\end{subequations} \newline \newline
\noindent
Capital letter $K$ is used for Euclidean four-wavevectors
\begin{subequations}
\begin{equation}\label{key}
K^{\mu} = (\omega_n, k_{i})  
\end{equation}
and since natural units are used the four-wave vectors is equal to the four-momentum: $p_\mu = k_\mu$.
A short-hand notation for Fourier transforms will now be introduced:
\end{subequations}
\begin{subequations}
\begin{equation}\label{key}
\phi(X) = \SumInt_{K} \tilde{\phi} \ e^{i K \cdot X},
\end{equation}
where
\begin{equation}\label{sumint}
\SumInt_{K} = T \sum_{\omega_n}
\int \frac{d^{3} k }{(2\pi)^{3}}
=
T \sum_{\omega_n}
 \int_{k},
\end{equation}
\begin{equation}\label{key}
\int_{\mathbf{k}} = \int \frac{d^{3} k }{(2\pi)^{3}}.
\end{equation}
The origin of the term $T \sum_{\omega_n}$, in which a summation over Matsubara frequencies is taken, is probably not clear to the reader at this point. 
The summation over Matsubara frequencies is due to the fact that energy is a discretized variable in the imaginary time formalism and I will refer to this short-hand notation again in section \ref{imagform} (where the imaginary time formalism is introduced) to prevent any confusion for the reader.
\end{subequations}

\section*{Acronyms}

\begin{table}[h!]
	\begin{tabular}{ l l }
		BE	:	& Bose-Einstein, \\
		DOF:	& Degrees Of Freedom, \\
		EFT: 	&Effective Field Theory, \\
		EH: 	&Einstein-Hilbert, \\
		FD		&Fermi-Dirac,		\\
		GR: 	&General Relativity, \\
		IDG: 	&Infinite Derivative Gravity, \\
		IR: 	&Infrared, \\
		ITF: 	&Imaginary Time Formalism, \\
		NLO		&	Next-to-Leading Order, \\
		NLO		&	Next-to-Next-to-Leading Order, \\
		RTF		&Real Time Formalism,	\\
				QED		&	Quantum ElectroDynamics, \\
		QCD		&	Quantum ChromoDynamics, \\
		QCP		&	Quark-Gluon Plasma,	\\ 
		QFT: 	&Quantum Field Theory, \\
		TFT: 	&Thermal Field Theory, \\
		UV: 	&Ultraviolet,  \\
		VEV		& Vacuum Expecation Value. \\
	\end{tabular}	
\end{table}

\section*{Introduction}
\addcontentsline{toc}{chapter}{Introduction}
\section{Why thermal field theory?}
Quantum field theories at zero temperature have been studied in great detail for several decades. Most features of zero temperature quantum field theories, such as QED and QCD, are now successfully understood, at least perturbatively. Perturbative calculations are carried out through the use of Feynman diagrams and corresponding Feynman rules, where
regularizations are fairly routine in order to tame either infrared (IR) and ultraviolet (UV) divergent radiative corrections.

QFTs have given rise to a plethora of predictions which are in exquisite accordance with experimental data obtained at particle colliders. This indicates that elementary particles can usually be understood sufficiently well within zero temperature field theories. One may however ask whether all finite temperature behavior can be encoded in QFTs and, if not, QFTs should not generically serve the role of sufficiently accurate approximations of the real world which is at finite temperature. 
\nnn
An important lesson to be learned from statistical physics  is that temperature tends to disorganize structures which are present at low temperatures. Examples are the Bose-Einstein condensate which bosons can only form at sufficiently low temperatures
and the Ising model in which a second order phase-transition occurs due to the disorganizing effect of temperature.
Consequently, it is desirable to have a quantum field theory at finite temperature at our disposal and, in fact, thermal field theory frameworks have already been introduced in the fifties when Matsubara constructed the \textit{imaginary time formalism} (ITF) \ref{refmatsubara}. 
This field theory allows the temperature $T$ and chemical potential $\mu$ to take on non-zero values and allows one to examine the induced effects on particle physics.\footnote{Calculations prove a lot more challenging than their zero temperature counterparts 
	and there is ``a continuous effort to simplify such calculations in order for thermal field theories to be embraced by the community at large'' \ref{das}.}
\nnn
In thermal field theory, the particle processes take place in a heat bath, \emph{i.e.} a background of particles at temperature $T$, rather than in an ordinary vacuum.
We will use the name “Thermal Field Theory" (TFT) to indicate
the modern version of quantum-mechanical many-body theory  which has been developed in the
fifties and sixties. TFT deals with the behavior of large assemblies of elementary particles
at non-zero temperature, and should be thought of as the unified framework of statistical mechanics
and elementary-particle physics (therefore it is the relativistic generalization of quantum statistical physics) with applications to the early universe, astrophysics
and ultra-relativistic nucleus-nucleus collisions. Moreover, it allows one to check whether predictions from QFT are robust at finite temperature, since the real world is at finite temperature.\footnote{From a philosophical point of view, thermal field theories (TFTs) ought to be more accurate theories than ordinary quantum field theories (QFTs). The real universe is at finite temperature and therefore one should always include the effects of the presence of a heat bath, however, quantum field theories are devoid of a temperature parameter and hence describe how particles behave at $T = 0$. Although quantum field theories at zero temperature give remarkably accurate results, TFTs provide more appropriate means for examining particle phenomena in general. A genuine necessity to use thermal field theory arises mostly at high temperatures, where the collective effects from, for instance, absorption of soft photons and electrons from the heat bath can no longer be neglected.}
\nnn
Thermal field theory allows one to study phase transitions involving symmetry
restoration in the context of theories with a  spontaneously broken symmetry such as the electroweak theory. Questions such as the chiral symmetry breaking phase transition \ref{refgross} and the confinement-deconfinement
phase transition in QCD \ref{wilczek} have drawn a lot much attention and are incredibly relevant as planned experiments
involving heavy ion collisions could either refute or verify these predictions. These experiments would improve our understanding of the quark-gluon plasma and its properties better. The main applications of TFT in high energy physics can roughly be sorted in three categories (with examples):
\begin{enumerate}[label=(\roman*)]
	\item {\textit{Cosmology:}} 
	The early universe is thought to have consisted of a quark-gluon plasma (QCP). The existence of this deconfined phase of QCD has been inferred from jet quenching in lead-lead (Pb-Pb) collisions at CERN (ATLAS \ref{atlas}, CMS \ref{cms} ALICE \ref{alice}). The consistence between experimental data and theoretical
	predictions of the Standard Model has to be verified for quark-gluon plasma and it would be appropriate to study this hot quark gluon plasma in thermal field theory. 
	
	In equilibrium thermal field field theory (typically carried out in the imaginary time formalism) one can study how the equation of state of QCD depends on temperature. One could study how fast (and with what kind of geometry) a hot plasma expands. These predictions can hopefully be tested at the RHIC (Relativistic Heavy Ion Collider) in Brookhaven where gold nuclei collide. In near-equilbrium physics one could study the production rate of dilepton pairs and photons from the plasma, both of which can be regarded analogues of the CMB from the primordial universe.
	The interested reader is referred to \ref{pisarki} for a detailed description of the QCD plasma (and more generally, gauge theories) in thermal field theory.
	
	\item{ \textit{High-energy physics}:
		The fundamental question
		of the asymmetry between matter and antimatter in the observable universe
		remains puzzling up to date, although many solutions to this question have been put forward. One of the first scientists who aimed to explain this baryonic asymmetry of the observable universe was Sakharov in 1967 who proposed that baryon number ($\mathcal{B}$) violations may occur at high temperatures \ref{sakharov}. 
		The work by Sakharov remained unnoticed for several
		years, because the key hypothesis of \emph{baryonic charge
			nonconservation}, which is required to generate an asymmetry between baryons and antibaryons, was initially
		not accepted. The Standard Model of particle physics is constructed in such a way that $\mathcal{B}$ is an accidental
		symmetry of the Lagrangian. This is true only at the classical level, however, quantum effects associated
		with the weak interactions violate baryon number non-perturbatively \ref{baryonviolation}.
		These violations arise because $\mathcal{B}$
		is anomalous with respect to the weak interactions.
		\footnote{Baryon number violation can also result from
			physics beyond the Standard Model theories such as sypersymmetry and Grand Unification Theories (GUTs).}
		The rate of baryon violation by the electroweak anomaly is, although mathematically challenging, a suitable problem to address within thermal field theory.
	}
	\item{\textit{Astrophysics}:
		The possiblility of the existence of a quark matter core
		in neutron stars can be studied in finite temperature field theory. Due to the close analogy with the QCP discussed in (i) \textit{Cosmology} this will not be discussed in more detail here.}
\end{enumerate}
These problems will not be addressed further in this thesis, but serve to convey a general outline of the kind of problems which can be solved using thermal field theory. Several aspects of quantum fields at finite temperature in the canonical ensemble will be discussed throughout this thesis, all of which can straightforwardly be generalized to a finite chemical potential in the grand canonical ensemble (since these ensembles are related through an inverse Laplace transformation on $\beta$) \ref{kapustabook}. Among the interesting results which will be discussed is the dynamically generated thermal mass of scalar particles caused by continuous interactions of the particle with the heat bath, which - in the context of the Higgs field - can potentially restore the spontaneously broken symmetry of the electroweak theory. 
\nnn
Subsequently I will discuss the thermal behavior of a nonlocal scalar field. I would like to stress that this is not a purely academic exercise - nonlocal models instead provide viable theories worth studying and therefore I have chosen to devote a whole chapter (the first chapter) to motivating nonlocal field theories, in particular in the gravity sector. The aim of my thesis is consequently twofold - 
\emph{1)} to motivate nonlocal theories of gravity to convey the relevance of studying such models at finite temperature (the nonlocal scalar model considered in chapter \ref{chapternonlocalfield} contains an exponential modification $a(\square) = \exp[-\square/M^2]$ whose origin lies in 
papers by Kuz'min and Tomboulis \ref{kuzmin} \ref{superrenormalizablegauge } but and is used in theories of gravity as well - namely in Infinite Derivative Gravity. And
\emph{2)} to discuss finite temperature in the context of both local and nonlocal models. I will now motivate why one would want to consider nonlocal field theories in the first place and discuss the meaning of nonlocality.
\nnn
 Readers who are \textit{already familiar with nonlocal field theories} and/or who are \textit{solely interested in local and nonlocal thermal field theory} can skip chapter \ref{chapterIDG}
and proceed to the chapters on thermal field theory (chapters \ref{chapterTFT}, \ref{chapternonlocalfield}) and the conclusions \ref{chapterconclusion}.
\section{Why nonlocal field theories?}
The Lagrangians which constitute the standard model are local and, in fact, locality is assumed in most undergraduate and graduate physics courses. Nonlocal Lagrangian are however ubiquitous in miscellaneous fields of physics.
As far back as the 1950s Hideki Yukawa considered nonlocal theories as an alternative to the usual point particle description. In a two paper series,
he investigated the possibility of removing divergences by taking into account the finite
size of elementary particles \ref{yukawa0}a,b.
Since then, nonlocal fields have been investigated in a variety of contexts. An excellent
example of this is work done by John Moffat and collaborators \ref{moffat} starting in the
nineties. He investigated the possibility of a nonlocal alternative to the Standard Model.
In much of his work he investigated an ultraviolet complete quantum field theory with
nonlocal interactions. Another interesting exploration by Evans, Moffat, Kleppe and
Woodard was the use of nonlocal regularizations of gauge theories \ref{moffatnonlocal}a-d.
\nnn
One of Moffat’s colaborators, Woodard, together with Deser subsequently explored a
nonlocal field theory of gravity model \ref{refdarkenergy1}. They managed to avoid some of the fine tuning which is necessary in the standard
cosmological models and moreover argued that nonlocal theories of gravity may prove fruitful to resolve
the black hole information problem.
But what exactly makes these theories interesting and which indications exist that nature may in fact be described by nonlocal field theories? We will motivate nonlocal field theories primarily in the gravity sector in both this introduction and more elaborately in chapter \ref{chapterIDG}.
\nnn
General relativity is plagued by singularities at both a cosmological (Big Bang) and astrophysical (singularities in Schwarzschild/Kerr metric) level.
The appearance of singularities in theories of nature
usually signals the breakdown of the theoretical description and motivates the need to go beyond the current framework. 
One of the key messages put forward in chapter \ref{chapterIDG} is that nonlocal extensions of the Einstein-Hilbert action (defining GR) are capable of circumventing typical problems which are exhibited by local generalizations of this action. 
\nnn
Finite higher derivative theories (in derivatives), for instance Fourth Order Gravity to be discussed in \ref{sectionstelle}, typically
invite ghosts into the theory – physical excitations with negative residue in the graviton propagator. This negative residue manifests itself as a negative kinetic energy, leading to instabilities even at a classical level \ref{luca} and a breakdown of unitarity in its quantum formulation. 
Higher-order derivative corrections to the Einstein-Hilbert action present a
\emph{unitary but non-renormalizable} \textbf{or} \emph{renormalizable but non-unitary} trade-off (as will be discussed elaborately in section \ref{sectionIDG}), whereas in fact
both options are unappealing from a theoretical point of view.
Interestingly, nonlocal theories of gravity can evade this problematic trade-off
and allow us to potentially construct a ghost-free (in order to preserve unitarity) and renormalizable theory of gravity.
We will discuss more captivating features which can be exhibited by nonlocal field theories, but not by local field theories.
\nnn
Examples of theoretical frameworks in which physicists aim to achieve a successful theory of quantum gravity are String Field Theory (SFT) and Loop Quantum Gravity (LQG).\footnote{SFT and LQC are often introduced as competing theories because their approaches for the quantization of gravity are fundamentally different.\footnote{This is not an exhaustive list of theoretical frameworks in which theories of quantum gravity are constructed. There is Causal Set approach,  dynamical triangulation, theories based on asymptotic safety and a lot more, with varying degrees of success.} 
	String theory aims to construct a unifying framework of all particle interactions and therefore unify gravity with the standard model. The quantum aspects of the gravitational
	field only emerge in a certain limit, where the different interactions can be
	distinguished from each other. All particles originate in excitations of
	fundamental strings according to string theory. The fundamental scale is given by the string length, which is
	supposed to be of the order of the Planck length \ref{siegelbook}. In loop quantum gravity, on the other hand, \emph{space and time are quantized}, analogously to how energy and momentum are quantized in quantum mechanics \ref{reflqc}.}
 A recurring feature of these theories is \emph{nonlocality}, since both theories exhibit nonlocality at a fundamental level.
Strings and branes are nonlocal by their very definition since, even classically, these objects interact over a region in spacetime rather than at a specific spatial point \ref{UVfiniteness}. The entire formulation of LQG is based on nonlocal
objects, such as Wilson loops and fluxes coming from the gravitational field \ref{UVinf}.
\nnn
Kuz'min \ref{kuzmin} and Tomboulis \ref{superrenormalizablegauge } concluded that non-polynomial
Lagrangians containing an infinite series in either the context of higher-derivative gauge theories and theories of gravitation can be rendered \emph{perturbatively super-renormalizable}. Their discovery had led to a great generation of interest in nonlocal theories of gravity - often constructed in String Field Theory or Quantum Loop Gravity. 
Inspired by the promising results of Kuz'min and Tomboulis,
Biswas, Mazumdar and Siegel have constructed an Infinite Derivative Theory of Gravity (IDG) \ref{UV} in \ref{bounce}. The authors concluded
 that the
absence of ghosts in the modified propagator coupled with asymptotic freedom can
\emph{only} be realized by considering an infinite set of higher derivative terms, while abiding general covariance. The corresponding action is called the BGKM action (named after its authors) and will be motivated elaborately in section \ref{sectionIDG}.
\nnn
Nonlocal gravity has been studied extensively in the literature, with theoretical work most notably on its renormalizability and black holes. Applications of nonlocal gravity have been studied in the contexts of inflation \ref{capozziello}, \ref{refstablebounce}, cosmology \ref{bounce}, \ref{singularityhigherderivative}, dark energy \ref{refdarkenergy1}, \ref{refdarkenergy2}, screening mechanisms \ref{refscreening1}, \ref{refscreening2}, cosmic structure formation \ref{refstructureform1}, \ref{refstructureform2} and dark matter \ref{refdarkmatter1}, \ref{refdarkmatter2}.  
Nonlocal models are, however, not exclusively used in the context of gravity but also in extensions of the Standard Model, noncommutative quantum field theory \ref{noncommutative} and in in effective field theories which are valid in low-energy limits when the effective theory has been derived from a more universal theory where degrees of freedom have been frozen out [\ref{feynmanwheeler}]. More general nonlocal theories exist as well, for instance models in which the derivatives do not necessarily appear in the Lorentz invariant combination $\square = -\partial_t^2
+ \nabla^2$, arise in non-commutative field
theories \ref{noncommutative}, field theories with a minimal length scale \ref{minimallengthmodel} (such as doubly special
relativity) and quantum algebras \ref{quantumalgebra}. 
 Nonlocal structures also appear in noncommutative geometry and SFT, p-adic
strings, zeta strings and strings quantized on a random lattice, for a review, see \ref{siegelbook}.
This raises the question whether nonlocality could be a fundamental feature of spacetime, this feature would become apparent only at short distances and high energies (both of which are UV features) \ref{conroy}.
\nnn
In the first chapter of this thesis we will discuss the aforementioned string-inspired nonlocal model of gravity called \textit{Infinite Derivative Gravity} (IDG) more elaborately. We discuss why this is a viable candidate for a renormalizable theory of gravity and exhibits many nice features which cannot be attained in local field theories of gravity.
Before addressing this model, it is illuminating to explain \textit{what a nonlocal model of gravity actually means (and what it does not imply)} - from  either a mathematical and a physical point of view.
\subsection{The meaning of nonlocality
	\label{sectionmeaningnonlocal}
}
\begin{figure}
	\centering
	\includegraphics[scale=0.5]{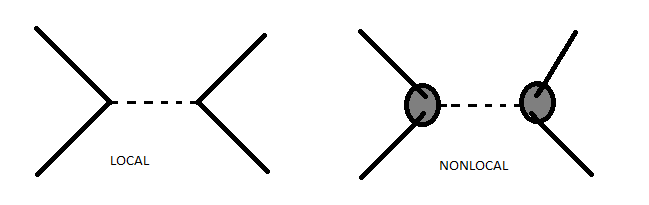}
	\caption{\ti{The first figure graphically represents a local interaction where the vertex can be thought of as pointlike. In nonlocal field theories, vertices are smeared over a region of space (see Eq. \ref{smearingvertex}) which renders the vertex nonlocal. In the nonlocal area (represented as a grey circle) one cannot know where the incoming particles interact and, in fact, acausality may occur in this region. This implies that the virtual particle may already be created before both incoming particles have been annihilated.
			\label{figurenonlocalfeynman}	
		}
	}
\end{figure}

A Lagrangian $\mathcal{L}[\phi]$ is said to be nonlocal if it contains terms which are nonlocal in the fields $\phi(x)$, \emph{i.e.} if they are not polynomials of fields  (and their derivatives) evaluated at a single point in space-time. An example of a nonlocal Lagrangians containing infinitely many derivatives is Lagrange's shift vector
\begin{equation}\label{key}
\phi(x + a) = \sum_{k = 0}^{\infty} \frac{a^{k}}{k!} \phi^{(k)}(x) = \exp(a \partial) \phi(x),
\end{equation}
where the occurrence of a spatial derivative in the exponent is a sufficient condition to introduce nonlocality, since the Taylor expansion of this function contains infinitely many derivatives \ref{conroy}. Infinite Derivative Theory of Gravity, which will soon be introduced, acquires a nonlocality by the same mechanism (but in a Lorentz invariant way).
The kinetic term contains an additional exponential of an entire function $\mathcal{K}(\square)= \exp[-\gamma(\square)]$ where often the Gaussian choice for the entire function is made \ref{UV}, \ref{UVfiniteness}, \ref{classicalproperties}
\begin{equation}\label{kbox}
\mathcal{K}(\square) = \exp\left[ -\square / M^2	\right],
\end{equation}
where $\square$ is the d'Alembertian and $M$ is the \textit{scale of nonlocality}.
\nnn
More degrees of freedom, on a classical level, means that more initial data needs to be provided in order to specify motion.  Nonlocal theories obviously have perturbation expansions containing higher derivatives, but these systems contain implicit constraints ensuring that, on the one hand, a lower-energy bound exists and on the other hand the degrees of freedom remains finite \ref{simon}. Nonlocal theories therefore evade the typical problems which are exhibited by polynomial higher-derivative theories.
\nnn
Cauchy's theorem on differential equations tells that, as you increase the number of
derivatives in your Lagrangian,  more initial data has to be  provided in order to find a solution ($\phi, \dot{\phi}, \ddot{\phi}, \cdots$). Whenever a Lagrangian contains an infinite number of derivatives (\textit{derivatives appearing
	non-polynomially} occur for instance when the kinetic term contains a ($1/\square$)-term), the situation is fundamentally different.\footnote{
	A series of infinite derivatives is a common feature of non-local theories, but it should be noted that this is \emph{not} a defining characteristic. Massive gravity
	theories which modify GR in the IR, for instance theories which have a kinetic correction $\mathcal{\square} \sim
	\frac{1}{\square^2}$ are, although 
	containing finite orders of the inverse d’Alembertian, indeed nonlocal \ref{conroy}. 
} 
Finding a solution for  a Lagrangian with second derivatives (such as the EH-action) requires providing both the
field and its first derivative at a specific point as \emph{initial data}. One does not need to
know the whole function, its value in the neighborhood of a point of its domain suffices, another way of saying that we solely need local information of the function. If the Lagrangian instead contains infinitely many derivatives, for
example if we think in terms of Taylor expansions around an initial value, then the complete function has to be specified. This is what is meant in the literature with \emph{nonlocal info}. But does this also imply that we need to provide infinitely many initial data points?
\nnn
We have discussed that the amount of initial data which needs to be specified to solve a $n$th order differential equation is $n$
terms ($\phi, \dot{\phi}, \cdots, \phi^{(n)}$) which raises the question whether one needs to assign an infinite amount of initial conditions in 
theories containing infinitely many derivatives. 
If that would be the case, one could never find a solution to the field equation and the theory would lose its physical predictability \ref{ghostinfinite}.
Fortunately, the Cauchy initial data problem
is more subtle in nonlocal models as has been pointed out in \ref{refinitialvalueinfinite}.
The amount of initial data which needs to be provided is uniquely determined by pole structure of the inverse kinetic operator $\mathcal{K}^{-1}(\square)$. If the nonlocal term does not contain any poles (and the nonlocal terms which will be considered in this thesis contain no poles)
one needs to specify the same number of initial data as when the nonlocal term is absent. This implies that, regardless of whether the following local or nonlocal kinetic term $\mathcal{K}(\square)$ is employed
\begin{equation}\label{key}
\mathcal{K}(\square) = \begin{cases}
\square - m^2 & \text{:local theory:} \\
\left[ \square - m^2 \right] \exp\left[-\square/M^2 \right] & \text{:nonlocal theory:}
\end{cases},
\end{equation}
the number of initial data which has to be provided 
to solve the homogeneous field equation
\begin{equation}\label{key}
\mathcal{K}(\square) \phi(\mathbf{x},t) = 0
\end{equation} 
is \emph{two}. 
The Wightman function, defined as a solution of the
homogeneous differential equation, is consequently neither affected by the aforementioned infinite derivative modification
$$
\exp\left[-\square/M^2\right] = \sum_{n=0}^{\infty} \frac{1}{n!} \left[ \frac{\square}{M^2} \right]^{n}
$$
nor by other nonlocal modifications \emph{provided that these introduce no new poles}. For the inhomogeneous (interacting) field equation one would have to provide four initial values - not infinitely many! \ref{ghostinfinite}
\nnn
From a physical point of view, interactions in nonlocal models can no longer be thought of as point-like interactions.
Nonlocal interactions result in an
exponential enhancement of the vertex factor, meaning that interactions do not take place actually at this point, as they would in local theories \ref{stelle}, \ref{refeinsteinuv}.
Interactions in nonlocal models are smeared out over a region of space-time - a feature which becomes most overt in the nonlocal vertex factors.\footnote{This nonlocal feature will be shown explicitly for the vertex factors of infinite-derivative theories of gravity in chapter \ref{chaptertoymodelfinitet}.}
Nonlocal theories of gravitation
yield nonlocal interactions and require a relaxation of \textit{the principle of locality}, whose unrelaxed version states that \emph{particles are no longer solely influenced by their immediate surroundings}.
\nnn
Let us show this explicitly for an interacting scalar theory with a nonlocal kinetic modification as given by Eq. \ref{kbox}, whose action is given by
\begin{equation}\label{nonlocalphi4}
S = \frac{1}{2} \phi 
\square
\exp\left[
-\square/M^2
\right]  \phi - \lambda 
\phi^4.
\end{equation}
This form of the action invites us to introduce the following field redefinition
\begin{equation}\label{key}
\phi \rightarrow \tilde{\phi} \exp \left[
\square/2M^2
\right]
\end{equation}
in order to remove the nonlocality from the kinetic term and obtain the equivalent action
\begin{equation}\label{smearingvertex}
S = \frac{1}{2} \tilde{\phi}  \square \tilde{\phi} - \lambda 
\tilde{\phi}^4 \exp \left[
2\square/M^2
\right].
\end{equation} 
This implies that the contribution from the free theory \emph{is exactly the same as the local free theory}, while the nonlocality does modify the contributions arising from interactions. The vertex factor 
\begin{equation}\label{key}
-\lambda
\exp \left[
-2k^2/M^2
\right]
\end{equation}
contains a Gaussian smearing term 
which implies that vertices are no longer pointlike as in local field theories. 
These phenomena are graphically represented in the nonlocal Feynman diagram of
Fig. \ref{figurenonlocalfeynman}.
\nnn
Nonlocal theories of the type studied here were known long ago to violate causality \ref{carone}. 
It has been argued by Coleman in 1970 \ref{coleman} that an observable acausality does not lead to logical paradoxes in
scattering experiments, since a unitary evolution of initial states to final states
is guaranteed by the existence of a
unitary
$S$-matrix.
Moreover, recent computations by Buoninfante, Mazumdar and Lambiase
in \ref{ghostinfinite} have shown that a nonlocal kinetic exponential modification given by
Eq. \ref{kbox} give rise to an acausalities which only emerge solely at microscopic
scales inside the nonlocal region defined by $t, r < M^{-1}$, where Heisenberg's uncertainty relation is operative. 
The authors of \ref{ghostinfinite} have therefore shown that 
acausality 
(and nonlocality) are inherently off-shell features and that
acausal effects only emerge at the microscopic scales $r \sim 1/M$, while physics remains causal outside this region.	
The acausality therefore does not pose a threat to the indispensable classical concept of causality!
\subsubsection*{Nonlocal thermal field theory}
Relatively little attention has been paid to nonlocal thermal field theory in the literature, although thermal corrections have been computed for several string-inspired models (see for instance computations for p-adic strings and open string tachyons \ref{p-adic}, \ref{sftmodels}). In this thesis I will discuss a nonlocal scalar theory (in chapter \ref{chapternonlocalfield}) which, for several reasons, is very interesting to study. It contains a kinetic exponential modification $\mathcal{K}(\square) = \exp[-\square/M^2]$ which is considered in Infinite Derivative Gravity as well. We will show that this nonlocal scalar model exhibits interesting features, such as Hagedorn phase transition at sufficiently high temperatures.
\nnn
A nonlocal scalar toy model of IDG has been constructed in the literature, the chapters on local and nonlocal field theory provide a thorough preparation for thermal field correction in this nonlocal scalar models. 
Several result for the nonlocal scalar field theory can be generalized, including the fact that the nonlocality does not contribute to the partition function in the free theory.
The biggest impediment in the computation of thermal correction in the context of the scalar IDG model will be the treatment of \emph{nonlocal derivative interactions}. Derivative interactions in the context of nonlocal thermal field theory has, to the best of my knowledge, not yet been considered in the thermal field theory literature so this will be a very captivating direction for future research.
\subsubsection*{Outline of the thesis}
The chapters of this thesis are structured as follows:
\begin{lyxlist}{00.00.0000}
\item [{Chapter \ref{chapterIDG}: \tb{Motivating nonlocal theories of gravity}}]
General relativity's classical space-time singularities non-renormalizability of its quantum formulation make quantum gravity an interesting research topic in contemporary science.  Following the discussion of nonlocal field theories of this introduction, we discuss why a fundamental theory of gravity ought to be nonlocal in order to have ameliorated UV-behavior (in comparison to GR) \emph{without introducing additional degrees of freedom}. In section \ref{sectionIDG} it is discussed that Infinite Derivative Gravity 
is endowed with these features and is, in fact, a suitable candidate for a renormalizable theory of gravity.
\item [{Chapter \ref{chapterTFT}}: \tb{Thermal field theory}]
The imaginary time formalism, also known as Matsubara formalism, is a powerful theory in which equilibrium systems at finite temperature can be studied. We firstly introduce the mathematical framework and 
discuss how to deal with expectation values, propagators, vertices and
 loop diagrams in the Matsubara formalism. The chapter concludes with several applications of thermal field theory: the partition function and corresponding thermodynamical functions are calculated for both a free scalar theory and an interacting scalar theory with a $\phi^4$ interaction term are computed and their thermal mass, which is the finite temperature contributions to the self-energy, will be calculated and interpreted. We will discuss why the thermal mass  generated by scalars bears importance to our understanding of the electroweak theory in the early universe in section \ref{lambdaphi4}.
\item [{Chapter \ref{chapternonlocalfield}: \tb{Nonlocal thermal field theory}}]
We will familiarize ourselves with nonlocal thermal field theory computations, firstly
by considering a free nonlocal scalar field
$\mathcal{L} = \frac{1}{2} \phi \square a(\square) \phi$,
where
$a(\square) = \sum_{n=0}^{\infty} \frac{1}{n!} \exp(-\square / M^2)^{n}$ where $\square$ is the d'Alembertian $g^{\mu \nu} \nabla_{\mu} \nabla_{\nu}$ and $M$ is the scale of nonlocality. The nonlocal term is inspired by Infinite Derivative Gravity and is a suitable choice for the kinetic term of the scalar theory since we aim to study the scalar IDG model with identical kinetic modification term in the future (the prelimary calculations of which can be found in appendix \ref{appendixidgfield}).
We will conclude that the nonlocality does not affect the partition function at the free level, while the nonlocality does affect the partition function of the nonlocal scalar theory when (arbitrary) interaction terms are included. This will be shown explicitly for a $\lambda \phi^4$ interaction term and we explain why the nonlocality only plays a physical role at the level of interactions. \nnn
In section \ref{sectiondiscussiontemperatureexpansion} we will
discuss our results and check whether these are in accordance with the conjectured stringy Hagedorn phase \ref{wittenatick}
and thermal duality \ref{hagedorn2}. We will conclude that thermal duality is absent in the nonlocal scalar theory, while Hagedorn behavior does emerge in the high temperature expansion. Explanations of why thermal duality may be violated in the nonlocal scalar model are provided in section \ref{sectionthermalduality}.
\item [{Chapter: \ref{chapterconclusion}: \tb{Conclusion and discussion}}] We summarize our results in this chapter and discuss interesting directions for future research.
\item [{Appendix \ref{appendixgr}}: \tb{General relativity}]
This appendix enumerates several important aspects of general relativity, all of which should successfully be recovered by a viable quantum gravity. We prove among other things that gravity is mediated by a massless, traceless spin-2 gauge boson and discuss the role of diffeomorphism invariance in general relativity.
\item [{Appendix \ref{appendixspinproject}}: \tb{Spin projection operators}] Spin projection operators are useful for the decomposition of a symmetric two-rank tensor such as the metric tensor into spin-2,-1 and -0 components. This formalism is introduced in \ref{appendixspinproject}.1 and is applied to general relativity in \ref{appendixspinproject}.2 to write the kinetic operator $\mathcal{O}^{\mu \nu \rho \sigma}$ in terms of spin projection operators.
\nnn
The graviton propagator in general relativity is obtained in this appendix using the spin projection operators formalism. Since the kinetic operator $\mathcal{O}^{\mu \nu \rho \sigma}$ cannot be inverted, a gauge fixing term (de Donder gauge) is included to obtain a kinetic operator with surplus structure, subsequently the gauge-invariant (so physical) part of the propagator is shown to be equal to $	\Pi_{\text{GR}} = \frac{1}{k^2} \left(\mathcal{P}^2 - \frac{1}{2} \mathcal{P}_{s}^{0} \right)$.
\item [{Appendix \ref{gravterms}}: \tb{Infinite derivative theories of gravity }]
The derivation of the most general quadratic action in section \ref{idgfeatures} has been deferred to this appendix due to its lengthy character. We will show that only one independent form-factor exist (called $a(\square)$) and motivate the Gaussian choice for this form factor. Among others, it is shown that this choice leads to a physical non-singular Newtonian potential. Lastly, we discuss how to yield the most general covariant \emph{quartic} action - which becomes important if one wants to include interactions.
\item [{Appendix \ref{toymod}}: \tb{The toy model IDG action}]
The scalar infinite derivative toy model of gravity will be introduced in \S\ref{infderscalar} and can be decomposed into a free and interacting action like $S = S_{\text{free}} + S_{\text{int}}$. The action is assumed to be invariant under scaling and shift symmetries $\phi \rightarrow (1+ \epsilon)\phi + \epsilon$. In this appendix, the intermediate mathematical steps are shown which one needs to derive the final expression for $S_{\text{int}}$ which was given in section \ref{infderscalar}.
\item [{Appendix \ref{freqsumappendix}: \tb{The spectral function and frequency sums}}] 
Firstly, the spectral intensity function is discussed which allows one to relate several Green functions, including the Euclidean Matsubara propagator and Minkowskian retarded, advanced propagators.
Frequency sums are  discussed elaborately in this appendix, with emphasis on how to compute frequency sums involving multiple bosonic propagators. The computation of an important recurring frequency sum involving two bosonic propagators is presented here.
\item [{Appendix \ref{appendixlocalnonlocal}}: \tb{Local and nonlocal $\phi^4$ theory}] 
It will be shown in this appendix that 
a naive perturbative expansion in $\lambda$ for scalar $ \phi^4$ theories break down at finite temperature due to infrared divergences, primarily but not exclusively due to Daisy (ring) diagrams.\nnn
In the second section we discuss that thermal duality, an important duality of string theory which relates physics at temperature $T$ to physics at temperature at $1/T$, is violated for 
open string tachyons (a nonlocal $\phi^4$ model). These computations are relevant for the discussion of thermal duality in section \ref{sectionthermalduality}.
\item [{Appendix \ref{chaptertoymodelfinitet}: \tb{Towards thermal field corrections in the scalar IDG model}}]
The prelimaries for thermal field corrections in the scalar toy IDG model will be introduced in this appendix. We have discussed how to obtain the vertex factors for theories with derivative interactions and written down the mathematical expressions for the lowest-order Feynman diagrams.
\end{lyxlist}

\setcounter{page}{1}
\pagenumbering{arabic}

\chapter{Motivating nonlocal theories of gravity
\label{chapterIDG}
}
The Newtonian theory of gravitation has been used throughout three centuries and yielded results which were in excellent accordance with gravitational phenomena on Earth. Newton himself however already realized that one aspects of his theory could not be philosophically justified and recognized the severe problems entailed by this  - gravitational effects from system $\mathcal{A}$ are instanteneously transferred onto another system, $\mathcal{B}$, without these bodies being in physical contact.\footnote{Newton wrote in ``Letters to Bentley'' (1692) that ``[i]t is inconceivable that inanimate Matter should, without the Mediation of something else, which is not material, operate upon, and affect other matter without mutual Contact.''}
Though sufficiently accurate for describing gravity on Earth, this nonlocal effect renders Newtonian mechanics useless in order to describe the grand structures in the universe, heavy objects such as neutron stars and, in general, situations in which relativistic effects should be taken into account.
The successive well-entrenched theory of gravitation - Einstein’s theory of General Relativity (GR) - eliminated this \textit{instantaneous action at a distance}. 
\nnn
General relativity, which is defined by the Einstein-Hilbert action \ref{wald}
\begin{equation}\label{einsteinhilbert}
S_{\text{EH}} = \int d^{4}x\sqrt{-g}\mathcal{R},
\end{equation}
(where $g = \det(g_{\mu \nu})$ and $\mathcal{R}$ is the Ricci scalar)\footnote{Let us note that the coupling constant $\kappa= M_{p}^{-2}$
	does not appear in the Einstein-Hilbert action.
	According to our conventions,
	this coupling constant is introduced when considering interactions with a matter source.}
is the most successful description of spacetime which has seen
numerous confirmations of observational tests at different length scales, albeit predominantly in the infrared (at large distances, far away
from the source, and at low energies). 
Its predicted existence of gravitational waves has recently
been confirmed by the advanced Laser Interferometer
Gravitational-Wave Observatory (LIGO), which has observed
a transient gravitational-wave signal and
tested the reliability of GR \ref{ligo}. In fact,
a vast amount of observational data has been collected all of which are in accordance with GR and have contributed to its remarkable success (see \ref{refeinstein} for an elaborate discussion of these tests). 
General relativity moreover recovers Newtonian mechanics in the IR and is therefore reconcilable with the successes of its predecessor at low energies and large distances.
General relativity has indubitably become a cornerstone of contemporary theoretical physics but, despite of its tremendous predictive power, general relativity has severe shortcomings pertaining to its predictive power arising mainly from a theoretical point of view.
 \nnn 
We will discuss that general relativity contains singularities at the classical level at the level of astrophysical singularities (in the Schwarzschild and Kerr metric, for instance) 
and at the cosmological level (the Big Bang) and that 
the quantum formulation of general relativity is non-renormalizable - and
why these features are problematic - in section \ref{modgrav}.
These problems have prompted physicists to construct alternative quantum theories which recover recover GR in the IR, where it is in perfect accordance with experimental data.\footnote{Since modified theories of gravity ought to recover GR in the infrared and therefore recover important aspects of this theory (the amount of DOFs of the GR graviton propagator, diffeomorphism invariance, Einstein Field Equations, \emph{et cetera}), important features of GR which should be recovered by modified theories of gravity
have therefore been derived in appendix \ref{appendixgr}.}
Interestingly, \emph{nonlocality} is a recurring feature in quantum gravities as has been discussed in the introduction of this thesis. This nonlocality is of a completely different nature compared to the nonlocal effects in Newtonian mechanics, since nonlocality is only noticeable in the quantum regime in the field theories which will henceforth be discussed.
Nonlocal field theories can ameliorate quantum loop divergences while
circumventing the emergence of ghosts and tachyons, which is impossible to achieve in finite higher derivative (thus local) theories of gravity. 
\section{Generalizing gravity \label{modgrav}}
General relativity is a remarkably successful theory but it does suffer from severe shortcomings at short distances and at small time scales (both of which are UV problems).  These singularities should be critically reflected upon since they may indicate that general relativity ceases to be predictive in the high-energetic regime of gravity and should be regarded an effective field theory. \nnn
The Einstein-Hilbert action predicts the occurrence of singular solutions - the cosmological singularity (Big Bang) in a time-dependent background and blackhole solutions in vacuum - both of which strongly suggest the incompleteness
of general relativity in the UV \ref{nonsingularsolution}.
The cosmic censorship forbids naked singularity in the context of a blackhole within general relativity \ref{penrose}, \ref{wald}. 
 The cosmological singularity is naked (not covered by an event horizon) and
introduces a problematic
uncertainty in the cosmological models at either the level of initial conditions for the big bang cosmology and for inflation as has been discussed in \ref{resolutionanisotropicsingularity}.

The cosmological singularity implies incomplete null/time
like-geodesics, entailing a breakdown of classical and quantum initial conditions  \ref{geroch}, \ref{waltbook}.
For these reasons, we could expect that nature avoids any kind of classical singularities (both naked and those covered by an
event horizon), which is a stronger version of the cosmic censorship hypothesis \ref{nonsingularsolution}, \ref{penrose}. \nnn
General relativity not only faces contentious problems on the classical level, but als on the quantum level. The Einstein-Hilbert action does not straightforwardly lead to quantization - when quantizing Einstein gravity one yields an infinitude of divergent 
Feynman diagrams which highlights the non-renormalizability of the quantum formulation, as will be discussed in \ref{GRsupdeg}. Removing these singularities at either a classical and quantum level is an outstanding problem in gravitational physics \ref{UVfiniteness}.
\subsection{Superficial degree of divergence in GR \label{GRsupdeg}}
The superficial degree of divergence $D$ in one-particle irreducible diagrams (1PI diagrams) is obtained by adding the contributions from \emph{loop integrals, internal propagators} and \emph{vertices} and study their combined dependence on momentum $k^{\mu}$ \ref{peskin}. In general relativity loop integrals $\int d^4k$ contribute $k^4$, propagators behave as $k^{-2}$ (as is shown in appendix \ref{grprop}) and vertices behave as as $k^{2}$. We consequently concludes that
\bse
\begin{equation}\label{superficialD}
 D = 4L -2I + 2V,
\end{equation}
where $L$ is the number of loops, $I$ is the number of internal propagators and $V$ is the number of vertices.
The superficial degree of divergence can be written in a more suggestive form by
exploiting the topological relation \ref{UV}
\begin{equation}\label{top}
L = 1 + I - V.
\end{equation} 
Now it can straightforwardly be deduced that 
\begin{equation}\label{key}
D = 4L -2I + 2(1+I-L) = 2(L + 1).
\end{equation}
\ese
so the degree of divergence increases linearly as the number of loops increases. 
This indicates that general relativity is non-renormalizable, in stark contrast with the field theories of the standard model, although the super-ficial degree of divergence is not always a reliable measure of (non-)renormalizability as can be deduced from examples given in \ref{peskin},  where $D$ predicts a linear divergence whereas it is actually logarithmic, or even worse, where $D$ falsely predicts that no divergences emerge. 
\nnn
For this reason, 't Hooft and Veltman have calculated the one-loop effective action of general relativity in 1972 \ref{hooft}. They concluded that gravity coupled to a scalar field is non-renormalizable and showed how  counter-terms can be introduced to make render GR finite at one-loop. The non-renormalizability of gravity coupled to various types of matter was established in the ensuing years. The decisive result that GR is non-renormalizable was provided by Goroff
and Sagnotti \ref{refeinsteinuv} and van de Ven \ref{ven} in 1986 and 1992 respectively, both of whom showed the existence of a divergent term which is cubic
in curvature at two loops in the action of Einstein gravity.
\nnn 
Renormalizability has become an essential criterion for quantum field theories to be considered viable, based on the conjection that fundamental field theories, as opposed to \emph{effective field theories}, ought to be
renormalizable. 
Infinities which arise in perturbative expansions are given a cut-off (integrating up to a hard cut-off $\Lambda$, called regularization) to yield finite results and are subsequently
renormalized \ref{kapustabook}. This means that the parameters of the Lagrangian and the corresponding cutoffs
are eliminated in favor of physical observables, such as mass and electric charge. 
 If a finite amount of primitively divergent diagrams occurs in a field theory, renormalizability could possibly be attained by taking into account \textit{counterterms} in the Lagrangian. 
There are good reasons to surmise that there are only a finite
number of independent parameters in the universe so if
the number of required cutoffs increases as the number
of loops increases (inexhaustibly) then
 an infinite number of observables has been to specified in order to define the theory, spoiling the predictivity of the theory. Such theories are called non-renormalizable and cannot be fundamental theories of nature, although these can still serve as useful effective field theories which for which it is not requisite to be renormalizable \ref{peskin}.\footnote{EFTs aim to describe phenomena accurately over a finite range of distances or momenta, so divergent contributions from the UV are to be expected here.} \nnn
The scattering matrix elements and, ultimately,
 the cross section can be obtained by
 studying scattering diagrams. Scattering diagrams hence are invaluable tools for QFT.
 Einstein gravity being non-renormalizable causes cross sections to blow up at high energies, since the cross section of a scattering diagram involving gravitons blows up
 for large external momenta  \ref{IDGscattering}.
 This nihilates the predictive power of this quantum theory. 
Physicists hence aim to devise an improved, renormalizable theory of gravitation for which general relativity would be an effective field theory. 
\subsection{Straying away from the Einstein-Hilbert action \label{awayEH}}
The non-renormalizability of GR has prompted worldwide research into quantum gravity in order to find a more fundamental quantum theory of gravity. 
Nature does not
forbid (neither locally nor globally) going beyond kinetic terms which go beyond two derivatives so there is no prohibition
in constructing higher derivative kinetic terms as long as these are Lorentz-invariant. 
Unifying general relativity with quantum field theory, both of
which are arguably the greatest advances of theoretical physics twentieth century,
is considered to be an unsolved problem of
paramount importance in the high-energy frontier of fundamental physics. \nnn
A common procedure to tame divergences in gravity is to generalize the Einstein-Hilbert action
$
S_{\text{EH}} = \frac{1}{2} \int \mathcal{R}\sqrt{-g} d^{4}x
$ by appending covariant terms to the action such as $\mathcal{R}^2$, $\mathcal{R}_{\mu \nu  \rho \sigma} \mathcal{R}^{\mu \nu}$, $\mathcal{R}_{\mu \nu} \mathcal{R}^{\mu \nu \rho \sigma}$ and $\mathcal{C}_{\mu \nu \rho \sigma} \mathcal{C}^{\mu \nu \rho \sigma}$, where $\mathcal{C}_{\mu \nu \rho \sigma}$ is the Weyl tensor (also known as the \emph{conformal tensor} - it represents tidal forces). 
The resulting action should reduce to the EH-action in the infrared, since this theory is well-supported by experimental results for large distances and low energies, implying that additional terms become negligible in this regime.
Modifications to the action however should lead to ameloriated UV-behavior in order to circumvent space-time singularities and (potentially) render the theory renormalizable.
\nnn
The modified action should moreover obey important theoretical principles. We for instance require that the action is devoid of (bad, as opposed to benign) ghost- and tachyon-like degrees of freedom in order to obtain a healthy action. The problem of ghosts is elaborated upon in footnote \ref{footnoteghost}, while the problem of tachyon is arguably more trivial since their faster than light propagation implies that these particles violate causality.\footnote{\label{footnoteghost} We should distinguish \textit{bad} and \textit{benign} ghosts.
A good ghost does not violate any fundamental principles
since this ghost never appears as observable physical state. One can think of  Fadeev-Popov ghosts, which are typically introduced in the path-integral formalism of gauge theories, but do not correspond to genuine, physical degrees of freedom.
A bad ghost, on the other hand, is associated with \textit{physical particles} and violates
	fundamental principles both at the classical and quantum level.
\begin{itemize}
	\item {}
\textit{At the quantum level}: ghosts correspond to states with
negative norm and could violate the unitarity condition.
\item {} \textit{At the classical
level}: the presence of ghost could cause instabilities of the vacuum since the energy is
not bounded from below. 
\end{itemize}
For an elaborate discussion of benign and bad ghosts, see appendix C of \ref{luca}.
} 
 Unfortunately, by straying away from the Einstein-Hilbert action, ghosts are easily invited into theories with (finitely many) higher-order derivative corrections.
A theory which is known to be able to resolve quantum divergences is Stelle's Fourth Order Gravity which incorporates $\mathcal{R}^2$ and $\mathcal{R}_{\mu \nu}\mathcal{R}^{\mu \nu}$ modifications to the Einstein-Hilbert action \ref{stelle}.\footnote{Although the theory is renormalizable, it does not eliminate classical singularties. Both the Schwarzschild and Kerr metric 
can be shown to contain singularities.}	
The spin-2 component is modified in this theory, however, a pole with wrong
residue sign (and therefore the infamous Weyl ghost) appears as will be discussed in section \ref{sectionstelle}. Firstly we will discuss 
Lovelock's theorem, 
which is an important theorem for both \emph{i)} the formulation of general relativity and \emph{ii)} concerning any valid
extension of the Einstein-Hilbert action is Lovelock's Theorem \ref{lovelock2}.
\subsection{Lovelock's theorem \label{sectionlovelock}}
Lovelock's theorem states that the only possible second-order Euler-Lagrange
expression obtainable in a four-dimensional space from a scalar density
with a Lagrangian dependent on the metric tensor (\emph{i.e.} $\mathcal{L} = \mathcal{L}(g_{\mu \nu})$ is \ref{lovelock}
\begin{equation}\label{key}
E^{\mu \nu} = \sqrt{-g} 
\left[
\alpha
\mathcal{R}^{\mu \nu} - \frac{1}{2} \alpha g^{\mu \nu} \mathcal{R}
+ \beta g^{\mu \nu}
\right]
= \sqrt{-g} 
\left[
\alpha G^{\mu \nu}
+ \beta g^{\mu \nu}
\right],
\end{equation}
where $G^{\mu \nu} = \mathcal{R}^{\mu \nu} - \frac{1}{2}  g^{\mu \nu} \mathcal{R}$ is the Einstein tensor and $\{\alpha, \beta\} \in \mathrm{R}$.
This is a remarkable result since the Einstein equation in the presence of the cosmological constant are retained when  considering $\alpha = 1, \beta = \Lambda$. What this theorem implies is that any gravitational
theory in a four-dimensional Riemannian space, whose corresponding field equations
are of second order or less, will be defined solely by the Einstein equation.
\nnn
The Einstein-Hilbert is not a unique action which yields the Einstein field
equations (EFEs) \ref{lovelock2}. The most generic form of the action
which achieves this is, in four dimensions, given by
\begin{equation}\label{lovelockeq}
\bal
S &= \int d^4x \sqrt{-g} \ \mathcal{L}_{\text{GB}},  \\
\text{where} \ \ \mathcal{L}_{\text{GB}} &= \left[ \alpha \mathcal{R} - 2\Lambda + \beta \left[ \mathcal{R}^2 - 4 \mathcal{R}^{\mu \nu} \mathcal{R}_{\mu \nu} + \mathcal{R}^{\mu \nu \rho \sigma} \mathcal{R}_{\mu \nu \rho \sigma} \right] \right] + \gamma 
\epsilon^{\mu \nu \lambda \sigma}
\mathcal{R}_{\ \ \ \mu \nu}^{\alpha \beta} \mathcal{R}_{\alpha \beta \lambda \sigma},
\eal
\end{equation}
and $\alpha, \beta, \gamma$ are real constants and $\epsilon^{\mu \nu \lambda \sigma}$ is the totally anti-symmetric Levi-Civita tensor. The
term 
\begin{equation}\label{gaussb}
\mathcal{R}^2 - 4 \mathcal{R}^{\mu \nu} \mathcal{R}_{\mu \nu} + \mathcal{R}^{\mu \nu \rho \sigma} \mathcal{R}_{\mu \nu \rho \sigma}.
\end{equation}
 is called the \emph{Gauss-Bonnet term} (hence the subscript ``GB'' in the Lagrangian) and
vanishes in four dimension due \textit{Euler's topological invariant} \ref{lovelock2}
\begin{equation}\label{eulerinvariant}
\mathcal{R}^{\mu\nu\rho\sigma}\mathcal{R}_{\mu\nu\rho\sigma}-4\mathcal{R}^{\mu\nu}\mathcal{R}_{\mu\nu}+\mathcal{R}^{2}=\nabla_{\mu}K^{\mu}
\end{equation}
which entails that the Gauss-Bonnet term can be written as a total derivative.
That being said, the Gauss-Bonnet term is in fact non-trivial in theories
containing more than 4 dimensions \ref{gaussbonnet}, \ref{conroy}.
\nnn
The implication of Lovelock's theorem for modified theories of gravity is the following. When aiming to describe a four-dimensional, generally covariant, metric tensor-based theory of gravity, while retaining the variational principle, we solely have
two options: 

\emph{1)}  Extend our approach into field equations that contain higher than second
order derivatives. 

\emph{2)}
 Allow for a degree of nonlocality to enter the system \ref{constantinos}. \nnn
 We will firstly discuss Stelle's Fourth Order Gravity and $f(\mathcal{R})$-models, both of which are based on possibility \emph{1)}. We will subsequently argue in section
 \ref{sectionIDG} that the second option should also be satisfied in order to obtain viable ghost- and singularity-free theories of gravity. 
\subsection{Stelle's fourth order gravity and $f(\mathcal{R})$ theories
\label{sectionstelle}
}
Stelle's Fourth Order Gravity\footnote{The theory is called ``Fourth Order Gravity'' because each term in the resulting
field equations contains four derivatives of the metric tensor.} is an extension of GR where all squared  curvature tensors have been appended to the EH-action
\begin{equation}\label{key}
S = \int d^{4}x \sqrt{-g} \left(
\alpha \mathcal{R} + \beta \mathcal{R}^2 + \gamma \mathcal{R}^{\mu \nu} \mathcal{R}_{\mu \nu}
\right).
\end{equation}
The squared Riemann tensor disappeared from Stelle's action
because of the existence of the  \textit{Euler topological invariant}
which was introduced in Eq. \ref{eulerinvariant}.
This topological invariant allows us to write the squared Riemann tensor in terms of squared Ricci scalar and tensors whenever the total derivative $\nabla_\mu K^{\mu}$ vanishes. Fourth Order Gravity is a rather
natural extension of gravity, since it corresponds to a generalization of the Gauss-Bonnet term
which appears in Lovelock gravity. 
\subsubsection*{Renormalizability at the cost of a Weyl ghost}
Because Fourth Order Gravity is fourth order in derivatives, the graviton propagator goes like $k^{-4}$, while vertices $V \sim k^{4}$ and loop integrals still provide $k^4$ contributions. Consequently the superficial degree of divergence $D$ (defined in Eq. \ref{superficialD}) 
is \emph{constant} since it satisfies 
\begin{equation}\label{key}
D = (L+I-V) = 4,
\end{equation}
where the  topological relation 
$
L = 1 + I - V
$
has been used
The degree of divergence, as a result, does not become more severe at higher loop orders. Interestingly, Stelle has proved that Fourth Order Gravity is actually perturbatively renormalizable, the modified graviton propagator implies a convergence of Feynman diagrams at 1-loop and beyond \ref{stelle}. This astounding discovery has led to a great generation of interest in quantum gravity in the seventies and eighties.
\nnn
The problem of non-renormalizability is resolved in Fourth Order Gravity, but the theory
unfortunately suffers from a negative-energy propagating DOF degree of freedom (a Weyl ghost) which renders the quantum theory non-unitary.
The graviton propagator is modified as follows \ref{stelle}, \ref{luca}
\begin{equation}\label{Pi}
\Pi = \Pi_{\text{GR}} - \frac{\mathcal{P}^{2}}{k^2 + m_2^2} + \frac{1}{2} \frac{ \mathcal{P}_s^0}{k^2 + m_0^2},
\end{equation}
where $\mathcal{P}^2$ and $\mathcal{P}_s^0$ are the spin-2 projection operator 
and a scalar projection operator respectively,
both of which are defined in appendix \ref{appendixspinproject} - (this appendix devoted to spin projection operators) and $\Pi_{\text{GR}}$ is the graviton propagator in GR, whose general form has been derived and expressed in spin projection operators in appendix \ref{grprop}. The second term in Eq. \ref{Pi} has a ``wrong sign'' causing this term to pick up a negative residue in the spin-2 component of the graviton
propagator which implies that the theory contains a spin-2 ghost.\footnote{Eq. \ref{Pi} entails that the theory contains a massive particle as well. This is a ``healty'' particle; it contain a plus sign in the propagator and is therefore not a ghost and for $m_0^2>0$ it is non-tacyonic.}
Loop integrals appearing in the Feynman diagrams of Stelle gravity are thus
convergent at 1-loop and beyond but at the cost of a  problematic, massive
spin-2 ghost \ref{stelle}. Can we circumvent the emergence of a ghost-like excitation by modifying the EH-action differently?
\nnn
Modifications involving scquared Ricci scalars and tensors, such as Stelle's Fourth Order Gravity, can improve the
UV-behavior but typically contain the Weyl ghost \ref{singfree}. 
The issue of ghost
persists for any finite-order, higher than 2-derivative theory for any spin when studying modifications including Ricci tensors \ref{IDGscattering}. 
We should get rid of this ghost-state, since the quantum theory is rendered non-unitary when interactions are included. In EFTs one may be tempted to ignore the problem of ghosts, but the presence of
ghosts almost always signal the presence of classical instabilities (such as Ostrogradsky instability).
Any higher derivative extension of gravity containing $\mathcal{R}_{\mu \nu}\mathcal{R}^{\mu \nu}$ and/or $\mathcal{R}_{\mu \nu \rho \sigma} \mathcal{R}^{\mu \nu \rho \sigma}$ terms inevitably suffer from either
ghost or classical instability due to the Ostrogradsky theorem, which cannot be
cured order by order in curvature corrections \ref{luca}.  
The Ostrogradsky
theorem tells that non-degenerate higher-derivative theories are bound
to contain ghost-like DOFs \ref{ostrogradsky}. 
These ghosts introduce non-unitarity into the theory and should therefore be absent from a viable theory of gravity. \nnn The only way to evade this in higher-derivative Lagrangians is by considering
degenerate theories \ref{roest}.\footnote{Higher-derivative theories, when unconstrained, exhibit features of a whole other nature than their lower-derivatives counterparts.  Higher-derivative Lagrangians  (i) contain more degrees of freedom and (ii) entail that the theory lacks a lower-energy bound. The latter poses a dramatic problem and \textit{always} occurs when higher-derivative terms are present in theories in which degeneracies are absent and no constraints are imposed \ref{simon}.} $f(\mathcal{R})$ models are interesting in this regard, this class of theories exhibits the astounding feature that no ghosts are introduced despite of the fact that the theory contains terms with higher-than-two derivatives of the metric \ref{f(r)ref}. 

\subsubsection*{Ghost-free $f(\mathcal{R})$ theories}
Perhaps the simplest generalization of the Einstein-Hilbert action comes in
the form of $f(\mathcal{R})$-gravity, where the curvature scalar $\mathcal{R}$ is replaced by an arbitrary
function of the curvature scalar. The
action is therefore given by
\begin{equation}\label{key}
S = \frac{1}{2} \int d^4x \sqrt{-g} f(\mathcal{R}).
\end{equation}
and should reduce to $\frac{1}{2} \int d^4x \sqrt{-g} f(\mathcal{R})$ in the infrared in order to be consistent with the successes of GR.
$f(\mathcal{R})$-type models are interesting models because it has been proven that such models can be ghost-free (see for instance \ref{luca}).\footnote{$f(\mathcal{R})$ theories do necessarily contain an additional scalar degree of freedom (in the $\mathcal{P}_s^0$ sector of the spin projection operators which are introduced in appendix \ref{appendixspinproject}), which need neither be ghostlike nor tachyonic. \ref{luca}}
These models have received much attention, especially in the context of dark energy models and inflationary scenarios in cosmology (see \ref{f(r)inflation1},\ref{f(r)inflation2} for an overview).\footnote{For instance, Starobinksy inflation is described by
$$\mathcal{L} \sim \mathcal{R} + \mathcal{R}^2.$$
}
It has however been shown that $f(\mathcal{R})$ theories cannot ameliorate the UV-behavior of Einstein gravity and therefore cannot resolve the classical singularities of GR \ref{bounce}. Although $f(\mathcal{R})$ models cannot be fundamental theories of gravitation, these can nontheless serve the role of fruitful effective field theories.\footnote{Perhaps gravity is better described by a $f(\mathcal{R})$ at higher energies, but this description ceases to be accurate at even higher energies.
Interestingly, Stelle's Fourth Order Gravity might also be a suitable EFT at higher energies, because the ghost is extremely massive ($m \sim M_p$) and would be integrated out of EFTs (whose validity would cease for high energies) \ref{stelle}.
Obviously, precision measurements in the future will have to shed light on the behavior of gravity at higher energy scales.
}.
\subsubsection*{Concluding remarks}
We want to modify the (GR) graviton propagator in order to obtain a potentially renormalizable quantum gravity, while no new degrees of freedom should be present in order circumvent the emergence of tachyonic and ghostlike DOFs in the theory of gravity. 
$f(\mathcal{R})$ type models can be ghost-free,
but cannot improve the UV behavior of the theory. Modifications of the action involving the Riemann tensor on the other hand can
improve the UV behavior but typically contain the Weyl ghost \ref{biswasstring},
Higher derivative theories \textit{always} face one or the other of aforementioned problems. Those which are free of ghosts are not renormalizable and vice versa \ref{pol}, \ref{stelle}.
A second theoretical principle which theories of modified gravity ought to abide is that tachyonic instabilities
are absent to avoid superluminal propagation in the theory.\footnote{Several
	dark energy motivated modified gravity theories are plagued by tachyonic instabilities, see \ref{biswasstring} for an overview.}
It becomes possible to avoid the ``dangerous ghosts'' by choosing special
combinations of the higher derivative corrections, as is shown in Gauss-Bonnet gravity \ref{nonperturbCMB}, \ref{gaussbonnet}, but these models typically exhibit other unphysical features. It has been shown in \ref{gaussbonnetanalysis} that particle instabilities, acausalities and violations of Lorentz invariance
 are fairly common features of Gauss–Bonnet gravity.
It is therefore
imperative that we explore alternative modifications to Einstein gravity, that is to say, we should examine \emph{nonlocal} theories of gravitation.
\section{Nonlocal theories of gravity \label{sectionIDG}}
We have now motivated why we want to delve into nonlocal theories of gravity and will do so in this section.
If you have not familiarized yourself with the meaning of nonlocality in field theory you are referred to the introduction of this thesis, where the meaning of nonlocality in field theories is discussed.
\nnn
Higher-order derivative corrections to the Einstein-Hilbert action present a
\emph{unitary but non-renormalizable} \textbf{or} \emph{renormalizable but non-unitary} trade-off, whereas
both options are undesirable 
since we aim to construct a ghost-free (therefore unitary) and renormalizable theory of gravity.
Interestingly, nonlocal theories of gravity can evade this problematic trade-off.
\begin{itemize}
	\item 
	Nonlocal models can circumvent the emergence of ghost-like degrees of freedom, maintaining unitarity in the quantum theory. 
	In fact, nonlocal models can prevent the introduction of \emph{any} additional degrees of freedom by choosing a suitable nonlocal modification of the theory. The existence of both features will be shown explicitly for Infinite Derivative Gravity in appendix \ref{gravterms}.
	This feature is not exhibited by $f(\mathcal{R})$ where one necessarily introduces one additional scalar degree of freedom.
	\footnote{That ghost-like DOFs are not necessarily absent in nonlocal models can easily be checked, for instance by choosing the kinetic term $\mathcal{K}(\square) = \frac{1}{\square}$ in Eq. \ref{nonlocaleuclidean}, which gives rise to an additional pole in the corresponding graviton propagator.} 
	\item{}
	It has been pointed out in the literature
	that nonlocal string theory models can ameliorate the UV-behavior of quantum loops and even allow the superrenormalizability
	of the theory \ref{UV}, \ref{superrenormalizablegauge }, \ref{moffat}.
	In 1989 and 1997 respectively, Kuz'min \ref{kuzmin} and Tomboulis \ref{superrenormalizablegauge } concluded that 
	Lagrangians containing an infinite series of higher derivatives gauge theories and theories of gravity
	can be made perturbatively super-renormalizable. Kuz'min and Tomboulis considered a Lorentz-invariant exponential modification of the kinetic term as given by
	\begin{equation}\label{key}
	\mathcal{K}(\square) = \exp\left[-\square/M^2\right]
	\end{equation}
	where $M$ is the scale of nonlocality.
	This nonlocal structure of
	quantum field theories is recurrent in many SFT models, including tachyonic actions in string field theory \ref{tachyon1}-\ref{tachyon4}, bulk
	fields localized on codimension-2 branes \ref{codimension} and various toy models of string theory such as p-adic strings \ref{p-adic}, \ref{p-adicmodel}, zeta strings \ref{zetastring}, and strings quantized on a random
	lattice \ref{regge}. 
\end{itemize}
We will now discuss the most general quadratic curvature action of gravity to examine what kind of modifications of gravity are, in fact, allowed. Subsequently, we will motivate  Infinite Derivative Gravity with the so-called BGKM action 
and discuss astounding results - including a non-singular Newtonian potential and ameliorated UV-behavior - which are yielded by this theory. We have primarily taken inspiration from Ref. \ref{pol} in the next subsection.
\subsection{Most general quadratic curvature action of gravity
\label{sectionbgkm}
}
We wish to construct the most general quadratic action for gravity around the simplest background metric, namely that of Minkowski space-time. When considering fluctations around the background metric $\eta_{\mu \nu}$ (expanding the metric as in Eq. \ref{metricexpansion}),
then we should only worry about the terms which are quadratic in $h_{\mu \nu}$. \nnn
Biswas, Gerwick, Koivisto and Mazumdar constructed in \ref{singfree} the most general four-dimensional covariant
metric-tensor-based gravitational action of quadratic-order (torsion-free and parity preserving) gravity with infinitely many derivatives around Minkowski background\footnote{The torsion free condition is imposed to ensure that the connection cannot be a separate field but
	is related to the metric. We are thus only dealing with the degrees of freedom associated with the metric.}
\begin{equation}\label{generalaction}
S_{\text{Q}} = \int \ d^{4}x \sqrt{-g} \left[
\mathcal{P}_0 + \sum_{n} \mathcal{P}_{n} \prod_{q} 
\left( 
\mathcal{O}_{nq} \mathcal{G}_{nq} 
\right)
\right],
\end{equation}
where $\mathcal{P}_{n}$ and  $\mathcal{G}_{nq}$ are functions of metric tensors and the Riemann tensor,
while the differential operator $\mathcal{O}_{nq}$ is made up exclusively of covariant operators in order to abide general covariance (and $\mathcal{O}_{nq}$ contains at least one covariant derivative). Since Minkowski space should be a legitimate vacuum solution
of the theory (and the Riemann tensor vanishes for the Minkowski space-time) it is
clear that all $P$ and $\mathcal{G}$ must be nonsingular for the Minkowski metric and therefore be expandable as a power series around $R_{\mu \nu \rho \sigma} = 0$. \nnn
The aforementioned quadratic action (defined in Eq. \ref{generalaction}) can be decomposed into the Einstein-Hilbert term adjoined with an additional term
\begin{subequations}
	\begin{equation}\label{key}
	S_{\text{Q}} = S_{\text{EH}} + S_{\text{UV}},
	\end{equation}
where $S_{\text{UV}}$ becomes important in the UV and ought to ameliorate the UV-behavior of the theory. The latter term may be recast, after expanding the metric like Eq. \ref{metricexpansion} and solely considering terms in the action that are quadratic in $h_{\mu \nu}$, into the following covariant form up to integration by parts [\ref{singfree}, \ref{bounce}, \ref{sitter}, \ref{genquad}]:
	\begin{equation}\label{key}
	S_{\text{UV}} = \int d^{4}x \sqrt{-g}
	\left(
	\mathcal{R}_{\mu_{1} \nu_{1} \lambda_{1} \sigma_{1}} 	
	\mathcal{O}_{\mu_{2} \nu_{2} \lambda_{2} \sigma_{2}}^{\mu_{1} \nu_{1} \lambda_{1} \sigma_{1}} 	
		\mathcal{R}^{\mu_{2} \nu_{2} \lambda_{2} \sigma_{2}} 
	\right).
	\end{equation}
The operator $ \mathcal{O}_{\mu_{2} \nu_{2} \lambda_{2} \sigma_{2}}^{\mu_{1} \nu_{1} \lambda_{1} \sigma_{1}} $ is a differential operator containing solely covariant derivatives and metrics $\eta_{\mu \nu}$. 
Note that a differential operator could act on the first tensor as well, but one can always recast that into the above expression using integrating by parts. \ref{pol}
For this purpose, the fact that $\mathcal{R}_{\mu \nu \lambda \sigma}$ vanishes around the Minkowski background, a consequence of which is that every Riemann
tensor contributes $\mathcal{O}(h)$ to the action has been used. This explains why all terms of interest for our purposes can be written
as a product of at most two curvature terms.
\nnn
It has been shown in Ref. \ref{pol} that $\mathcal{R}_{\mu_{1} \nu_{1} \lambda_{1} \sigma_{1}} 	
\mathcal{O}_{\mu_{2} \nu_{2} \lambda_{2} \sigma_{2}}^{\mu_{1} \nu_{1} \lambda_{1} \sigma_{1}} 	
\mathcal{R}^{\mu_{2} \nu_{2} \lambda_{2} \sigma_{2}} $ corresponds to fourteen terms, allowing one to write $S_{\text{UV}}$ as
\begin{equation}\label{genaction}
\begin{aligned}
S_{\text{UV}} = \frac{1}{2} \int d^{4}x \sqrt{-g}
& \ \Big[
\mathcal{R} \mathcal{F}_1(\square) \mathcal{R}
+ \mathcal{R} \mathcal{F}_2(\square)
\nabla_{\nu} \nabla_{\mu} \mathcal{R}^{\mu \nu}
+ \mathcal{R}_{\mu \nu} \mathcal{F}_3(\square) \mathcal{R}^{\mu \nu}
+ \mathcal{R}_{\mu}^{\nu} \mathcal{F}_4(\square)
\nabla_{\nu} \nabla_{\lambda} \mathcal{R}^{\mu \lambda} \\
&+ \mathcal{R}^{\lambda \sigma} \mathcal{F}_5(\square)
\nabla_{\mu} \nabla_{\sigma}
\nabla_{\nu} \nabla_{\lambda} \mathcal{R}^{\mu \lambda} 
+ \mathcal{R} 
\mathcal{F}_6(\square)
\nabla_{\mu} 
\nabla_{\nu} 
\nabla_{\lambda}
\nabla_{\sigma} \mathcal{R}^{\mu \nu \sigma \lambda} 
+ \mathcal{R}_{\mu \nu} \mathcal{F}_{7}(\square) 
\nabla_{\nu}
\nabla_{\sigma}
\mathcal{R}^{\mu \nu \lambda \sigma} \\
&+ \mathcal{R}_{\lambda}^{\rho} \mathcal{F}_{8}(\square) 
\nabla_{\mu}
\nabla_{\sigma}
\nabla_{\nu}
\nabla_{\rho}
\mathcal{R}^{\mu \nu \lambda \sigma}
 + \mathcal{R}^{\mu_{1} \nu_{1}} \mathcal{F}_{9} (\square) \nabla_{\mu_{1}} \nabla_{\nu_{1}} \nabla_{\mu} \nabla_{\nu} \nabla_{\lambda} \nabla_{\sigma} \mathcal{R}_{\mu \nu \lambda \sigma}\\
 &+
 \mathcal{R}_{\mu \nu \lambda \sigma} \mathcal{F}_{10} (\square) \mathcal{R}^{\mu \nu \lambda \sigma} 
 +
 \mathcal{R}_{\mu \nu \lambda}^{\rho} \mathcal{F}_{11}(\square) \nabla_{\rho} \nabla_{\sigma} \mathcal{R}^{\mu \nu \lambda \sigma} 
 +
 \mathcal{R}_{\mu \rho_{1} \nu \sigma_{1}} \mathcal{F}_{12} (\square) \nabla^{\rho_{1}} \nabla^{\sigma_{1}} \nabla_{\rho} \nabla_{\sigma} \mathcal{R}^{\mu \rho \nu \sigma} \\
 &+
 \mathcal{R}_{\mu}^{\nu_{1} \rho_{1} \sigma_{1}} \mathcal{F}_{13} (\square) \nabla_{\rho_{1}} 
 \nabla_{\sigma_{1}} 
 \nabla_{\nu_{1}} 
 \nabla_{\nu} 
 \nabla_{\rho} 
 \nabla_{\sigma}
 \mathcal{R}^{\mu \nu \lambda \sigma} \\
&+
   \mathcal{R}^{\mu_{1} \nu_{1} \rho_{1} \sigma_{1}} \mathcal{F}_{14} (\square) \nabla_{\rho_{1}} 
  \nabla_{\sigma_{1}} 
  \nabla_{\nu_{1}}
  \nabla_{\mu_{1}}
  \nabla_{\mu} 
  \nabla_{\nu} 
  \nabla_{\rho} 
  \nabla_{\sigma}
  \mathcal{R}^{\mu \nu \lambda \sigma}
\Big].
\end{aligned}
\end{equation}
\end{subequations}
The most
general quadratic action is captured by 14 arbitrary functions, the $\mathcal{F}_i(\square)$'s, all of which are
functions of the d'Alambertian operator
\begin{equation}\label{Fi}
\mathcal{F}_{i} (\square) = \sum_{n=0}^{\infty}	f_{i_{n}} \left( \square / M^2
\right)^{n}.
\end{equation}
which is divided by $M^2$ to obtain a dimensionless variable.
The coefficients $f_{i_{n}}$ are, for the time being, left unconstrained to ensure
that these are arbitrary infinite derivative functions, although their form will later on be fundamental for our results. We will firstly eliminate eleven out of fourteen aforementioned functions.
\nnn
It will now be argued, following Refs. \ref{pol} and \ref{conroy}, that several terms occurring in Eq. \ref{genaction} are in fact dependent. Consequently eight out of the fourteen terms can be omitted from $S_{\text{UV}}$. The Riemann tensor
is antisymmetric on its first and last pair of indices \bse
\begin{equation}\label{riemannantisymmetric}
\mathcal{R}_{\{\mu\nu\}\sigma \rho} = \mathcal{R}_{\mu\nu\{\sigma \rho\}} = 0.
\end{equation}
and the Jacobi identity is satisfied, \textit{i.e}.
\begin{equation}\label{bianchi}
\nabla_{\omega} \mathcal{R}_{\mu \nu \lambda \sigma} 
+ \nabla_{\sigma} \mathcal{R}_{\mu \nu \omega \lambda} 
+ \nabla_{\lambda} \mathcal{R}_{\mu \nu \sigma \omega} = 0,
\end{equation}
\ese
the combination of which greatly simplifies the action which was introduced in Eq. \ref{genaction}. \nnn
We will prove that 
$\mathcal{R} \mathcal{F}_2(\square)
\nabla_{\nu} \nabla_{\mu} \mathcal{R}^{\mu \nu}$
and $\mathcal{R}_{\mu}^{\nu} \mathcal{F}_4(\square)
\nabla_{\nu} \nabla_{\lambda} \mathcal{R}^{\mu \lambda}
$ can be omitted from the action, because these terms can be recast as $\mathcal{R}\mathcal{F}_1(\square)\mathcal{R}$ due to the Bianchi identity.
A contraction of the Bianchi identity in Eq. \ref{bianchi} yields 
\begin{equation}\label{key}
\nabla_{\mu} \mathcal{R}^{\mu \nu} = \frac{1}{2} \nabla^{\nu} \mathcal{R}.
\end{equation}
It can therefore be shown that the following combination of terms can be rewritten as one single term
\begin{equation}
\bal
& \ \ \ \ \mathcal{R}\mathcal{F}_1(\square)\mathcal{R} 
+
\mathcal{R} \mathcal{F}_2(\square)
\nabla_{\nu} \nabla_{\mu} \mathcal{R}^{\mu \nu}
+ \mathcal{R}_{\mu}^{\nu} \mathcal{F}_4(\square)
\nabla_{\nu} \nabla_{\lambda} \mathcal{R}^{\mu \lambda} \\
&= \mathcal{R}\mathcal{F}_1(\square)\mathcal{R} 
+ \frac{1}{2} \left[
\mathcal{R} \mathcal{F}_2(\square) \square
\mathcal{R}
+ \mathcal{R}_{\mu}^{\nu} \mathcal{F}_4(\square)
\nabla_{\nu} \nabla^{\mu} \mathcal{R}
\right] \\
& \overset{!}{=} \mathcal{R}\mathcal{F}_1(\square)\mathcal{R} 
+ \frac{1}{2} \left[
\mathcal{R} \mathcal{F}_2(\square) \square
\mathcal{R}
+ \nabla^{\mu} \nabla_{\nu}  \mathcal{R}_{\mu}^{\nu} \mathcal{F}_4(\square)
\mathcal{R}
\right] \\
&= \mathcal{R}\mathcal{F}_1(\square)\mathcal{R} 
+ \frac{1}{2} 
\mathcal{R} \mathcal{F}_2(\square) \square
\mathcal{R}
+ \frac{1}{4}   \mathcal{R} \mathcal{F}_4(\square)
\square
\mathcal{R}\\
&= \mathcal{R}\mathcal{F}_1(\square)\mathcal{R}.
\eal
\end{equation}
Integration by parts has been used to obtain the expression with the exclamation mark (`$\overset{!}{=}$') and $\mathcal{F}_1(\square)$ has been redefined in the last equality such that it includes $\mathcal{F}_2(\square)$ and $\mathcal{F}_4(\square)$. 
We have actually gotten rid of 8 out of 14 terms, the proofs for the other terms are reminiscent to the one just provided and consequently will not be provided here (the interested reader is referred to \ref{conroy}).
\nnn
After eliminating superfluous terms from the action one obtains the following action for the quadratic gravity:
\begin{equation}\label{simpaction2}
\begin{aligned}
S _{\text{UV}}= \frac{1}{2} \int d^{4}x \sqrt{-g}
& \ \Big[
\mathcal{R} \mathcal{F}_1(\square) \mathcal{R} \\
& \ \ + \mathcal{R}_{\mu \nu} \mathcal{F}_2(\square) \mathcal{R}^{\mu \nu} \\
& \ \ + \mathcal{R}_{\mu \nu \lambda \sigma} \mathcal{F}_{3}(\square) \mathcal{R}^{\mu \nu \lambda \sigma} \\
& \ \ +  \mathcal{R} 
\mathcal{F}_4(\square)
\nabla_{\mu} 
\nabla_{\nu} 
\nabla_{\lambda}
\nabla_{\sigma} \mathcal{R}^{\mu \nu \sigma \lambda} \\
& \ \ +
\mathcal{R}_{\mu}^{\nu_{1} \rho_{1} \sigma_{1}} \mathcal{F}_{5} (\square) \nabla_{\rho_{1}} 
\nabla_{\sigma_{1}} 
\nabla_{\nu_{1}} 
\nabla_{\nu} 
\nabla_{\rho} 
\nabla_{\sigma}
\mathcal{R}^{\mu \nu \lambda \sigma} \\
& \ \ +
\mathcal{R}^{\mu_{1} \nu_{1} \rho_{1} \sigma_{1}} \mathcal{F}_{6} (\square) \nabla_{\rho_{1}} 
\nabla_{\sigma_{1}} 
\nabla_{\nu_{1}}
\nabla_{\mu_{1}}
\nabla_{\mu} 
\nabla_{\nu} 
\nabla_{\rho} 
\nabla_{\sigma}
\mathcal{R}^{\mu \nu \lambda \sigma}\Big],
\end{aligned}
\end{equation}
where the dummy indices ``$i$'' which occur in the form factors $\mathcal{F}_i(\square)$'s of Eq. \ref{simpaction2} have been relabelled.
The resulting action can be simplified more rigorously by keeping in mind that covariant derivatives are replaced by partial derivatives in a Minkowski background, which entails that covariant derivatives can freely commute. This causes the last three terms in Eq. \ref{simpaction2} to vanish. \nnn 
Because of the vanishing value of
the symmetric-antisymmetric products between partial derivatives and the Riemann tensor,
whose first and last index pairs are antisymmetric,
the terms with $F_4(\square)$, $F_5(\square)$ and $F_6(\square)$ in the quadratic action do not contribute in the limit of a flat background .
The proof for $\mathcal{F}_4(\square)$ is as follows
\begin{equation}\label{key}
\begin{aligned}
\mathcal{R} 
\mathcal{F}_4(\square)
\nabla_{\mu} 
\nabla_{\nu} 
\nabla_{\lambda}
\nabla_{\sigma} \mathcal{R}^{\mu \nu \sigma \lambda}
&= \frac{1}{2} \left[
\mathcal{R} 
\mathcal{F}_4(\square)
\nabla_{\nu} 
\nabla_{\mu} 
\nabla_{\lambda}
\nabla_{\sigma} \mathcal{R}^{\mu \nu \sigma \lambda}
+
\underbracket[0.5pt][7pt]{
\mathcal{R} 
\mathcal{F}_4(\square)
\nabla_{\mu} 
\nabla_{\nu} 
\nabla_{\lambda}
\nabla_{\sigma} \mathcal{R}^{\mu \nu \sigma \lambda} }_{\text{relabel dummy indices: }\mu \leftrightarrow \nu}
\right] \\
&= 
 \left[ \mathcal{R} 
\mathcal{F}_4(\square)
\nabla_{\nu} 
\nabla_{\mu} 
\nabla_{\lambda}
\nabla_{\sigma} \mathcal{R}^{\mu \nu \sigma \lambda}
+
\mathcal{R} 
\mathcal{F}_4(\square)
\nabla_{\nu} 
\nabla_{\mu} 
\nabla_{\lambda}
\nabla_{\sigma} \mathcal{R}^{\nu \mu \sigma \lambda} \right] \\
&= \frac{1}{2}  \mathcal{R} 
\mathcal{F}_4(\square)
\nabla_{\nu} 
\nabla_{\mu} 
\nabla_{\lambda}
\nabla_{\sigma} \mathcal{R}^{ \{\mu \nu \}  \sigma \lambda} \\
&= 0,
\end{aligned}
\end{equation} 
which vanishes due to the antisymmetric properties of the Riemann tensor (Eq. \ref{riemannantisymmetric}).
Proofs that $\mathcal{F}_5(\square)$ and $\mathcal{F}_6(\square)$ also vanish become trivial when using the antisymmetry of $\mathcal{R}_{\mu \nu \rho \sigma}$ in its first and second pair of indeces.
We have therefore yielded the most general form of a quadratic, covariant metric based gravitational action 
\begin{align}
S_{\text{Q}} &= \frac{1}{2} \int d^4x \sqrt{-g} \mathcal{R} + S_{\text{UV}}, \\
\text{where } \ S_{\text{UV}} &= \frac{1}{2} \int d^{4}x \sqrt{-g} \bigg[ \mathcal{R} \mathcal{F}_1 (\square) \mathcal{R} + \mathcal{R}_{\mu \nu}  \mathcal{F}_{2} (\square) \mathcal{R}^{\mu \nu}  + \mathcal{R}_{\mu \nu \lambda \sigma} \mathcal{F}_{3} (\square) \mathcal{R}^{\mu \nu \lambda \sigma}		\bigg] \label{SUV}.
\end{align}
The action contains only three independent, arbitrary dimensionless functions $\mathcal{F}_1(\square)$, $\mathcal{F}_2(\square)$ and $\mathcal{F}_3(\square)$.\footnote{Well-known theories of gravitation can be retrieved by making appropriate choices for $\mathcal{F}_1(\square), \mathcal{F}_2(\square), 
	\mathcal{F}_3(\square)$. Stelle gravity, for instance, corresponds to an action where the functions are non-zero constants and the Starobinsky model (a model for inflation) corresponds to $\mathcal{F}_1(\square) = f_0 > 0$ while $\mathcal{F}_2(\square) = \mathcal{F}_3(\square) = 0$.} 
We can actually get rid of the Riemann-squared term as well in Minkowski background,
since the covariant derivatives become simple
partial derivatives and the functions $\mathcal{F}_i(\square)$ can be placed on the left of the curvature term by means of
integrating by parts.
By subsequently implementing the Euler topological invariant relation (which was already introduced in our discussion of Stelle gravity but repeated here for the reader's convenience)
\begin{equation}\label{key}
\mathcal{R}^{\mu\nu\rho\sigma}\mathcal{R}_{\mu\nu\rho\sigma}-4\mathcal{R}^{\mu\nu}\mathcal{R}_{\mu\nu}+\mathcal{R}^{2}=\nabla_{\mu}K^{\mu},
\end{equation}
where $\nabla_{\mu}K^{\mu}$ is a four-divergence (surface term) which does not contribute to variation of the the action, it can be concluded that
the product of Riemann
tensors disappears since this can be rewritten in terms of  [scalar curvature]$^2$ + [Ricci]$^2$ terms in Eq. \ref{SUV}.
Similar constructions are also possible around any other constant-curvature
backgrounds such as anti-deSitter (AdS) and deSitter (dS) backgrounds, see \ref{stablebackground}. \nnn 
The three form factors 
$\mathcal{F}_{i}(\square)$ satisfy
\bse
\begin{equation}\label{key}
2\mathcal{F}_{1}(\square) + \mathcal{F}_{2}(\square) + 2\mathcal{F}_3(\square) = 0 .
\end{equation}
Without loss of generality we can set $\mathcal{F}_3(\square) = 0$,
which implies that\footnote{That one can set $\mathcal{F}_3(\square) = 0$ without loss of generality only holds true up to the perturbative $\mathcal{O}(h^2)$ level, because Euler's topological invariant can be used to write the squared Riemann tensor in terms of squared Ricci scalars and tensors.}
The form factors are now related by
\begin{equation}\label{key}
2\mathcal{F}_1(\square) = -\mathcal{F}_2(\square) = \frac{a(\square) - 1}{\square}.
\end{equation}
as is shown explicitly in appendix \ref{appendixidgfield}.\footnote{Similar constraints
	on $\mathcal{F}_i$ can be derived for dS and AdS backgrounds, see \ref{stablebackground}. }
Both of the remaining form factors can thus be written in terms of \emph{one kinetic modification term} $a(\square)$.
\ese
The analytic function $a(-k^2)$ should be such that it does not contain any poles in the finite complex plane in order to circumvent the emergence of additional degrees of freedom in the propagator. Furthermore, we require $a(-k^2) \rightarrow 1$ for $k^2 \rightarrow 0$ in order to recover GR in the IR. 
By requiring that 
\textit{any suitable infinite derivative modification can be expressed by an exponential of an entire function},
it is possible to obtain a ghost-free graviton propagator with
no additional poles other than the familiar two massless degrees of freedom of $S_{\text{EH}}$ (the derivation of which is provided in appendix \ref{grprop}). 
Stringy kinetic modifications
combine to an exponential of an entire function and these functions do not have poles in the complex plane (they are holomorphic over the whole complex plane) and vanish only at infinity.
Consequently exponential of entire functions do not introduce any new states; no ghosts, tachyons, `healthy particles' or otherwise \ref{UVinf}. This
property has been exploited to construct various nonlocal infinite derivative field theories. 
Interestingly, the nonlocality arising from infinite derivative extensions of GR has been shown to
play a pivotal role at the classical level by resolving the cosmological
singularity \ref{singfree}, \ref{bounce}, \ref{dimitri}--\ref{craps}. 
This has been shown for a Minkowski background but also for more generic backgrounds,  for other constant-curvature backgrounds \ref{singfree}, \ref{stablebackground}. The previous requirements are satisfied by employing a Gaussian choice for the analytic function $a(\square)$
\begin{equation}\label{key}
a(\square) = \exp[-\square/M^2] = \sum_{n=0}^{\infty} \frac{1}{n!} \left[-\frac{\square}{M^2} \right]^n.
\end{equation}
and the corresponding action is called the BGKM action
\begin{equation}\label{key}
\bal
S &= \int d^4x \left[ \frac{\mathcal{R}}{2} + \mathcal{R}\mathcal{F}_{1}(\square) 
\mathcal{R} 
+
\mathcal{R}_{\mu \nu }
\mathcal{F}_{2}(\square)
\mathcal{R}^{\mu \nu}
\right]
 \\
&= \int d^4x \left[ \frac{\mathcal{R}}{2} 
+  \mathcal{R} 
\left[ \frac{\exp[-\square/M^2] - 1}{\square} \right]
\mathcal{R} 
-
2
\mathcal{R}_{\mu \nu }
\left[ \frac{\exp[-\square/M^2] - 1}{\square} \right] \mathcal{R}^{\mu \nu} 
\right].
\eal
\end{equation}
Exponentials of entire functions contain all orders of the d'Alembertian operator,
where each operator has been modulated by the scale of nonlocality $M$ (indicating the scale where nonlocal modifications become important) to ensure that these
functions are dimensionless.  
Other choices for the form factors however cannot, a priori, be excluded. It will be discussed in section \ref{idgfeatures} that the Gaussian choice yields physical results, lending support to the viability of the corresponding quantum gravity.
It is proved in appendix \ref{gravterms} that the graviton propagator in the BGKM theory is given by
\begin{equation}\label{key}
\Pi(-k^2) = \frac{1}{a(-k^2)}\Pi_{\text{GR}}(-k^2) = e^{-\frac{k^2}{M^2}} \Pi_{\text{GR}}(-k^2),
\end{equation}
where the graviton propagator from GR (the derivation of which can be found in Appendix \ref{grprop}) is given by
\begin{equation}\label{key}
\Pi_{\text{GR}}(-k^2) =  \frac{1}{k^2 } \Bigg[ \mathcal{P}^2 - \frac{1}{2} \mathcal{P}_{s}^{0}		\Bigg].
\end{equation}
We notice that the only difference with the GR graviton propagator is that the IDG graviton propagator
is modified by a multiplicative trascendental function.
We note that the propagator becomes exponentially suppressed in the UV which implies that the theory becomes asymptotically free in the UV, in close analogy with QCD. Infinite Derivative Gravity consequently has ameliorated UV-behavior in comparison with general relativity, both at a
classical and at a quantum level (a discussion of the implications on the classical level can be found in  \ref{luca}, for quantum aspects see  \ref{UVinf}).
Applications pertaining to Regge behavior and the stringy conjectured Hagedorn transition can be found in  \ref{regge} and \ref{hagedorn1}, \ref{hagedorn2}, \ref{hagedorn3}, respectively. 
\subsection{Features of the BGKM action \label{idgfeatures}}
It was discussed in section \ref{modgrav}
that a viable theory of gravity is devoid of cosmological and blackhole singularities.
It has been argued in the literature that the BGKM action
avoids these singularities at the linearized level around the Minkowski background, a conclusion which has been drawn by several research groups both in the context of static configurations \ref{singfree}, \ref{genquad}, \ref{newtonianpotential}, \ref{luca} and in the context of a rotating blackhole
case  \ref{rotating}.
Ensuing studies have shown that the cosmological singularity
can successfully be resolved at the non-linear level as well \ref{bounce}, \ref{nonperturbCMB}, \ref{singularityhigherderivative}. \nnn
Another remarkable result of the BGKM model is that it predicts a modified Newtonian potential which can better be justified than the usual Newtonian potential whose $1/r$-dependence predicts a pole for $r = 0$. 
In fact, the Newtonian $1/r$-fall in the gravitational potential $\Phi(r)$ has been tested only up to a distance of $5.6 \cdot 10^{-5} \text{ m}$ in torsion
balance experiments, which has been tested up to energies of only 0.004 eV \ref{newtonpotentialtest}. This implies that there is a huge desert of roughly 30 orders of magnitude (between 0.004 eV and the Planck mass) where the nature of
gravitational interactions is not yet properly understood. Consequently, deviations from the $1/r$-fall could occur for sufficiently small distances and, in fact, the BGKM action predicts that an error function is appended to the Newtonian potential. The error function counteracts the ubiquitous infrared $1/r$-divergence Newtonian potential of \emph{local theories of gravity}.  \nnn
By working in the weak-field regime (the Newtonian approximation) in the presence
of a static point-source of mass $m$,
the energy-momentum tensor is described by
\begin{equation}\label{key}
\tau_{\mu \nu} = \rho \delta_{\mu}^{0} \delta_{\nu}^{0} = m\delta^{(3)}(\mathbf{x}) \delta_{\mu}^{0} \delta_{\nu}^{0}
\end{equation} 
and the linearized metric can be expressed in
isotropic coordinates as
\begin{equation}\label{key}
ds^{2} = -(1 + 2 \Phi)dt^2 + (1 - 2\Phi) \left[dr^2 + r^2 d\Omega^2 \right], 
\end{equation}
where $d\Omega^2$ is the squared solid angle given by $d\Omega^2 = d\theta^2 + \sin^2\theta d\phi^2$. The metric potential has been derived in Appendix \ref{gravterms} and reads
\begin{equation}\label{modifiednewtonianpotential}
\Psi(r) = \Phi(r) 
=
- {m_g  \over 16 \pi M_p^2} \frac{1}{r}  \erf \left(
{r M \over 2}
\right).
\end{equation}
The error function satisfies $\erf(\frac{rM}{2}) \rightarrow 1$ for $r \rightarrow \infty$ which implies that the correct Newtonian potential with $1/r$ dependence is recovered for GR in the infrared \ref{bounce}, \ref{singfree}, \ref{luca}.
Moreover, the error function satisfies the interesting property that $\erf\left(\frac{rM}{2} \right)  \rightarrow \frac{rM}{2} $ for $r \rightarrow 0$ which entails that the Newtonian potential no longer diverges to $-\infty$ in the UV:
\begin{equation}\label{key}
\lim_{r \rightarrow 0}
- {m_g  \over 16 \pi M_p^2} \frac{1}{r}  \erf \left(
{r M \over 2}
\right) = - {m_g M \over 32 \pi M_p^2} 	\sim - {m_g M \over  M_p^2}
\end{equation}
Tn the non-local UV-regime where $Mr < 2$ the potential becomes non-singular and converges to a finite negative constant \ref{newtonianpotential}. Interestingly, although the matter source contains a delta function singularity the Newtonian potentials remain finite. 
Having obtained a probable modified Newtonian potential lends support to our particular \textit{choice} of the entire function $a(\square)$ (remember that the Gaussian function $a(-k^2) = \exp[k^2/M^2]$ is not the only allowed option for $a(\square)$). 
One is actually free to choose any entire function for the nonlocal term $a(\square)$, but in order to recover the propagator of the usual GR the following constraint needs to be satisfied: $a(k^2) \rightarrow 1$ for IR momenta
$k \rightarrow 0$. Additionally, the sign of $a(\square)$ is crucial in order to recover a physical Newtonian potential:
 one can show that $a(\square) = \exp(\square/M^2)$ does not recover the Newtonian potential of general relativity and can consequently be ruled out as a suitable choice for an exponential of an entire function \ref{pol}, \ref{singfree}.
The behavior of the Newtonian potential in IDG for other choices of $a(\square)$
was studied  elaborately in \ref{IDGNewtonianpotential}, where a more general ghost-free form factor
\begin{equation}\label{key}
a(\square) = e^{-\gamma \left(\square/M^2 \right)}
\end{equation}
has been studied, where $\gamma\left(\square/M^2 \right)$ is restricted to \emph{entire function} but  need not be Gaussian. The authors concluded that a nonsingular Newtonian potential is obtained for those cases and the potential converges to approximately the same values as for the Gaussian choice of the form factor.
What this astounding result implies is that a universal class of entire
functions $\gamma(\square/M^2)$ exists which is singularity-free in the UV. 
\nnn
Furthermore, the time-dependent solution of the BGKM action (obtained by conjoining Eq. \ref{SUV} with the EH-action) yields a \textit{non-singular bouncing cosmology in a vacuum} for time $t < M^{-1}$.
This solitonic solution is absent in general relativity. 
The temperature at the bounce has been computed to be close to (or even higher than) the
Hagedorn temperature, depending on the scale of the nonperturbative 
corrections to the Einstein-Hilbert action \ref{nonperturbCMB} which depend on the value for $M$. 
\nnn
Given that Infinite Derivative Gravity is asymptotically free, could the theory be renormalizable?
The main problem with gravitational theories,
contrary to scalar field theories, is that these constitute gauge theories. One of the key features of gauge theories is that its free kinetic action is related (via gauge symmetry) to the interaction terms.
Exponential suppressions in propagators inevitably give rise to exponential
enhancements in the vertex factors - as was shown explicitly for the nonlocal scalar particle with $\phi^4$ interaction term in section \ref{sectionmeaningnonlocal}. 
This compensating interplay between propagators and vertices is actually exhibited by \emph{all covariant theories of gravity}, including
Einstein’s theory and Fourth Order Gravity \ref{stelle}, see also \ref{sagnotti1}, \ref{sagnotti2}.
Also in infinite derivative gravity, the exponential suppressions in the propagators  inevitably give rise to exponential enhancements in the corresponding vertex factors \ref{UVinf}. Whether this interplay of exponential enhancements and suppressions either ameliorates of deteriorates the UV-behavior of IDG compared to that of general relativity
would be a relevant question to ask at this point. The computation of the superficial degree of divergence in IDG differs from how this is computed in the case of GR.
The counting is now based on the pre-factors of the exponents rather than on the degree of polynomial divergence, due to the fact that exponentials dominate every polynomial growth in the UV. 
\nnn
The superficial degree of divergence however suggests that diagrams
with more than one loop should be finite for the BGKM action \ref{UVinf}, \ref{modesto}, \ref{tomboulis}.  Although this is a propitious result it does not constitute a full-fledged proof for UV-finiteness of the theory, but highlights that this nonlocal theory of gravity is a suitable candidate for a renormalizable theory of gravity which is guaranteed to be ghostfree.\footnote{This naive expectation about the convergence properties of the Feynman diagrams has been verified for a scalar toy model resembling IDG, where
	explicit calculations involving quantum loop calculations revealed its UV-completeness when \emph{dressed propagators and vertices are used}. This scalar toy model has been studied extensively in the literature and we will introduce it in section \ref{infderscalar}. }
We will further motivate the BGKM action by discussing several successes which have been achieved for Infinite Derivative Gravity.
\subsubsection*{Arduous computations}
Now that we have introduced and motivated Infinite Derivative Gravity, let us proceed to the second goal of this thesis - the computation of thermal field corrections. Is it feasible to study this theory in nonlocal thermal field theory within the scope of this thesis?
Thermal field theory in the context of gravity is incredibly complicated for several reasons and is therefore beyond the scope of this thesis, although preliminary steps are shown in appendix \ref{chaptertoymodelfinitet}.
Gravitational theories are all order theories and thus contain interactions
of all orders in $h_{\mu \nu}$. Computing all these interactions is well beyond the scope of the current paper and even a computation of the complete cubic interactions is already difficult to compute at zero temperature (see \ref{UVinf}) and becomes significantly more challenging at finite temperature. Thermal field theory computations for a BGKM action thus turn out to be incredibly complicated, but we can nevertheless obtain useful information by considering a reminiscent (but mathematically speaking considerably more tractable) scalar toy model which captures imporant features of the BGKM action. 
\nnn
 This scalar theory is based upon identical (diffeomorphism) symmetries as general relativity which implies that the scalar theories obeys \emph{scaling and shifting symmetries}. Interestingly, quantum aspects of this toy model have been studied in \ref{UVinf} where it was shown that the  1-loop 2-point function is divergent, while all other 1-loop functions and 2-loop 2-point functions do remain finite. Talaganis, Biswas and Mazumdar discussed how these computations and arguments can be generalized to arbitrary loops, strongly indicating that this toy model contains solely one divergent Feynman diagram.
\nnn
The finite temperature behavior of this temperature will provide useful insights into the gauge quantum gravity, but we need to reflect on the robustness of the results when replacing the scalar toy model would by its more realistic brother - the full gauge theory described by the BGKM action.
We therefore aim to derive the thermodynamic behavior of this scalar toy model and introduce local and nonlocal thermal field theory (using scalar fields) in the ensuing chapters
to acquire invaluable skills for the ensuing computation of thermal corrections in the scalar toy model.\footnote{This model is more complicated than nonlocal $\phi^4$ theory because the scalar toy model contains derivative acteractions, which have not been studied before in the context of finite temperature field theory!}
\subsection{Infinite derivative scalar toy model \label{infderscalar}}
Canonical examples of scalar infinite derivative actions which appear in string literature\footnote{Additionally, this nonlocality in the form of an
infinite set of derivatives has been discussed in the context of RG arguments within the context of assymptotic safety \ref{asymptoticsafety}.} can all be written as
\begin{equation}\label{action}
S_{\text{inf der}} = \int \ d^{D}x \ \left[	\frac{1}{2} \phi \mathcal{K}(\square) \phi - \mathcal{V}_{\text{int}}	\right], 
\end{equation}
where the kinetic operator $\mathcal{K}(\square)$ contains an infinite series of higher-derivative terms. This action describes, among others, p-adic toy models for strings ($\mathcal{K}=-e^{\square/M^2}$), random lattices ($\mathcal{K}=-(\square + m^2)e^{\square/M^2}$) and bosonic stringy tachyons, where $m^2 < 0$ and $M^2 >0$ are proportional to the string tension \ref{UV}, \ref{p-adic}, \ref{sftmodels}. 
\nnn
A scalar toy model IDG with a string-inspired free action based on Eq. \ref{action} ($D=4$) which captures essential features of the
infinite-derivative BGKM action was derived in Ref. \ref{UV}. The free action reads
\begin{subequations}
\begin{equation}\label{sfreee}
S_{\text{scalar}} = \frac{1}{2} \int \ d^{4}x \left( \phi \square a(\square) \phi \right),
\end{equation}
 and the free theory is accompanied by the following interaction term
\begin{equation}\label{key}
S_{\text{int}} = \frac{1}{M_p} \int \ d^{4}x \Big( 
\alpha_1 \phi \partial_{\mu} \phi \partial^{\mu} \phi 
+
\alpha_2 \phi \square \phi a(\square) \phi 
+
\alpha_3 \phi \partial_{\mu} \phi \partial^{\mu} a(\square) \phi 
\Big).
\end{equation}
 Following \ref{UVinf} we choose the identical Gaussian function
\begin{equation}\label{key}
a(\square) = \exp \left[ -\square/M^2 \right] 
\end{equation}
which is present in the BGKM action.
\end{subequations}
The field equations of GR obey a global scaling symmetry
\begin{subequations}
\begin{equation}\label{key}
g_{\mu \nu} \rightarrow \lambda g_{\mu \nu}
\end{equation}
and this translates to a symmetry for the $h_{\mu \nu}$ when expanding the metric around Minkowski space. The infinitesimal version of this symmetry is given by
\begin{equation}\label{key}
h_{\mu \nu} \rightarrow (1 + \epsilon)h_{\mu \nu} + \epsilon \eta_{\mu \nu}.
\end{equation}
Though the scaling symmetry is not expected to be an unbroken fundamental symmetry of nature, the symmetry relates the free and interaction terms in a reminiscent fashion to how
gauge symmetries relate these terms. The symmetry thus plays an important role for our scalar theory of gravity and consequently the shift-scaling symmetry
\begin{equation}\label{symmetry}
\phi \rightarrow (1+ \epsilon)\phi + \epsilon
\end{equation}
\end{subequations}
is imposed in order to obtain a scalar toy model whose vertices and propagators preserve the compensating nature exhibited by the full-fledged BGKM gravity. \nnn
Equation \ref{symmetry} (uniquely) fixes the cubic interaction term $S_{\text{int}}$ as
\begin{equation}\label{intaction}
S_{\text{int}} = \frac{1}{4M_{p} }\int d^{4}x \Bigg( 	\phi \partial_{\mu} \phi \partial^{\mu} \phi + \phi \square \phi a (\square) \phi - 
 \phi \partial_{\mu} \phi a(\square)\partial^{\mu} \phi	\Bigg),
\end{equation}
the full derivation of which can be found in Appendix  \ref{toymod}. \nnn 
It has been pointed out in Ref. \ref{UV} that the infinite-derivative scalar
toy model given by Eqs. \ref{sfreee} and \ref{intaction} is renormalizable at 1-loop order. This was done by first constructing suitable Feynman rules for this action with derivative interactions. Subsequently, integration of the loop momentum variable $k^\mu$ at 1-loop (for the 1-loop, 2-point function) turned out to lead to a UV-divergent diagram, however, the counterterm has also been computed in Ref. \ref{UV}.
Scattering amplitudes for the scalar toy model are superficially
convergent for $L > 1$, where L is the number of loops occurring in the Feynman diagram. Additionally, the highest divergence
of the 1-loop, 2-point function with nonvanishing external momenta $p$ and
$-p$ is $\Lambda^4$ (where $\Lambda$ is a hard cutoff) and the highest divergence of the 2-loop, 2-point function with vanishing external momenta is also $\Lambda^4$. \ref{UV}, \ref{UVfiniteness}. This result poses a striking contrast with the result from GR,
where the 2-loop diagrams diverge as $\Lambda^6$ \ref{UV} and  proves to be a huge advancement over GR in this regard.\nnn
The dressed propagator is stronger exponentially suppressed compared to the bare propagator and dressed vertices behave as exponentials of external momenta
when the external momenta are large. By employing dressed propagators and
dressed vertices, $n$-loop, 2- \& 3-point diagrams constructed out of lower-loop,
2- \& 3-point diagrams become finite in the UV with respect to the internal
loop momentum $k^\mu$. The promising result is that no additional UV-divergences arise (other than the two divergent diagrams from the 1-loop 2-point and 2-loop 2-point functions) so no new counterterms are required, guaranteeing normalizability of the scalar theory. The exponential suppression of the graviton propagator renders the theory asymptotically free and therefore provides promising prospects for a resolution of various classical
and quantum divergences which were discussed in section \ref{modgrav}. \nnn
The UV-behavior of scattering diagrams within the context of an
infinite-derivative scalar toy model has also been investigated, see \ref{IDGscattering}. The authors concluded that the external
momentum ($p^\mu$) dependence of the scattering diagrams is convergent for sufficiently large $p^\mu$. 
The bare vertices of the scattering diagrams were replaced by dressed
vertices by considering the renormalized propagator and by computing vertex loop corrections to the bare vertices. Interestingly, the exponents in the dressed vertices successively decrease as the loop order increases and even become negative at sufficiently high loop-order.
\nnn
Subsequently, the UV finiteness
with respect to both \textit{internal loop momenta and external momenta} for a general class of Feynman diagrams within the context of infinite-derivative scalar field theories has been studied in \ref{UVfiniteness}. 
The results for the 1-loop, 2-point function with \textit{non-vanishing external momenta} $p^\mu$,
where the bare propagators were replaced by the appropriate dressed propagators, have shown that these Feynman integrals are actually convergent. By employing dressed vertices and propagators, $n$-loop,
$N$-point diagrams which are constructed out of lower-loop ($2$- \& $3$-point functions) are UV-finite. In general, $N_i$-point
diagrams have been shown to be UV-finite as well \ref{UVfiniteness}.
The most
general 1PI Feynman diagrams are consequently finite in the UV within the framework of this infinite-derivative scalar field theory. Establishing
that an infinite-derivative theory of gravity  is  finite (or even renormalizable)
at higher loop-orders would be a huge achievement and would facilitate a theory of quantum gravity which is successful on all fronts.
These arduous computations have not yet appeared in the literature. \nnn
We will contribute to our understanding of the scalar toy model by computating thermal corrections for the scalar IDG toy model action. 
The thermal properties of the scalar toy model defined by Eqs. \ref{sfreee} and \ref{intaction} will be derived in future work, but the prelimaries on how to compute relevant Feynman diagrams in this particular nonlocal model have been included in appendix \ref{chaptertoymodelfinitet} of this thesis.
As has already been discussed, this theory resembles the full gauge theory in many aspects and the results from thermal corrections would consequently give useful insights pertaining to the predictions which infinite-derivative theories of gravity make for the early universe and in other hot environments such as neutron stars.

\chapter{Thermal field theory
\label{chapterTFT}
}
 In this chapter, the mathematical machinery of thermal field theory is introduced with a particular emphasis on the imaginary time formalism (ITF). 
 The formalism has been developed by many authors, but is usually named
 after Matsubara (1953) who was the first to set up a diagrammatic perturbation theory
 for the partition function on a field-theoretic basis in \ref{refmatsubara}. 
 The imaginary time formalism (or \textit{Matsubara formalism}) is the oldest TFT framework and is well-suited to describe equilibrium behavior at finite temperature.
 The major advantage of ITF
 is that the Feynman rules are very similar to those of the vacuum
 theory in Fourier language, except that the energies in the propagators are imaginary and discrete which entail that certain Feynman rules ought to be modified, among others $\int \frac{d^4k}{(2\pi)^4} \rightarrow T \sum_n\int \frac{d^3k}{(2\pi)^3}$.  
 Dynamical problems, on the other hand, necessitate an analytic continuation of
 the Matsubara Green functions to the real axis (the notion of time ordering does not exist in Euclidean field theory due to the unphysical representation of time; time evolution should therefore be introduced by a suitable choice of the analytic continuation to Minkowski space-time) but are generally better evaluated in the real time formalism (RTF) in which time-evolution can be introduced straightforwardly. 
\nnn
Firstly it will be discussed how computations can be carried out in the imaginary time formalism and, for this purpose, the relevant mathematical machinery will be introduced in section \ref{imagform} which allows us, among other things, to compute corrections to the partition function and compute \emph{thermal masses} (contributions to the self-energy arising from continuous interactions with the heat bath).
The partition function for a non-interacting scalar theory will be calculated in section \ref{freescalartheory} and is  ensued by a calculation of the dominant contribution from a $\phi^4$ interaction term in section \ref{lambdaphi4}. 
We conclude that this interaction term provides a \emph{negative contribution} to the partition function and corresponding thermodynamic state functions such as energy, pressure and entropy - a result which can be found in many textbooks on TFT (for instance \ref{bellac},  \ref{kapustabook}, \ref{weert}) but an interpretation of this effect is usually absent and will therefore be given in section \ref{lambdaphi4}. Subsequently, we will prove that scalar particles acquire a thermal mass in addition to their vacuum mass, even if the latter is zero. The topics of this chapter pave the way to an understanding of \emph{nonlocal} thermal field theory, 
which will be discussed in the ensuing chapter \ref{chapternonlocalfield}.\footnote{
Nonlocal thermal field theory is best understood if one is already familiar with local field theory since one can then adjust appropriate aspects of the local theory, for instance replace the local propagator by the nonlocal propagator.
Most of the  mathematical machinery which is introduced in this chapter however remains applicable in nonlocal field theory. Those who are primarily interesting in nonlocal thermal field theory are therefore advised to firstly read this chapter.}
\section{The imaginary time formalism \label{imagform} }
Thermal field theory is based on the formal analogy, first noted by
F. Bloch in 1932, between inverse temperature $\beta = 1/(k_BT)$ and the time evolution operator in imaginary time $\tau = it$, \emph{i.e.}
\begin{equation}\label{key}
\exp \left[ - i t H / \hbar \right]   \leftrightarrow \exp\left[	- \beta H	\right], \ \ t = - i \hbar \beta
\end{equation}
This leads to so-called
temperature Green functions with purely imaginary time arguments in the imaginary time formalism. That is, one deals with Euclidean field theories. The framework is not equivalent to the Euclidean formalation of quantum field theory, since fields at finite temperature obey periodicity constraints (as will be discussed elaborately) which constrain the imaginary time variable to the regime $[0,\beta]$. These statements will be proved and elaborated upon in the following sections, but we will firstly provide a brief review of quantum statistical mechanics (as thermal field theory should be conceived of as a relativistic generalization of quantum statistical mechanics).
\subsection{Review of quantum statistical mechanics}
Since thermal field theories aim to describe quantum systems in a statistical fashion, a logical starting point would be to choose any of the canonical ensembles known from statistical mechanics. In both the canonical ensemble and grand canonicle ensemble ``$\beta = T^{-1}$'' should be thought of as a Lagrange mulitiplier which determines the mean energy of the system. In the grand canonical ensemble, the chemical potential $\mu$ plays the role of a second Lagrange multiplier and determines the average number of particles which are present in the system.\footnote{Particles are continually created and annihilated in relativistic field theories, entailing that the grand canonical ensemble is often more appropriate. In this case the density matrix generalizes to
	\begin{equation}
	\rho = \exp[- \beta [H - \mu Q]],
	\end{equation}
	where $Q$ are conserved quantum numbers, such as baryon number $\mathcal{B}$, and the corresponding chemical potentials are their Lagrange multipliers. 
	In addition to the thermodynamic state functions defined in Eq. \ref{thermquant} one can obtain
	the expectation value for the particle number by differentiation to the chemical potential:
	\begin{equation}
	\begin{cases}
	\text{Particle number} & N_{i} = \frac{1}{V} \frac{ \partial \mathz }{ \partial \mu_i}.
	\end{cases}
	\end{equation}
Calculations in this thesis are performed in the canonical ensemble, but it should be kept in mind that	one may always pass over to the grand canonical ensemble by performing an \emph{inverse Laplace transform} on the variable $\beta$ \ref{kapustabook}.
 } 
The statistical properties of a system in thermodynamic equilibrium are described by the partition function, the canonical partition function is defined as
\begin{equation}\label{parttt}
\bal
\mathz(\beta) &= \Tr \exp\left[-\beta H \right] \\
&= \sum_{n}  \bra{ n } e^{- \beta H} \ket{ n},
\eal
\end{equation}
where we sum over a complete basis of states $n$ and $H$ is the Hamiltonian of the system. All the macroscopic properties of the system in \emph{thermal equilibrium} may be calculated from $\mathz$.
\nnn
Thermodynamic state functions are simply
derivatives of $\ln \mathz$ multiplied by constants\footnote{
The thermodynamic potential $\Omega$ can be calculated via the equation
\begin{equation}\label{key}
V \Omega(\beta) = -\ln \mathz(\beta)
\end{equation}	
and loop corrections to $\Omega$ are often calculated in the literature, for instance in \ref{bellac}.
	We have chosen to express the thermodynamic state function in terms of (derivatives of) $\ln \mathz$, a convention which is rather arbitrary since $\ln \mathz$ and $\Omega$ differ only in a respective factor of $(-V)$. Loop corrections to the  partition function will therefore be computed in this thesis rather than corrections to the thermodynamic potential.}
\begin{equation}\label{thermquant}
\begin{cases}
\text{Pressure} &	P(\beta) = \frac{T}{V} \ln \mathz, \\	
\text{Energy density} &	\epsilon(\beta) = - \frac{1}{V} \frac{\partial}{ \partial \beta} \ln \mathz, \\
\text{Entropy density} &	s(\beta) = \frac{1}{V} \frac{\partial }{ \partial T} \left( T \ln \mathz \right),
\end{cases}
\end{equation}
where $V$ is the volume of the system.
The vacuum expectation value of an observable $[\cdots]$ is determined by
\begin{equation}\label{key}
\expval{\cdots}_{0} = \sum_{n} \bra{n} [\cdots] \ket{n}
\end{equation}
in quantum field theory.
Thermal averages are calculated in a reminiscent fashion:
vacuum expectation values now have to be replaced
by quantum-statistical expectation values
\begin{equation}\label{exptemp}
\expval{\cdots}_{\beta} 
= \Tr \left( \left[ \cdots \right] \rho(\beta)  \right)
= {1 \over \mathz} \sum_{n} \bra{n} [\cdots] e^{-\beta H} \ket{n} 
= {1 \over \mathz} \sum_{n} \bra{n} [\cdots]  \ket{n} e^{-\beta E_n},
\end{equation}
where  
$H$ is the Hamiltonian of the system\footnote{Although $`H'$ should be interpreted as $(H - \mu Q)$ when working in the grand canonical ensemble. \emph{We will however perform calculations in the canonical ensemble in the rest of this thesis}, but one can relatively straightforwardly generalize these results for the grand partition function $\Xi$.} 
and the canonical density matrix is defined as
\begin{equation}\label{key}
\rho(\beta) = \mathz^{-1} \exp \left( -\beta H \right).
\end{equation}
Note that all operators are the familiar zero-temperature ones and that the temperature enters only in the exponential $\exp[-\beta H]$ where it characterizes the particular ensemble of states used to calculate expectation values of the operators $\expval{\cdots}$.
\subsubsection*{Example: bosonic harmonic oscillator }
Let us now explicitly compute the thermodynamic quantities provided in Eq. \ref{thermquant} for the bosonic harmonic oscillator in the canonical ensemble using Eq. \ref{parttt}.
The bosonic harmonic oscillator is given by the Hamiltonian
\begin{subequations}
\begin{equation}\label{key}
H = K + V, 
\end{equation}
where the kinetic and potential terms are
\begin{equation}\label{key}
K = \frac{p^2}{2m} \ \ \ \ \ \text{and} \ \ \ \ \ \ V = \frac{1}{2}m \omega^2 x^2.
\end{equation}
The non-degenerate energy eigenvalues $\ket{n}$ are given by
\begin{equation}\label{key}
E_n = \hbar \omega \left(n + \frac{1}{2} \right),
\end{equation}
where $\frac{\hbar \omega}{2}$ is the zero-point energy of the vacuum.
\end{subequations}
The partition function can be obtained in a trivial fashion when written in the energy basis, i.e.
\begin{subequations}
\begin{equation}\label{key}
\bal
\mathz(\beta) = \Tr e^{-\beta H} &= \sum_{n=0}^{\infty} \bra{n} e^{-\beta H} \ket{n}  = \sum_{n=0}^{\infty} \bra{n} e^{-\beta \hbar \omega \left(n + \frac{1}{2} \right)} \ket{n}  \\
&=  
\frac{ e^{-\beta \hbar \omega /2 } }{
1 - e^{-\beta \hbar \omega }
} =  
\frac{ 1 }{
	2 \sinh \left( \frac{\hbar \omega \beta}{2 } \right) }.
\eal
\end{equation}
The logarithm of the partition function is thus given by
\begin{equation}\label{bosonicharmonicoscillator}
\ln \mathz (\beta) =
\ln \left( \frac{ e^{-\beta \hbar \omega /2 } }{
	1 - e^{-\beta \hbar \omega }
} \right) = -\frac{\beta \hbar \omega}{2} - \ln\left(
	1 - e^{-\beta \hbar \omega}
\right)
\end{equation}
\end{subequations}
and the thermodynamic quantities are
\begin{subequations}
\begin{align}
S &= - \frac{\partial}{\partial T} \left(
\frac{ \hbar \omega}{2} +
T \ln \left(1 - e^{ \hbar \omega / T} \right) \right) =  - \ln \left(1 - e^{ \hbar \omega / T} \right) + \frac{\hbar \omega }{T}
\frac{1}{e^{ \hbar \omega / T} - 1 } \approx \begin{cases}
 \frac{\hbar \omega }{T}
 e^{- \hbar \omega / T}   & \text{for } T \ll \hbar \omega \   \\
1 + \ln \frac{T}{\hbar \omega}  & \text{for } T \gg \hbar \omega \   \\
\end{cases}, \\
	P &= \frac{T}{V} \left[
\ln \left(1 - e^{ \hbar \omega / T} \right) - \frac{\hbar \omega}{2}
	 \right] \approx 
	\begin{cases}
	\frac{ \hbar \omega T}{2V }  & \text{for } T \ll \hbar \omega \   \\
	\frac{T}{V} \exp\left(\hbar \omega / 2T \right) & \text{for } T \gg \hbar \omega \   \\
	\end{cases}, \\
E &= \frac{\partial}{\partial \beta}
\left(
\frac{\beta \hbar \omega}{2} + \ln\left(
1 - e^{-\beta \hbar \omega}
\right)
\right) = \hbar \omega \left[ \frac{1}{2} + \frac{1}{e^{-\beta \hbar \omega} - 1}
\right] \approx
\begin{cases}
\frac{\hbar \omega}{2} & \text{for } T \ll \hbar \omega \   \\
T & \text{for } T \gg \hbar \omega \   \\
\end{cases}.
\end{align}
\end{subequations}
Note how one can separate the contribution of the ground state, which dominates at low
temperatures $T \ll \hbar \omega$, from that of the thermally excited states. The latter are characterized by the appearance
of the Bose-Einstein distribution $n_{\text{BE}}(\hbar \omega) \equiv \left[\exp(\beta \hbar \omega) - 1 \right]^{-1}$. 
\subsection{KMS condition. 
\label{sectionKMS}
}
It was mentioned in the introduction that Green
functions 
in the Matsubara formalism
are periodic in $\beta$ and
 are solely defined at its
so-called \emph{Matsubara frequencies}, which constitute a
discrete set of points on the imaginary axis of the complex energy plane. We will first prove these statements before delving into the mathematical machinery which is required for computations in thermal field theory (such as frequency sums) in the ensuing subsection. \nnn
The expression for finite temperature $N$-point correlation function follow follows straightforwardly from the definition of thermal averages, the definition of which can be found in Eq. \ref{exptemp}. 
Because the partition function involves a trace, it leads to an interesting identity relation for correlation function following from the cyclicity of the trace.
The finite temperature \emph{two-point correlator} satisfies
\begin{equation}\label{twopoint}
\bal
\expval{
	\phi(\mathbf{x}, t), \phi(\mathbf{y}, t')
}_{\beta} &=
\mathz^{-1} \Tr \left[ 
 e^{- \beta H}
	\phi(\mathbf{x}, t) \phi(\mathbf{y}, t')
\right] \\
&= 
\mathz^{-1} \Tr \left[ e^{- \beta H} \phi(\mathbf{y}, t') e^{- \beta H}   \phi(\mathbf{x}, t) 
e^{ \beta H} 
\right] \\
&= 
\mathz^{-1} \Tr \left[
e^{- \beta H}
\phi(\mathbf{y}, t')   \phi(\mathbf{x}, t + i\beta) \right] \\
&= \expval{
	\phi(\mathbf{y}, t' ), \phi(\mathbf{x}, t + i\beta)
}_{\beta},
\eal
\end{equation}
where the quantum-mechanical time-evolution relation in the Heisenberg picture
has been used with an imaginary time variable $\tau = i\beta$, \emph{i.e.}
\begin{equation}\label{key}
\phi(\mathbf{x}, \tau) = \exp(-H \beta) \phi(\mathbf{x}, 0) \exp( H\beta).
\end{equation}
This relation is known as the KMS	(Kubo-Martin-Schwinger (KMS) condition \ref{refkubo} which generalizes to
all statistical ensemble averages and plays a crucial role in the study of thermal field
theories \ref{kapustabook}. The KMS condition is responsible for nontrivial differences between Euclidean (zero-temperature) field theory and thermal field theory. 
\nnn
It can be inferred from Eq. \ref{twopoint} that the field $\phi(\mathbf{y}, t')$ evolves into imaginary time $t'+ i \beta$. 
Eq. \ref{twopoint} still holds true for purely imaginary time coordinates $t$, $t'$, so the equilibrium two-point correlation functions satisfy \ref{weert}
\begin{equation}\label{KMS}
\expval{
	\phi(\mathbf{x}, \tau), \phi(\mathbf{y}, \tau')
}_{\beta} =  
\expval{
	\phi(\mathbf{y}, \tau' + \beta), \phi(\mathbf{x}, \tau)
}_{\beta}
\end{equation}
and imply a periodic boundary condition
\begin{equation}\label{periodicity}
\phi(\mathbf{y}, \tau') = \pm \phi(\mathbf{y}, \tau' + \beta),
\end{equation}
where the equality holds true for arbitrary $\tau'$ 
and implies that initial and the final states must be the identical.
It should be noted that we interchanged the fields $	\phi(\mathbf{x}, \tau)$, $\phi(\mathbf{y}, \tau')$
in the first step of Eq. \ref{KMS}.
Therefore one yields a plus-sign whenever these fields commute, that is to say, for bosonic fields, and a minus-sign whenever these anticommute, so for fermionic fields. The imaginary time Green's function thus exhibit either periodicity or antiperiodicity in imaginary time with period $\beta$.
This (anti)periodicity relation plays a pivotal role in the imaginary time formalism, responsible for non-equivalent Feynman rules in the IFT as compared to the Euclidean formulation of quantum field theory.
\nnn
That we are using an imaginary time variable is unsurprising when realizing that
inverse temperature is related to imaginary time (as was noted by Bloch in 1958 \ref{refbloch}) by\footnote{We will subsequently omit $\hbar$ from our expressions and thereby abide natural units ($c = \hbar = k_B = 1$) which are used throughout the rest of the thesis. For more info, read ``\hyperlink{conventions}{Conventions and notation}''.}
\begin{equation}\label{key}
\exp \left[ - i t H / \hbar \right]   \leftrightarrow \exp\left[	- \beta H	\right], \ \ t = - i \hbar \beta
\end{equation}
For this reasons it is convenient to introduce an imaginary time variable via $\tau = it$,  where 
$t \in \mathrm{R}$ so
$\tau \in \mathrm{C}$. 
Bloch realized \ref{refbloch} that the partition function can therefore be written like a time evolution operator into the imaginary axis.
Writing the partition function in the coordinate $x$ basis
\begin{equation}
\mathz(\beta) = \int dx \bra{x} \exp[-\beta H] \ket{x}
\end{equation}
we can straightforwardly show that it can be given the following path integral representation\footnote{
Note that the transition amplitude has $t_i =0, t_f = -i\beta$ and $x_i = x_f = x$.
}
\begin{equation}\label{pathintegral}
\mathz(\beta) = \int \mathcal{D}\phi \exp\left[-S_E[\phi]\right]
\end{equation}
where 
\begin{equation}\label{key}
S_E(\phi) = \int_0^\beta d\tau L_{E}(\phi),
\end{equation}
 subscript $\text{E}$ stands for ``Euclidean" and $L_{E}$ is the the Euclidean (Wick-rotated onto the negative imaginary axis) Lagrangian which is related to the Minkowskian $\mathcal{L}(\phi)$ by 
\begin{equation}\label{key}
L_{E} =  \int d^3x \mathcal{L}(-i \tau, \mathbf{x}).
\end{equation}
Bloch therefore discovered that the partition function can be computed using reminiscent field-theoretic methods as in QFT (path integrals, Green functions, \emph{etc}) provided that the time coordinate is interchanged for an imaginary (equilibrium) temperature variable.
Also the expectation value for $\expval{\cdots}_{\beta}$ can consequently be formulated in terms of path-integrals
\begin{equation}\label{key}
\expval{\cdots}_{\beta} 
= \frac{
	\int \mathcal{D}\phi 	\left[ \cdots \right] \exp\left[-S_{\text{E},0}\right]
}{
	\int \mathcal{D}\phi  \exp\left[-S_{\text{E},0}\right],
}
\end{equation}
This path integral representation for expectation values will prove fruitful in the calculations of section \ref{sectionapplications} and chapter \ref{chapternonlocalfield}.\footnote{A historical note is appropriate here: Matsubara  had originally constructed the imaginary time formalism in an operator description, it has only later been formulated in path integral formulation \ref{das}.} 
Note that the imaginary time formalism is a Euclidean field theory so if one is interested in Minkowskian Green functions
one needs to perform an analytic continuation.
\nnn
The KMS condition implies a periodic boundary condition which identifies the fields at $\phi(\tau = 0) = \phi_1$ and $\phi(\tau = \beta) = \phi_2$. The Euclidean path integral
\begin{equation}\label{key}
\bra{\phi_2} \exp\left[ - \beta H \right] \ket{\phi_1} = \int_{\phi_1}^{\phi_2} \mathcal{D} \phi \ e^{-S_{E}[\phi]}
\end{equation}
involves a split into space and (Euclidean) time, where $\phi_{1,2}$
is a boundary condition that
specifies data at fixed times $\tau = 0$ and $\tau = \beta$.
The meaning of this
this path integral depends on the
topology of space.
This identification of fields yields an infinitely long cylinder for field theories on a line and for field theories on a circle this yields a torus, since torii are in fact the Euclidean product of two $S^1$ circles (both of which may be familiar results from SFT models with compactified higher dimensions which satisfy similar boundary conditions).
The field $\phi_1$ in the following diagrams evolves vertically in Euclidean time and `splits' are made at $\tau = 0$ and $\tau = \beta$

\noindent\begin{minipage}{.5\linewidth}
	\begin{equation}
	\bal
	\mathz(\beta) &= \Tr \exp\left[- \beta H \right] \\
	&= \sum_{\phi_1} \bra{\phi_1} e^{-\beta H} \ket{\phi_1} \\
	&= 
	\sum_{\phi_1} 
	\vcenter{
		\begin{tikzpicture}[baseline=(v1), node distance=0.6cm and 0.6cm]
		\begin{feynman}[inline = (v1)]
		\coordinate[] (v1)	; 
		\coordinate[right=of v1] (v2); 
		\coordinate[above=of v1] (v3);
		\coordinate[right=of v3] (v4); 
		\coordinate[above=of v3] (v5);
		\coordinate[right=of v5] (v6);
		\coordinate[right=0.2cm of v2] (v7);
		\vertex[above=of v7] (v8) {\(\beta\)};
		\coordinate[right=0.2cm of v6] (v9);
		\diagram* {
			(v1) -- [ghost, edge label'=\(\phi_1\)] (v2),
			(v5) -- [ghost, edge label=\(\phi_1\)] (v6),
			(v1) -- (v5);
			(v4) -- (v2);
			(v4) -- (v6);
			(v8) -- [fermion](v7);
			(v8) -- [fermion](v9);
		};
		\end{feynman}.
		\end{tikzpicture}
	} \\
	&= 
	\vcenter{
		\includegraphics[scale=0.2]{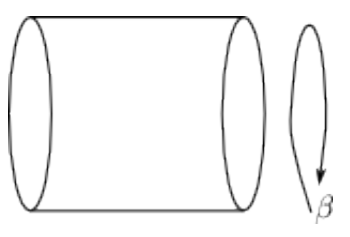}
	}
	\eal
	\end{equation}
\end{minipage}%
\begin{minipage}{.5\linewidth}
	\begin{equation}
	\bal
	\mathz(\beta) &= \Tr \exp\left[- \beta H \right] \\
	&= \sum_{\phi_1} \bra{\phi_1} e^{-\beta H} \ket{\phi_1} \\
	&= 
	\sum_{\phi_1} 
	\vcenter{
		\includegraphics[scale=0.2]{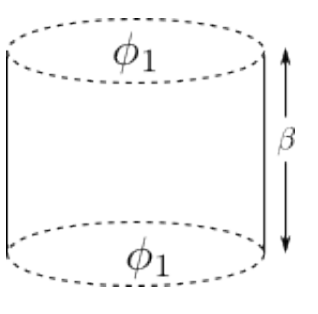}
	} \\
	&= 
	\vcenter{
		\includegraphics[scale=0.2]{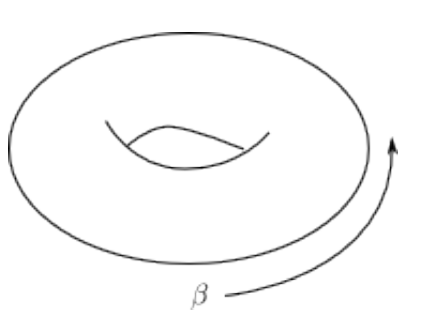}
	}
	\eal
	\end{equation}
\end{minipage}	
\nnn
In the penultimate steps of the previous equations, by summing over $\phi_1$
we are actually imposing periodic boundary conditions on the 2D plane (cylinder). This glues together the two ends of the plane (cylinder), represented by the vertical ``$\leftrightarrow$'', producing
a cylinder (torus). The thermal partition function for a line
is equal to a path integral on a cylinder and  the thermal partition function for a 
2D theory on a circle is equal to a path
integral on a torus.
\subsubsection*{Matsubara frequencies}
Since every periodic function can be expanded in a Fourier series, the field
 $\phi(\mathbf{x}, \tau)$ satisfies
\begin{equation}\label{fourierexp}
\phi(\mathbf{x}, \tau) = T \sum_{n = -\infty}^{\infty} \phi_{n}(\mathbf{x}, \omega_n) \ e^{i \omega_n \tau},
\end{equation}
where $\omega_n$ are called the Matsubara frequencies \ref{hagedorn1}.
Since reality\footnote{More precisely, the condition $\phi^{*}(\mathbf{x}, \tau) = \phi(\mathbf{x}, \tau)$ is imposed.} is imposed on the fields $\phi(\mathbf{x}, \tau)$ the Fourier series implies that the Fourier modes satisfy
\begin{equation}\label{reality}
\left[\tilde{\phi}_{n}(\mathbf{k}, \omega_n) \right]^{*} = \tilde{\phi}_{n}(-\mathbf{k}, -\omega_{n})
\end{equation}
which implies that only half of the Fourier modes are independent.
This property will  recurringly be used in the computions of section \ref{lambdaphi4}.
Moreover, because $\phi(\mathbf{x}, 0) = \pm \phi(\mathbf{x}, \beta)$, a discretized set of frequencies emerges in the imaginary time formalism. Equations \ref{periodicity} and \ref{fourierexp} imply that the Matsubara frequencies satisfy
\begin{equation}\label{key}
\omega_n = 
\begin{cases}
2  \pi n \beta^{-1}	& \text{for bosons}, \\ (2n + 1) \pi \beta^{-1}	& \text{for fermions}.
\end{cases}
\end{equation}
Matsubara frequencies can be thought of as the possible \emph{frequency modes on a circle} (or torus, depending on the topology of space), where $n= 0$ provides the lowest possible frequency $\omega_n = \pi T$ for fermionic thermal modes while the fundamental frequency for bosons is given by $n = \pm 1$ and reads $\omega_n = 2\pi T$. The fact that bosons can have $\omega_n = 0$ is responsible for many infrared divergences in bosonic thermal field theories, we will encounter examples of such divergences in appendix \ref{appendixbreakdownperturbation}.
\nnn
Since energy now takes discrete values, we have to replace the integration over the loop energies to summation over internal Matsubara frequencies in loops 
\begin{equation}\label{key}
\int {d^4p \over (2\pi)^4 } \rightarrow T \sum_n \int \frac{d^3p}{(2\pi)^3}.
\end{equation}
Summation over these integers at higher loops presents a nontrivial challenge in the imaginary time formalism.
Thermal field computations will be computed exclusively for scalar theories in this thesis, so only bosonic TFT will be discussed henceforth. 
\subsubsection*{Imaginary versus real time formalisms}
\begin{figure}[h!]
	\centering
	\includegraphics[scale=0.5]{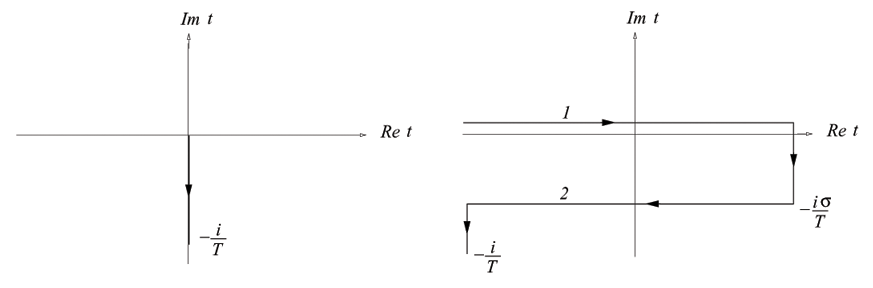}
		\caption{
		\textit{The left contour shows the contour which is used in the Matsubara formalism. In this formalism, energy is discrete and imaginary and a summation over Matsubara frequencies has to be taken. The right contour shows an alternative contour which leads to the \textit{real-time formalism}. One is free to choose any value between 0 and 1 for the real parameter $\sigma$, but two choices have proven to be specifically fruitful. For $\sigma = 1$, one obtains the Keldysh-Schwinger (or closed time path) formalism which can used to describe non-equilibrium phenomena \ref{das}.
	For $\sigma = 1/2$, the real time formalism corresponds to thermofield dynamics of Umezawa et. al \ref{umezawa}, which is particularly well-suited to examine the Hilbert space structure of a theory. The unphysical parameter $\sigma$ never appears in the final results.} Figures were taken from \ref{das}
	}
\label{contourfigure}
\end{figure}

The imaginary contour of the Matsubara formalism corresponds to a straight line from the origin to $-i\beta$ (shown in Fig. \ref{contourfigure}) and follows directly from the definition of the partition function.
What it is fact important is that the \emph{contour starts at $t =0$} and \emph{terminate the contour at $t = -i\beta$} so one is free to choose whichever contour one prefers as long as they satisfy these constraints. The second contour shown in Fig. \ref{contourfigure} lies at the heart of real time formalisms (or in-in formalisms).
Time takes values over the entire real interval in real-time formalisms (hence the name) and, for most applications, the contributions from imaginary branches decouple because these lie at $\text{Re } t = \pm \infty$ (but see \ref{refrealtime} for exceptions where the imaginary branches \emph{do} contribute).
\nnn
 Energy in real time formalisms thus has a continuous spectrum much like in the zero temperature theory, but one should take into account the fact that \emph{two} real time contours contribute to this spectrum \ref{das}. A necessity to double the amount of degrees of freedom arises and therefore a more intricate formalism than the imaginary time formalism arises when choosing the real-time contour in Fig. \ref{contourfigure} which is known as Umezawa's \emph{thermofield dynamics} \ref{umezawa}. Alternatively, one can use the Keldysh-Schwinger formalism, where the contour goes from $(-\infty)$ to ($\infty$), then to $(\infty - i\beta)$ and then to $(-\infty - i\beta)$ \ref{bellac}.  We may consequently write the propagator for scalar fields as a 2 by 2 matrix (see \ref{das})
\begin{equation}\label{key}
\mathcal{D}(p) = \begin{pmatrix}
\frac{1}{p^2 - m^2 +i\epsilon} + 2\pi n_\text{BE}(\abs{p_0}) \delta(k^2 - m^2) 
&  2\pi 
\left(\theta(-p_0) + n_\text{BE}(\abs{p_0}) \delta(k^2 - m^2) \right) \\
  2\pi 
\left(\theta(p_0) + n_\text{BE}(\abs{p_0}) \delta(k^2 - m^2) \right)
& -\frac{1}{p^2 - m^2 +i\epsilon} + 2\pi n_\text{BE}(\abs{p_0}) \delta(k^2 - m^2) \\
\end{pmatrix},
\end{equation}
where $n_\text{BE}(\abs{p_0}) = \frac{1}{e^{\beta \abs{p_0}} - 1}$ is the Bose-Einstein distribution (note that it is the absolute value of $p_0$, this absolute value actually encumbers computations in the RTF significantly). I do not want to go into the details of these real time formalisms here, I mainly wanted to convey that multiple formalisms exist in which  thermal corrections can be computed. Interestingly, real time formalism are more appropriate to describe non-equilibrium physics because one has a real time variable and can consequently introduce a time-evolution into the system \ref{umezawa}. In the imaginary time formalism one can only study equilibrium physics, although one can study \emph{near-equilibrium physics} by introducing a time-evolution in the analytic continuation.
\nnn
Both the imaginary and the real time formalisms (RFTs) are well-suited to describe equilibrium physics, but one can choose whether one prefers to avoid (possibly tedious) Matsubara frequency summations and hence use a RTF where the amount of DOFs is doubled, or whether one prefers fewer degrees of freedom and use the ITF. We have chosen to work exclusively in the imaginary time formalism in this thesis, since  we will not go beyond second order loops while the computation of Matsubara summations typically becomes cumbersome only for higher-order loop contributions.

\subsection{Propagators
\label{sectionpropagators}
}
The presence of particles in the background heat bath modifies propagators at finite temperature from their zero-temperature counterparts, even at the tree level \ref{renorm}.
For this reason,
we devote a section to finite temperature propagators and discuss how different kinds of propagators (e.g. the imaginary time, real time, retarded and advanced propagator) can be obtained in TFT, all of which are defined in terms of the \ti{spectral intensity function} which is introduced in appendix \ref{freqsumappendix}. We will however focus primarily on how the most important propagator in the imaginary time formalism can be derived, the Matsubara propagator.
\nnn
This imaginary time propagator and one cannot formulate alternative propagators, such as retarded and advanced propagators, in the imaginary time formalism.
This is because the notion of time-ordered products, defined in Minkowskian space as
\begin{equation}\label{key}
\frac{
	\expval{T(\phi(x_1) \cdots \phi(x_n) \exp\left( i \int \mathcal{L} dx \right) }
}{
	\expval{T \exp \left( i \int \mathcal{L} dx \right) }
},
\end{equation}
where $T$ denotes time ordering, $\phi$ are free fields in Minkowski space-time, $\mathcal{L}$
is the interaction Lagrangian, and $\expval{\cdots}$ are vacuum averages, does not exist in Euclidean space-time. One can introduce an analytic continuation to the Schwinger
points (the details of which can be found in \ref{euclideantimeorder}) and yield the following
expression for the analytic continuation of time-ordered field products (now called \emph{Schwinger functions})
\begin{equation}\label{key}
S(x_1,x_2, \cdots,x_n) = \frac{
	\expval{\phi(x_1) \cdots \phi(x_n) \exp\left(- S_E \right) }
}{
	\expval{ \exp\left(- S_E \right) }
},
\end{equation}
where $x_1,x_2, \cdots,x_n$ are points in Euclidean space and the fields $\phi$ are now Euclidean fields. Since the fields are commutative, \emph{the chronological time ordering disappears} and one can no longer speak of a distinguished ``time'' direction in the Euclidean space \ref{euclideantimeorder}.
 \nnn
Let us first recall how propagators are derived in ordinary (Minkowskian) QFT. The free (massive) scalar theory will be considered, whose action is
\begin{equation}\label{massivescalar}
S = \int d^4x \phi ( \square + m^2) \phi.
\end{equation}
The same differential equation may have many different Green's functions depending on the boundary conditions, leading to inequivalent retarded, advanced, Feynman propagator et cetera. Let's consider a generic Green's function of the
Klein-Gordon equation, that is, some function $G(x - y)$ satisfying
\begin{equation}\label{key}
\left[\square + m^2\right]G(x-y) = -i\delta^{(4)}(x-y).
\end{equation}
Fourier transforming the previous equation using
\begin{equation}\label{key}
G(x-y) = \int \frac{d^4k}{(2\pi)^4} e^{-ik(x-y)} \cdot \tilde{G}(k) 
\end{equation}
yields
\begin{equation}\label{green}
(-k^2 + m^2) \tilde{G}(k) = -i,
\end{equation}
which naively indicates that the propagator, in fact, equals $i$ times the inverse of the kinetic operator, \emph{i.e.}
\begin{equation}\label{key}
G(k) = \frac{i}{k^2 - m^2} \ \ \ \ \ G(x-y) = \int \frac{d^4k}{(2\pi)^4} \frac{i\ e^{-ik(x-y)}}{k^2 - m^2}.
\end{equation}
Note that $G(k)$ contains poles at $k = \pm m$
and,
in general, an integral of a singular
function over its pole is ill-defined and must be regularized in order to get a definite answer. There are many different ways to regularize an integral,
while regulators yield different regularized integrals. Many inequivalent
Green's functions of the same Klein–Gordon equation emerge consequently. 
The (time-ordered) Feynman Green function reads
\begin{equation}\label{key}
 G(x-y) = \int \frac{d^4k}{(2\pi)^4} \frac{i\ e^{-ik(x-y)}}{k^2 - m^2 + i \epsilon},
\end{equation}
where $\epsilon$ is an infinitesimally small real parameter, but depending on the choices of sign of ``$i \epsilon$'' one can obtain different kinds of propagators. Following our previous discussion on the role of time in Euclidean field theories, ``$i \epsilon$'' will not play a role in thermal field theory.
\nnn
We can use the similar methodology in Euclidean space, where the RHS of Eq. \ref{green} is multiplied by $i$ due to the Wick rotation underlying the imaginary time formalism. The propagator for massless scalars is naively given by
\begin{equation}\label{key}
 -k^2  \tilde{G}(k) = 1 \rightarrow \tilde{G}(k) = \frac{1}{-k_0^2 + \mathbf{k}^2}.
\end{equation}
We should take into account the fact that frequencies are Wick rotated and moreover \emph{discretized} due to the KMS condition, implying that $k_0 \rightarrow i\omega_n$ where $\omega_n$ are the bosonic Matsubara frequencies. This implies that the Matsubara (imaginary time) propagator in frequency momentum space is given by
\begin{equation}\label{key}
\mathcal{D}_0(i\omega_n, \mathbf{k}) = \frac{1}{\omega_n^2 + \mathbf{k}^2},
\end{equation}
while the propagator in configuration space is
\begin{equation}\label{fourierprop}
\mathcal{D}(\mathbf{x}, \tau) = T \sum_{n=-\infty}^{\infty} \int \frac{d^3p}{(2\pi)^3} \frac{1}{\omega_n^2 + \mathbf{p}^2}.
\end{equation}
It should be kept in mind that this is the \emph{bare propagator} - the \emph{physical propagator} $\mathcal{D}(i\omega_n, \mathbf{k})$ is obtained only after including the self-energy contributions arising from interactions (see section \ref{sectionselfenergy})
and modifies the propagator due to the existence of virtual particles surrounding the propagator (due to vacuum fluctuations, as in QFT) and due to interactions of the propagator with particles in the surrounding heat bath. \nnn
The Matsubara propagator can be analytically continued to Minkowskian retarded, advanced and Feynman propagators through the \emph{spectral intensity function} which is introduced in appendix \ref{freqsumappendix}. 
The retarded and advanced propagators find  many applications in linear response theory, including the possibility to describe screening effects, collective excitations and dispersion relations of particles in media \ref{bellac}.\footnote{The retarded and advanced propagators ($D_R(k_0)$ and $D_A(k_0)$ respectively) can be obtained through \ref{bellac}
	$$D_R(k_0) = -i \mathcal{D}(p_0 + i \epsilon) , \ \ \ \ \ 
	D_A(k_0) = i \mathcal{D}(p_0 - i \epsilon).$$}
Through analytic continuation  one can introduce a (small) time-evolution into the system and therefore study \emph{near-equilibrium physics}.
This is not possible in the Euclidean imaginary-time formalism where the time variable has been traded for an inverse temperature parameter \ref{kapustabook}.
\subsection{Frequency sums \label{freqsumsection}}
\begin{figure}[h!]
	\centering
	\includegraphics[scale = 0.5]{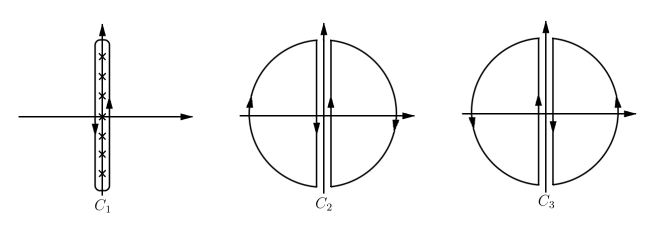}
	\caption{
		\textit{Frequency sums can be expressed as complex contour integrals because $\frac{\beta}{2} \coth \left( \beta p^0 / 2 \right)$ produces a collection of poles with unit residue at  $p^0 = i \omega_n$. A subsequent contour deformation 
		allows one to employ either contour $C_2$ or $C_3$, both of which cover the entire complex plane except the imaginary axis and differ in the sign of the residues only. The $C_3$  contour will  be used whenever frequency sums are calculated in this thesis.} Figure taken from \ref{kapustabook}} 
	\label{contourfig}
\end{figure}

Eq. \ref{fourierprop} implies that a summation over Matsubara frequencies needs to be taken when Fourier transforming fields from frequency-momentum to configuration space. It will now be explained how frequency sums can be carried out (generically)  within bosonic field theories (where the possibility of having fermionic exterior lines is left open in appendix \ref{appendixfreqsum}).
\nnn
For bosonic Matsubara frequencies $\omega_n = 2\pi n T$, the frequency sum has a generic form
	\begin{equation}\label{freq}
	T \sum_{n = -\infty}^{\infty} f(p^0 = i\omega_n = 2 \pi n T i).
	\end{equation}
Bosonic frequency sums can conveniently be rewritten as contour integrals in the complex $p^0$ plane, because hyperbolic cotangent function $\frac{\beta}{2} \coth \left( \beta p^0 / 2 \right)$ produces a collection of poles with unit residue at exactly $p^0 = i \omega_n$ and is everywhere else bounded and analytic \ref{kapustabook}. The frequency sum of Eq. \ref{freq} is therefore equivalent to
\begin{equation}\label{freqeval}
\beta^{-1} \sum_{n} f(p^0 = i\omega_n) = \beta^{-1} \int_{C_1} {\dd p^0 \over 2 \pi i} f(p^0) \frac{\beta}{2} 
\coth \left( {\beta p^0 \over 2 } \right),
\end{equation}
where $C_1$ is a contour with infinitesimal width around the imaginary axis (an illustration of which is provided in Fig. \ref{contourfig}). Caveat: this prescription works only whenever  the meromorphic function$f(p^0)$  is regular on the $\Re p^0  = 0$ line \footnote{In the grand canonical ensemble $f(p^0)$ should be regular on the line $\Re p^0 = \mu$.} and decreases faster than $p_0^{-1}$ for $\abs{p_0} \rightarrow \infty$ \ref{bellac} (this asymptotic behavior guarantees that the function vanishes at infinity, allowing one to close complex contours using large half-circles).	
The contour $C_1$ can be deformed to $C_2$ in Fig. \ref{contourfig}, which is a com contour which covers the whole complex plane except for the imaginary axis. Since the hyperbolic cotangent solely produces poles on the imaginary axis, all contributions to the frequency sum originate in the residuals of the poles which are created by the meromorphic function $f(p^0)$. This contour enables us to calculate frequency sums through complex analysis.
For mathematical convenience, the direction of $C_2$ is reversed in order to obtain positive residues (where a minus-sign is added to the RHS of Eq. \ref{freqeval} to take this into account), this new contour is referred to as `$C_3$'.\footnote{A similar prescription exists for fermionic frequency sums with one crucial difference: that the hyperbolic cotangent function is replaced by a hyperbolic tangent function $\frac{\beta}{2} \tanh \left( \beta p^0 / 2 \right)$ in order to generate poles at $p^0 =  (2n + 1)\pi / \beta$. Since only bosonic theories are discussed in this thesis, the interested reader is referred to \ref{kapustabook} for more details about fermionic frequency sums. }
\nnn
In section \ref{freescalartheory} (Eq. \ref{referfrequencysum}), a bosonic frequency sum 
with $
f(p^0) = \left[ (\beta \omega_n)^2 + x^2 \right]^{-1} = \left[
- (\beta p^0)^2 + x^2
\right]^{-1} $
where $p^0 = 2 \pi n T i$ needs to be evaluated. This frequency sum will now be converted to a contour integral representation (using $C_3$)
\begin{equation}\label{C3eq}
	\frac{1}{\beta} \sum_{n = -\infty}^{\infty} f(p_0) 
=
\frac{1}{\beta}
\sum_{n} \frac{1}{ (2 \pi n)^2 + x^2} =
{1 \over 2}  \int_{C_3} \frac{\dd p^0}{2 \pi i} \frac{1}{(\beta p^0)^2 - x^2} \coth \left(
\frac{\beta p^0}{2}
\right)
\end{equation}
which can straightforwardly be evaluated.\footnote{The frequency sum will evaluated using a different method as well in section \ref{freescalartheory} to check the result.} 
The fact that $f(p^0)$ has two poles $\left(p^0 = \pm x/\beta \right)$ implies that solely two residuals give contributions, namely:
\begin{equation}\label{freqq1}
\bal
{1 \over \beta}  \sum_{n=-\infty}^{\infty} f(p^0)
 &=
\frac{1}{2 } \left[
\frac{1}{2  x} \coth \left(x / 2 \right) - \frac{1}{ 2 x} \coth(-x / 2)
\right]
= 
\frac{1}{2 x} \coth \frac{x}{2} 
= \frac{1}{2 x} \left[
1 + \frac{2}{e^x - 1}
\right].
\eal
\end{equation} 
Subsequently, in section \ref{lambdaphi4}, a reminiscent bosonic frequency sum will be encountered where $f(p^0) = \left[-(p^0)^2 + \mathbf{p}^2\right]^{-1}$
and whose poles are thus located at $p^0 = \pm p$. The frequency sum yields
\begin{equation}\label{freqq}
\bal
{1 \over \beta}  \sum_{n=-\infty}^{\infty} f(p^0)
&=
\frac{1}{2 } \left[
\frac{1}{2  p} \coth \left(\beta p / 2 \right) - \frac{1}{ 2 p} \coth(-\beta p / 2)
\right]
= 
\frac{1}{2 p} \coth \frac{\beta p}{2} 
= \frac{1}{2 p} \left[
1 + \frac{2}{e^{\beta p} - 1}
\right],
\eal
\end{equation} 
which \emph{separates into a vacuum contribution and a thermal contribution} (the latter of which contains a BE-distribution function $n_{\text{BE}}(p) = \frac{1}{e^{\beta p} - 1}$). It will now be shows that this trait is exhibited by every frequency sum.
\subsubsection*{Vacuum and thermal contribution}
That frequency sums always result in a (vacuum + matter) contribution can be seen
by rewriting $C_3$. Rearranging the (exponentials of) the hyperbolic cotangent in Eq. \ref{freqeval} yields
\begin{equation}\label{contourform}
\bal
T \sum_{n = -\infty}^{\infty} f(p^0 = i\omega_n) &=
\frac{ 1 }{ 4 \pi i}
\int_{C_3} \dd p^0  f(p^0)  
\coth \left( {\beta p^0 \over 2} \right) \\
&=
\frac{ 1 }{ 2 \pi i}
\int_{C_3} \dd p^0  f(p^0)  
\left[ \frac{1}{2} + \frac{1}{e^{\beta p^0} - 1} \right] \\
&=
\frac{ 1 }{ 2 \pi i} \left[
\int_{-i \infty + \epsilon}^{i \infty + \epsilon} \dd p^0  f(p^0)  
\left[ \frac{1}{2} + \frac{1}{e^{\beta p^0} - 1} \right] 
+  \int_{i \infty - \epsilon}^{- i \infty - \epsilon} \dd p^0  f(p^0)  
\left[ - \frac{1}{2} - \frac{1}{e^{\beta p^0} - 1} \right] \right]. \\
\eal\end{equation}
Now setting $p^0 \rightarrow -p^0$ in the first term yields
\begin{equation}
\bal
T \sum_{n = -\infty}^{\infty} f(p^0 = i\omega_n)
&= \frac{1}{2 \pi i} 
\left[
\int_{- i \infty}^{i \infty} \ dp_0 [f(p_0) + f(-p_0) ] + \int_{- i \infty + \epsilon}^{i \infty + \epsilon} \ dp_0 [f(p_0) + f(-p_0) ] n_{\text{BE}}(p^0) \right],
\eal
\end{equation}
where $n_{\text{BE}}(p^0) = \left[e^{\beta p^0} - 1\right]^{-1}$ is the BE-distribution function.
As was already disussed, the frequency sum then naturally separates into a
temperature-independent part (the vacuum part) and a thermal part containing a Bose–Einstein or Fermi-Dirac distribution.\footnote{One obtains a Fermi-Dirac distribution if one replaces $\coth \left( \frac{\beta p^0}{2} \right)  $ by $\tanh \left( \frac{\beta p^0}{2} \right)$ as is appropriate for fermions, whose Matsubara frequencies are antisymmetric in period $\beta$ as was already discussed in section \ref{sectionKMS}.}
\nnn
Frequency sums involving multiple propagators frequencly recur in thermal field theory (we will encounter these in the scalar toy IDG model in appendix \ref{chaptertoymodelfinitet}), an example of which is
A recurring frequency sum in this thesis involves two bosonic propagators in a loop diagram, exhibiting the generic form
\begin{equation}\label{loopfreq}
\begin{aligned}
S(i \omega_n, \mathbf{p}) &= \frac{1}{\beta} \sum_{n} \mathcal{D}(i \omega_n, \mathbf{k}) \mathcal{D}(i (\omega_n - \omega_m), \mathbf{k} - \mathbf{p}) \\
&= \frac{1}{\beta} \sum_{n} \frac{1}{\omega_n^2 + E_{\mathbf{k}}^2} \frac{1}{(\omega_n - \omega_m)^2 + E_{\mathbf{p}-\mathbf{k}}^2}.
\end{aligned}
\end{equation}
Frequency sums involving multiple propagators can be evaluated using the aforementioned contour integral methodology as well, this procedure is elaborated upon in appendix \ref{freqsumappendix}. 
\subsection{Self-energy
\label{sectionselfenergy}
}
Bare propagators do not give a realistic description of propagators, since quantum mechanics (Heisenberg's uncertainty principle) entails that particle-antiparticle pairs can temporarily be created even from the vacuum. For instance, the physical photon propagator should take into account this this particle creates electron/positron pairs which undergo electromagnetic interactions with the photon and hence modify the propagator. In thermal field theory, we should also take into account the phenomenon that fields propagate through a \emph{heat bath} and can consequently undergo interactions with this heat bath \ref{renorm} (think of absorption of an electron or photon, for instance).
Comparable situations occur for scalar particles (of which the SM Higgs is the only fundamental particle), we will now describe how self-energies can be calculated in order to obtain the physical propagator
\begin{equation}\label{physicalpropagator}
\mathcal{D}(i\omega_n, \mathbf{p}) = \left[\mathcal{D}_0^{-1}(i\omega_n, \mathbf{p}) + \Pi(i \omega_n, \mathbf{p}) \right]^{-1},
\end{equation}
where $\mathcal{D}_0(i \omega_n, \map)$ and $\mathcal{D}(i \omega_n, \map)$ are the bare and physical propagator, respectively. We will now derive the formulae for the first- and second-order self-energy, closely following Kapusta's treatment of this topic in \ref{kapustabook}.
\nnn
Eq. \ref{physicalpropagator} implies that
\begin{equation}\label{key}
\Pi(i \omega_n, \map) = \mathcal{D}^{-1}(i \omega_n, \map) - \mathcal{D}_{0}^{-1}(i \omega_n, \map),
\end{equation}
Consequently we can write
\begin{equation}\label{D0}
\mathcal{D}(i \omega_n, \map) = \left(1 + \mathcal{D}_0 \Pi \right)^{-1} \mathcal{D}_0
\end{equation}
The free propagator $\mathcal{D}(i \omega_n, \mathbf{p})$ can be written as
\begin{equation}\label{key}
\mathcal{D}(i \omega_{n}, \mathbf{p}) = 
\expval{\beta^2 \phi_{n}(\mathbf{p}) \phi_{-n} (-\mathbf{p})} 
		=
		\frac{
\beta^2 \int \mathcal{D} \phi \ \phi_{n}(\mathbf{p})
\phi_{n}^{*}(\mathbf{p}) \exp[-S_E(\phi)]
}{
\int \mathcal{D} \phi \ \exp[-S_E(\phi)] }.
\end{equation}
By using the path integral representation of the partition function and the fact that the free scalar action can be written as (this will be proved in Eq. \ref{s0})  
\begin{equation}\label{key}
S_0 = -\frac{1}{2} \sum_{n} \sum_{\mathbf{p}} \mathcal{D}_0^{-1} \phi_n(\mathbf{p}) \phi_n^{*}(\mathbf{p}),
\end{equation}
we conclude that
\begin{equation}\label{D02}
\bal
\mathcal{D}(i \omega_{n}, \mathbf{p}) 
&= 		\frac{\beta^2}{\mathz} \int \mathcal{D} \phi \ \phi_{n} (\mathbf{p})
	\phi_{n}^{*}(\mathbf{p})\exp[-S_E(\phi)] \\ 
	&= -2 \frac{\delta \ln \mathz}{\delta \mathcal{D}_0^{-1}} \\
	&= 2\mathcal{D}_0^{2} \frac{\delta \ln \mathz}{\delta \mathcal{D}_0}.
	\eal
\end{equation}
Eqs. \ref{D0}, \ref{D02} entail that
\begin{equation}\label{pi0}
\left(1 + \mathcal{D}_0 \Pi \right)^{-1} = 2\mathcal{D}_0 \frac{\delta \ln \mathz}{\delta \mathcal{D}_0}.
\end{equation}
The following expression is obtained when solving Eq. \ref{pi0} up to first order
\begin{equation}\label{key}
\ln \mathz_0 = \frac{1}{2} \sum_n \sum_{\mathbf{p}}
\ln \left[\beta^{-2}
\mathcal{D}_0(i \omega_n, \mathbf{p})
\right],
\end{equation}
which entails that
\begin{equation}\label{key}
\frac{ \delta \ln \mathz_0}{\delta \mathcal{D}_0} = \frac{1}{2} \mathcal{D}_0^{-1}(i \omega_n, \map).
\end{equation}
Moreover, since $\left(1 + \mathcal{D}_0 \Pi_1 \right)^{-1} 
\approx 1 - \mathcal{D}_0 \Pi_1$ we yield
$$
\left(1 + \mathcal{D}_0 \Pi_1 \right)^{-1} 
\approx 
 1 + 2 \mathcal{D}_0 \frac{ \delta \ln \mathz_1 }{\delta \mathcal{D}_0},
$$
yielding the first-order self-energy
\begin{equation}\label{selfenergyy}
 \Pi_1(\omega_n, \map) = - 2 \frac{ \delta \ln \mathz_1 }{\delta \mathcal{D}_0}.
\end{equation}
This equality entails that we should consider all Feynman diagrams which contribute to the partition function and examine all topologically inequivalent Feynman diagrams which arise after \textit{cutting one line} (due to $\delta/\delta \mathcal{D}_0$).
It is useful to consider the formal expansion of $\Pi$ in a power series in the coupling constant $\lambda$ 
\begin{equation}\label{key}
\Pi(\omega_n, \map) = \sum_{l = 1}^{\infty} \Pi_l(i \omega_n, \map)
\end{equation}
to obtain the physical propagator
\begin{equation}\label{key}
\mathcal{D} = 2\mathcal{D}_0^2 \sum_{n = 0}^{\infty} 
\left[
\frac{\delta \ln \mathz_n}{\delta \mathcal{D}_0}
\right]
= \mathcal{D}_0 + 2 \mathcal{D}_0^2 
\left[
\frac{\delta \ln \mathz_1}{\delta \mathcal{D}_0}
+
\frac{\delta \ln \mathz_2}{\delta \mathcal{D}_0}
+
\cdots
\right].
\end{equation}
In real life, one however rarely includes higher than second-order contributions because the computations become successively more arduous, while the corrections successively decrease at higher orders.
Differentating the second and third term in Eq. \ref{selfenergyy} and retaining terms of order $\lambda^2$ yields
\begin{equation}
- \mathcal{D}_0 \Pi_2 + \mathcal{D}_0 \Pi_1 \mathcal{D}_0 \Pi_1 = 2 \mathcal{D}_0 \frac{\delta \ln \mathz_2}{\delta \mathcal{D}_0},
\end{equation}
so the second contribution to the self-energy reads
\begin{equation}\label{selfenergy2}
 \Pi_2  =  \Pi_1 \mathcal{D}_0 \Pi_1 - 2  \frac{\delta \ln \mathz_2}{\delta \mathcal{D}_0}.
\end{equation}
In section \ref{lambdaphi4} first-order thermal self-energies will be calculated for the scalar $\lambda \phi^4$ model and the second-order thermal mass will be computed in appendix \ref{appendixbreakdownperturbation} where it will be shown that infrared divergent diagrams emerge beyond the leading-order contribution to the thermal mass. Finite temperature Feynman rules are indispensable to compute self-energies (and thermal correction to the partition function) efficiently.
\section{Feynman rules in the imaginary time formalism \label{feynmanrules}}
Matsubara was the first to set up diagrammatic techniques in order to compute the partition function using field-theoretic methods \ref{refmatsubara}. 
Feynman diagrams are the preferred means to visualize particle processes and, more importantly, compute radiative corrections efficiently, in analogy with the important role played by Feynman diagrams and rules in ordinary QFT. Let us therefore enumerate the Feynman rules for the imaginary time formalism, which are in fact remarkably similar to the Feynman rules for zero temperature field theories \ref{kapustabook}.
\begin{enumerate}
	\item{} Draw all topologically inequivalent connected Feynman diagrams to the desired order in perturbation theory.
	\item {} Assign a factor of $$T \sum_n \int \frac{d^3p}{(2\pi)^3} \mathcal{D}_0(i \omega_n, \mathbf{p})$$ to each line, where $\mathcal{D}_0(i \omega_n, \mathbf{p})$ is the Matsubara propagator.\footnote{One will often have to replace this bare propagator by a more realistic propagator in which (at least the lowest contributions to) the self-energy has been included, most notably to tame IR divergent diagrams. For a more detailed discussion of IR-divergent diagrams	and its relation to the self-energy, the reader is referred to appendix \ref{appendixbreakdownperturbation}.}	 
	\item {In $\lambda \phi^4$ theory, assign a factor 
	$$(-\lambda) \beta (2\pi)^3 
	\delta \left(
	\sum_{i=1}^{4} \omega_i
	\right)	
	\delta^{(3)} \left(
	\sum_{i=1}^{4} \mathbf{p}_i
	\right)	$$
	to every vertex where the Kronecker delta arise due to energy-momentum
	conservation. This can easily be generalized; simply replace $(-\lambda)$ by the appropriate vertex factor 
	and $\delta \left(
	\sum_{i=1}^{4} \omega_i
	\right)	
	\delta^{(3)} \left(
	\sum_{i=1}^{4} \mathbf{p}_i
	\right) \rightarrow \delta(\omega_{\text{in}} - \omega_{\text{out}})
	\delta^{(3)}(\map_{\text{in}} - \map_{\text{out}})$
	when dealing with other types of interactions. }
	\item{} Determine the overall combinatoric symmetry factor in the same fashion as in QFT and multiply every Feynman diagram by the appropriate symmetry factor.
	\item {} There will be an overall factor of $\beta(2\pi)^3\delta(0) = \beta V$.
\end{enumerate}
One must not be misled by these superficial similarities and keep in mind that the ITF differs from ordinary QFT in many regards, for instance in its unphysical representation of energy and time.
The Euclidean $S^1$-topology is responsible for several non-equivalent features of (Minkowskian) QFT and IFT which are not reflected in the similarities of the Feynman rules. It should also be kept in mind that one may still have to perform an
 analytic continuation of one's results, for instance to introduce a time-dependency or to obtain real-time Green functions which find applications in linear response theory (for instance to study screening effects, dispersion relations and quasiparticles in media) \ref{kapustabook}.

\section{Applications
\label{sectionapplications}
}
Simple applications of thermal field theory will be discussed in this section to familiarize the reader with local finite temperature field theory. Firstly, the partition function for a free scalar theory is calculated in the Matsubara formalism.  Subsequently, the first-order thermal loop correction to the partition function in an interacting scalar $\phi^4$ model will be computed. The aim of this section is two-fold: \emph{1)} it serves the role of a warm-up exercise for the significantly more challenging loop corrections for a nonlocal scalar theory 
and \emph{2)} the results for the local scalar theory will be compared with those for a nonlocal scalar theory in the next chapter.\footnote{Moreover, the results of the latter should corroborate those for the local scalar theory at low temperatures, since the nonlocal Lagrangian reduces to Eq. \ref{freescalarlagrangian}) in that regime. Having computed thermodynamic state functions in the local field theory allows us to check the validity of the results in the nonlocal regime, because the results from local field theory should be recovered in the appropriate regime.}
\subsection{Free scalar theory \label{freescalartheory}}
The free scalar theory is defined by the following Lagrangian density
\begin{equation}\label{freescalarlagrangian}
\mathcal{L}_{\phi} = \frac{1}{2} \phi(\mathbf{x}, \tau) \square \phi(\mathbf{x}, \tau),
\end{equation}
where the d'Alembertian $\square$ in the ITF reads
\begin{equation}\label{key}
\square = \frac{\partial^2}{\partial \tau^2} + \nabla^2.
\end{equation}
The finite temperature scalar action is therefore given by
\begin{equation}
S_{0} = \int_{0}^{\beta} d\tau \int d^{3}x \	\mathcal{L}_{\phi} = \frac{1}{2}  \int_{X} 
\phi(\mathbf{x}, \tau) \left[ \frac{\partial^{2}}{\partial \tau^2} + \nabla^2   \right] \phi(\mathbf{x}, \tau),
\end{equation}
where $\int_X$ is short-hand notation for the recurring term
$$
\int_{X}  \equiv  \int_{0}^{\beta} d\tau \int d^{3}x.
$$
As usual, a Fourier expansion for the scalar fields
\begin{equation}\label{fourier}
\phi(\mathbf{x}, \tau) = \sqrt{ \frac{\beta}{V} } \sum_{n = -\infty}^{\infty} \sum_{\mathbf{p}} e^{i (\mathbf{p} \cdot x + \omega_n \tau  )} \phi_n(\mathbf{p}).
\end{equation}
Note that we have a summation over discrete momenta rather than an integral over a continuous momentum variable because we start with a \emph{lattice field theory} and subsequently recover the continuum limit by taking $ \frac{1}{V} \sum_{\mathbf{p}} \rightarrow \frac{d^3p}{(2\pi)^3}$.
The finite temperature action will now be written as (henceforth dropping the subscript ``E'' in the action as the reader knows that we work in Euclidean space and will be warned if we decide to go back to Minkowski space time)
\begin{equation}
\begin{aligned} \label{s0}
S_{0}  &=  \frac{1}{2}   \cdot \frac{\beta}{V} \int_{0}^{\beta} d\tau  \int d^{3}x \
\sum_{n = -\infty}^{\infty} \sum_{\mathbf{p}} \phi_n(\mathbf{p}) \left[ -\omega_n^2 - \mathbf{p}^2  \right] \phi_{-n}(- \mathbf{p}) \\
&= - \frac{1}{2}  \beta^2 
\sum_{n} \sum_{\mathbf{p}}
 \left[ \omega_n^2 + \mathbf{p}^2    \right] \abs{\phi_n(\mathbf{p})}^2 \\
 &= - \frac{1}{2}  \beta^2 
 \sum_{n} \sum_{\mathbf{p}}
 \mathcal{D}_0^{-1}(i\omega_n, \mathbf{p}) \abs{\phi_n(\mathbf{p})}^2,
\end{aligned}
\end{equation}
where reality of the field has been used in the last step.
We now want to calculate the partition function by using the path integral representation of $\mathz_0$ in Eq. \ref{pathintegral}.
The integrand of the partition function $\int \mathcal{D} \phi_n(\mathbf{p}) \exp\left[-S_0 \right]$ depends on the \emph{magnitude} of the field $\phi_n(\mathbf{p})$, so we introduce 
$A_n(\mathbf{p}) = |\phi_{n}(\mathbf{p})|$
\begin{equation}\label{mathz0}
\begin{aligned}
\mathz_{0} &= N'  \int_{- \infty}^{\infty} d A_n(\mathbf{p}) \exp \left(	- \frac{\beta^2 }{2}  
\sum_{n} \sum_{\mathbf{p}}
 \left[  \omega_n^2 + \mathbf{p}^2    \right] A_n^2(\mathbf{p})	\right) \\
&= N' \prod_{n} \prod_{\mathbf{p}} \int_{- \infty}^{\infty} d A_n(\mathbf{p}) \exp \left(	- \frac{\beta^2 }{2}  
 \left[  \omega_n^2 + \mathbf{p}^2    \right] A_n^{2}(\mathbf{p})	\right) \\
 &= N' \prod_{n} \prod_{\mathbf{p}}  (2\pi)^{1/2} \left( \beta^2 
 \left[  \omega_n^2 + \mathbf{p}^2    \right] 	\right)^{-1/2}.
\end{aligned}
\end{equation}
The thermodynamical functions are independent of constants in the partition function and the factors $\sqrt{2\pi}$ and $N'$ can consequently be omitted without loss of generality \ref{kapustabook}. The logarithm of the partition function can easily be derived from Eq. \ref{mathz0}
\begin{equation}\label{key}
\ln \mathz_{0} = -\frac{1}{2} \sum_{n} \sum_{\mathbf{p}}   \ln \left( \beta^2 
\left[  \omega_n^2 + \mathbf{p}^2    \right] 	\right) = -\frac{1}{2} \sum_{n} \sum_{\mathbf{p}}   \ln \big( (2 \pi n )^2 + \beta^2 
\mathbf{p}^2	\big),
\end{equation}
where the following equality is useful
\begin{equation}\label{key}
\ln \big( (2 \pi n )^2 + \beta^2 \mathbf{p}^2 \big) = \int_{1}^{\beta^2 \mathbf{p}^2} \frac{d \theta^2}{\theta^2 + (2 \pi n)^2 } + \ln
\left(
	1 + (2 \pi n)^2
\right).
\end{equation}
We yield the following expression for the logarithmic partition function
\begin{equation}\label{referfrequencysum}
\begin{aligned}
\ln \mathz_{0} &= -\frac{1}{2} \sum_{n} \sum_{\mathbf{p}} \ln \big( (2 \pi n )^2 
+ \beta^2 \mathbf{p}^2 \big)  \\
 &= 
 -\frac{1}{2} 
  \sum_{n} \sum_{\mathbf{p}}  
 \Bigg[\int_{1}^{\beta^2 \mathbf{p}^2}
 \frac{d \theta^2}{\theta^2 + (2 \pi n)^2 } 
 {\color{magenta}
 +  \ln
\left(
1 + (2 \pi n)^2
\right) }
\Bigg], \\
\end{aligned}
\end{equation}
where the term in magenta ($ -\frac{1}{2} \sum_n \sum_{\mathbf{p}}\ln
\left(1 + (2 \pi n)^2 \right)$) will be omitted (it is, just like $N'$ and $\sqrt{\pi}$, independent of $T$ and $V$). 
The remaining frequency sum has already been evaluated using complex contour integral methods in 
 section \ref{freqsumsection} (Eq. \ref{freqq1})
where the following result was obtained\footnote{In order the check the result:
	the frequency sum can alternatively be performed by utilizing the equality
	$$
	\sum_{n = -\infty}^{\infty} \frac{1}{n^2 + (\theta / 2)^2 } = \frac{ 2 \pi^2}{ \theta}  \left[
	1 + \frac{2}{e^\theta - 1}
	\right],
	$$
	which entails
	$$
	\frac{1}{\theta^2 + (2\pi n)^2}
	= \frac{1}{4\pi^2} \cdot \frac{1}{n^2 + (\theta/2)^2} = \frac{1}{2\theta} \left[
	1 + \frac{2}{e^\theta - 1}
	\right]. 
	$$}
\begin{equation}\label{key}
T \sum_{n=-\infty}^{\infty} \frac{1}{-(\beta p^0)^2 + \theta^2}= \frac{1}{2 \theta} \left[
1 + \frac{2}{e^{\theta} - 1}
\right].
\end{equation}
Now, all we  need to do is compute the $d^3p$ integral
$$
\begin{aligned}
\ln \mathz_0 &= 
-\frac{1}{2} 
\sum_{n} \sum_{\mathbf{p}}  
\int_{1}^{\beta^2 \mathbf{p}^2}
\frac{d \theta^2}{\theta^2 + (2 \pi n)^2 } \\
&= 
-\frac{1}{4}
\sum_{\mathbf{p}} 
 \int_{1}^{\beta^2 \mathbf{p}^2}  \
\frac{ \dd \theta^2 }{ \theta}  \left[
1 + \frac{2}{e^\theta - 1}
\right] \\
&= 
-
\sum_{\mathbf{p}} 
 \int_{1}^{\beta p} \dd \theta
\left[
\frac{1}{2} + \frac{1}{e^\theta - 1}
\right]  \\
&
\overset{!}{=}  
\sum_{\mathbf{p}} 
\left[
- \frac{\beta p}{2}  -  \ln ( e^{\beta p} - 1 )	
\right], 
\end{aligned}
$$
where ``$\overset{!}{=}$'' in the last step should be interpreted as ``omitting terms independent of temperature.'' Taking the thermodynamic limit $\sum_{\mathbf{p}} \rightarrow V \int \ \frac{d^3p}{(2\pi)^3} $ we yield  
\begin{equation}\label{freescalarpartition}
\bal
\ln \mathz_0 &= 
V \int \ \frac{d^3p}{(2\pi)^3}
\left[
-\frac{\beta p}{2}  -  \ln ( e^{\beta p} - 1 )  
\right] \\
&= 
 \frac{V}{2\pi^2}  \int_{0}^{\infty} dp  
\left[
{ \color{red}
- \frac{1}{2}\beta p^3 
}
- p^2 \ln ( 1 - e^{- \beta p} )
\right].
\eal 
\end{equation}
We have successfully the partition function for non-interacting bosonic harmonic oscillators, Eq. \ref{bosonicharmonicoscillator}.
The divergent term (in red) is nothing but the familiar \emph{zero-point energy} from QFT \ref{zee}, which diverges since
it sums over an infinitude of zero-point modes. Moreover, this term diverges in the $T \rightarrow 0$ ($\beta \rightarrow \infty$) limit, which indicates that the partition function contains an infinite constant contribution at $T = 0$. A $1/T$ contribution in the partition function corresponds to a constant contribution to the thermodynamic potential $\Omega$ and we typically omit this irrelevant constant.\footnote{Adding a constant to $\Omega$ corresponds to adding a constant to $P$, which does not change the entropy density $s$.  Usually the constant is chosen so that $P=0$ at $T=0$. This also implies that the energy density $e=0$ at $T=0$ due to the identity $e = -P+Ts$.}
It should be kept in mind that the energy and pressure of physical configurations are measured with respect to the energy and pressure of the vacuum, we can usually tractably normalize this term to zero. 
The zero-point energy cannot generically be ignored - the ground states of vacua may change due to a background
fields - for instance due to an electric field (\emph{Casimir effect}) or gravitational field \ref{peskin}. One should include the zero-point energy in those cases, but these effects play no role within the scope of this thesis so we can safely set the zero point energy to zero. 
\nnn
We therefore yield the following (logarithmic) partition function for free massless scalars
\begin{equation}\label{key}
\ln \mathz_0 = - \frac{V}{2\pi^2}  \int_{0}^{\infty} \ dp \ p^2 
 \ln \left(1 - e^{- \beta p} \right) = -\frac{V}{2\pi^2} \left[- \frac{ \pi^4 }{45 \beta^3} \right]
 = \frac{V \pi^2}{90 \beta^3}
\end{equation}
and the corresponding thermodynamical functions (which were defined in Eq. \ref{thermquant}) read
\begin{equation}\label{scalarthermo}
\boxed{E(T,V) = - \frac{\partial}{\partial \beta} \ln \mathz_0 = \frac{\pi^2 }{30 } VT^4, \ \
P(T) = \frac{T}{V} \ln \mathz = \frac{ \pi^2}{90 }T^4,
\ \
S(T,V) = \frac{\partial }{\partial T} \left( T \ln \mathz \right) = \frac{2 \pi^2 }{45 } VT^3 }  .
\end{equation}
We note that the pressure and energy are proportional to $T^4$, while the entropy is proportional to $T^3$ both of which comply with what one should expect for pointlike particles based on dimensional analysis. Moreover, these thermodynamic state functions (including their prefactors) share remarkable similarities with those for a blackbody photon gas, whose thermodynamic functions are given by
\begin{equation}\label{key}
E(T,V) =  \frac{\pi^2 }{15 } VT^4, \ \
P(T)  = \frac{ \pi^2}{45 }T^4,
\ \
S(T,V) = \frac{4 \pi^2 }{45 } VT^3
\end{equation}
and we conclude that the thermodynamic state functions of Eq. \ref{scalarthermo} are those for an ultra-relativistic ideal gas of spinless particles. 
The thermodynamic state functions for a blackbody photon gas are twice as large as those for free scalars and this difference is well-understood. Scalar particles have only one degree of freedom, while massless vectors have two transversal polarization modes. Due to the additional degree of freedom, the partition function (and hence every thermodynamic function) is multiplied by a factor two.
\nnn
 Let us now proceed to the next logical step for understanding TFT, namely
 perturbation theory in the imaginary time formalism. In the presence of a coupling constant $\lambda > 0$, one can no longer determine the partition function of the system
 exactly, neither through a path integral approach nor  in the canonical formalism. Approximation schemes are therefore required and, just like in ordinary quantum field theory, perturbation theory comes to the rescue!
\subsection{Scalar $\lambda \phi^4$ theory \label{lambdaphi4}}
Perturbational methods are at our disposal whenever the action can be decomposed as $S = S_0 + S_{\text{I}}$, where $S_0$ is at most quadratic and $S_{\text{I}}$ is the interacting part of the action satisfying the weak-coupling condition ($\lambda \ll 1$ is thus assumed). Now the result for the observable in question can then be represented as a generalized Taylor series in $\lambda$.
The partition function can be expanded in terms of $S_{\text{I}}$:
\begin{subequations}
	\begin{equation}\label{key}
	\bal
	\mathz &= 
	N' \int \mathcal{D} \phi \exp \left(-S_0 - S_{\text{I}} \right) \\
	&=
	N' \int \mathcal{D} \phi \exp\left(-S_0\right)
	\cdot \left[1 -  S_{\text{I}} + {1 \over 2}  S_{\text{I}}^2 - \frac{1}{6} S_{\text{I}}^3
	+ \cdots \right] \\
	&=
	N' \int \ \mathcal{D} \phi \ \exp\left(-S_0\right) \sum_{l = 0}^{\infty} \frac{(-1)^{l}}{l!} S_{\text{I}}^{l}.
	\eal
	\end{equation}
The \emph{logarithm of the partition function} is the quantity of our interest when calculating thermodynamical quantities, since the thermodynamic functions $\epsilon, P, s$ are defined in terms of derivatives of $\ln \mathz$. This term reads
	\begin{equation}\label{partfunc}
	\begin{aligned}
	\ln \mathz &= \ln \left(	N'\int \ \mathcal{D} \phi \exp\left(-S_0\right) \sum_{l = 0}^{\infty} \frac{(-1)^{l}}{l!} S_{I}^{l}	\right)  \\
	&= \ln \left(	N'\int \ \mathcal{D}\phi \exp\left(-S_0\right) \right)  
	+
	\ln \left( 1 + \sum_{l = 1}^{\infty} \frac{(-1)^{l}}{l!} S_{I}^{l}	\right) \\
	&= \ln \mathz_0 + \ln \mathz_{\text{I}},
	\end{aligned}
	\end{equation}
\end{subequations}
from which it can be inferred that this extensive quantity nicely separates into an ideal gas contribution ($\ln \mathz_{0}$, which was already calculated in the previous section) and a contribution arising from interactions ($\ln \mathz_{\text{I}}$). 
The following Taylor expansion
\begin{equation}\label{taylor}
\ln(1 + x) = \sum_{n = 1}^\infty \frac{(-1)}{n}^{n+1} \  x^n 
\end{equation}
is very useful in order to further expand $\ln \mathz_{\text{I}}$.\footnote{Regarding the employed notation: contributions from Feynman diagrams which are first-order in the coupling parameter will be combined in $\mathz_1$, contributions from $\lambda^2$ in $\mathz_2$, et cetera, the subscript of $\mathz$ thus denotes the order of the series expansion from which the corresponding contributions have arisen.}
\nnn
A good starting point for thermal perturbation theory is to include a $\phi^4$ interaction term, \emph{i.e.} the full action is defined by
$$S = S_{0} + S_{\text{I}} =  \int  d^4x \left[ \frac{1}{2} \phi(\tau, \mathbf{x}) \square \phi(\tau, \mathbf{x}) - \lambda \phi(\tau, \mathbf{x})^4 \right],$$
where $\lambda$ is the coupling constant. 
The contribution from the free theory was derived in the previous subsection, 
 We will now compute the lowest-order correction (in $\lambda$) to $\ln \mathz$ corresponding to a 2-loop Feynman diagram. Before using the aforementioned Feynman rules at finite temperature (discussed in section \ref{feynmanrules}), \emph{this lowest-order correction will be computed explicitly} since this will improve our understanding as to why the Feynman rules at finite temperature actually differ from those at zero temperature and show where symmetry factors come from.
\nnn
What requires evaluation is the following term
\begin{equation}\label{key}
\begin{aligned}
\ln \mathz_1 = \frac{ \int_{0}^{\beta} \dd \tau \int\ \dd^{3}x \int \mathcal{D}\phi \ e^{S_0} 	   \mathcal{L}_{\text{int}} 	 	}{\int \mathcal{D} \phi \ e^{S_0} } 
&= 
\frac{ -\lambda \int_{X} \int \mathcal{D} \phi e^{S_0} 	  \  \phi^{4} 	 	}{\int \mathcal{D} \phi \ e^{S_0} }, \\
\end{aligned}
\end{equation}
Expressing $\phi(\mathbf{x}, \tau)$ as a Fourier series (Eq. \ref{fourier}) yields
\begin{subequations}
\begin{equation}\label{key}
\ln \mathz_1 = - \lambda \int_{X} \sum_{n_{1}, \cdots , n_{4}}
\sum_{\mathbf{p}_{1}, \cdots , \mathbf{p}_{4}}
\frac{\beta^2}{V^2}
\exp[i (\mathbf{p}_{1} + ... + \mathbf{p}_{4}) \cdot \mathbf{x} ]
\exp[i (\omega_{n_{1}} + ... + \omega_{n_{4}}   ) \tau ] \frac{A}{B},
\end{equation}
where
\begin{align}
\label{eqA}
A &= \prod_{l} \prod_{\mathbf{q}} \int \mathcal{D} \phi_{l}(\mathbf{q}) 
\exp \left[
-\frac{\beta^2}{2}  (\omega_l^2 + \mathbf{q}^2)
\abs{\phi_{l}(\mathbf{q})}^{2}
\right]
\phi_{n_{1}}(\mathbf{p}_{1}) 
\cdots
\phi_{n_{4}}(\mathbf{p}_{4}) , \\
B &= \prod_{l} \prod_{\mathbf{q}} \int \mathcal{D} \phi_{l}(\mathbf{q}) 
\exp \left[
-\frac{\beta^2}{2}  (\omega_l^2 + \mathbf{q}^2)
 \abs{ \phi_{l}(\mathbf{q}) }^{2}
 \right].
\end{align}
\end{subequations}
The scalar fields are represented in energy-momentum space, where the allowed discrete energies are given by the bosonic Matsubara frequencies. The $\int d\tau$ and $\int d^3 x$ integral integrals contribute
$\beta \delta(\omega_{n_{1}} + \cdots + \omega_{n_{4}} )$ and  $ V \delta^{(3)}(\mathbf{p}_{1} + \cdots + \mathbf{p}_{4})$ respectively.
 \nnn
 The numerator vanishes for the grand majority of possible configurations, thereby giving no contribution to $\ln \mathcal{Z}_1$. This is caused by the fact that the expectation value of an odd number of fields, just like it does at zero temperature, yields zero expectation value, in particular $\expval{\phi } = 0$ causes many field configurations, which are in fact allowed by 
 $\delta(\omega_{n_{1}} + \cdots + \omega_{n_{4}} ) \delta^{(3)}(\mathbf{p}_{1} + \cdots + \mathbf{p}_{4})$,
 to vanish. 
 
 An example of a non-vanishing contribution is when
$n_4 = -n_1$, $\mathbf{p}_4 = - \mathbf{p}_1$ and  $n_3 = -n_2$, $\mathbf{p}_3 = - \mathbf{p}_2$. These conditions satisfy the constraints posed by the Kronecker deltas and, for every constraint, the exponent in Eq \ref{eqA} factorizes. We conclude that \emph{three identical contributions} to the partition survive the conditions posed by the Kronecker deltas, all of which have the property that the exponential functions
 factorize.\footnote{
 	\label{constraintsscalar}
 	The other permutations are
 	\begin{itemize}
 		\item {}
 		$n_3 = -n_1$, $\mathbf{p}_3 = -\mathbf{p}_1$ and $n_4 = -n_2$, $\mathbf{p}_4 = -\mathbf{p}_2$,
 		\item {
 		$n_2 = -n_1$, $\mathbf{p}_2 = -\mathbf{p}_1$ and $n_4 = -n_3$, $\mathbf{p}_4 = -\mathbf{p}_3.$}
 	
 \noindent
 These three allowed permutations correspond to the following exhaustive set of allowed Bloch-de Dominicis contractions for $\phi^4$ theory:
 $$\wick{ 
\c1 \phi_1
\c2 \phi_2
\c2 \phi_3
\c1 \phi_4}, \ \  	
\wick{
\c1 \phi_1
\c2 \phi_2
\c1 \phi_3
\c2 \phi_4} \ \ \text{and} \ \ 
\wick{
\c1 \phi_1
\c1 \phi_2
}
\wick{
\c1 \phi_3
\c1 \phi_4
},
 $$
which shows that the combinatoric factor $\mathcal{C}=3$ is correct.
 	\end{itemize} 
}
\nnn
Additionally, contributions from the numerator are cancelled by (identical) contributions from the denominator, except in a few exceptional cases where additional constraints are satisfied. When considering the numerator corresponding to the aforementioned constraints, 
 this contribution is not cancelled by the denominator when $l = n_1$, $\mathbf{q} = \mathbf{p}_1$ and $l = n_2$, $\mathbf{q} = \mathbf{p}_2$. Similar constraints exist for the remaining two permutations which comply with the constraints of footnote \ref{constraintsscalar}.
 Consequently,
 three identical contributions to $\ln \mathz_{1}$ should be taken into account. We will compute the contribution from the aforementioned constraint ($n_4 = -n_1$, $\mathbf{p}_4 = - \mathbf{p}_1$ and  $n_3 = -n_2$, $\mathbf{p}_3 = - \mathbf{p}_2$) and multiply this by the  appropriate combinatorial factor $\mathcal{C} = 3$, \emph{i.e.}
\begin{equation}\label{key}
\bal
 \ln \mathcal{Z}_1 &= - \underset{(=\mathcal{C})}{3} \lambda \beta V \cdot \frac{\beta^2}{V^2} 
 \sum_{n_1,n_2}
 \sum_{\mathbf{p}_1, \mathbf{p}_2}
 \prod_{l} \prod_{\mathbf{q}} \int \mathcal{D} \tilde{\phi}_{l}(\mathbf{q}) \ 
  \frac{
	\exp \left[
	-\frac{1}{2} \beta^2 (\omega_{l}^2 + \mathbf{q}^2 )
	 \tilde{\phi}_{l}^2(\mathbf{q})
 \right]
 \tilde{\phi}_{n_{1}}^2(\mathbf{p}_{1}) 
 \tilde{\phi}_{n_{2}}^2(\mathbf{p}_{2}) 
}{ 
	\exp \left[
	-\frac{1}{2} \beta^2 (\omega_l^2 + \mathbf{q}^2 )
		\tilde{\phi}_{l}^2(\mathbf{q})
		\right],
} \\
\eal
\end{equation}
where reality of the fields (Eq. \ref{reality}) has been used to simply the expression (\emph{inter alia} $ \tilde{\phi}_{n_{1}}(\mathbf{p}_{1}) \tilde{\phi}_{- n_{1}}(- \mathbf{p}_{1}) = 
\tilde{\phi}_{n_{1}}^2(\mathbf{p}_{1})$).
Subsequently imposing the secondary set of constraints ($l = n_1$, $\mathbf{q} = \mathbf{p}_1$ and $l = n_2$, $\mathbf{q} = \mathbf{p}_2$) yields
\begin{equation}
\bal
\ln \mathz_1
 &= -\frac{3 \lambda \beta^3}{V} \Bigg[
 \sum_{n_1} \sum_{\mathbf{p}_1}
 \frac{ \int \mathcal{D} \tilde{\phi}_{n_{1}}(\mathbf{p}_{1}) 
	\exp \left[
	-\frac{1}{2} \beta^2 (\omega_{n_1}^2 + \mathbf{p}_{1}^2 )
 \tilde{\phi}_{n_{1}}(\mathbf{p}_{1}) \tilde{\phi}_{- n_{1}}(- \mathbf{p}_{1}) \right]
	\cdot
	\tilde{\phi}_{n_{1}}(\mathbf{p}_{1}) \tilde{\phi}_{ - n_{1}}( - \mathbf{p}_{1}) 
}{  \int \mathcal{D} \tilde{\phi}_{l}(\mathbf{q}) 
	\exp \left[
	-\frac{1}{2} \beta^2 (\omega_{n_1}^2 + \mathbf{p}_{1}^2)
	 \tilde{\phi}_{n_{1}}(\mathbf{p}_{1}) \tilde{\phi}_{- n_{1}}(- \mathbf{p}_{1})
\right]
} \\
&   \ \ \ \ \ \ \ \ \ \ \ \ \  \times 
\sum_{n_2} \sum_{\mathbf{p}_2}
\frac{  \int \mathcal{D} \tilde{\phi}_{n_{2}}(\mathbf{p}_{2}) 
	\exp \left[
	-\frac{1}{2} \beta^2 (\omega_{n_2}^2 + \mathbf{p}_{2}^2 )
	\tilde{\phi}_{n_{2}}(\mathbf{p}_{2}) \tilde{\phi}_{- n_{2}}(- \mathbf{p}_{2}) \right]
	\cdot
	\tilde{\phi}_{n_{2}}(\mathbf{p}_{2}) 
	\tilde{\phi}_{ - n_{2}}( - \mathbf{p}_{2})
}{  \int \mathcal{D} \tilde{\phi}_{l}(\mathbf{p_{2}}) 
	\exp \left[
	-\frac{1}{2} \beta^2 (\omega_{n_2}^2 + \mathbf{p}_{2}^2 )
 \tilde{\phi}_{n_{2}}(\mathbf{p}_{2}) \tilde{\phi}_{- n_{2}}(- \mathbf{p}_{2})
\right]
}
\Bigg] \\
&= -  \frac{ 3 \lambda \beta^{3}}{V} \left( \sum_{n} \sum_{\mathbf{p}}  \frac{1}{ \beta^{2} (\omega_n^2 + \mathbf{p}^2 )} \right)^{2} \\
&= - 3 \lambda \beta V \left( T \sum_{n}  \int \frac{d^{3}p}{(2 \pi)^3 } \frac{1}{  (\omega_n^2 + \mathbf{p}^2 ) } \right)^{2}, 
\eal
\end{equation}
where we have used 
\begin{equation}\label{key}
 \frac{\int_{-\infty}^{\infty} dx \ x^2 e^{-\alpha x^2 / 2} }
{	\int_{-\infty}^{\infty} dx \  e^{-\alpha x^2 / 2}
}
= \frac{1}{ \alpha}
\end{equation}
twice in the penultimate step with $\alpha = (\omega_n^2 + \mathbf{p}^2)$.
The first order correction to the partition function is consequently
\begin{equation}\label{lnZ1}
\begin{aligned}
\ln \mathz_1
&= 
- 3 \lambda \beta V \left(
\SumInt_{p} \mathcal{D}_0(i \omega_n, \mathbf{p})
\right)^{2},
\end{aligned}
\end{equation}
where the Matsubara propagator in frequency-momentum space is given by \begin{equation}\label{key}
\mathcal{D}_0(i \omega_n, \mathbf{p}) = \left( \omega_n^2 + \mathbf{p}^2  \right)^{-1}
\end{equation} 
and the following short-hand notation has been used 
\begin{equation}\label{key}
\SumInt = T \sum_{n}  \int \frac{d^{3}p}{(2 \pi)^3 }
\end{equation} 
which will often recur in computations.
Note that the loop corrections can be represented pictorially as
\begin{equation}\label{key}
\ln \mathz_1 = \chi \left(
\begin{tikzpicture}[node distance=0.25cm and 0.25cm]
\coordinate[vertex] (v1);
\coordinate[right=of v1] (v2);
\coordinate[above=of v2] (v3);
\coordinate[below=of v2] (v4);
\coordinate[left=of v1] (v5);
\coordinate[above=of v5] (v6);
\coordinate[below=of v5] (v7);
\draw  (v1) -- (v3);
\draw  (v1) -- (v4);
\draw  (v1) -- (v6);
\draw  (v1) -- (v7);
\end{tikzpicture}
\right)
=
3	\ \begin{tikzpicture}[baseline=(v3), node distance=1cm and 1cm]
\begin{feynman}[inline = (v3)]
\coordinate[] (v3);
\coordinate[vertex, right=of v3] (v4);
\coordinate[right=of v4] (v5);
\semiloop[fermion]{v3}{v4}{0}[$(\mathbf{p}_1, \omega_{n_{1}})$];
\semiloop[fermion]{v4}{v3}{180};
\semiloop[fermion]{v4}{v5}{0};
\semiloop[fermion]{v5}{v4}{180}[$(\mathbf{p}_2, \omega_{n_{2}})$][below];
\end{feynman}
\end{tikzpicture}
\end{equation}
according to Bloch-de Dominicis' theorem (see footnote \ref{constraintsscalar}) where $\chi\left(\cdots\right)$ represents all possible contractions of the diagram $(\cdots)$.
The contribution to the partition function of this Feynman diagram can be computed by means of the Feynman rules which were introduced in section. \ref{feynmanrules}, which would straightforwardly have yielded Eq. \ref{lnZ1}. We will henceforth always use the Feynman rules in order to efficiently compute loop diagrams. \nnn
The frequency sum of Eq. \ref{lnZ1} has already been computed in the section on frequency sums (Eq. \ref{freqq}) where we obtained
\begin{equation}\label{partitionfunctionsplit}
T \sum_n \mathcal{D}(i \omega_n, \mathbf{p})  
= \frac{1}{2p} \left[
1 + \frac{2}{e^{\beta p} - 1}
\right]
\end{equation}
where $p = \abs{\mathbf{p}}$. Consequently
\begin{equation}\label{key}
 \SumInt_p \ \mathcal{D}(i \omega_n, \mathbf{p})  
= \frac{1}{4\pi^2} \int   \left[
p + \frac{2 p}{e^{\beta p} - 1}
\right]dp,
\end{equation}
where the first integral diverges quadratically. This term does not contain a BE-distribution function and, in fact, can be identified as the vacuum term. This yields, also in ordinary quantum field theory, a diverging $\int p \ dp$ integral which can nonetheless be set to zero since we can only measure differences with respect to the vacuum (an obvious exception to this rule is the \emph{Casimir effect} \ref{peskin} but this phenomenon does not play a role for the purposes of this thesis). We consequently choose to normalize this to zero and hence only have to evaluate the second integral in order to evaluate $\ln \mathz_1$ (defined in Eq. \ref{lnZ1})
\begin{equation}
\bal
\ln \mathz_1 &= - 3 \lambda \beta V \left[
\SumInt_{p} \mathcal{D}_0(i \omega_n, \mathbf{p})
\right]^{2} \\
&= - 3 \lambda \beta V \left[
\frac{1}{2\pi^2} \int \frac{p dp}{e^{\beta p} - 1} 
\right]^{2} \\
&= - 3 \lambda \beta V \left[
\frac{1}{2\pi^2} \frac{1}{\beta^2} 
\int \frac{x dx}{e^{x} - 1} 
\right]^{2} \\
&= - 3 \lambda \beta V \left[
\frac{1}{12\beta^2} 
\right]^{2} \\
&= - \frac{ \lambda  V}{48 \beta^3}
\eal
\end{equation}
where we have defined $x \equiv \beta p$ and subsequently used 
\begin{equation}\label{key}
\int_{0}^{\infty}  {x^{n-1}  dx \over e^x - 1} = (n-1)! \zeta(n), \ \ \therefore \int_{0}^{\infty}  {x \ dx \over e^x - 1} = \zeta(2)= {\pi^2 \over 6}.
\end{equation}
The partition function up to first order in $\lambda$ is therefore given by 
\begin{equation}\label{totalpartition}
\ln \mathz(\beta, V) = \frac{V \pi^2}{90 \beta^3} - \frac{\lambda  V}{48 \beta^3},
\end{equation}
and we can now calculate the thermodynamic quantities (which were defined in Eq. \ref{thermquant}
\subsubsection*{Thermodynamic functions}
\bse
\begin{align}
P(T) &= \frac{T}{V} \ln \mathz =  \left(
\frac{\pi^2}{90} - \frac{\lambda}{48} 
\right) T^4,\\
\epsilon(T) &=  -\frac{1}{V} \frac{\partial}{\partial \beta} \ln \mathz
= \left(\frac{ \pi^2}{30} - \frac{\lambda}{12} \right)T^4, \\
s(T) &= 
\frac{1}{V} \frac{\partial}{\partial T}
\left( T \ln \mathz \right)  =   \left(
\frac{2 \pi^2}{45} - \frac{\lambda}{48}
\right)T^3. 
\end{align}
\ese
Note that $\phi^4$ interactions decrease the thermodynamic state functions as compared to in the \emph{free theory}. This can be envisaged as follows. The $\phi^4$ interaction term corresponds to self interactions of the scalar gas and consequently lowers the outwards pressure because the scalar can no longer travel freely. A close analogy exists with the ideal gas law $P = nRT/V$ and the van der Waals equation, where the \emph{attraction between molecules} (which lowers the pressure and is encoded in $a$) and the finite diameter of molecules (which increases the pressure and is encoded in $b$) have been taken into account: 
\begin{equation}\label{key}
P = \frac{nRT}{V - nb} - \frac{a n^2}{V^2} \approx  \frac{nRT}{V} - \frac{a n^2}{V^2} \ \ \text{(large volume limit)}.
\end{equation}
Although this law has originally been devised for molecules (which attract one another through electromagnetic and van der Waals forces) one can in fact justify its applicability to scalar gases  by noting that 
$$
\begin{tikzpicture}[node distance=0.25cm and 0.25cm]
\coordinate[vertex] (v1);
\coordinate[right=of v1] (v2);
\coordinate[above=of v2] (v3);
\coordinate[below=of v2] (v4);
\coordinate[left=of v1] (v5);
\coordinate[above=of v5] (v6);
\coordinate[below=of v5] (v7);
\draw  (v1) -- (v3);
\draw  (v1) -- (v4);
\draw  (v1) -- (v6);
\draw  (v1) -- (v7);
\end{tikzpicture} \ \ \ \ (\phi^4 \text{ interactions})
$$
are mediated by an attractive force and effectively correspond to $a>0$ in the van der Waals equation. The pressure therefore decreases for identical reasons as why interacting molecules lead to a lower pressure in comparision with the case of non-interacting molecules.
Scalar particles now have a \emph{finite mean free path} and are scattered after each interaction, which, overall, leads to a lowered pressure because particles can be scattered in a different outgoing than ingoing direction. 
\nnn
Due to the $\phi^4$ attractions, the amount of inhomogeneities in the system increases as well compared to the local cause which because there are local overdensities (and therefore also underdensities) of particles which lowers the overall entropy. 
This is not a standard explanation in the literature (in fact, I have not found any explanation as to why these thermodynamic function decrease as a  consequence of interactions) and this is my own interpretation which I deem rather plausible, but the validity of this explanation should of course be studied more thoroughly.
\subsubsection*{Self-energy}
Let us now calculate the lowest-order contribution to the self energy, \emph{i.e.} $\Pi_1$, recalling from Eq. \ref{selfenergyy} that 
\begin{equation}\label{selfenergy}
\Pi_1 = -2 \left( \frac{\delta \ln \mathz_1 }{\delta \mathcal{D}_0} \right)_{\text{1PI}}.
\end{equation}
Differentating to a propagator corresponds, pictorially speaking, to ``cutting a line'', where the to-be-cut line is is graphically represented by ``$ \mathbf{\cdots}$'' in the ensuing Feynman diagrams. We obtain
\bse
\begin{equation}\label{key}
\bal
\Pi_1 &=
-2 \cdot 3
 \frac{\delta}{\delta \mathcal{D}_0}
	\Bigg[ \ 
	\begin{tikzpicture}[baseline=(v3), node distance=1cm and 1cm]
	\begin{feynman}[inline = (v3)]
	\vertex [] (v3) ;
	\vertex [right=of v3] (v4);
	\vertex [right=of v4] (v5);
	\semiloop[fermion]{v3}{v4}{0};
	\semiloop[fermion]{v4}{v3}{180};
	\semiloop[fermion]{v4}{v5}{0};
	\semiloop[fermion]{v5}{v4}{180};
	\end{feynman}
	\end{tikzpicture}
	\Bigg] \\
&= -6
\Bigg[ \ 
\begin{tikzpicture}[baseline=(v3), node distance=1cm and 1cm]
\begin{feynman}[inline = (v3)]
\vertex [] (v3) {\(  \mathbf{\cdots} \)};
\vertex [right=of v3] (v4);
\vertex [right=of v4] (v5);
\semiloop[fermion]{v3}{v4}{0};
\semiloop[fermion]{v4}{v3}{180};
\semiloop[fermion]{v4}{v5}{0};
\semiloop[fermion]{v5}{v4}{180};
\end{feynman}
\end{tikzpicture}
+
\begin{tikzpicture}[baseline=(v3), node distance=1cm and 1cm]
\begin{feynman}[inline = (v3)]
\vertex [] (v3);
\vertex [right=of v3] (v4);
\vertex [right=of v4] (v5) {\( \mathbf{\cdots} \)};
\semiloop[fermion]{v3}{v4}{0};
\semiloop[fermion]{v4}{v3}{180};
\semiloop[fermion]{v4}{v5}{0};
\semiloop[fermion]{v5}{v4}{180};
\end{feynman}
\end{tikzpicture}
\Bigg] \\
&= -12 \ \
\begin{tikzpicture}[baseline=(v3), node distance=1cm and 1cm]
\begin{feynman}[inline = (v3)]
\vertex [] (v1);
\vertex [above=of v1] (v2);
\vertex [left=of v1] (v3);
\vertex [right=of v1] (v4);
\semiloop[fermion]{v1}{v2}{90};
\semiloop[fermion]{v2}{v1}{270};
\draw (v3) -- (v4);
\end{feynman}
\end{tikzpicture} \\
&= -12 (-\lambda ) \frac{1}{\beta} \sum_n \int \frac{d^3p}{(2\pi)^3} \frac{1}{\omega_n^2 + \mathbf{p}^2} \\
&= 12 \lambda  \int_{C_3} \frac{dp^0}{2\pi i} \int \frac{d^3p}{(2\pi)^3} \frac{1}{(p^0)^2 - \mathbf{p}^2} \coth\left( \frac{\beta p^0}{2} \right).
\eal
\end{equation}  
This frequency sum has already been evaluated in section \ref{freqsumsection}, the result is
\begin{equation}\label{key}
\Pi_1 = 12 \lambda \int \frac{d^3p}{(2\pi)^3} \frac{1}{2p} \left[
1 + \frac{2}{e^{\beta p} - 1}\right],
\end{equation}
\ese
from which we conclude that the lowest-order self-energy
of a free scalar field at finite temperature splits into two parts 
\begin{equation}\label{key}
\Pi_{1} = \Pi_{1}^{\text{vac}} + \Pi_1^{\text{matter}}.
\end{equation}
 in a reminiscent fashion to  the partition function in Eq. \ref{partitionfunctionsplit}.
\nnn
The temperature independent (vacuum) part reads
\bse
\begin{equation}\label{key}
\Pi_1^{\text{vac}} = 12 \lambda \int \frac{d^3p}{(2\pi )^3} \frac{1}{2p}
= 12 \lambda \int \frac{d^3p}{(2\pi)^3} \int \frac{dp^0}{2 \pi i} \frac{-1}{(p^0)^2 - \mathbf{p}^2},
\end{equation}
where the $dp^0$ integral covers the imaginary axis and is closed in the left half of the $\Re(c)<0$ complex plane and consequently only picks up a residue from the $p^0 = -p$ pole. This integral has been studied in many quantum field theory
textbooks (also in one-loop self energy), where a Wick rotation of $p^0$ to the imaginary axis has been performed \ref{peskin}, \ref{zee}. Defining $p^0 = ip^4$ and $d^4p = d^3p dp_4$ we obtain 
\begin{equation}\label{selfenergydivergent}
\bal
\Pi_1^{\text{vac}} 
&= 12 \lambda \int \frac{d^3p}{(2\pi )^3} \int \frac{dp^4}{2 \pi i} \frac{-i}{-(p^4)^2 - \mathbf{p}^2} \\
&= 12 \lambda \int \frac{d^4p}{(2\pi )^4} \frac{1}{(p^4)^2 + \mathbf{p}^2},
\eal
\end{equation}
which is quadratically divergent in the UV and therefore needs to be regulated by introducing a high-momentum cut-off $\Lambda_p$. Let us define $p \equiv \sqrt{p_4^2 + \mathbf{p}^2}$ and use the fact that the solid angle $\Omega_D$ subtended by a hypersphere in $D$ dimensions is 
\begin{equation}\label{key}
\Omega_D = 2\pi^{D/2} \left[
\Gamma(D/2)
\right]^{-1}
\end{equation} (so $\Omega_4 = 2\pi^2$) to obtain the vacuum contribution to $\Pi$:
\begin{equation}\label{key}
\bal
\Pi_{1}^{\text{vac}} 
&= \frac{24 \pi^2 \lambda}{(2\pi)^4}  \int_0^{\Lambda_p} \frac{p^3 dp}{p^2} 
= 
\frac{3 \lambda}{2\pi^2}  \int_0^{\Lambda_p} \frac{p^3 dp}{p^2} \\
&=
\frac{3 \lambda}{4\pi^2}   \Lambda_p^2 
\eal
\end{equation}
\ese
We note that the propagator to first order in $\lambda$ is given by
\begin{equation}\label{key}
\mathcal{D}^{-1}(p_4, \mathbf{p}) = p_4^2 + \mathbf{p}^2 + \Pi_{1}^{\text{vac}}, 
\end{equation}
and we want to counteract the divergent mass by adding an appropriate counterterm to the Lagrangian: $-\frac{1}{2} \delta m^2 \expval{\phi^{2} }_{0}$. 
This corresponds to an additional interaction
\begin{equation}\label{key}
-\frac{1}{2} \delta m^2 \expval{\phi^{2} }_{0}
=
-\frac{1}{2} \
\begin{tikzpicture}[baseline=(v3), node distance=1cm and 1cm]
\begin{feynman}[inline = (v3)]
\vertex [] (v3);
\vertex [right=of v3] (v4);
\vertex [right=of v4] (v5) {\( \mathbf{\cross} \)};
\semiloop[fermion]{v4}{v5}{0};
\semiloop[fermion]{v5}{v4}{180};
\end{feynman}
\end{tikzpicture}
\end{equation}
which needs to be taken into account in the action since it contributes to the partition function.
We yield the following contribution to the self-energy $(\Pi_{1,\cross}^{\text{vac}})$
\begin{equation}\label{key}
\Pi_{1,\cross}^{\text{vac}} = 
\begin{tikzpicture}[baseline=(v3), node distance=1cm and 1cm]
\begin{feynman}[inline = (v3)]
\vertex [] (v1);
\vertex [right=0.5cm of v1] (v2) {\( \mathbf{\cross} \)};
\vertex [right=0.8cm of v2] (v3);
\draw (v1) -- (v3);
\end{feynman}
\end{tikzpicture}
= \delta m^2.
\end{equation}
We thus obtain
\begin{equation}\label{key}
\Pi_{1}^{\text{vac, ren}}
= \frac{3  \lambda}{2 \pi^{2}} \Lambda_c^{2} + \delta m^2,
\end{equation}
where the counterterm $\delta m^2$ will be chosen conveniently such that the scalar remains massless. 
\nnn
As a matter of fact, theories which are renormalizable at zero temperature are necessarily well-defined at finite temperature as well. We note from Eq. \ref{contourform} that  frequency sums generically give rise to both a vacuum and thermal contribution. The vacuum contribution corresponds to the self-energy which can be obtained in QFT (this may be UV-divergent), while the thermal contribution on the other hand is heavily suppressed in the UV because it contains a distribution function (either Bose-Einstein or Fermi-Dirac) which is proportional to $\exp\left[
- \beta E
\right]
$ and therefore ameliorates the UV-behavior at finite temperature. 
In fact, no UV-divergences arises in the thermal part of both the partition function and the self-energy \ref{bellac}, \ref{kapustabook}, \ref{weert}.
\nnn
The thermal part can easily be evaluated (using $ \int   \frac{x dx}{  e^{x} - 1 } = \frac{\pi^2 }{6}$ in the penultimate step):
\begin{equation}\label{thermalselfenergy}
\bal
\Pi_{1}^{\text{matter}}
&= 12 \lambda \int \frac{d^3p}{(2\pi)^3}  \frac{1}{p \left( e^{\beta p} - 1 \right)} \\
 &= \frac{6 \lambda}{\pi^2} \int   \frac{p dp}{  e^{\beta p} - 1 } \\
 &= \frac{6 \lambda}{\pi^2 \beta^2} \int   \frac{x dx}{  e^{x} - 1 } \\
 &=  \lambda T^2. \\
\eal
\end{equation}
Since the physical and bare imaginary-time propagators ($\mathcal{D}(i\omega_n, \mathbf{p})$ and $\mathcal{D}_0(i\omega_n, \mathbf{p})$, respectively) are related by
\begin{equation}\label{key}
\mathcal{D}(i\omega_n, \mathbf{p}) = \left[\omega_n^2 + \mathbf{p}^2 + \Pi(i \omega_n, \mathbf{p}) \right]^{-1} = \left[\mathcal{D}_0(i\omega_n, \mathbf{p})^{-1} + \Pi(i \omega_n, \mathbf{p}) \right]^{-1},
\end{equation}
we conclude that the thermal background introduces a dynamically generated \textit{thermal mass} for scalar particles of order 
\begin{equation}\label{thermalmass}
m_{\text{eff}} = \Pi_{1}^{\text{matter}} = \lambda T^2
\end{equation}
This thermal mass is the effective mass that particles acquire through their \emph{continuous
interaction with the heat bath}.
It has been noted in \ref{yang} that the dynamically generated thermal mass may have profound implications for our understanding of the (role played by the) electroweak theory in the early universe. \nnn
The Standard Model contains one elementary scalar particle, the Higgs particle, which contains a \emph{negative squared mass} $\mu^2 < 0$ and is described by following Lagrangian
\begin{equation}\label{key}
\mathcal{L}_{\text{Higgs}} = \frac{1}{2} 
\left[ 
\partial_{\mu} \phi \partial^{\mu} \phi - \mu^2 \phi^2
\right]
- \lambda \phi^4.
\end{equation}
It is well-known that the Higgs Lagrangian has a spontaneously broken gauge symmetry at $T = 0$, because a nonzero vev  $\expval{\phi_0}  = \sqrt{\mu / 2\lambda}$ is generated which results in the Mexican hat potential. 
If the magnitude of the dynamically generated \emph{positive} is larger than that of the the vacuum mass, the minimum of the effective
potential be located at the origin (\emph{i.e.} $\phi_G = 0$) and the spontaneously broken symmetry ceases to exist above a critical temperature $T_{\text{c}}$.
A phase transition between the high and low temperature domain,
which is an important consequence for our understanding of the early universe, in which the Higgs mechanism could potentially not yet generate masses for elementary particles.\footnote{It should be kept in mind that elementary particles not only acquire mass through interactions with the Higgs field but also through chiral symmetry breaking, which is in fact responsible for approximately 99\% of the mass generation within nuclei \ref{peskin}.} 
\nnn
The existence of an electroweak phase transition would be very interesting in the  context of the \emph{baryon/antibaryon asymmetry in the universe}, because
baryon $\mathcal{B}$-violating processes may become unsuppressed under those circumstances according to several theories \ref{baryonviolation}, \ref{sakharov}.
I would however like to stress that Eq. \ref{thermalselfenergy} and the discussion above should \emph{not be interpreted as an intended proof of an electroweak phase transition}. The discussion is merely intended to convey a general idea of how temperatures could influence the
behaviors of the Higgs particle and we will now discuss which problems have to be overcome if one wants to compute higher order corrections to the thermal mass of the Higgs boson. 
\nnn
A first objection would be that that the Higgs particle is \emph{massive} 
whereas the scalar particle which we have studied is not.
The integral in Eq. \ref{thermalselfenergy} therefore changes accordingly
$$\frac{6 \lambda}{\pi^2} \int   \frac{p dp}{  e^{\beta p} - 1 } \rightarrow \frac{6 \lambda}{\pi^2} \int   \frac{p dp}{  e^{\beta \sqrt{\mu^2 + \mathbf{p}^2}} - 1 }.$$
At very high temperatures the original integral is recovered since $\sqrt{m^2 + \mathbf{p}^2} \approx |\mathbf{p}| = p$ in those circumstances and one may conclude that this thermal mass is robust for applications pertaining to the hot, early universe. A more severe criticism would be that
 higher order corrections to the self-energy have to be computed to check whether the combined contributions (the \emph{total thermal mass}) is larger than the vacuum mass.
\nnn
Following our discussion of self-energies of section
\ref{sectionselfenergy}
one can relatively straightforwardly expand the self-energy up to second order, resulting in
\begin{equation}\label{key}
\bal
\Pi_2 &=    \Pi_1 \mathcal{D}_0 \Pi_1 - 2  \frac{\delta \ln \mathz_2}{\delta \mathcal{D}_0} \\
&=
\left( -12 \begin{tikzpicture}[baseline=(v3), node distance=1cm and 1cm]
\begin{feynman}[inline = (v3)]
\coordinate[] (v1); 
\coordinate[vertex, right=of v1] (v2); 
\coordinate[right=of v2] (v3);
\coordinate[above=of v2] (v4);
\draw   (v1) -- (v3);
\semiloop[fermion]{v2}{v4}{90};
\semiloop[fermion]{v4}{v2}{270};
\end{feynman}.
\end{tikzpicture} 
\right)
\cross
\left( -12 \begin{tikzpicture}[baseline=(v3), node distance=1cm and 1cm]
\begin{feynman}[inline = (v3)]
\coordinate[] (v1); 
\coordinate[vertex, right=of v1] (v2); 
\coordinate[right=of v2] (v3);
\coordinate[above=of v2] (v4);
\draw   (v1) -- (v3);
\semiloop[fermion]{v2}{v4}{90};
\semiloop[fermion]{v4}{v2}{270};
\end{feynman}.
\end{tikzpicture} \right)
-2 \frac{\delta}{\delta \mathcal{D}_0}
\left(
+ 36 \
\begin{tikzpicture}[baseline=(v3), node distance=1cm and 1cm]
\begin{feynman}[inline=(v3)]
\coordinate[] (v3);
\coordinate[vertex, right=of v3] (v4);
\coordinate[vertex, right=of v4] (v5);
\coordinate[right=of v5] (v6);
\semiloop[fermion]{v3}{v4}{0};
\semiloop[fermion]{v4}{v3}{180};
\semiloop[fermion]{v4}{v5}{0};
\semiloop[fermion]{v5}{v4}{180};
\semiloop[fermion]{v5}{v6}{0};
\semiloop[fermion]{v6}{v5}{180};
\end{feynman}
\end{tikzpicture} 
+ 12 
\right) \\
&= 144 \ \begin{tikzpicture}[baseline=(v3), node distance=1cm and 1cm]
\begin{feynman}[inline = (v3)]
\coordinate[] (v1); 
\coordinate[vertex, right=of v1] (v2); 
\coordinate[above=of v2] (v4);
\coordinate[vertex, right=1.2cm of v2] (v5); 
\coordinate[right=of v5] (v6);
\coordinate[above=of v5] (v7);
\draw   (v1) -- (v6);
\semiloop[fermion]{v2}{v4}{90};
\semiloop[fermion]{v4}{v2}{270};
\semiloop[fermion]{v5}{v7}{90};
\semiloop[fermion]{v7}{v5}{270};
\end{feynman}.
\end{tikzpicture}
-144
\begin{tikzpicture}[baseline=(v3), node distance=1cm and 1cm]
\begin{feynman}[inline = (v3)]
\coordinate[] (v1); 
\coordinate[vertex, right=of v1] (v2); 
\coordinate[right=of v2] (v3);
\coordinate[vertex, above=of v2] (v4);
\coordinate[above=of v4] (v5); 
\draw   (v1) -- (v3);
\semiloop[fermion]{v2}{v4}{90};
\semiloop[fermion]{v4}{v2}{270};
\semiloop[fermion]{v4}{v5}{90};
\semiloop[fermion]{v5}{v4}{270};
\end{feynman}.
\end{tikzpicture}
- 144 \ \begin{tikzpicture}[baseline=(v3), node distance=1cm and 1cm]
\begin{feynman}[inline = (v3)]
\coordinate[] (v1); 
\coordinate[vertex, right=of v1] (v2); 
\coordinate[above=of v2] (v4);
\coordinate[vertex, right=1.2cm of v2] (v5); 
\coordinate[right=of v5] (v6);
\coordinate[above=of v5] (v7);
\draw   (v1) -- (v6);
\semiloop[fermion]{v2}{v4}{90};
\semiloop[fermion]{v4}{v2}{270};
\semiloop[fermion]{v5}{v7}{90};
\semiloop[fermion]{v7}{v5}{270};
\end{feynman}.
\end{tikzpicture}
-96 \ 
\begin{tikzpicture}[baseline=(v3), node distance=1cm and 1cm]
\begin{feynman}[inline = (v3)]
\coordinate[] (v1); 
\coordinate[vertex, right=of v1] (v2); 
\coordinate[vertex, right=of v2] (v3); 
\coordinate[right=of v3] (v4); 
\draw   (v1) -- (v4);
\semiloop[fermion]{v2}{v3}{0};
\semiloop[fermion]{v3}{v2}{180};
\end{feynman}.
\end{tikzpicture} \\
&= -144
\begin{tikzpicture}[baseline=(v3), node distance=1cm and 1cm]
\begin{feynman}[inline = (v3)]
\coordinate[] (v1); 
\coordinate[vertex, right=of v1] (v2); 
\coordinate[right=of v2] (v3);
\coordinate[vertex, above=of v2] (v4);
\coordinate[above=of v4] (v5); 
\draw   (v1) -- (v3);
\semiloop[fermion]{v2}{v4}{90};
\semiloop[fermion]{v4}{v2}{270};
\semiloop[fermion]{v4}{v5}{90};
\semiloop[fermion]{v5}{v4}{270};
\end{feynman}.
\end{tikzpicture}
-96 \ 
\begin{tikzpicture}[baseline=(v3), node distance=1cm and 1cm]
\begin{feynman}[inline = (v3)]
\coordinate[] (v1); 
\coordinate[vertex, right=of v1] (v2); 
\coordinate[vertex, right=of v2] (v3); 
\coordinate[right=of v3] (v4); 
\draw   (v1) -- (v4);
\semiloop[fermion]{v2}{v3}{0};
\semiloop[fermion]{v3}{v2}{180};
\end{feynman}.
\end{tikzpicture}
.
\eal
\end{equation}
It has however been shown \ref{kapustabook} \ref{bellac} that this
perturbative expansion breaks down beyond first-order corrections in $\lambda$. 
The next-to-leading order (NLO) correction to the partition function is not $\mathcal{O}(\lambda^2)$ but in fact of the order $\lambda^{3/2}$, which can be ascribed to infrared divergences of the so-called Daisy diagrams - this is discussed elaborately in appendix \ref{appendixbreakdownperturbation} . 
The close relationship between $\ln \mathz$ and $\Pi$ entails that both perturbative expansions cease to be meaningful beyond the leading-order term.
We were however expanding the perturbative series with free scalar propagators, where the dominant contribution to the thermally generated mass $(\lambda T^2)$ has not been included in the propagator.
\nnn
We have shown that a thermal mass $m_{th} =  \sqrt{\lambda}T$ is generated by the lowest contribution of the self-energy.  Upon closer examination of Eq. \ref{thermalselfenergy}, primarily momenta whose order of magnitude is $p \sim T$ have contributed to this integral - these momenta are called \emph{hard momenta}, while \emph{soft momenta} are defined to be of the order $\sqrt{\lambda T}$ \ref{kapustabook}. 
Interestingly, when using the more realistic estimate of the physical propagator 
\begin{equation}\label{key}
\mathcal{D}(i\omega_n, \mathbf{p}) = \frac{1}{\omega_n^2 + \mathbf{p}^2 + \lambda T^2},
\end{equation}
in our perturbative expansion we find that hard momenta provide negligible contributions to higher-order self-energy terms. Soft momenta, on the other hand, generate a thermal mass  of approximately the same magnitude as inverse bare propagator! Consequently, one typically wants to separate the hard and soft momenta and utilize an effective field theory in which \emph{only soft momenta are retained}. Systematic calculations of amplitudes involving only soft lines can be achieved if one includes the combined effect of all possible \emph{hard thermal loops}, which means that \emph{effective propagators and vertices} have to be used in the soft-momenta EFT. The theoretical framework called ``Hard Thermal Loop Resummation'' has been devised by Pisarski and Braaten and the interested reader is referred to their orginal treatmeant of this topic \ref{pisarski1}, \ref{pisarki} for more info. 
Calculations for the Higgs NLO and NNLO contributions become rather complicated and, to the best on my knowledge, no
decisive calculations have been published up to date which show whether or not the dynamically generated mass is higher than the vacuum Higgs mass.

\chapter{Nonlocal thermal field theory \label{chapternonlocalfield}}
We have motivated an interesting scalar toy model which mimics the BGKM action for gravity at the end of the first chapter. 
We aim to examine its
thermal properties by computing thermal loop corrections in this theory,  firstly have to get acquainted with simpler nonlocal thermal field theories. 
In this chapter, we employ finite temperature methods which were developed for nonlocal theories (see \ref{hagedorn1}, \ref{hagedorn2}, \ref{bluhm}) in order to obtain the partition function for a nonlocal scalar model, firstly for the free theory.
We will prove that nonlocalities do not modify the partition function of the free theory, provided that these introduce no additional poles in the propagator.\footnote{This is an important result and implies that we, in fact, already know what the partition function in the context of the \emph{free} scalar IDG toy model is, namely $\mathz = \frac{V\pi2}{90}T^3$.}
\nnn
Subsequently we compute the leading order contribution to the partition function from $\phi^4$ interactions. Astounding results will be obtained which comply the conjectured existence of a stringy Hagedorn phase at temperatures above the scale of nonlocality, where stringy excitations contribute to the partition function. The nonlocality radically modifies the partition function as compared to in local field theories when interactions are included.
 \section{Nonlocal scalar $\mathbf{\lambda \phi^4}$ theory \label{nonlocallambda4} }
The action for our infinite derivative scalar toy model mimicking a gauge theory gravity with BGKM action was motivated and deduced in section \ref{infderscalar}. The free action
\begin{equation}\label{S0scalar}
S_0 = {1 \over 2} \int d^{4}x \ \phi \square a(\square) \phi,
\end{equation} 
reduces to that of simple scalar theory for temperatures below the scale of nonlocality, since $$a(-k^2) = e^{ \left( \frac{k}{M} \right)^{2}} \approx e^{0} = 1 \ \ \ \text{for  } \ \ \frac{T}{M} \ll 1.$$
The theory is local in this regime and the graviton propagator is given by the ordinary scalar propagator
\begin{equation}\label{key}
\mathcal{D}_{0}(i\omega_n, \mathbf{p}) = (\omega_n^2 + \mathbf{p}^2)^{-1}
\end{equation}
since the propagator is equal to the inverse of the kinetic operator in the Lagrangian.
For sufficiently high energies and temperatures the free action is however given by Eq. \ref{S0scalar} and the corresponding nonlocal free  propagator (in frequency-momentum space) can trivially be derived to be
\begin{equation}\label{nonlocalprop}
\mathcal{D}_{0}(i\omega_n, \mathbf{p}) = \frac{ e^{- (\omega_n^2 + \mathbf{p}^2) M^2} }{\omega_n^2 + \mathbf{p}^2}
 =  \left(\omega_n^2 + \mathbf{p}^2 \right)^{-1} e^{- \bar{p}^{2}} ,
\end{equation}
where barred quantities denote four-vectors divided by the scale of nonlocality (so $\bar{p} = p / M$). 
We will now prove that the nonlocality does not affect the partition function of the free theory. Consequently we will recover the partition function for the \emph{local free theory} whose action is given by $S_0 = \frac{1}{2} \int d^4x \phi \square  \phi$ and which has been discussed in length in section \ref{freescalartheory}. That the partition functions for local and nonlocal scalar theories are equivalent can be understood from a physical point of view - this will be discussed in length at the end of the next subsection.
\subsection{Contribution from the  free theory}
The free action is straightforwardly calculated from Eq. \ref{nonlocallambda4} (where $\int d^4x \rightarrow \int_0^\beta d\tau \int d^3x$) 
\begin{equation}\label{freeactionsolve}
\bal
S_0 &= 
-\frac{1}{2} \beta^2 \sum_{n} \sum_{\mathbf{k}} \mathcal{D}_0^{-1}(i\omega_n, \mathbf{k})\phi_{n}(\mathbf{k}) \phi_{n}^{*}(\mathbf{k}) \\
&-\frac{1}{2} \beta^2 \sum_{n} \sum_{\mathbf{k}} (\omega_n^2 + \mathbf{k}^2) 
e^{(\omega_n^2 + \mathbf{k}^2) / M^2} \abs{\phi_{n}(\mathbf{k})}^2 ,
\eal
\end{equation}
and is reminiscent to the calculation for the local scalar theory in Eq. \ref{s0}, except that the propagator now contains an additional exponential modification.
Realizing that $\ln \mathz_0 = \int \mathcal{D}\phi_n(\mathbf{k}) \exp\left[S_0 \right] $ depends only on the magnitude of $\phi_n(\mathbf{k})$, we reintroduce the notation $A_n(\mathbf{k}) \equiv \abs{ \phi_n(\mathbf{k}) }$.
Eq. \ref{freeactionsolve} entails that
\begin{align}
\ln \mathcal{Z}_0 &=
\ln \left( N' \int  \mathcal{D} \phi \ e^{S_0}  \right) \\
&= \ln \left(
N' \prod_{n} \prod_{\mathbf{k}} 
\left[
\int_{-\infty}^{\infty} dA_n(\mathbf{k}) e^{-\frac{1}{2} \beta^2 \mathcal{D}_0^{-1} (i \omega_n, \mathbf{k}) A_n^2 (\mathbf{k})}
\right]
\right) \nonumber  \\
&= \ln \left(
N' \prod_{n} \prod_{\mathbf{k}} 
\left[
2 \pi \beta^{-2} \mathcal{D}_0 (i \omega_n, \mathbf{k})
\right]
\right)^{1/2} \nonumber \\
&=  \sum_{n} \sum_{\mathbf{k}}
\ln \left( \beta^{2} \mathcal{D}_0^{-1}(i\omega_n, \mathbf{k}) \right)^{-1/2} \nonumber \\
&=
\label{z0}
-\frac{1}{2} V \sum_n \int_{\mathbf{k}} \Big[ 
(\omega_n^2 + \mathbf{k}^2)/M^2
+\ln \beta^2 \left[\omega_n^2 + \mathbf{k}^2 \right]
\Big].
\end{align}
In the penultimate step, we have used the fact that 
$\ln \left(
N' \prod_{n} \prod_{\mathbf{k}} 
\left[
2 \pi \mathcal{D}_0 (i \omega_n, \mathbf{k})
\right]
\right)^{1/2} = \ln 
N' + \frac{1}{2} \ln (2 \pi) \sum_{n} \sum_{\mathbf{k}}$ 
$+ \text{ Eq.} (\ref{z0}),
$
and we can choose $N'$ conveniently such that only Eq. \ref{z0} remains. Note that the second term within the brackets is exactly the contribution to the partition function we obtained for the local scalar theory, while the first term originates from the nonlocality. 
\nnn
The first term within the brackets often appears in p-adic string field theories. That this term vanishes has been shown in Ref. \ref{sftmodels} by (i) expressing the frequency sum as a contour integral
as in \S \ref{freqsumsection} and (ii) utilizing a dimensional regularization scheme, both of which are discussed below.
Since we need to use dimensional regulatization in the context of thermal corrections to the scalar IDG model later on, this dimensional regularization will now be discussed in detail, while the more familiar contour integral method is only briefly touched upon in this footnote.\footnote{That the vacuum contributions  to $\ln \mathz$ is zero can be understood by expressing the sum as a contour integral:
	\begin{equation}
	T \sum_{n = -\infty}^{\infty} f(k_0 = i\omega_n) = \frac{1}{4 \pi i} \int_{- i \infty}^{i \infty} \ dk_0 [f(k_0) + f(-k_0) ] 
	+ \frac{1}{2\pi i} \int_{- i \infty + \epsilon}^{i \infty + \epsilon} \ dk_0 [f(k_0) + f(-k_0) ] \frac{1}{e^{\beta k_0} - 1},
	\end{equation}
	where the vacuum and thermal contributions are neatly separated.
	The vacuum contribution
	$
	\int {\dd k_4 \ \dd^3k \over (2 \pi)^4} \left(
	\frac{k_4^2 + \mathbf{k}^2 }{M^2}
	\right)
	$
	diverges but vanishes within regularization schemes. This is remarkably similar to the divergent vacuum contribution to the self-energy in $\lambda \phi^4$ theory (see Eq. \ref{selfenergydivergent}), which after regularization and adding a term $-\frac{1}{2}\delta m^2 \expval{\phi_0^2}$ to the Lagrangian is actually zero.  
	\nnn
	What about the matter contribution?
	In this frequency sum, the meromorphic function $f(k_0)$ is given by $f (k_0) = (-k_0^2 + \mathbf{k}^2) / M^2$ (where $k_0 = ik_4$) which obviously has no poles in the complex plane. It is therefore analytic everywhere, the contour can be pushed to infinity without picking up residues and the contribution consequently vanishes.}
\subsubsection*{Dimensional regularization}
We aim to show that the contribution to the partitition function \emph{from the nonlocality} ($\ln \mathz_{0, \text{NL}}$) vanishes in Eq. \ref{z0} (we thus need to prove that $ \ln \mathz_{0, \text{NL}} = - \frac{1}{2} V \sum_n \int \frac{d^3k}{(2\pi)^3} (\omega_n^2 + \mathbf{k}^2)/M^2= 0$).
 This term reads
\bse
\begin{equation}\label{lnmathz0}
\ln \mathz_{0, \text{NL}} = -{V\over 2 M^2}\sum_n\int \frac{d^3k}{(2\pi)^3}\ (k^2+(2\pi nT)^2).
\end{equation}
This integral can be solved using dimensional regularization, for this purpose we introduce two small parameters $\epsilon$, $\delta \ll 1$ and express $\ln \mathz_0$ as 
\begin{equation}\label{key}
\ln \mathz_{0, \text{NL}} = -{V\over 2 M^2}\sum_n I_n (\epsilon, \delta),
\end{equation}
where
\begin{equation}\label{I_n}
I_n(\epsilon,\delta)  =\int
\frac{d^{3+\epsilon}k}{(2\pi)^{3 + \epsilon} }
{1 \over (k^2+(2\pi nT)^2)^{-1+\delta}} .
\end{equation}
\ese
Although the integral in Eq. \ref{lnmathz0} diverges, we will show that it actually \emph{vanishes} when computing integral Eq. \ref{I_n} which corresponds to the same integral in the limit $\epsilon, \delta \rightarrow 0$.
\nnn
In order to solve Eq. \ref{I_n}, let us utilize the standard formula
\begin{equation}\label{key}
\int
\frac{d^Dk}{(2\pi)^D}\frac{k^A}{(k^2+m^2)^B}=\frac{\Gamma(B-A-D/2)\Gamma(A+D/2)m^{2A-2B+D}}{(4\pi)^{D/2}\Gamma(B)\Gamma(D/2)},
\end{equation}
where  $B=-1+\delta$, $D=3+\epsilon$, $A=0$ and $m=2\pi nT$, \emph{i.e.} 
\begin{equation}\label{key}
I_n(\epsilon,\delta) = \frac{\Gamma(-5/2+\delta-\epsilon/2)\Gamma(3/2+\epsilon/2)(2\pi nT)^{5-2\delta+\epsilon}}{(4\pi)^{3/2+\epsilon/2}\Gamma(-1+\delta)\Gamma(3/2+\epsilon/2)} \nonumber \\
= \frac{\Gamma(-5/2+\delta-\epsilon/2)(2\pi nT)^{5-2\delta+\epsilon}}{(4\pi)^{3/2+\epsilon/2}\Gamma(-1+\delta)} \, .
\end{equation}
The logarithmic partition function can now be written as
\begin{equation}\label{key}
\begin{aligned}
\ln \mathz_{0, \text{NL}} &= 
-{V\over 2 M^2} \sum_{n} I_n(\epsilon, \delta) \\
&= -{V  (2\pi)^{2-2\delta+\epsilon}  T^{5-2\delta+\epsilon} \over 2 (4\pi)^{3/2+\epsilon/2} M^2  }
\frac{\Gamma(-5/2+\delta-\epsilon/2)}{\Gamma(-1+\delta)} \sum_{n} n^{5-2\delta+\epsilon},
\end{aligned}
\end{equation}
where $\sum_{n} n^{5-2\delta+\epsilon}$ can be rewritten in terms of the Riemann-Zeta function:
\begin{equation}\label{key}
\sum_{n} n^{5-2\delta+\epsilon} = 2 \zeta(5-2\delta+\epsilon).
\end{equation}
We now want to retain our partition function in 4D (taking $\epsilon, \delta \rightarrow 0$). 
Since $\Gamma(x) = \int_0^\infty e^{-t}t^{x - 1} \dd t$ is not defined for its singular points, so for negative integers, we conclude that the only non-regular term in our expression is $\Gamma (-1 + \delta)$. Consequently, this term is replaced by its Laurent series expansion
\begin{equation}\label{key}
\Gamma(-n+\delta)=\frac{(-1)^n}{n!} \left[ \frac{1}{\delta}-\gamma+\psi(n)+ \mathcal{O}(\delta) \right],
\end{equation}
where $\gamma \approx 0.5772$ is the Euler-Mascheroni constant and $\psi(n)$ is defined as
\begin{equation}\label{key}
\psi(n)=\sum_{a=1}^n \ { 1 \over a}.
\end{equation}
Dimensional regularization therefore yields (displaying `$\delta$' in magenta in order for the reader to immediately see where these terms occur in the expression)
\begin{equation}\label{key}
\begin{aligned}
\ln \mathz_{0, \text{NL}} &= -V \lim_{
{\color{magenta}	
	\delta \rightarrow 0 }}   \left[ 
{  2\pi^{2}   \over  (4\pi)^{3/2}   } \frac{T^{5}}{M^2} \
\frac{\Gamma(-5/2) \zeta( 5 )}{\frac{(-1)^1}{1!} \left[ \frac{1}{{\color{magenta}\delta}}-\gamma+\psi(1)+ \mathcal{O}({\color{magenta}\delta}) \right] }  \right] \\
&\approx V \lim_{
	{\color{magenta}	
		\delta \rightarrow 0 }}   \left[ 
{  \sqrt{\pi}   \over 4   } \frac{T^{5}}{M^2} \
\frac{\Gamma(-5/2) \zeta( 5 )}{  \frac{1}{{\color{magenta}\delta}} + 0.4228 + \mathcal{O}({\color{magenta}\delta)}  }  \right] \\
&= V \lim_{
	{\color{magenta}	
		\delta \rightarrow 0 }}    \left[ 
{ \sqrt{\pi} \over 4   } \frac{T^{5}}{M^2} \
\Gamma(-5/2) \zeta( 5 )  \right] \left[{\color{magenta}\delta} - \left[0.4228 + \mathcal{O}({\color{magenta}\delta}) \right]{\color{magenta}\delta^2} + \left[0.4228 + \mathcal{O}({\color{magenta}\delta}) \right]^2{\color{magenta}\delta^3} \right] \\
&= 0,\\
\end{aligned}
\end{equation}
where we have used the Maclaurin expansion up to third order in $\delta$
\begin{equation}\label{key}
\frac{1}{\frac{1}{\delta}+C} \simeq \delta -C \delta^2 + C^2 \delta^3,
\end{equation}
where $C = -\gamma + \psi(1) + \mathcal{O}(\delta) \approx 0.4228 + \mathcal{O}(\delta)$.\footnote{We conclude that the nonlocality in Eq. \ref{z0} does not contribution to the partition function \emph{in the free theory}.
\nnn
The finite temperature contribution
from $\ln \mathz = -\frac{1}{2} V \sum_n \int_{\mathbf{k}} 
\ln \beta^2 \left[\omega_n^2 + \mathbf{k}^2 \right]$ has already been computed in section \ref{lambdaphi4} so the partition function for the free non-local scalar theory at $T>M$ reads
\begin{equation}\label{nonlocalpartitition}
\ln \mathcal{Z}_0 = - 4 \pi^3 V \beta \SumInt_k \ln \beta \left[\omega_n^2 + \mathbf{k}^2 \right] = \frac{V \pi^2}{90 \beta^3} + \text{[vacuum]},
\end{equation}
which is exactly the same result as in the local field theory where the nonlocality does not play any role.
It will be shown in the next section that the nonlocality, although not modifying the free partition function, does actually modify the contributions which arise due to interactions (this will be shown for $\lambda \phi^4$ theory, where the partition function differs from local $\lambda \phi^4$ theory).}
\subsubsection{Interpretation of result}
We could have expected that the nonlocality does not contribute to the free theory, because $a(\square)$ does not affect the pole-structure of the theory.
The free field equation is homogeneous (\emph{i.e.} contains no source $J(\tau, \mathbf{x})$) and therefore the nonlocal field equation reads 
\begin{equation}\label{nonlocalkleingordon}
\square
\exp\left[
-\square/M^2
\right]  \phi(\tau, \mathbf{x}) = 0.
\end{equation}
The absence of a source enables us to remove the nonlocality by a field redefinition $\phi \rightarrow \tilde{\phi}\exp\left[\square/M^2\right]$ which entails that
 the dynamics for $J(\tau, \mathbf{x}) = 0$ are, in fact, identical to the solutions of the
 local \emph{Klein-Gordon equation} (for a massless scalar) $\square \tilde{\phi}(\tau, \mathbf{x}) = 0$.\footnote{See \ref{ghostinfinite} for a proof that the Wightman function, defined as a solution to the homogeneous field equation,
 	remains the same for local and for infinite derivative generalization of such theories in which the kinetic operator $\mathcal{K}(\square)$ contains an additional $\exp[\square/M^2]$ term.}
 Since the homogeneous solution obeys the local equations of motion (\emph{i.e.} the
 plane wave solution of the local field theory)
  the nonlocal term $a(\square)$ does not modify the partition function of the free theory.
\nnn
Since nonlocalities cannot always be removed by field redefinitions one may not generalize the previous statement to the claim that nonlocal terms in the kinetic term generically cannot modify the partition function. What one may conclude is that the nonlocality can \emph{always} be removed from the homogeneous field equations by a suitable field redefinitions (or by multiplying both sides with the inverse of the nonlocal operator).
In fact, whether the partition function is modified by the nonlocality hinges completely on \emph{whether or not the nonlocal term introduces new poles} as was pointed out in
\ref{refinitialvalueinfinite}.\footnote{Caveat: this only holds true for Lorentz invariant nonlocal term and therefore not for term such as $a(\nabla) = \exp\left[-\nabla^2/M^2\right]$), in which case the solutions to the field equations are modified. Again the interested reader who is interested in a full-fledged proof is referred to \ref{refinitialvalueinfinite}.}
\nnn
A useful way to think of this is that nonlocalities \emph{which do not introduce any poles} do not change the dynamics of the theory, while bringing about modifications in interacting theories as compared to in local theories. 
Let us show this explicitly for the scalar theory, where the action is given by
\begin{equation}\label{nonlocalphi4}
S = \frac{1}{2} \phi 
\square
\exp\left[
-\square/M^2
\right]  \phi - \lambda 
\phi^4.
\end{equation}
This form of the action invites us to introduce the following field redefinition
\begin{equation}\label{key}
\phi \rightarrow \tilde{\phi} \exp \left[
\square/2M^2
\right]
\end{equation}
in order to remove the nonlocality from the kinetic term and obtain the equivalent action
\begin{equation}\label{smearingvertex}
S = \frac{1}{2} \tilde{\phi}  \square \tilde{\phi} - \lambda 
\tilde{\phi}^4 \exp \left[
2\square/M^2
\right].
\end{equation} 
This implies that the contribution from the free theory \emph{is exactly the same as the local free theory}, while the nonlocality does modify the contributions arising from interactions. The vertex factor 
\begin{equation}\label{key}
-\lambda
\exp \left[
-2k^2/M^2
\right]
\end{equation}
contains a Gaussian smearing term 
which implies that vertices are no longer pointlike as in local field theories. 
These phenomena are graphically represented in the nonlocal Feynman diagram of
Fig. \ref{figurenonlocalfeynman}.
\nnn
The nonlocal term $a(\square)$ therefore brings about modifications to the interactions of fields. Different thermodynamic behavior of nonlocal scalar fields (compared to their local equiparts) may therefore not be not be precluded and, in fact, several field theories (including p-adic string toy models) and string field theories which contain a nonlocal modification in the kinetic sector $a(\square)$ exhibit thermodynamic behavior which differs from that of local field theories. These include a more gradual increase of the partition function (Hagedorn phase), a thermal duality of the partition function and a characteristic asymptotic behavior of the partition function, all of which will be discussed elaborately in the ensuing section. What I aim to convey here is that interesting thermodynamic characteristics may emerge due to the presence of $a(\square)$ in the scalar field theory. We will now delve into the corresponding calculations. 
\section{Perturbation theory}
Perturbative techniques will be used in the nonlocal regime of aforementioned free action with a $\lambda \phi^4$ interaction term\footnote{It should be kept in mind that we aim to include the following interaction term for the scalar IDG model (which was derived in Eq. \ref{intaction}) 	\begin{equation}\label{S_int}
	S_{\text{int}} = \frac{1}{4M_{p} }\int d^{4}x \Bigg( \phi \partial_{\mu} \phi \partial^{\mu} \phi + \phi \square \phi a (\square) \phi -  \phi \partial_{\mu} \phi a(\square)\partial^{\mu} \phi	\Bigg)
	\end{equation}
	and subsequently compute thermal corrections in this theory. Since this theory contains three interaction terms all of which have derivative interactions, this proves to be an arduous task.
	Also, from a pedagogical point of view, it is fruitful if we first familiarize ourselves with \emph{nonlocal} thermal field theory so let us choose a simpler interaction term and deal with Eq. \ref{S_int} in the ensuing chapter.}
\begin{equation}\label{intact}
S =  \int_0^{\beta} d\tau \int d^{3}x \left[ {1 \over 2} \phi \square a(\square) \phi - \lambda \phi^4 \right].
\end{equation} 
The first-order contribution (in the coupling constant $\lambda$) to the partition function for the nonlocal scalar theory will be calculated and subsequently juxtaposed with the results of the local field theory which have already been obtained in section \ref{lambdaphi4}.
One can deduce from Eq. \ref{nonlocalphi4} that
the propagator is modified compared to in the local case (now containing an additional exponential enhancement factor)
\begin{equation}\label{nonlocalprop2}
\mathcal{D}_{0}(i\omega_n, \mathbf{p}) = \frac{ e^{- (\omega_n^2 + \mathbf{p}^2) M^2} }{\omega_n^2 + \mathbf{p}^2}
\end{equation}
 while the vertex factor remains $(- \lambda)$.
The mathematical machinery which has been developed in section \ref{lambdaphi4} in order to evaluate 
\begin{equation}\label{nonlocalfeynman}
\begin{aligned}
\ln \mathcal{Z}_1 &= 
3 \begin{tikzpicture}[baseline=(v3), node distance=1cm and 1cm]
\begin{feynman}[inline=(v3)]
\coordinate[] (v3);
\coordinate[vertex, right=of v3] (v4);
\coordinate[, right=of v4] (v5);
\semiloop[fermion]{v3}{v4}{0}[$(\mathbf{p}_1, \omega_{n_{1}})$];
\semiloop[fermion]{v4}{v3}{180};
\semiloop[fermion]{v4}{v5}{0};
\semiloop[fermion]{v5}{v4}{180}[$(\mathbf{p}_2, \omega_{n_{2}})$][below];
\end{feynman}
\end{tikzpicture} = 3 (- \lambda) \beta V \Bigg[
\SumInt_{p} \mathcal{D}_0(i \omega_n, \mathbf{p})
\Bigg]^{2},
\end{aligned}
\end{equation}
thus remains applicable, although the calculations becomes more challenging with the nonlocal propagator. 
\nnn
Note that the propagator is IR-divergent for the zero thermal mode $(n=0)$. A Schwinger parametrization based on the equality
\begin{equation}\label{schwinger}
 \frac{ e^{ -
		(
		\omega_n^2 + \mathbf{p}^2 
		) / M^2} }{ \omega_n^2 + \mathbf{p}^2  } 
=
\frac{1}{M^2} 
\int_{1}^{\infty} d\alpha \ e^{- \alpha(\omega_n^2 + \mathbf{p}^2 ) / M^2}
\end{equation}
will be utilized in order to deal with this divergence, though the Schwinger parameter ($\alpha$ in Eq. \ref{schwinger}) is replaced by $\alpha^2$ since this facilitates an easy evaluation of the integral. We obtain
\bse
\begin{equation}\label{solve}
\bal
\sum_{n} \frac{ e^{ -
		(
		\omega_n^2 + \mathbf{p}^2 
		) / M^2} }{ \omega_n^2 + \mathbf{p}^2  } 
&=
\frac{1}{M^2} \sum_{n}
\int_{1}^{\infty} d\alpha^{2} e^{- \alpha^2(\omega_n^2 + \mathbf{p}^2 ) / M^2} \\
&=
\frac{2}{M^2} \sum_{n}
\int_{1}^{\infty} d\alpha \  \alpha  e^{- \alpha^2(\omega_n^2 + \mathbf{p}^2 ) / M^2}.
\eal
\end{equation}
The term within brackets in Eq. \ref{nonlocalfeynman} then reduces to
\begin{equation}\label{sumintp}
\begin{aligned}  
\SumInt_{p} \mathcal{D}_0(i \omega_n, \mathbf{p})
&=
\SumInt_{p}
\frac{ e^{ -
		(
		\omega_n^2 + \mathbf{p}^2 
		) / M^2} }{ \omega_n^2 + \mathbf{p}^2  }  =
\frac{ T }{  \pi^2 M^2} 
\sum_{n}
\int_{1}^{\infty} d\alpha \ \alpha  
\int_0^\infty dp \ p^2   
e^{- \alpha^2(\omega_n^2 + \mathbf{p}^2) / M^2}. \\	
\end{aligned}
\end{equation}
\ese
We can easily evaluate these integrals, since the second integral resembles the following integral, which can straightforwardly be solved: 
\begin{equation}\label{standardintegral}
\int_0^{\infty} dp \ p^2  
e^{- C p^2 }  = - \frac{\partial }{\partial C} \int_0^\infty dp \ e^{-C p^2}   
= - \frac{\partial }{\partial C} \left[\frac{1}{2} \sqrt{\frac{\pi}{C}} \right]
= \frac{1}{4} \sqrt{ \frac{\pi}{C^{3}}}.
\end{equation}
Plugging in $C = \alpha^2 / M^2$ in Eq. \ref{standardintegral} allows one to rewrite the second integral in Eq. \ref{sumintp} as
\begin{equation}\label{usefulequality}
\int_0^{\infty} dp \ p^2  
e^{- \alpha^2 (\omega_n^2 + \mathbf{p}^2)/M^2 }  =  
\frac{M^3}{4 \alpha^3} \sqrt{\pi} e^{- \frac{\alpha^2 \omega_n^2}{M^2} }
\end{equation}
so Eq. \ref{sumintp} reduces to
\begin{subequations}
	\begin{equation}\label{key}
	\begin{aligned}
	\SumInt_{p} \frac{ e^{ -
			(
			\omega_n^2 + \mathbf{p}^2 
			) / M^2} }{ \omega_n^2 + \mathbf{p}^2  }  
	&=
	\frac{M T}{4 \pi \sqrt \pi }
	\sum_{n}
	\int_{1}^{\infty}  \frac{d \alpha}{\alpha^2}
	e^{-  \left(\frac{2 \pi T \alpha }{M} \right)^2 n^2 } \\
	&=
	\frac{M T}{4 \pi \sqrt \pi }
	\int_{1}^{\infty}  \frac{d \alpha}{\alpha^2}
	\zeta\left(\frac{2 \pi T \alpha }{M} \right) \\
	&=
	\frac{M T}{4 \pi \sqrt \pi }
	f(T / M),
	\end{aligned}
	\end{equation}
	\begin{equation}\label{zeta}
\text{where }	\ \ \boxed{ \zeta(s) = \sum_{n = -\infty}^{\infty} e^{-s^2 n^2},
		\ \ \ 
		f(T / M) = \int_1^\infty \frac{ d \alpha}{\alpha^2} \zeta\left(\frac{2 \pi T \alpha }{M} \right) }. 
	\end{equation}
\end{subequations}
The first equality for $\zeta$ (which is \emph{not} the Riemann-zeta function) naively suggests that higher thermal modes are strongly suppressed. The zeroth ($n = 0$) and first thermal modes ($ n = \pm1$) determine the thermodynamical functions accurately, with minute corrections arising due to higher thermal modes for ``high'' $T/M$.\footnote{This is easily verified by taking the limit $\frac{T}{M} \rightarrow \infty$, where \emph{solely} the $n = 0$ mode contributes to $\zeta\left(\frac{2 \pi T \alpha }{M} \right)$.}. On the other hand, a plethora of thermal modes provide non-negigible contributions in the $\frac{T}{M} \rightarrow 0$ limit. 
A solution to this problem (inspired by \ref{sftmodels}), where only a few thermal modes need to be taken into account which facilitate a convenient evaluation of the infinite series, will now be discussed. \nnn
The zeta function can be written as a third (Jacobi) elliptic theta function
\begin{equation}\label{key}
\zeta(s) = \vartheta_3(0, \exp\left(-s^2\right) ),
\end{equation}
where
\begin{equation}\label{key}
\vartheta_{3}(u, z) = \sum_{n=-\infty}^{\infty} z^{n^2} e^{2\pi i u}.
\end{equation}
This elliptic theta function satisfies the following useful identity
\begin{equation}\label{vartheta}
\vartheta_3(u, e^{-x^2}) = { \sqrt \pi \over x} e^{-u^2 / x^2} \vartheta_3 \left( {i \pi u \over x^2}, e^{- \pi^2 / x^2} \right)
\end{equation}
which allows us to expand the $\zeta$ function differently in both the low and high temperature regimes.
More precisely, by combining Eqs. \ref{zeta} and \ref{vartheta},
	the zeta function can be written alternatively as 
\begin{equation}
\zeta(s) = \sum_{n = -\infty}^{\infty} e^{-  n^2 s^2 }
=
\frac{ \sqrt{\pi} }{s}
\sum_{m = -\infty}^{\infty} e^{- \pi^2 m^2 / s^2 }
=
\frac{ \sqrt{\pi} }{s} \zeta \left(
\frac{ \pi }{ s }
\right),
\end{equation}
which implies that the zeta function in Eq. \ref{zeta} satisfies (up to constants of $2\pi$)
\begin{equation}\label{zetas}
\zeta \left(\frac{T}{M} \right) = \sum_{n = -\infty}^{\infty} e^{-  n^2 T^2 /M^2 }
=
\sqrt{\pi}
\frac{  M }{T}
\sum_{m = -\infty}^{\infty} e^{- \pi^2 m^2 M^2 / T^2 }
=
\sqrt{\pi}
\frac{  M }{T} \zeta \left(
\pi
\frac{ M}{ T }
\right).
\end{equation}
Interestingly, Eq. \ref{zetas} suggests that
the partition function can alternatively be interpreted as the addition
of contributions of a  set of modes $m=1/n$ which are not proportional to the temperature, but are \emph{proportional to
the inverse of the temperature}.
\nnn
Note that both expansions yield identical series for $s = 2 \pi T \alpha/M = \sqrt \pi$
\begin{equation}
\zeta(\sqrt \pi) = \sum_{n=-\infty}^{\infty} e^{-\pi n^2} = \sum_{m=-\infty}^{\infty}  
 e^{-\pi m^2}.
 \end{equation}
  A natural question to ask at this point is which formula for $\zeta$ is more useful in the regimes $s < \sqrt \pi$ and $s > \sqrt \pi$. Let us introduce the following abbreviations
  $\mathcal{N} \equiv \sum_{n = -\infty}^{\infty} e^{-  n^2 s^2 }$ and $
  \mathcal{M} \equiv
  \frac{ \sqrt{\pi} }{s}
  \sum_{m = -\infty}^{\infty} e^{- \pi^2 m^2 / s^2 }$.
 \begin{itemize} 
 	\item{\tb{High temperature expansion}}
 \end{itemize}
 For high temperatures (high $s \equiv \frac{2 \pi T \alpha}{M }$) the summation over $n$-modes in $\mathcal{N}$ is mathematically easier to deal with. In this case, higher order modes are strongly exponentially suppressed and therefore negligible, while a plethora of higher-order invsere thermal modes modes $m$ would have to be taken into account in $\mathcal{M}$.
 \nnn
 The high temperature expansion is used when  $s  > \sqrt{\pi}$,
 where it should be kept that the Schwinger parameter $\alpha$ is integrated over and thus vanishes from the exponent. The high temperature expansions should consequently be used for temperatures $T > M/2\sqrt \pi$.
 \begin{itemize}
\item{\tb{Low temperature expansion}}
\end{itemize}
 On the other hand, for low temperatures the function $\mathcal{M}$ makes the mathematics more tractable - the $M^2/T^2$ term in the exponential provides a strong suppression of higher $m$-modes for low $T/M$. This becomes strikingly apparent in the $T \rightarrow 0$ limit, where the leading and only
 term is given by the \textit{inverse zero mode} $m = 0$, while subdominant contributions  from the other lowest-order inverse modes emerge for small but nonzero $T$. 
 This expansion is thus useful for $T<M/2\sqrt \pi$.
\nnn
We have thus yielded the following low and high temperature expansions
\begin{equation}\label{lowTexp}
\boxed{T < M/2\sqrt{\pi}: \zeta\left(\frac{2 \pi T \alpha }{M} \right) 
	= \frac{ M}{2  T \alpha \sqrt{\pi}}
	\sum_{m=-\infty}^{\infty}
	e^{- m^2 \pi^2 \big/ \left(\frac{ 2 \pi T \alpha }{M }\right)^{ 2}}} \ ,
\end{equation}
\begin{equation}
\boxed{T > M/2\sqrt{\pi}: \zeta\left(\frac{2 \pi T \alpha }{M} \right) 
	= 
	\sum_{n=-\infty}^{\infty}
	e^{- n^2 \left(\frac{ 2 \pi T \alpha }{M }\right)^{ 2}}}
\end{equation}
and these series \emph{are identical} at the self-dual temperature $T = M/2\sqrt{\pi}$.\footnote{Similar expansions were used for p-adic models 
where the duality of $\zeta$ implied a duality of the partition function
$$
\mathz(T) = \mathz \left( \frac{T_c^2}{T} \right),
$$
where $T_c = M/2\sqrt{\pi}$ \ref{hagedorn1} \ref{hagedorn2}. The existence of this stringy thermal duality
has been verified in stringy computations including \ref{hagedornpolchinski}, \ref{hagedornsathiapalan}, \ref{hagedorndienes} and it would be interesting to study whether the nonlocal scalar field exhibits this thermal duality as well. This is deferred to section \ref{sectionthermalduality}
where an \emph{interpretation} of this  thermal duality will be given as well.
}
\nnn
In summary, the first-order loop correction to the partition function is given by
\begin{equation}\label{nonlocalphi4partition}
\bal
\begin{aligned}
\ln \mathcal{Z}_1 &= -3  \lambda \beta V \Bigg[
\SumInt_{p} \frac{e^{-(\omega_n^2 + \mathbf{p}^2) / M^2})}{\omega_n^2 + \mathbf{p}^2}
\Bigg]^{2} \\
&=
- \frac{3\lambda V}{T}
\left[	\frac{M T}{4 \pi \sqrt \pi }
f(T / M) \right]^2,
\eal
\end{aligned}
\end{equation}
where the function $f( T / M)$ will be derived in both the low and high temperature regimes of the nonlocal field theory. We utilize the fact that the zeta function is best understood as an exponential of inverse thermal modes $m=1/n$ for ``low'' temperatures $T < M/2\sqrt{\pi}$ (``low'' should thus be interpreted as low compared to the scale of nonlocality)
and as an exponential of thermal modes $n$ for high temperatures at high temperatures $T > M/2\sqrt{\pi}$. 
\subsection{Low temperature expansion
\label{subsubsectionlowtemperatureexpansion}
}
The appropriate low-temperature expansion for the $\zeta$ function reads
\begin{equation}\label{key}
\zeta\left(\frac{2 \pi T \alpha }{M} \right) 
= \frac{ M}{2  T \alpha \sqrt{\pi}}
\sum_{m=-\infty}^{\infty}
e^{- m^2 \pi^2 / \left(\frac{ 2 \pi T \alpha }{M }\right)^{ 2}}
=
\frac{ M}{2  T \alpha \sqrt{\pi}}
\left[
1 
+2 \sum_{m =1 }^{\infty}
e^{- m^2  \left(\frac{M }{ 2 T \alpha } \right)^{ 2}}
\right],
\end{equation}
where we used the fact that the exponential is symmetric in $m$ and the $m=0$ mode can be taken outside of the summation.
In order to compute the partition function we firstly have to compute
(see Eq. \ref{nonlocalphi4partition})
\begin{equation}\label{key}
\bal
f(T / M) 
&= \frac{ M}{2  T  \sqrt{\pi}} \int_1^\infty \frac{d\alpha}{\alpha^3}
\left[
1 
+2 \sum_{m =1 }^{\infty}
e^{- m^2  \left(\frac{M }{ 2 T \alpha } \right)^{ 2}}
\right]. \\
&= \frac{ M}{2  T  \sqrt{\pi}}  
\left[
\frac{1}{2} 
+2 
\sum_{m=1}^{\infty}
\int_1^\infty \frac{d\alpha}{\alpha^3}
\exp \left(-   \left( \frac{m M}{2T} \right)^{2} 
\frac{1}{\alpha^2} \right) 
\right] \\
&= \frac{M}{2T \sqrt{\pi}} \left[
\frac{1}{2} + \frac{4T^2}{M^2} \sum_{m=1}^{\infty}
\left(
 \frac{1}{m^2} - \frac{ \exp \left( -m^2 M^2/4T^2 \right) }{m^2}
 \right)
\right],
\eal
\end{equation}
where 
\begin{equation}
\bal
\int_1^\infty \frac{d\alpha}{\alpha^3} e^{-C^2/\alpha^2} 
&= \frac{1}{2C^2} \frac{\partial}{\partial \alpha}\int_1^\infty d\alpha \ e^{-C^2/\alpha^2} \\
&= \frac{1}{2C^2}  \ e^{-C^2/\alpha^2} \Big\rvert_{\alpha = 1}^{\infty} \\
&= \frac{1}{2C^2} \left[1 - \exp(-C^2) \right] 
 \eal
\end{equation}
has been used in the penultimate step with  $C= \frac{mM}{2T}$ . Now using\footnote{This series is known as the 	``Basel Problem'' and was solved by Leonhard Euler in 1734. The proof of this equality can be found in many textbooks.}
\bse
\begin{equation}\label{key}
\sum_{m=1}^{\infty} \frac{1}{m^2} = \frac{\pi^2}{6}
\end{equation}
and realizing that
$$
\sum_{m=1} \frac{e^{-m^2\cdot K^2}}{m^2},
$$
 (where $K = M^2/4T^2$)
 can neither be solved analytically nor numerically using Wolfram Mathematica 11.2, however, due to the strong exponential suppresion in $ \left(\frac{M^2}{T^2} \right) \gg 1$ primarily the first inverse thermal modes $(m= \pm1)$ provide a contribution (since higher modes are significantly stronger suppressed) and we can take the approximation
 \begin{equation}
 \sum_{m=1}^{\infty} \frac{e^{-  m^2 \cdot \frac{M^2}{4T^2}  }  }{m^2} \approx  e^{-  \frac{M^2}{4T^2}}.
 \end{equation}
\ese
$f(T/M)$ can thus be written as
\begin{align}
\bal
f(T/M) &= \frac{M}{2 \sqrt{\pi} T} \left[
\frac{1}{2} + \frac{2T^2 \pi^2}{3M^2}
- \frac{4T^2}{M^2} e^{- M^2/4T^2}
\right] \\
&= \frac{M}{4 \sqrt{\pi} T} 
+  \frac{\pi \sqrt{\pi} T}{3M }
-  \frac{2T}{M \sqrt \pi} e^{- M^2/4T^2}
 \\ 
&= \frac{1}{4 \sqrt{\pi} } \frac{M}{T} 
+  \frac{ \pi \sqrt \pi }{3 } \frac{T}{M}
-  \mathcal{O} \left( \frac{T}{M} \exp \left(- M^2/4T^2 \right) \right). \label{f(T/M)}
\eal
\end{align}
The corresponding contribution to the logarithmic partition function is given by
\begin{equation}\label{key}
\bal
\ln \mathz_1 &= - \frac{3\lambda V}{T}
\left[	\frac{M T}{4 \pi \sqrt \pi }
f(T / M) \right]^2 \\
&=
- \frac{3\lambda V}{T}
\left[	\frac{M^2 }{16 \pi^2  } 
+  \frac{  T^2}{12 }
-  \mathcal{O} \left( T^2 \exp \left(- M^2/4T^2 \right) 
\right) \right]^2 \\
&=
- \frac{3\lambda V}{T}
\left[	\frac{M^4 }{16^2 \pi^4} 
+  \frac{  T^4}{9 \cdot 16 }
+ \frac{M^2 T^2}{6 \cdot 16 \pi^2}
 \right] \\
&
-{3 \lambda V  \over 16 \pi^3 }
\left[
{M^4  \over 16 \pi T} + {\pi^3 T^3 \over 9 } + \frac{\pi M^2 T}{6 }
\right],
\eal
\end{equation}
where the $ \mathcal{O}
\left(
T^2 \exp\left(- M^2 / 4 T^2  \right) \right)$ term was ignored after the penultimate step. The partition function now reads
\begin{equation}\label{vacuumpartition}
\ln \mathz(V,T) = 
\frac{V \pi^2}{90 }T^3
-{3 \lambda V  \over 16 \pi^3 }
\left[
{\color{red}
{  1 \over 16 \pi} \frac{M^4}{T} 
}
+ {\pi^3  \over 9 } T^3 
+ \frac{\pi }{6} M^2 T
\right].
\end{equation}
The $M^4/T$ contribution bears reminiscence to the $1/T$ contribution 
to the partition function which occurred in the local scalar field theory (see Eq. \ref{freescalarpartition}). There, we realized that there is simply a constant contribution to $\Omega$ (the thermodynamical potential) at $T=0$.
This constant does not affect the thermodynamic properties and is consequently normalized to zero. 
\nnn
We should recover the results from local field theory in the $\frac{T}{M} \rightarrow 0$ limit. Although the $M^2T$ term seemingly diverges in this limit, we will now discuss why this term in fact vanishes in the local field theory limit.
The $M^2T$ contribution gives a divergent contribution for the local field theory but this term is in fact meaningless in the local regime, where the scale of nonlocality is an unphysical parameter. We consequently have to omit this term from the partition function.
The partition function is thus given by
\begin{equation}\label{key}
\ln \mathz(V,T) = 
\frac{V \pi^2}{90 }T^3
-{3 \lambda V  \over 48  }
 T^3,
\end{equation}
which is the same results which was obtained for local scalar $\phi^4$-theory and therefore complies with our expectation for the thermodynamic behavior of the nonlocal scalar in the $T \ll M$ regime. The $T \rightarrow M/2\sqrt{\pi}$ limit is discussed in appendix \ref{appendixTM}. We will now compute the thermal mass for the low-temperature regime and conclude that it can be reconciled with the thermal mass from the local field theory.
\subsubsection*{Thermal mass}
Now that the function $f(T/M)$ has been derived, we can easily obtain the lowest-order contribution to the thermal mass.
\begin{equation}\label{key}
\bal
\Pi_1 &= -12 \ \
\begin{tikzpicture}[baseline=(v3), node distance=1cm and 1cm]
\begin{feynman}[inline = (v3)]
\vertex [] (v1);
\vertex [above=of v1] (v2);
\vertex [left=of v1] (v3);
\vertex [right=of v1] (v4);
\semiloop[fermion]{v1}{v2}{90};
\semiloop[fermion]{v2}{v1}{270};
\draw (v3) -- (v4);
\end{feynman}
\end{tikzpicture} \\
&= 12 \lambda  \left[
	\frac{M T}{4 \pi \sqrt \pi }
f(T / M) \right] \\
&= 	\frac{3 M T \lambda}{4 \pi \sqrt \pi }
\left[
{\pi \sqrt \pi T \over 3M } 
- \frac{2T}{M \sqrt \pi}
e^{- \frac{M^2 }{4 T^2  }}
\right] \\
&= 	
\lambda T^2 \left[1 
- \frac{3}{2\pi^2}e^{- \frac{M^2 }{4 T^2  }}
\right].
\eal
\end{equation}  
Let us note that the thermal mass which was derived for the local scalar field, \emph{i.e.} $Pi_1 = \lambda T^2$ is recovered in the $\frac{T}{M} \rightarrow 0$ limit.
Interestingly, the dynamically generated thermal mass becomes slightly smaller than for the local scalar field when approaching the scale of nonlocality - it would be interesting to study the physical origin of this phenomenon in the future. 
\subsection{High temperature expansion
\label{sectionhightemperatureexpansion}
}
We are interested in the thermodynamic quantities at temperatures roughly above the scale of nonlocality $M$ (more precisely $T > M/2\sqrt{\pi}$). 
The exponential kinetic modification term does not vanish in this high-$T$ regime 
and the thermodynamical functions will consequently differ from those in the high temperature regime of the local field theory. 
\nnn
Inspired by literature on p-adic strings and SFT tachyons we expect
that modified thermodynamic behavior (as compared to in the low-$T$ limit) emerges at sufficiently high temperatures since all string states are expected to contribute to the partition function at temperatures above $M$ \ref{p-adic}, \ref{sftmodels}, \ref{hagedorn1}, \ref{hagedorn2}, \ref{hagedorn3}. 
In fact, Atick and Witten have predicted in 1988 in their seminal paper \ref{wittenatick} the existence of a \ti{Hagedorn phase} emerges at sufficiently high temperatures above the scale of nonlocality, which is, among others, reflected in a more gradual increase in free energy and pressure than at low temperatures (\ti{inter alia} $P(T) \sim T^2$ instead of $P(T) \sim T^4$). 
Since the nonlocal kinetic modifications often appears in SFT, we would expect that the Hagedorn phase comes about in the nonlocal scalar theory. 
An interesting characteristic of strings is that these become \emph{elongated two-dimensional strings} at sufficiently high temperatures, implying a more gradual increase of the partition function than at lower temperatures, where contributions arise solely from ``pointlike'' particles.
The features of the Hagedorn phase and its physical interpretation 
will be described in detail in section \ref{sectionhagedorn}, where the  connections between our results and this Hagedorn phase are discussed too.
\nnn
As was discussed in section \ref{nonlocallambda4}, the appropriate expression for the $\zeta$ function at temperatures $T\geq M/2\sqrt{\pi}$ is
\begin{equation}\label{key}
\zeta\left(\frac{2 \pi T \alpha }{M} \right) = 1 +2 \sum_{n=1}^{\infty} 
e^{- n^2 \left(\frac{2 \pi T \alpha }{M} \right)^{2}},
\end{equation} 
which entails that
\begin{equation}\label{nonlocalscalarhighT}
\begin{aligned}
\SumInt_{p} \mathcal{D}_0(i \omega_n, \mathbf{p}) 
&= \frac{M T}{4 \pi \sqrt{\pi} }
f(T / M) \\
&= \frac{M T}{4 \pi \sqrt{\pi} }
\int_1^\infty \frac{d\alpha}{\alpha^2} 
\left[
1 +2\sum_{n = 1}^{\infty}
e^{- \frac{\omega_n^2 \alpha^2}{M^2}  }
\right] \\
&= \frac{M T}{4 \pi \sqrt{\pi} }
+
\frac{M T}{2 \pi \sqrt{\pi} }
\int_1^{\infty}
\frac{d\alpha}{\alpha^2}
\left[
\sum_{n = 1}^{\infty}
e^{- \frac{\omega_n^2 \alpha^2}{M^2}  }
\right].
\end{aligned}
\end{equation}
The following standard integral is used  to evaluate the integral occurring in Eq. \ref{nonlocalscalarhighT}
\begin{equation}\label{key}
\int_1^{\infty}
\frac{d\alpha}{\alpha^2}
e^{-C^2 \alpha^2 }
=
e^{-C^2} - C \sqrt{\pi} \left[1 - \erf(C) \right],
\end{equation}
where $C = 2 \pi n T / M$ and $\erf(C)$ is the error function:
\begin{equation}\label{error}
\erf(C) = \frac{2}{\sqrt{\pi}} \int_0^{C} dt \ e^{-t^2}.
\end{equation}
We therefore obtain
\begin{equation}\label{keykey}
\frac{M T}{4 \pi \sqrt{\pi} }
f(T / M)
=
\frac{M T}{4 \pi \sqrt{\pi} }
\left[
1 + 
2
\sum_{n = 1}^{\infty}
\left\{
e^{-4 \pi^2 n^2 T^2  / M^2} 
{\color{magenta}
- \frac{2 \pi  n T \sqrt{\pi}}{M}
\left[
1 - \erf \left(
\frac{2 \pi  n T}{M}
\right)
\right]
}
\right\}
\right],
\end{equation}
where the terms represented in magenta will now be expanded around infinity\footnote{We thus use the large argument limit of the error function. The error function asymptotically reaches infinity $\erf(x) \rightarrow 1$ for $x \rightarrow \infty$ so we can Taylor expand the residual $[1-\erf(x)]$. 
	The following expression was given by Wolfram Mathematica 11.2: Series[x $*$ (1 - Erf[x]), \{x, Infinity, 2\}]
	=
	$e^{-x^2 - \mathcal{O}(\frac{1}{x})^4} \left[
\frac{1}{\sqrt \pi} - \frac{1}{2\sqrt{\pi} x^2} + \mathcal{O}(\frac{1}{x^3}) \right]$,
where $x = \frac{2\pi n T}{M}$. Note that this expression still needs to be multiplied by $-\sqrt \pi$ to obtain the terms which are shown in magenta.
}
\begin{equation}\label{key}
-  \frac{2 \pi  n T \sqrt{\pi} }{M}
\left[
1 - \erf \left(
\frac{2 \pi  n T}{M}
\right)
\right]
\simeq e^{-4 \pi^2 n^2 T^2  / M^2}  \left[-1 + \frac{1}{2}  \left( \frac{M}{2 \pi n T} \right)^2 - \mathcal{O}\left( \frac{M}{2 \pi n T} \right)^3 \right].
\end{equation}
This simplifies the term between square brackets in Eq. \ref{keykey}
\begin{equation}\label{key}
\bal
&1 + 
2
\sum_{n = 1}^{\infty}
\left\{
e^{-4 \pi^2 n^2 T^2  / M^2} 
	- \frac{2 \pi  n T \sqrt{\pi}}{M}
	\left[
	1 - \erf \left(
	\frac{2 \pi  n T}{M}
	\right) \right] \right\} \\
=   & 1 +   \frac{M^2}{4 \pi^2 T^2} \sum_{n=1}^{\infty} \frac{ e^{-4 \pi^2 n^2 T^2  / M^2}  }{n^2}    - \mathcal{O}\left( \frac{M^3}{T^3}   e^{-4 \pi^2 n^2 T^2  / M^2} \right) ,
\eal
\end{equation}
where we omit the $\mathcal{O}\left( \frac{M^3}{ T^3} e^{-4 \pi^2 n^2 T^2  / M^2} \right)$ term since higher modes are strongly exponentially suppressed,
and likewise we only maintain the $n=1$ thermal mode in the summation. 
We have thus obtained
\begin{equation}\label{key}
f(T/M) = 1 + \frac{M^2}{4 \pi^2 T^2 } e^{- 4 \pi^2 T^2 / M^2}
\end{equation}
and as a result the leading-order contribution in $\lambda$ to the partition function is
\begin{equation}\label{correctionexpT^2/M^2}
\bal
\ln \mathz_1 &=
- \frac{ 3 \lambda V }{T}
\left[	
	{MT \ \over 4 \pi \sqrt{\pi} } \left[ 1 + \frac{M^2}{4 \pi^2 T^2 } e^{- 4 \pi^2 T^2 / M^2} \right] \right]^2  \\
&= - \frac{ 3 \lambda  V }{16 \pi^3 } M^2 T
- \frac{ 3 \lambda  V }{32 \pi^5 } \frac{M^4}{T}
\left[
e^{- 4 \pi^2 T^2 / M^2 }
+ \frac{1 }{8 \pi^{2} } \frac{M^2}{T^2} e^{- 8 \pi^2 T^2 / M^2 } 
\right].
\eal
\end{equation}
The partition function can now, up to the leading-ording contribution in $\lambda$, be expressed as
\begin{equation}\label{nonlocalpartitionfunc}
\bal
\ln \mathcal{Z}(V,T) &= 
\frac{V \pi^2 }{90 } T^3
- \frac{ 3 \lambda  V }{16 \pi^3 } M^2 T
- \frac{ 3 \lambda  V }{32 \pi^5 } \frac{M^4}{T}
\left[
   e^{- 4 \pi^2 T^2 / M^2 }
+ \frac{1 }{8 \pi^{2} } \frac{M^2}{T^2} e^{- 8 \pi^2 T^2 / M^2 } 
\right].
\eal
\end{equation}
\subsubsection*{$\mathbf{T \gg M/2\sqrt{\pi}}$}
In the very high temperature regime $T \gg M/2\sqrt \pi$ we can neglect the exponentials in the partition function and yield\footnote{See appendix \ref{appendixTM} for the $T \rightarrow M/2\sqrt{\pi}$ limit.}
\begin{subequations}	
\begin{align}
\ln \mathz(V,T) &= \frac{V \pi^2 }{90 } T^3
- \frac{ 3 \lambda  V }{16 \pi^3 } M^2 T  \ \ \ \ \ \ \left(\frac{T}{M} \gg \frac{1}{2\sqrt{\pi}}\right),	\\
\epsilon(T)	 &= \label{scalarnonlocalenergy2}
 \frac{ \pi^2}{30} T^4 -
	{3\lambda   \over 16 \pi^3} M^2 T^2, \\
	P(T) &= 
	\frac{ \pi^2}{90} T^4
	- \frac{3 \lambda }{16 \pi^3} M^2 T^2, \\
	s(T)  &= 
	\frac{2 \pi^2 }{45 }T^3
	-
	\frac{3 \lambda}{8 \pi^3} M^2 T.
	\end{align}
\end{subequations}	
\mathcenter
The thermodynamic behavior at temperatures above $M$ is indeed fundamentally different from that that of the local scalar field. the partition function satisfies $\mathcal{Z}(T) \sim T^3$ at low temperatures while it satisfies $\mathz(T) \sim (T^3 - T)$ at high temperatures. 
The corresponding thermodynamic behavior ties in beautifully with the conjectured \textit{Hagedorn phase transition} \ref{wittenatick}  and will be discussed elaborately in section \ref{sectionhagedorn}.
\section{Discussion of results
\label{sectiondiscussiontemperatureexpansion}
}
It is interesting to check whether the results of the previous section can be reconciled with predictions made in string field thermodynamics. 
We have briefly alluded to the fact that the nonlocal scalar model may exhibit a thermal duality at leading order in $\lambda$ in the previous section. 
The prediction in string field theory is that \emph{a thermal duality of the partition function should exist when including the leading-order contributions in $\lambda$} and that this duality is broken when one includes NLO, NNLO and, in fact, arbitrary higher order contributions. This feature has been verified for p-adic strings and several other SFT models \ref{hagedorn2}, \ref{hagedornpolchinski}. After a (very) brief introduction of p-adic strings in the following section we will verify that the nonlocal scalar theory does not exhibit the thermal duality which is obeyed by p-adic models (toy models for strings) and several string field theories \ref{p-adic}, \ref{hagedorn1}.
\nnn
We will subsequently  discuss the conjectured Hagedorn phase which was briefly alluded to in the introduction of section \ref{sectionhightemperatureexpansion}. Atick and Witten predicted the existence of this stringy state of matter in their seminal 1988 paper \ref{wittenatick} which is, among other things, characterized by a more gradual increase of the partition function with temperature because fewer degrees of freedom are activated in the stringy gas than in a gas of elementary particles. 
Our results for the nonlocal scalar field are reconcilable with this Hagedorn phase as will be discussed in section \ref{sectionhagedorn}, where we also briefly allude to potential implications of such a Hagedorn phase.
\subsection{Thermal duality
\label{sectionthermalduality}
}
It has been argued by Atick and Witten in their seminal paper \ref{wittenatick}
that the existence of an intriguing thermal duality is entailed by string thermodynamics.
The existence of this thermal duality
($T \leftrightarrow 1/T$) bears resemblance to $T$-duality which states that a string which is compactified on a
torus with radius $R$ is equivalent to a string which is compactified on a torus with
radius $\alpha'/R$ where $1/\alpha'$
is the string tension. 
Note that $T$-duality thus corresponds to a duality which relates physics at large compactification radii to physics at small compactification radius radii.
A thermal duality would relate the
physics at temperature $T$ to the physics at inverse temperature $T_c^2/T$ via
\begin{equation}
\mathz(T)  = \mathz\left({T_\text{c}^2 \over T}\right).
\end{equation}
 where the \emph{self-dual} or \emph{critical temperature} $T_c$ (depending on whether or not a phase transition occurs at $T_c$) is related to the string scale $m_s$.
\nnn
This duality this has in fact been verified for several string field theories \ref{hagedornpolchinski}, \ref{hagedornsathiapalan}, \ref{hagedorndienes}, but also for p-adic string toy strings which are, in fact, quantum field theories \ref{p-adic} \ref{hagedorn2}. p-adic field theories which will be discussed in more detail below because the thermal duality in those theories emerges because $\zeta(T/M)$ allows for a dual description $\zeta(M/T)$ which relates the physics at temperature $T$ to physics at temperature $1/T$. The partition function for the nonlocal scalar field theory contains the same $\zeta$-function, but we will show that, and explain why, this theory does \emph{not obey thermal duality}. Let us first briefly discuss p-adic field theories and show how thermal duality of the partition function emerges there.
\subsubsection*{Thermal duality in p-adic string models}
\begin{figure}
 	\centering
 	\includegraphics[scale=0.35]{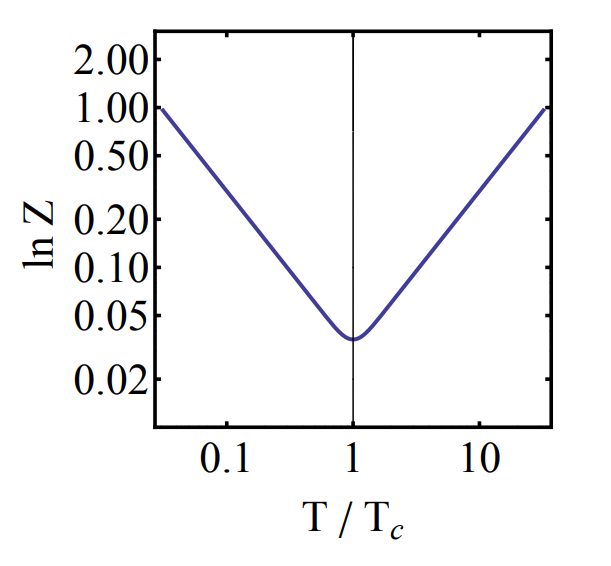}
 	\caption{
 		\ti{Partition function of the p-adic string model with $p = 3$ and $D = 4$ 
 	in arbitrary units. The symmetry of $\mathz_1(T)$ with respect to $T_c$ shows the exact realization
 		of the thermal duality at leading order in $\lambda$. Our nonlocal scalar model exhibits the same thermal duality at first order in $\lambda$, where the thermal duality arose because the third Jacobi elliptic function satisfies has inequivalent expansions for low and high temperature regimes.}
 	Figure was taken from \ref{p-adic}.}
 	\label{thermaldualityfig}
 \end{figure}

The finite-temperature action for p-adic strings 
is given by
\begin{equation}\label{ppp}
S = \frac{m_s^D}{g_p^2} \int_0^\beta d\tau \int d^{D-1}x \left[-\frac{1}{2} \varphi e^{- \left( \partial_\tau^2 + \nabla^2 \right) /M^2} \varphi - \frac{1}{p+1}\varphi^{p+1}\right],
\end{equation}
where $m_s$ is the string mass scale (defined by $m_s^2 = 1/2\alpha'$ where $\alpha'$ is the string tension) and 
$\frac{1}{g_p^2} = \frac{1}{g_0^2} \frac{p^2}{p - 1}$, where
$g_0$ is the open string coupling constant.  
The aforementioned action of Eq. \ref{ppp} resembles the
action for our nonlocal scalar model when choosing $D=3$, $p=3$ (this theory is called ``3-adic'') since the following rescalings of the fields and vertex factor
\begin{equation}\label{key}
\phi = \frac{m_s^2}{g_3} \varphi, \ \ \ \ \lambda = - \frac{1}{18} \frac{g_0^2}{m_s^4},
\end{equation}
yield the equivalent action for 3-adic theory
\begin{equation}\label{padicaction}
S = \int_0^\beta d\tau \int d^3x \left[-\frac{1}{2} \phi e^{-\left( \partial_\tau^2 + \nabla^2 \right)/M^2} \phi - \lambda\phi^{4}\right].
\end{equation}
This action differs only from the nonlocal scalar action considerd in Eq. \ref{intact} in that \emph{i)} the kinetic term contains no poles  which implies that p-adic strings have no perturbative
excitations \ref{hagedorn2}
and  \emph{ii)}
the dimensions of fields and vertices are different in p-adic string models ($\left[\phi\right]=E^2$ and consequently $\left[\lambda\right]=E^{-4}$). 
The most important difference between our nonlocal scalar theory and p-adic strings is that the propagator of the latter does not contain a pole-structure 
and is given by
\begin{equation}\label{padicpropagator}
\mathcal{D}_{0}(i\omega_n, \mathbf{p}) =  e^{- (\omega_n^2 + \mathbf{p}^2) M^2}.
\end{equation}
Because the propagator contains no pole, the free theory is empty and all the thermal properties are generated
by interactions, even if the model is weakly coupled.
The 3-adic propagator as defined in Eq. \ref{padicpropagator} modifies the thermal corrections significantly as compared to for nonlocal scalar fields. The contribution from the two-loop diagram
$$\begin{tikzpicture}[baseline=(v3), node distance=1cm and 1cm]
\begin{feynman}[inline=(v3)]
\coordinate[] (v3);
\coordinate[vertex, right=of v3] (v4);
\coordinate[, right=of v4] (v5);
\semiloop[fermion]{v3}{v4}{0}[$(\mathbf{p}_1, \omega_{n_{1}})$];
\semiloop[fermion]{v4}{v3}{180};
\semiloop[fermion]{v4}{v5}{0};
\semiloop[fermion]{v5}{v4}{180}[$(\mathbf{p}_2, \omega_{n_{2}})$][below];
\end{feynman}
\end{tikzpicture}$$ 
has already been computed in section \ref{nonlocallambda4} for the nonlocal scalar fields and, in the context of p-adic strings, the contribution reads
\begin{equation}\label{padicz}
\bal
\ln \mathz_1 &= 
3(-\lambda)\beta V \left[
T \sum_{n} \int \frac{d^3p}{(2\pi)^3}
\mathcal{D}(i \omega_n, \mathbf{p})
\right]^2 \\
=
&-3\lambda\beta V \left[
\frac{T}{2\pi^2} \sum_{n}  \int_0^{\infty} dp
\ e^{- (\omega_n^2 + \mathbf{p}^2) M^2}
\right]^2 \\
\overset{(\text{Eq. }\ref{usefulequality}, \alpha =1)}{=} &
-3\lambda\beta V \left[
\frac{M^3 T}{8\pi\sqrt{\pi}}   \zeta \left( \frac{2\pi T}{M}\right)
\right]^2.
\eal
\end{equation}
The partition function for p-adic strings is directly proportional to $\zeta^2 \left(\frac{2 \pi T }{M} \right)$
and, as a result, inherits its thermal duality from the dual description of the $\zeta$ function!
\nnn
The self-dual temperature $T_c$ of $\zeta(s)$  is given by $T_c = M/2
\sqrt{\pi}$ (as can easily be deduced from Eq. \ref{zetas})
which straightforwardly yields\footnote{That the thermal duality ceases to exist when higher-order interactions are included has been argued in the string field theory literature - this was done firstly by Atick and Witten \ref{wittenatick} and subsequent calculations by by Sathiapalan and Sircar \ref{hagedornsathiapalan} have lent further support to this statement. The same result was found in computations for p-adic strings (with $p=3$) \ref{hagedorn2}.
	There is an easy way to check whether the duality is broken in those Feynman diagrams.
	Note that the existence of the thermal duality hinges on the dual description
	\begin{equation}\label{vartheta3}
	\vartheta_3(u, e^{-s^2}) = { \sqrt \pi \over s} e^{-u^2 / s^2} \vartheta_3 \left( {i \pi u \over s^2}, e^{- \pi^2 / s^2} \right),
	\end{equation}
	where $u = 0$. In that case, a straightforward alternative interpration of $\vartheta_3 \left(0, \exp \left(\frac{2\pi  n  T}{M} \right) \right)$ exists in which thermal modes $m=1/n$ are summed over. Whenever $u \neq 0$ one can still use a dual description, but since Eq. \ref{vartheta3} implies that inverse
	modes need to be weighted differentently in this case, thermal duality would be broken.
	It has been shown in \ref{hagedorn2} that the sunset diagram (a second-order diagram) provides a
	$\vartheta_3 \left(\frac{1}{2}\phi, \exp(-x^2) \right)$ contribution to the partition function and all higher order terms provide similar contributions with $u \neq 0$. Thermal duality is thus broken beyond the leading order contributions to the partition function.}
\begin{equation}
\mathz(T)  = \mathz\left({T_\text{c}^2 \over T}\right).
\end{equation}
Interestingly, the thermal duality emerged because the leading-order Feynman diagram could be expressed in terms of $\zeta(s)$ which is known to have a \emph{dual description} - a mathematical property which we have exploited to generate non-equivalent low- and high-temperature expansions in the context of the nonlocal scalar field.
Nonetheless, Eq. \ref{padicz} poses a strong contrast with the partition function for our nonlocal scalar model
\begin{equation}\label{key}
\ln \mathz_1 
=
- 3\lambda \beta V
\left[	\frac{M T}{4 \pi \sqrt \pi }
\int_1^\infty \frac{ d \alpha}{\alpha^2} \zeta\left(\frac{2 \pi T \alpha }{M} \right) \right]^2
\end{equation}
in which a nontrivial integral $f(T / M) = \int_1^\infty \frac{ d \alpha}{\alpha^2} \zeta\left(\frac{2 \pi T \alpha }{M} \right) $ over the Schwinger parameter $\alpha$ has to be taken (where the $\zeta$ function also contains the Schwinger parameter). That is why thermal duality is violated in the nonlocal scalar field. 
\nnn
Note that thermal duality implies the following \emph{asymptotic behavior of the partition function}.
\begin{equation}\label{asymptoticbehavior}
\ln \mathz \rightarrow
\begin{cases}
- \Lambda V / T 	& \text{for } T \ll M \\ 
-4 \pi\Lambda  T V / M^2 	& \text{for } T \gg M \\
\end{cases},
\end{equation}
where $\Lambda$ is the cosmological constant.
This behavior has been verified for p-adic (again with $p=3)$ strings in  \ref{hagedorn2}, but is \emph{not satisfied} for the nonlocal scalar theory.\footnote{Note that thermal duality implies that the partition function at high temperatures is equal to the partition function at low temperatures multiplied by $4\pi T^2/M^2$. In the context of the nonlocal scalar model it, instead, has to be multiplied by $\frac{9 M^2}{\pi^2 T^2}$ where the $M^2/T^2$ is responsible for the more gradual increase of the partition function with temperature in the high-T expansion - a result which is reconcilable with the Hagedorn phase.} This makes sense because both theories are governed by the exponential at high temperatures, but are fundamentally different at low temperatures.\footnote{The thermodynamic behavior of the nonlocal scalar field is determined by both the low-$T$ expansion of $\zeta$ \emph{and the two derivatives from the d'Alembertian}, while in the context of 3-adic fields this is completely determined by the low-$T$ expansion of $\zeta$.} 
But what is the physical reason underlying the existence of such a duality?
We will conclude that the absence of thermal duality is  quite sensible once we can \emph{interpret the physical origin of this duality}.
\nnn
Due to
the compact nature of one dimension in the
canonical thermal computation of string models, not only
the ordinary Matsubara thermal modes $\omega_n$ contribute to the partition function. The contributions from strings wrapped on the circle\footnote{Or torus, depending on the topology of space as was discussed in section \ref{sectionKMS}.} $S^1$ with circumference
$1/T$ should also be included - these are known as \emph{topological winding modes}, see Fig. \ref{thermaldualityfig}.
\begin{figure}[h!]
	\centering
	\includegraphics[scale=0.7]{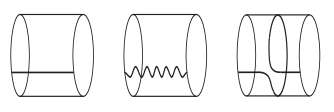}
	\caption{\ti{The first and second schematic description show Matsubara modes, the $n=0$ (ground) thermal mode is shown , juxtaposed by the $n=11$ thermal mode in the middle. On the right, the first topological winding mode ($\omega_{\text{ST}} = 1$), where strings are wrapped on the torus $S^1$ of circumference
		$\beta$, is shown. String theory predicts that winding modes contribute to the partition function at high temperatures.}
		Figure was taken from \ref{hagedorn2}.
	}
\end{figure}
The additional thermal modes $\omega_{\text{ST}}$ have been argued to satisfy (see \ref{hagedorn2})
\begin{equation}\label{omegastringtheory}
\omega_{\text{ST}} = \frac{\abs{n_{\text{ST}}}m_s^2}{\pi T},
\end{equation}
where the winding states $n_{\text{ST}}$ are labelled by the first homotopy group
$\pi_1(S^1) = \bf{Z}$ (keep in mind that $S^1$ is the topology of Euclidean space) and the string scale $m_s^2 = 1/2\alpha'$, where $\alpha'$ is related to the string tension $T_s$ by $T_s = 1/\sqrt{\alpha'}$  \ref{hagedorn2}. For bosonic string theories, the following symmetry is implied by Eq. \ref{omegastringtheory} and the bosonic Matsubara frequencies $\omega_n = 2\pi n T$:
\begin{equation}\label{thermalmodessymmetry}
2\pi T \leftrightarrow \frac{m_s^2}{\pi T}, \ \ \ \text{and} \ \ \ n \leftrightarrow n_{\text{ST}}.
\end{equation}
The straightforwardly leads to a thermal duality 
\begin{equation}\label{thermalduality}
\mathz(T) \sim \mathz\left( {m_s^2 \over 2\pi^2 }\frac{1}{T} \right) = \mathz\left({T_\text{c}^2 \over T}\right),
\end{equation}
where the critical temperature $T_{\text{c}} = \frac{m_s}{\sqrt{2}\pi}$ is related to the Hagedorn temperature via $T_{\text{H}} = T_{\text{c}} / a$, where $a$ depends on the particular string theory.\footnote{For instance, $a = 2$ for
bosonic strings, while $a =
\sqrt{2}$ for the Type II superstring and
$a = 1 + 1/
\sqrt2$ for the heterotic string \ref{hagedorn2}, \ref{hagedorndienes}.}
The thermal duality
can be expected to hold true if Eq. \ref{thermalmodessymmetry} is a genuine symmetry exhibited by nature and, indeed, compelling arguments have been put forward that thermal duality should be placed on equal footing with
other fundamental dualities of string theory such as $S$-duality and $T$-duality \ref{dienes3}.
\nnn
The physical reason why thermal duality emerges is thus the phenomenon that strings can have topological winding mode excitations. It is therefore unsurprising that string field theories have verified this thermodynamic behavior, especially given that the thermal duality symmetry of SFT follows immediately
from T-duality and Lorentz invariance and this symmetry, as a result, appears to be equally profound as the dualities which occur at zero temperature. What is however startling is that p-adic toy models, which are in fact quantum field theories rather than string field theories, could reproduce a thermal duality. Biswas and Kapusta themselves seemed surprised by their result
as well and wrote that ``[i]t is remarkable that this non-local quantum field
theory is able to reproduce the fundamental properties of
the thermodynamics of string theories,'' 
primarily prompted by their observation that 
``in p-adic models there are no obvious topological
counterparts [of the topological winding modes] because we are working with a quantum
field theory.'' \ref{hagedorn2}. From a physical point of view the thermal duality is not to be expected
and the behavior seems to have come about (accidentally) by the form of the 3-adic propagator, namely $\mathcal{D}_0(i\omega_n, \mathbf{p}) = \exp[p^2/M^2]$, the origin of which \emph{does lie in SFT} where it has been studied rigorously by Kuz'min and Tomboulis \ref{kuzmin} \ref{superrenormalizablegauge }. That the nonlocal scalar theory does not abide thermal duality may be due to the following possibilities:

\emph{i)} The nonlocal scalar theory should not be considered a full-fledged string field theory (like p-adic string theories) and consequently thermal duality can only emerge accidentally (for instance because the partition function is proportional to $\zeta(s)$) because the propagator \emph{happens to have an appropriate form this to happen}. The form of the nonlocal scalar propagator facilitates the introduction of a Schwinger parameter $\alpha$ which spoils the thermal duality.

\emph{ii)} Thermal duality is not as fundamental a symmetry as is often assumed. In the literature one finds statements such as ``it seems more \emph{natural}
to consider thermal duality as a fundamental property of a consistent string theory'' (my italics) but physicists do not unanimously agree upon the existence of this duality. In fact, the stringy bosonic tachyon whose thermodynamic behavior is derived in \ref{sftmodels} does \emph{not abide thermal duality either}, although this may perhaps be ascribed to the fact that their action is what one obtains in
the simplest level of truncation in the String Field Theory approach \ref{witten1986} where
only the tachyon field is kept and therefore should not be considered to be a genuine String Field Theory. This last option seems the most plausible to me, as will now be discussed, and would imply that the nonlocal field theory should not be considered to be a full-fledged SFT either.
\nnn
Since thermal duality is entailed by the combination of $T$-duality and Lorentz invariance, it seems very reasonable to expect that the duality should be satisfied by SFTs. Arguments have been put forward that the (bosonic) tachyonic strings from \ref{sftmodels} may not obey this duality because the action is too simple a truncation of the full stringy action and, in fact, their action is reminiscent to the nonlocal action which we have considered. Open string tachyons are described by
\begin{equation}\label{tachyoncond}
S = \int \ d^{4}x \ \left[	\frac{1}{2} \phi \left[\square - m^2 \right] e^{-\square / M^2} \phi - \lambda \phi^4	\right]
\end{equation}
which differs only in that the nonlocal scalar field is massless $m^2 = 0$. The absence of thermal duality in the context of open-string tachyons is shown explicitly in appendix \ref{appendixthermal}.\footnote{Unfortunately the authors of \ref{sftmodels} have not addressed the absence of thermal duality in their paper, it would be interesting to study more throroughly why this duality is not present for open-string tachyons in the near future.}
Interestingly, the reminisence in Lagrangians  allows us to check our expression for the partition function of the nonlocal scalar field in both temperature expansions (it should be identical to the partition function for open-string tachyons in the limit $m^2 \rightarrow 0$) and this cross-check confirms that our partition functions are, indeed, correct and thermal duality is consequently broken in the nonlocal scalar theory.
\subsection{Hagedorn phase
\label{sectionhagedorn}
}
Atick and Witten predicted that there should be a phase transition 
occurs for string
gases in thermal equilibrium
near the so-called Hagedorn temperature $T_H$ (at finite coupling). 
The Hagedorn phase is beyond the scope of my thesis, but a brief introduction of the Hagedorn phase is appropriate to give the reader a general idea of what the Hagedorn phase amounts to.
This new phase is dominated by the fundamental degrees of freedom which
underly string theory - namely the DOFs of strings -
and has become a well-entrenched cornerstone of string thermodynamics.
What happens near the Hagedorn temperature is that, as you put more energy into the system, you basically excite a \emph{single string} rather than change the thermal distribution of many strings. You can see the breakdown of the thermodynamic description because the specific heat $c_v$ of the system (which is a measure of the strength of the thermal fluctuations) diverges near $T_H$ \ref{dienes}.
\nnn
Above the Hagedorn energy density closed fundamental strings form a long string
phase. 
The exponential growth density of states of strings (as a function
of mass) makes the canonical partition function
\begin{equation}\label{key}
\mathz(\beta) = \Tr \left[\exp[-\beta H]\right]
\end{equation}
diverge at the Hagedorn temperature. A free string gas therefore becomes ill-defined as one approaches the so-called Hagedorn temperature. Different physical interpretations of this fact have been offered, including the interpretation that the Hagedorn
temperature defines an \emph{absolute limiting temperature in physics} or that it \emph{signals a transition to an unknown high temperature phase where strings should be replaced by even more
fundamental degrees of freedom} \ref{hagedornsathiapalan}.
\nnn
As the temperature
approaches
$T_H$, it becomes possible to excite the string oscillatory and winding mode
degrees of freedom. 
The energy density $\epsilon$ at high energy densities is found
to favor the formation of a single long string which carries most of the available energy
\ref{refstringsoup}.
The energy is in
the radiative degrees of freedom, the momentum modes of the strings.
In fact, close to the Hagedorn temperature, most of the energy in the
string gas is contained in winding strings.
According to Atick and Witten the free energy has to grow more slowly than for ordinary radiation \ref{wittenatick2} and they showed that
the high temperature behavior of the free energy $\mathcal{F}$ of strings presents a dependence
on the temperature characteristic of \emph{two-dimensional field theories}, \emph{i.e.} \ref{wittenatick} 
\begin{equation}\label{key}
\mathcal{F}(T) \propto T^2
\end{equation}
as compared to the usual behavior in local field theories where $\mathcal{F}(T) \propto T^4$. 
 Interestingly, the partition function of the nonlocal scalar field no longer goes as $\ln \mathz(T) \propto T^3$ but instead as 
 $$
 \ln \mathz = \frac{V \pi^2 }{90 } T^3
 - \frac{ 3 \lambda  V }{16 \pi^3 } M^2 T
 $$
 at high temperatures. This behavior has actually been predicted by
Atick and Witten, who conjectured that a Hagedorn transition occurs at high temperatures beyond which fewer degrees of freedom are activated than would be the case in quantum field theories.
The stringy system contains
many fewer degrees of freedom than one would  expect from the zero-temperature
string spectrum, or even in point-like particle field theories \ref{wittenatick2}. 
In the Hagedorn phase the energy density is dominated
by \emph{the most massive string states} and as a result behaves as a pressureless fluid \ref{p-adic}. It is quite surprising that the nonlocal scalar theory \emph{does exhibit Hagedorn behavior, but does not satisfy thermal duality}, especially because both effects can be ascribed to stringy topological winding modes. One of the arguments which were put forward in the previous section was that thermal duality might be broken because the action arises from the lowest order of truncation from a SFT action and consequently cannot capture all stringy phenomena. 
This entails that we should carefully examine (in future work)
whether the $M^2T$ term in the partition function has really arisen due to a Hagedorn phase transition or \emph{merely resembles} Hagedorn behavior. This is in close analogy to the p-adic theory which satisfies thermal duality, where there are no obvious topological
counterparts because it is in fact a quantum
field theory rather than a string field theory \ref{hagedorn2}. In my interpretation, one should be careful in those cases because the characteristic behavior may have arisen due to string-unrelated reasons. It is, however, remarkable that the nonlocal scalar theory reproduces
a fundamental property of
the thermodynamics of string theories, namely
 Hagedorn behavior.
\subsubsection*{Future work: thermal corrections in the infinite derivative scalar theory of gravity}
Now that we have computed the partition function and corresponding thermodynamical quantities up to first order for a relatively simple nonlocal Lagrangian, the computations however proved to be significantly more challenging than in the local scalar field theory. The next step is to compute these quantities for
the scalar toy model IDG action which, I remind you, is a nonlocal field theory which contains derivative interactions. Due to the close analogy with the BGKM gauge theory, thermal corrections of this scalar toy model will provide useful insights into the thermal behavior of infinite derivative theories of gravity. 
It would be interesting to study whether a Hagedorn phase occurs in that theory, especially because a Hagedorn phase in the early universe would have a plethora of implications for cosmology.
The stringy Hagedorn phase has found several cosmological applications, especially
in the context of cyclic/bouncing cosmologies \ref{hagedornsoup} and thermal structure formation scenarios \ref{hagedornbrandenberger}.
Interestingly, if matter is ultimately described by String Field Theory, massive string states could have been
excited in the early universe which would have corresponded to a Hagedorn phase for string matter.
In this case, cosmological fluctuations
could have been generated via the string gas mechanism which has been proposed by Nayeri, Branderberger and Vafa in \ref{hagedornbrandenberger} where a scale-invariant spectrum of perturbations would be obtained in the Hagedorn phase of string cosmology. In this era of the universe, the metric would have been almost static, and the Hubble radius would have been almost infinite \ref{nonperturbCMB}. 
These
topics are beyond the scope of my thesis and will not be discussed here, however, I wanted to convey that \emph{i)} the Hagedorn phase of matter could have grand implications for our understanding of the universe and \emph{ii)} that a Hagedorn phase could leave imprints in the CMB which make it a very interesting and relevant topic to study.
\nnn
It has been argued by Carone \ref{carone} that the acausal ordering of production and decay vertices for ordinary resonant particles in nonlocal models may provide a phenomenologically distinct signature for those models. It is worth studying which signatures would be present in IDG and whether these could be observed in the CMB data. Thermal field corrections for the BGKM are of crucial importance to study which predictions the theory makes for the early universe - one could subsequently test these predictions by comparing these with CMB data in order to (potentially) verify or refute this quantum gravity. 
\nnn
The preliminary steps for these thermal corrections (which will be finished in future work) can be found in appendix \ref{chaptertoymodelfinitet}.

\newpage

\chapter{Conclusions and discussion
\label{chapterconclusion}
}
In this thesis we have firstly motivated why nonlocal field theories are theories worth studying -nonlocal models have primarily been motivated in the gravity sector. Kuz'min \ref{kuzmin} and Tomboulis \ref{superrenormalizablegauge } concluded (in 1989 and 1997, respectively) that
gauge theories and theories of gravity can in fact be made perturbatively super-renormalizable if one considers a non-polynomial \emph{Lagrangian which contains an infinite series in higher derivatives}.
Biswas, Mazumdar and Siegel \ref{bounce} have subsequently argued  
that the absence of propagating ghosts and the renormalizability of the gauge theory can \emph{only}
be realized
when considering an infinite number of derivative terms. Following \ref{pol} we have further motivated the use of the exponential function 
$a(\square) = \exp[-\square/M^2] = \sum_{n=0}^{\infty} \frac{1}{n!} \left[-\square/M^2 \right]^{n} $
(where $M$ is the scale of nonlocality)
in the kinetic sector, which is allowed by general covariance and yields promising results including a non-singular Newtonian potential. Moreover, recent computations suggest that the theory is indeed a suitable candidate for a renormalizable theory of gravity \ref{UV} \ref{UVfiniteness} and that the theory may remove singularities at a classical level \ref{bounce}.  Biswas, Gerwick, Koivisto and Mazumdar concluded that theories of gravity whose kinetic term contains the infinite series of derivatives 
could be asymptotically free in the UV regime without violating fundamental physical principles such as unitarity, that is to say, the theory remains ghostfree \ref{singfree}.
\nnn
Subsequently we have motivated why finite temperature field theory is an interesting theoretical framework, which allows one to compute thermal corrections to zero-temperature field theories. Finite temperature self-energies correspond to thermal masses and are acquired because fields undergo interactions with the surrounding heat bath. 
Using the (Euclidean) Matsubara formalism we studied local scalar $\phi^4$ theory, whose free partition function is identical to that of a photon black body spectrum\footnote{Up to a factor of two originating in the fact that photons have two DOFs, while scalars have one. 
This means that the partition function for an ultrarelativistic scalar gas has been obtained successfully.} and has been ensued by
a computation of
the leading-order loop correction. The magnitudes of $\epsilon$, $P$ and $s$ have been shown to decrease due to self-interactions, an interpretation of which has been presented in section 
\ref{lambdaphi4}.
\nnn
We have afterwards shown in section  that a scalar (Higgs) particle acquires a thermal mass $\Pi_1 = \lambda T^2$ (the leading-order contribution) which  contributes a \emph{positive mass-squared to the propagator}. 
If the thermal mass of the Higgs field has been larger than its vacuum mass in the early universe, an electroweak symmetry restoration phase would have been realized in which elementary particles could not have acquired masses through the Higgs mechanism.
Higher-order contributions to the thermal mass would most tractably be computed in Braaten's and Pisarski's Hard Thermal Loop (HTL) effective field theory, as has been argued in section \ref{lambdaphi4} and appendix \ref{appendixbreakdownperturbation}. 
\nnn
We then delved into nonlocal thermal field theory and, following Biswas, Mazumdar and Siegel \ref{bounce} we modified the kinetic operator by adding an exponential term $a(\square) = \exp[-\square/M^2]  $ to the kinetic sector of the scalar theory in chapter \ref{chapternonlocalfield}.  We have shown that the nonlocality does not modify the partition function of \emph{the free theory} and could explain this feature by showing that the Wightman function of this nonlocal field theory is identical to that of the local scalar field theory.
The nonlocality does contribute to the partition function when interactions are included - this was shown explicitly for $\phi^4$ interactions. The local field theory results from chapter \ref{lambdaphi4} were recovered in the $M \rightarrow \infty$ limit.
\nnn
New physics emerges at temperatures $T > M/2\sqrt{\pi}$ since the scale of nonlocality becomes important in this regime. The partition function is no longer proportional to $T^3$ but to ($T^3 - M^2 T$), which can be understood from a stringy point of view. The partition function can be reconciled with the  conjectured stringy Hagedorn phase \ref{wittenatick}, \ref{wittenatick2}, \ref{hagedornsathiapalan} where elongated stings provide the dominant contributions to the partition function in this regime and the partition function increases more gradually with temperature, see section \ref{sectionhagedorn}. Moreover, we have concluded that stringy thermal duality is not satisfied in the nonlocal scalar field. Several reasons why this well-entrenched symmetry could be violated in the nonlocal scalar theory have been discussed in section \ref{sectionthermalduality}.
\nnn
An interesting direction for future research is the computation of thermal corrections to the scalar IDG model, whose nonlocal free action is identical to the one considered in chapter \ref{chapternonlocalfield} and contains derivative interactions
$$
S_{\text{int}} = \frac{1}{4M_{p} }\int d^{4}x \Bigg( 	\phi \partial_{\mu} \phi \partial^{\mu} \phi + \phi \square \phi a (\square) \phi - 
\phi \partial_{\mu} \phi a(\square)\partial^{\mu} \phi	\Bigg).
$$
To the best of my knowledge, thermal field corrections have not yet been computed for nonlocal models with derivative interactions and it would be illuminating to study the thermal properties of such models. 
It is an interesting research project to study whether the scalar IDG theory exhibits Hagedorn behavior at temperatures above the scale of nonlocality $T>M$ and study which consequences would be entailed for the hot, early universe. 
Thermal corrections in models involving derivative interactions will be calculated in future work.
 \nnn
In the more remote future thermal properties of the BGKM action will have to be researched, the calculations of which are unfortunately encumbered by the gauge structure of the theory.  
Once thermal field corrections have been successfully obtained for the scalar IDG model, one could study how the calculations should be modified in a gauge theory.\footnote{Remember that this generalization is quite straightforward (only the propagator is modified), but the computation of the nonlocal Feynman diagrams is, typically, significantly more arduous than in local field theories.}
 Thermal field corrections for the BGKM are of crucial importance to find out which predictions for the early universe are entailed by this action and subequently test these predictions (for instance, by examining its compatibility with CMB data) in order to potentially verify or refute this quantum gravity. Although Infinite Derivative Gravity is remarkably successful from a theoretical point of view, experimental tests will eventually be decisive for our understanding of whether or not nature is described by this kind of nonlocal theory!

\newpage

\appendix

\chapter{General relativity
\label{appendixgr}
}
General relativity is defined by the Einstein-Hilbert action
\begin{equation}\label{einsteinhilbert}
S_{\text{EH}} = \int d^{4}x\sqrt{-g}\mathcal{R},
\end{equation}
where $g = \det(g_{\mu \nu})$ and $\mathcal{R}$ is the Ricci scalar is the action describing Einstein gravity. Let us note that the coupling constant $\kappa= M_{p}^{-2}$
does not appear in the Einstein-Hilbert action. According to our conventions,
the coupling constant is introduced when considering an interaction term with a matter source. \nnn
We will firstly discuss \emph{linearized gravity} since linearized field equations will be obtained later on (for Infinite Derivative Gravity) in a reminiscent fashion. 
The action will now be varied with respect to the metric in order to obtain the corresponding equation of motion. Plugging in the definition of the Ricci scalar, \emph{i.e.} $\mathcal{R} = g^{\mu \nu} \mathcal{R}_{\mu \nu}$, one yields
\begin{equation}\label{key}
\delta S_{\text{EH}} = \int d^{4}x 
\left[
\sqrt{-g} g^{\mu \nu} \delta \mathcal{R}_{\mu \nu} 
+ \sqrt{-g} \mathcal{R}_{\mu \nu} \delta g^{\mu \nu} 
+  \mathcal{R} \delta \sqrt{-g} 
\right]
\end{equation}
That the first contribution vanishes is shown explicitly in \ref{carroll}. The second contribution can be calculated trivially, for the third contribution we have to understand how to deal with $\delta(\sqrt{-g})$ in a variational approach.
\nnn
For every square matrix $M$ with nonvanishing determinant, the following equality holds true
\begin{equation}\label{key}
\ln(\det M) = \Tr(\ln M).
\end{equation}
By varying this equality one obtains
\begin{equation}\label{key}
(\det M)^{-1} \delta (\det M) = \Tr (M^{-1} \delta M)
\end{equation}
and consequently
\begin{equation}\label{deltag}
\delta g = \delta \det(g_{\mu \nu}) = g(g_{\mu \nu}^{-1} \delta g_{\mu \nu}) = g(g^{\mu \nu} \delta g_{\mu \nu}).
\end{equation}
By employing the equality
\begin{equation}\label{key}
\delta g_{\mu \nu} = - g_{\mu \rho} g_{\nu \sigma} \delta g^{\rho \sigma}
\end{equation}
and substituting this in Eq. \ref{deltag} one finds 
\begin{equation}\label{finaldeltag}
\delta g  = g(- g_{\rho \sigma}  \delta g^{\rho \sigma}).
\end{equation}
Using Eq. \ref{finaldeltag}, $\delta \sqrt{-g}$ can be evaluated straightforwardly
\begin{equation}\label{variationmetric}
\bal
\delta \sqrt{-g} &= - \frac{\delta g}{2 \sqrt{-g}}  =  \frac{g}{2 \sqrt{-g}}  g_{\rho \sigma}  \delta g^{\rho \sigma} =  - \frac{1}{2} \sqrt{-g}  g_{\mu \nu}  \delta g^{\mu \nu}.
\eal
\end{equation}
We thus have yielded\footnote{We could have obtained the same result by performing the perturbation around Minkowski background
	\begin{equation}\label{sehexpansion}
	S_{\text{EH}}(g_{\mu\nu})=S_{\text{EH}}(\eta_{\mu\nu}+\delta g_{\mu\nu})=S_{\text{HE}}(\eta_{\mu\nu})+\frac{\delta S_{\text{EH}}}{\delta g^{\mu\nu}}\delta g^{\mu\nu}+\mathcal{O}\left((\delta g)^{2} \right),
	\end{equation}
	where cubic terms in the perturbation $h_{\mu\nu}$ are neglected. Since $\mathcal{R}(\eta_{\mu\nu})=0$ also the first terms in the taylor expansion in Eq. \ref{sehexpansion} vanishes, i.e. $S_{\text{EH}}(\eta_{\mu\nu})=0$.
	Moreover, terms which are linear in the perturbation $\delta g^{\mu\nu}$ do
	not appear. The linearized Einstein-Hilbert action, quadratic in $\delta g^{\mu\nu}$,
	is thus given by
	\begin{equation}
	S_{\text{EH}}=\frac{\delta S_{\text{HE}}}{\delta g^{\mu\nu}}\delta g^{\mu\nu}= \int d^{4}x \left[\mathcal{R}_{\mu\nu}-\frac{1}{2}\eta_{\mu\nu}\mathcal{R}\right] \delta g^{\mu\nu}.
	\end{equation}
}
\begin{equation}\label{EHaction}
S_{\text{EH}} = \int d^{4}x \sqrt{-g} 
\left[
\mathcal{R}_{\mu \nu} - \frac{1}{2} g_{\mu \nu} \mathcal{R}
\right] \delta g^{\mu \nu}
\end{equation}
\section{Lagrangian and its kinetic operator}
Let us now introduce a perturbation of the metric tensor
\begin{equation}\label{metricexpansion}
g_{\mu\nu}(x) = \eta_{\mu\nu} + h_{\mu\nu}(x).
\end{equation}
It is most convenient to employ the linearized
forms of the Riemann tensor, Ricci tensor and scalar tensor when doing perturbation theory, these are given by
\begin{equation}
\begin{array}{rl}
\mathcal{R}_{\mu\nu\lambda\sigma}= 
& {\displaystyle \frac{1}{2}}\left[\partial_{\nu}\partial_{\lambda}h_{\mu\sigma}+\mathcal{\partial_{\mu}\partial_{\sigma}}h_{\nu\lambda}-\mathcal{\partial_{\sigma}\partial_{\nu}}h_{\mu\lambda}-\partial_{\mu}\partial_{\lambda}h_{\nu\sigma}\right],\\
\\
\mathcal{R_{\mu\nu}}=g^{\alpha\rho}\mathcal{R}_{\alpha\mu\rho\nu}= & {\displaystyle \frac{1}{2}}\left[\partial_{\rho}\partial_{\nu}h_{\mu}^{\rho}+\partial_{\rho}\partial_{\mu}h_{\nu}^{\rho}-\partial_{\mu}\partial_{\nu}h-\square h_{\mu\nu}\right],\\
\\
\mathcal{R}= & \partial_{\mu}\partial_{\nu}h^{\mu\nu}-\square h.
\end{array}\label{eq:97linearized curv}
\end{equation}
Noting that $\delta g^{\mu\nu}=-h^{\mu\nu}$, 
the perturbed action in Eq. \ref{EHaction} becomes 
\begin{equation}
\begin{array}{rl}
S_{\text{EH}}= & {\displaystyle \int}d^{4}x(-h^{\mu\nu})\left(\mathcal{R}_{\mu\nu}-\frac{1}{2}\eta_{\mu\nu}\mathcal{R}\right)\\
\\
= & - \int d^{4}x \ h^{\mu\nu}\left[\frac{1}{2}\left(\partial_{\rho}\partial_{\nu}h_{\mu}^{\rho}+\partial_{\rho}\partial_{\mu}h_{\nu}^{\rho}-\partial_{\mu}\partial_{\nu}h-\square h_{\mu\nu}\right) -\frac{1}{2}\eta_{\mu\nu}\left(\partial_{\alpha}\partial_{\beta}h^{\alpha\beta}-\square h\right)\right]\\
\\
= & - \int d^{4}x\left(h_{\sigma}^{\mu}\partial^{\sigma}\partial^{\nu}h_{\mu\nu}-h\partial^{\mu}\partial^{\nu}h_{\mu\nu}-\frac{1}{2}h_{\mu\nu}\square h^{\mu\nu}+\frac{1}{2}h\square h\right)\\
\\
\equiv & \int d^{4}x \ \mathcal{L}_{\text{EH}},
\end{array}\label{eq:40.1}
\end{equation}
thus the Lagrangian for any symmetric two-rank tensor is
\begin{equation} \label{LaggHE}
\mathcal{L}_{\text{EH}} = h_{\sigma}^{\mu}\partial^{\sigma}\partial^{\nu}h_{\mu\nu}-h\partial^{\mu}\partial^{\nu}h_{\mu\nu}-\frac{1}{2}h_{\mu\nu}\square h^{\mu\nu}+\frac{1}{2}h\square h.
\end{equation}
We want to recast this Lagrangian into the following form
\begin{equation}
\mathcal{L}_{\text{EH}}=\frac{1}{2}h_{\mu\nu}\mathcal{O}^{\mu\nu\rho\sigma}h_{\rho\sigma},\label{eq:41}
\end{equation}
so that the kinetic operator $\mathcal{O}^{\mu \nu \rho \sigma}$ can be obtained. By raising and lowering the indices with the metric tensor $\eta_{\mu\nu},$ the Lagrangian density can be rewritten in the following way

\begin{equation}
\begin{aligned}
\mathcal{L}_{\text{EH}} &=  h_{\mu\nu}\left(\partial^{\nu}\partial^{\sigma}\eta^{\mu\rho}\right)h_{\rho\sigma}-h_{\mu\nu}\left(\partial^{\mu}\partial^{\nu}\eta^{\rho\sigma}\right)h_{\rho\sigma}  -\frac{1}{2}h_{\mu\nu}\left(\eta^{\mu\rho}\eta^{\nu\sigma}\square\right)h_{\rho\sigma}+\frac{1}{2}h_{\mu\nu}\left(\eta^{\mu\nu}\eta^{\rho\sigma}\square\right)h_{\rho\sigma}\\
&= \frac{1}{2}h_{\mu\nu}\left[2\partial^{\nu}\partial^{\sigma}\eta^{\mu\rho}-2\partial^{\mu}\partial^{\nu}\eta^{\rho\sigma}-\eta^{\mu\rho}\eta^{\nu\sigma}\square+\eta^{\mu\nu}\eta^{\rho\sigma}\square\right]h_{\rho\sigma} \\
&=
\frac{1}{2}h_{\mu\nu}\bigg[-\left(\frac{1}{2}\eta^{\mu\rho}\eta^{\nu\sigma}+\frac{1}{2}\eta^{\mu\sigma}\eta^{\nu\rho}-\eta^{\mu\nu}\eta^{\rho\sigma}\right)\square-\eta^{\mu\nu}\partial^{\rho}\partial^{\sigma}-\eta^{\rho\sigma}\partial^{\mu}\partial^{\nu} \\
&
\qquad \qquad \qquad \qquad \quad \ \ \
+\frac{1}{2}\left(\eta^{\nu\rho}\partial^{\mu}\partial^{\sigma}+\eta^{\nu\sigma}\partial^{\mu}\partial^{\rho}+\eta^{\mu\rho}\partial^{\nu}\partial^{\sigma}+\eta^{\mu\sigma}\partial^{\nu}\partial^{\rho}\right)\bigg]h_{\rho\sigma}.\\
\end{aligned}
\end{equation}
The kinetic operator $\mathcal{O}^{\mu\nu\rho\sigma}$ is hence given by 
\begin{equation}\label{eq:42}
\begin{array}{rl}
\mathcal{O}^{\mu\nu\rho\sigma} = & -\left(\frac{1}{2}\eta^{\mu\rho}\eta^{\nu\sigma}+\frac{1}{2}\eta^{\mu\sigma}\eta^{\nu\rho}-\eta^{\mu\nu}\eta^{\rho\sigma}\right)\square-\eta^{\mu\nu}\partial^{\rho}\partial^{\sigma}-\eta^{\rho\sigma}\partial^{\mu}\partial^{\nu}\\
\\
& +\frac{1}{2}\left(\eta^{\nu\rho}\partial^{\mu}\partial^{\sigma}+\eta^{\nu\sigma}\partial^{\mu}\partial^{\rho}+\eta^{\mu\rho}\partial^{\nu}\partial^{\sigma}+\eta^{\mu\sigma}\partial^{\nu}\partial^{\rho}\right)
\end{array}
\end{equation}
and it can rather straightforwardly be shown that following properties hold
\begin{equation}\label{eq:43}
\mathcal{O}^{\mu\nu\rho\sigma}=\mathcal{O}^{\nu\mu\rho\sigma}=\mathcal{O}^{\mu\nu\sigma\rho}=\mathcal{O}^{\rho\sigma\mu\nu}.
\end{equation}
\section{Field equations}
By varying the linearized action, we can obtain the Euler-Lagrange
equations%
\footnote{From the geometrical point of view we can obtain the same field equations
	by linearizing Einstein equations.%
} for the symmetric two-rank tensor $h_{\mu\nu}.$ In order to do this,
it is convenient to rewrite the Lagrangian given in Eq. \ref{LaggHE} solely
in terms of the first derivatives of $h_{\mu\nu}$ by means of integration
by parts:
\begin{equation}\label{eq:104L}
\mathcal{L}_{\text{EH}}=-\partial_{\rho}h_{\alpha}^{\rho}\partial_{\beta}h^{\alpha\beta}+\partial_{\alpha}h\partial_{\beta}h^{\alpha\beta}+\frac{1}{2}\partial_{\rho}h^{\alpha\beta}\partial^{\rho}h_{\alpha\beta}-\frac{1}{2}\partial_{\rho}h\partial^{\rho}h.
\end{equation}
The field equations are given by 
\begin{equation}\label{eulerlagrange}
\partial_{\sigma}\frac{\partial\mathcal{L}_{\text{EH}}}{\partial(\partial_{\sigma}h_{\mu\nu})}=\frac{\partial\mathcal{L}_{\text{EH}}}{\partial h_{\mu\nu}}
\end{equation}
so the derivatives of the Lagrangian in Eq. \ref{eq:104L} with respect to 
$\partial_{\sigma}h_{\mu\nu}$ and $h_{\mu\nu}$ have te be computed. Since
\[
\begin{array}{rl}
{\displaystyle \frac{\partial\mathcal{L}_{\text{EH}}}{\partial(\partial_{\sigma}h_{\mu\nu})}}= & {\displaystyle -\eta^{\mu\sigma}\partial_{\rho}h^{\nu\rho}-\eta^{\nu\sigma}\partial_{\rho}h^{\rho\mu}+\partial^{\sigma}h^{\mu\nu}}  {\displaystyle +\eta^{\mu\nu}\partial_{\rho}h^{\sigma\rho}+\eta^{\nu\sigma}\partial^{\mu}h-\eta^{\mu\nu}\partial^{\sigma}h}
\end{array}
\]
the LHS of Eq. \ref{eulerlagrange} equals
\begin{equation}
\begin{array}{rl}
\partial_{\sigma}\frac{\partial\mathcal{L}_{\text{EH}}}{\partial(\partial_{\sigma}h_{\mu\nu})}= & {\displaystyle -\partial^{\mu}\partial_{\rho}h^{\nu\rho}-\partial^{\nu}\partial_{\rho}h^{\rho\mu}+\square h^{\mu\nu}} {\displaystyle +\eta^{\mu\nu}\partial_{\rho}\partial_{\sigma}h^{\rho\sigma}+\partial^{\mu}\partial^{\nu}h-\eta^{\mu\nu}\square h.}
\end{array}\label{fieldLHS}
\end{equation}
The RHS of Eq. \ref{eulerlagrange} vanishes
\begin{equation}
\frac{\partial\mathcal{L}_{\text{EH}}}{\partial h_{\mu\nu}}=0\label{fieldRHS}
\end{equation}
since every metric tensor is acted upon by a partial derivative in Eq. \ref{eq:104L}.
\begin{subequations}
	Eqs. \ref{fieldLHS} and \ref{fieldRHS} imply that the field equations in vacuum are given by\footnote{The equality $\eta_{\mu \nu} \eta^{\nu \rho} = \delta_{\mu}^{\rho}$ is used repeatedly.}
	\begin{equation}\label{eq:47}
	\partial^{\mu}\partial_{\rho}h^{\nu\rho}+\partial^{\nu}\partial_{\rho}h^{\rho\mu}-\square h^{\mu\nu}+\eta^{\mu\nu}\partial_{\sigma}\partial_{\rho}h^{\sigma\rho}-\partial^{\mu}\partial^{\nu}h+\eta^{\mu\nu}\square h=0.
	\end{equation}
	Lowering the indices with $\eta_{\mu\nu}$ gives the equivalent expression
	$$
	\left(\eta_{\mu\rho}\eta_{\nu\sigma}\square+\eta_{\rho\sigma}\partial_{\mu}\partial_{\nu}-\eta_{\mu\sigma}\partial_{\nu}\partial_{\rho}-\eta_{\nu\sigma}\partial_{\mu}\partial_{\rho}+\eta_{\mu\nu}\partial_{\sigma}\partial_{\rho}-\eta_{\mu\nu}\eta_{\rho\sigma}\square\right)h^{\rho\sigma}=0
	$$
	and a subsequent multiplication with $\eta^{\mu\nu}$ gives
	$$ 2\left(\square\eta_{\rho\sigma}-\partial_{\rho}\partial_{\sigma}\right)h^{\rho\sigma}+4\left(\partial_{\rho}\partial_{\sigma}-\square\eta_{\rho\sigma}\right)h^{\rho\sigma}=0,
	$$
	from which it can be inferred that the following combination of terms vanishes
	\begin{equation}\label{vanish}
	\left(\partial_{\rho}\partial_{\sigma}-\square\eta_{\rho\sigma}\right)h^{\rho\sigma} =  \partial_{\rho}\partial_{\sigma}h^{\rho\sigma}-\square\eta_{\rho\sigma}h =0.
	\end{equation}
	Eq. \ref{vanish} entails that the vacuum field equations can be written in
	a simplified form%
	\footnote{From the geometrical point of view it means that $\mathcal{R}=0$
		in the vacuum. In fact the linearized form of the Ricci scalar is
		$\mathcal{R}=\partial_{\rho}\partial_{\sigma}h^{\rho\sigma}-\square h=\left(\partial_{\rho}\partial_{\sigma}-\eta_{\rho\sigma}\square\right)h^{\rho\sigma}$ so the Ricci scalar should vanish in the vacuum.%
	}
	\begin{equation}\label{fieldequationsvacuum}
	\left(\eta_{\mu\rho}\eta_{\nu\sigma}\square+\eta_{\rho\sigma}\partial_{\mu}\partial_{\nu}-\eta_{\mu\sigma}\partial_{\nu}\partial_{\rho}-\eta_{\nu\sigma}\partial_{\mu}\partial_{\rho}\right)h^{\rho\sigma}=0.
	\end{equation}
\end{subequations}
In presence of a source $\tau^{\mu\nu}$ the term $-\kappa h_{\mu\nu}\tau^{\mu\nu}$ has to be appended to the RHS of $\mathcal{L}_{\text{EH}}$.\footnote{The minus sign arises because starting from GR in a generic background
	we have: $S=S_{\text{HE}}+S_{m},$ where $S_{\text{EH}}$ was already examined in
	Eq. \ref{eq:40.1}  and the matter part whose variation with respect
	to the metric is $$\delta S_{m}=\kappa\int d^{4}x \ \delta g^{\mu\nu}\tau_{\mu\nu}=\kappa\int d^{4}x \ (-g^{\mu\alpha}g^{\nu\beta}\delta g_{\alpha\beta})\tau_{\mu\nu}.$$
	Since $\delta g_{\alpha\beta}=h_{\alpha\beta},$ from the point
	of view of the field theory approach we consider $-\kappa h_{\alpha\beta}\tau^{\alpha\beta}.$
}
The variation of the Lagrangian density with respect to $h_{\mu \nu}$ no longer vanishes, since
\begin{equation}
\frac{\partial\mathcal{L}_{\text{EH}}}{\partial h_{\mu\nu}}=-\kappa\tau^{\mu\nu}
\end{equation}
so the full field equations thus become
\begin{equation}
\square h^{\mu\nu}-\partial^{\mu}\partial_{\rho}h^{\nu\rho}-\partial^{\nu}\partial_{\rho}h^{\rho\mu}+\eta^{\mu\nu}\partial_{\sigma}\partial_{\rho}h^{\sigma\rho}+\partial^{\mu}\partial^{\nu}h-\eta^{\mu\nu}\square h=-\kappa\tau^{\mu\nu},\label{fullfieldequations}
\end{equation}
We can introduce spin projector operators in the space of the symmetric two-rank tensors. We can decompose a symmetric
two-rank tensor in terms of spin-$2,$ spin-$1$ and two spin-$0$
components under the rotation group $SO(3),$ i.e. $h_{\mu\nu}\in\mathbf{0}\oplus\mathbf{0}\oplus\mathbf{1}\oplus\mathbf{2},$
by introducing a set of projection operators. 
Using the spin projection procedure, the kinetic operator can be shown to be
\begin{equation}\label{OO}
\mathcal{O}_{\mu\nu\rho\sigma}=k^{2}\left(\mathcal{P}^{2}-2\mathcal{P}_{s}^{0}\right)_{\mu\nu\rho\sigma},
\end{equation}
the full computation of which can be found in appendix \ref{grprop}.
\nnn
We conclude that the graviton exhibits both spin-$2$ and spin-$0$ degrees of freedom, while the spin-$1$ component is absent.
Thus, in total we have six degree of freedom: five spin-$2$ and one
spin-$0$ components. Four degrees of freedom drop out due to gauge invariance exhibited by the Lagrangian as will be proved in the next section. What this implies is the following: \emph{off-shell gravitons have 6 degrees of freedom, while on-shell gravitons have only 2 degrees of freedom}.

\section{Invariance under diffeomorphisms}
We should recall
that a diffeomorphism $\phi$ is a (bijective) $C^{\infty}$ map between manifolds which has
a $C^{\infty}$ inverse. In other words, the diffeomorphism group is formed by the set of mappings
$\phi: \mathcal{M} \rightarrow \mathcal{M}$ which preserves the structure of the manifold. There are two geometrical interpretations which are commonly referred to as \emph{passive} and \emph{active}
diffeomorphisms.
Passive diffeomorphism invariance refers to invariance under change of coordinates, \emph{i.e.}
the same object is represented in different coordinate systems, whereas active diffeomorphisms on the other hand relate different objects in $\mathcal{M}$ in the same
coordinate system. This means that $f$ is viewed as a map associating one point in the
manifold to another one \ref{carroll}. 
\nnn
General relativity is distinguished from other dynamical field theories by its invariance
under active diffeomorphisms. Any theory can be made invariant under passive diffeomorphisms.
Passive diffeomorphism invariance on the one hand is a property of the formulation of a dynamical
theory, while active diffeomorphism invariance on the other hand is a \emph{property of the dynamical theory} \ref{gaul}. If we displace all dynamical objects in the manifold at once, we
generate nothing but an equivalent mathematical description of the same physical state,
because localization with respect to the manifold is, in fact, irrelevant due to invariance under diffeomorphism.
\nnn
We could ask what is the gauge symmetry
for $\mathcal{L}_{\text{EH}}$ and its field equations from Eq. \ref{fieldequationsvacuum}. Let us consider the
one-parameter group of diffeomorphisms $x'^{\mu}\equiv x'^{\mu}(x)$
and its infinitesimal form in particular
\begin{equation}
x'^{\mu} = x^{\mu} + \xi^{\mu}(x)\label{eq:62}.
\end{equation}
We are interested in how the metric tensor $g_{\mu\nu}(x)$ transforms
under diffeomorphisms:
\begin{equation}
\begin{array}{rl}
g'_{\mu\nu}(x')= & {\displaystyle \frac{\partial x^{\alpha}}{\partial x'^{\mu}}\frac{\partial x^{\beta}}{\partial x'^{\nu}}g_{\alpha\beta}}(x)=\left(\delta_{\mu}^{\alpha}-\partial_{\mu}\xi^{\alpha}\right)\left(\delta_{\nu}^{\beta}-\partial_{\nu}\xi^{\beta}\right)g_{\alpha\beta}(x)\\
\\
= & g_{\mu\nu}(x)-\left(\partial_{\mu}\xi^{\alpha}\right)g_{\alpha\nu}-\left(\partial_{\nu}\xi^{\alpha}\right)g_{\alpha\mu}.
\end{array}\label{eq:63}
\end{equation}
Moreover, by expanding in Taylor series $g'_{\mu\nu}(x')$ one gets%
\footnote{Note that we are using the fact that the transformation is infinitesimal,
	namely $\partial_{\alpha}g_{\mu\nu}'(x)=\partial_{\alpha}g{}_{\mu\nu}(x)+\mathcal{O}(\xi^{2})$
	and $\partial'_{\mu}\xi^{\alpha}(x')=\partial_{\mu}\xi^{\alpha}(x)+\mathcal{O}(\xi^{2}).$%
}
\begin{equation}
g'_{\mu\nu}(x')\simeq g'_{\mu\nu}(x)+\xi^{\alpha}\partial_{\alpha}g{}_{\mu\nu}(x).
\end{equation}
Substituting the latter equation in Eq. \ref{eq:63}, we obtain
$$
\bal
\delta g_{\mu\nu}(x)\equiv g'_{\mu\nu}(x)-g_{\mu\nu}(x)=&-\left(\partial_{\mu}\xi^{\alpha}\right)g_{\alpha\nu}-\left(\partial_{\nu}\xi^{\alpha}\right)g_{\mu\alpha}-\xi^{\alpha}\partial_{\alpha}g_{\mu\nu}\\
=&-\partial_{\mu}\xi_{\nu}-\partial_{\nu}\xi_{\mu}+\xi^{\alpha}\left(\partial_{\mu}g_{\alpha\nu}+\partial_{\nu}g_{\mu\alpha}-\partial_{\alpha}g_{\mu\nu}\right)\\
=&-\partial_{\mu}\xi_{\nu}-\partial_{\nu}\xi_{\mu}+2\xi_{\alpha} \Gamma_{\mu\nu}^{\alpha}.
\eal
$$
This expression can be further simplified by noting that
\begin{equation}
\delta g_{\mu\nu}(x)=-\nabla_{\mu}\xi_{\nu}-\nabla_{\nu}\xi_{\mu}\label{eq:64}
\end{equation}
and that the covariant derivatives are replaced by partial derivatives within a Minkowski background, entailing that
\[
\delta h_{\mu\nu}=-\partial_{\mu}\xi_{\nu}-\partial_{\nu}\xi_{\mu}.
\]
Additionally, we redefine $-\xi_{\mu}\rightarrow\xi_{\mu}$ so the
variation of $h_{\mu\nu}$ now reads 
\begin{equation}
\delta h_{\mu\nu}=\partial_{\mu}\xi_{\nu}+\partial_{\nu}\xi_{\mu}.\label{gaugetrans}
\end{equation}
We can easily verify that the transformation from Eq. \ref{gaugetrans} is a gauge
symmetry for $\mathcal{L}_{\text{HE}}$ and its associated
field equations. Substituting Eq. \ref{gaugetrans} into the field equation from Eq. \ref{fieldequationsvacuum} gives
\[
\square\partial_{\mu}\xi_{\nu}+\square\partial_{\nu}\xi_{\mu}+2\partial_{\mu}\partial_{\nu}\partial_{\alpha}\xi^{\alpha}-\partial_{\mu}\partial_{\alpha}\partial^{\alpha}\xi_{\nu}-\partial_{\nu}\partial_{\alpha}\partial^{\alpha}\xi_{\mu}-2\partial_{\nu}\partial_{\mu}\partial_{\alpha}\xi^{\alpha}=0,
\]
i.e. the field equations are invariant under the gauge
transformation from Eq. \ref{gaugetrans}. This allows us to transform the tensor field $h_{\mu\nu},$ so that we can choose a special gauge
in which the field equations simplify. A suitable choice of gauge for our symmetric tensor field is the $\mathit{De}$ $\mathit{Donder}$ $\mathit{gauge}$: 
\begin{equation}
\partial_{\alpha}h_{\mu}^{\alpha}-\frac{1}{2}\partial_{\mu}h= \partial_{\alpha}\left[h_{\mu}^{\alpha}-\frac{1}{2}\delta_{\mu}^{\alpha}h\right]=0.\label{eq:66}
\end{equation}
The field equations in vacuum, which were obtained in Eq. \ref{fieldequationsvacuum} and repeated here for the reader's convenience
\begin{equation}
\square h_{\mu\nu}+\partial_{\mu}\partial_{\nu}h-\partial_{\mu}\partial_{\alpha}h_{\nu}^{\alpha}-\partial_{\nu}\partial_{\alpha}h_{\mu}^{\alpha}=0,
\end{equation}
reduce to those for photons when working in De Donder gauge, in other words,
the wave equation is obtained:
\begin{equation}\label{wave}
\square h_{\mu\nu}-\partial_{\mu}\left(\partial_{\alpha}h_{\nu}^{\alpha}-\frac{1}{2}\partial_{\nu}h\right)-\partial_{\nu}\left(\partial_{\alpha}h_{\mu}^{\alpha}-\frac{1}{2}\partial_{\mu}h\right) =  \square h_{\mu\nu} = 0.
\end{equation}
The wave equation in momentum space satisfies $-k^{2}h_{\mu\nu}=0$, since $k^{2}=0$
the field $h_{\mu\nu}$ thus describes a massless graviton.
\subsubsection*{Concluding remarks}
We thus have found that the graviton is a massless particle according to general relativity. It is shown explicitly in Appendix \ref{appendixfieldeqGR} (using the spin projection formalism) that the graviton has 6 degrees of freedom, one due to its spin 0 mode and five due to its spin 2 mode.
Theories of modified gravity should also have a single gauge boson, massless and having spin-2 and spin-0 modes, as will be discussed soon.

\chapter{Spin projection operators \label{appendixspinproject}}
One can
introduce a useful set of spin projector operators in the space of the symmetric two-rank tensors. Spin projection operators enable a decomposition of symmetric
two-rank tensors in terms of spin-$2,$ spin-$1$ and two spin-$0$
components under the rotation group $SO(3)$. One can also decompose other Lorentz tensors under the $SO(3)$ group, but since this is irrelevant for the purposes of the thesis the reader is referred to Refs. \ref{luca}, \ref{maggiore} for generalizations of the spin projection methodology to tensors which are not necessarily of the ``symmetric two-rank'' type.
\section{Spin projection operator formalism
\label{appendixspinproj1}
}
Let us consider a symmetric part metric tensor $h_{\mu\nu}\in\mathbf{0}\oplus\mathbf{0}\oplus\mathbf{1}\oplus\mathbf{2}.$
Viewing $h_{\mu\nu}$ as a symmetric tensor product of two four-vectors $h_{\mu \nu} \equiv U_\mu \otimes V_\mu$,
the decomposition in projection operators can be performed as follow: 
\begin{equation}
\begin{array}{rl}
h_{\mu\nu}= & \eta_{\mu \rho} \eta_{\nu \sigma} h^{\rho\sigma} = \left(\theta_{\mu\rho}+\omega_{\mu\rho}\right)\left(\theta_{\nu\sigma}+\omega_{\nu\sigma}\right)h^{\rho\sigma}\\
\\
= & \left(\theta_{\mu\rho}\theta_{\nu\sigma}+\theta_{\mu\rho}\omega_{\nu\sigma}+\omega_{\mu\rho}\theta_{\nu\sigma}+\omega_{\mu\rho}\omega_{\nu\sigma}\right)h^{\rho\sigma}\\
\\
= & \frac{1}{2}\left(\theta_{\mu\rho}\theta_{\nu\sigma}+\theta_{\mu\sigma}\theta_{\nu\rho}\right)h^{\rho\sigma}-\frac{1}{3}\theta_{\mu\nu}\theta_{\rho\sigma}h^{\rho\sigma} +\frac{1}{3}\theta_{\mu\nu}\theta_{\rho\sigma}h^{\rho\sigma}+\omega_{\mu\nu}\omega_{\rho\sigma}h^{\rho\sigma}\\
\\
& +\frac{1}{2}\left(\theta_{\mu\rho}\omega_{\nu\sigma}+\theta_{\mu\sigma}\omega_{\nu\rho}+\theta_{\nu\rho}\omega_{\mu\sigma}+\theta_{\nu\sigma}\omega_{\mu\rho}\right)h^{\rho\sigma},
\end{array}\label{hdecomposition}
\end{equation}
where the transverse projector $\theta_{\mu\nu}$ and longitudinal projector $\omega_{\mu\nu}$ are defined as\footnote{These projection operators are used to project the spin-0 (longitudinal) and spin-1 (transverse) components from four-vectors $V_\mu$ like
$$
	V_\mu = \omega_{\mu \nu} V^{\nu} + \theta_{\mu \nu} V^{\nu}.
$$
We can immediately conclude for photons with 4-momentum $k^\mu$ that $k^\mu \omega_{\mu \nu} = k_\nu$, \emph{i.e.} $\omega_{\mu \nu}$ projects the longitudinal component and likewise $\theta_{\mu \nu}$ projects the transversal component since $k^\mu \theta_{\mu \nu} = 0$.
One can prove that the longitudinal component has spin 0
$$\eta^{\mu \nu} \omega_{\mu \nu} =  1 = 2(0) + 1$$
and that the transversal component has spin 1
$$\eta^{\mu \nu} \theta_{\mu \nu} = 4 - 1 = 2(1) + 1. $$
These spin projection operators play a useful role for the decomposition of 2-rank tensors as well, since we can employ the identity $\eta_{\mu nu} = \theta_{\mu \nu} + \omega_{\mu \nu}$ and therefore obtain spin projection operators for tensors which are defined in terms of $\{\omega, \theta\}$. \label{fourvecdecomp}
}
	
\begin{equation}
\{\omega,\theta\}:\,\,\,\,\,\begin{cases}
{\displaystyle \omega_{\mu\nu} = \frac{\partial_{\mu}\partial_{\nu}}{\square}}\\
&  \\
{\displaystyle \theta_{\mu\nu} = \eta_{\mu\nu} - \omega_{\mu \nu}}\\
\end{cases}\label{eq:4}
\end{equation}
The following properties can easily be proved and entail that the set $\{\omega, \theta\}$ is orthogonal and complete.
\begin{equation}
\begin{array}{rl}
&
\theta_{\mu\nu}+\omega_{\mu\nu}=\eta_{\mu\nu}
\Leftrightarrow \theta+\omega= I,\\
\\
& \omega^{2}=\omega,\,\,\,\theta^{2}=\theta,\,\,\,\theta\omega=0,\\
\\
 & \theta_{\mu\nu}\theta_{\rho}^{\nu}=\theta_{\mu\rho},\,\,\,\omega_{\mu\nu}\omega_{\rho}^{\nu}=\omega_{\mu\rho},\,\,\,\theta_{\mu\nu}\omega_{\rho}^{\nu}=0.
\end{array}\label{eq:5}
\end{equation}
Eq. \ref{hdecomposition} allows us to introduce the following spin projection operators:
\begin{subequations}
	\begin{equation}\label{spinproj}
\begin{aligned}
	\mathcal{P}_{\mu\nu\rho\sigma}^{2}&={\displaystyle \frac{1}{2}\left(\theta_{\mu\rho}\theta_{\nu\sigma}+\theta_{\mu\sigma}\theta_{\nu\rho}\right)-\frac{1}{3}\theta_{\mu\nu}\theta_{\rho\sigma}} && \text{(spin 2)} ,\\
	\mathcal{P}_{\mu\nu\rho\sigma}^{1}&={\displaystyle \frac{1}{2}}\left(\theta_{\mu\rho}\omega_{\nu\sigma}+\theta_{\mu\sigma}\omega_{\nu\rho}+\theta_{\nu\rho}\omega_{\mu\sigma}+\theta_{\nu\sigma}\omega_{\mu\rho}\right) && \text{(spin 1)},\\
	\mathcal{P}_{s,\,\mu\nu\rho\sigma}^{0}&={\displaystyle \frac{1}{3}}\theta_{\mu\nu}\theta_{\rho\sigma},\,\,\,\,\,\,\,\,\,\,\mathcal{P}_{w,\,\mu\nu\rho\sigma}^{0}=\omega_{\mu\nu}\omega_{\rho\sigma} && \text{(spin 0)}.
\end{aligned}
	\end{equation}
	The set 
$
\left\{ \mathcal{P}^{2},\mathcal{P}_{m}^{1},\mathcal{P}_{s}^{0},\mathcal{P}_{w}^{0}\right\}
$
	forms a complete set of spin projector operators in terms of which
	a symmetric two-rank tensor can be decomposed. In fact, one can easily
	verify that
	the spin projection operators are orthogonal
	\footnote{We are suppressing the additional indices, the ``proper'' orthogonality relation is given by $\mathcal{P}_{a,\,\mu\nu\alpha\beta}^{i}\mathcal{P}_{b,\,\alpha\beta\rho\sigma}^{i}=\delta_{ij}\delta_{ab}\mathcal{P}_{a,\,\mu\nu\rho\sigma}^{i}.$
	}
	 	\begin{equation}\label{orthogonality}
	\mathcal{P}_{a}^{i}\mathcal{P}_{b}^{j}=\delta_{ij}\delta_{ab}\mathcal{P}_{a}^{i}
	\end{equation}
and form a \emph{complete set of spin projector operators}
	\[
	\begin{array}{rl}
	\mathcal{P}^{2}+\mathcal{P}^{1}+\mathcal{P}_{s}^{0}+\mathcal{P}_{w}^{0}= & \frac{1}{2}\left(\theta_{\mu\rho}\theta_{\nu\sigma}+\theta_{\mu\sigma}\theta_{\nu\rho}\right)+\omega_{\mu\nu}\omega_{\rho\sigma} +\frac{1}{2}\left(\theta_{\mu\rho}\omega_{\nu\sigma}+\theta_{\mu\sigma}\omega_{\nu\rho}+\theta_{\nu\rho}\omega_{\mu\sigma}+\theta_{\nu\sigma}\omega_{\mu\rho}\right)\\
	\\
	= & \frac{1}{2}\eta_{\nu\sigma}\theta_{\mu\rho}+\frac{1}{2}\eta_{\nu\rho}\theta_{\mu\sigma}+\frac{1}{2}\theta_{\nu\rho}\omega_{\mu\sigma}+\frac{1}{2}\theta_{\nu\sigma}\omega_{\mu\rho}+\omega_{\mu\nu}\omega_{\rho\sigma}\\
	\\
	= & \frac{1}{2}\left(\eta_{\mu\rho}\eta_{\nu\sigma}+\eta_{\nu\rho}\eta_{\mu\sigma}\right)+\frac{1}{2}\eta_{\nu\rho}\omega_{\mu\sigma}+\frac{1}{2}\eta_{\nu\sigma}\omega_{\mu\rho} -\frac{1}{2}\eta_{\nu\sigma}\omega_{\mu\rho}-\frac{1}{2}\eta_{\nu\rho}\omega_{\mu\sigma}\\
	\\
	= & \frac{1}{2}\left(\eta_{\mu\rho}\eta_{\nu\sigma}+\eta_{\nu\rho}\eta_{\mu\sigma}\right)= I.
	\end{array}
	\]
Symmetric two-rank tensor can thus be decomposed completely
in terms of projector operators, an example of which is given for
$h_{\mu\nu}:$ 
\begin{equation}
\begin{array}{rl}
h_{\mu\nu}= & \mathcal{P}_{\mu\nu\rho\sigma}^{2}h^{\rho\sigma}+\mathcal{P}_{\,\mu\nu\rho\sigma}^{1}h^{\rho\sigma}+\mathcal{P}_{s,\,\mu\nu\rho\sigma}^{0}h^{\rho\sigma}+\mathcal{P}_{w,\,\mu\nu\rho\sigma}^{0}h^{\rho\sigma}\\
\\
= & \left(\mathcal{P}^{2}+\mathcal{P}^{1}+\mathcal{P}_{s}^{0}+\mathcal{P}_{w}^{0}\right)_{\mu\nu\rho\sigma}h^{\rho\sigma}.
\end{array}\label{eq:b.356sym}
\end{equation}
Additionally, two additional terms are required to close the algebra (these mix the two scalar components
$s$ and $w$ and are therefore cannot be viewed as genuine projection operators) and form a basis
	of symmetric four-rank tensors in terms of which the operator space of the gravitational
	field equations can be spanned
	\begin{equation}
	\mathcal{P}_{sw,\,\mu\nu\rho\sigma}^{0}={\displaystyle \frac{1}{\sqrt{3}}\theta_{\mu\nu}\omega_{\rho\sigma},\,\,\,\,\,\,\,\mathcal{P}_{ws,\,\mu\nu\rho\sigma}^{0}=\frac{1}{\sqrt{3}}}\omega_{\mu\nu}\theta_{\rho\sigma}. \label{closealgebra}
	\end{equation}
\end{subequations}
The operators defined in Eqs. \ref{spinproj},d  satisfy the completeness property and the following orthogonality
relations 
\begin{equation}
\begin{array}{lll}
\mathcal{P}_{a}^{i}\mathcal{P}_{b}^{j}=\delta_{ij}\delta_{ab}\mathcal{P}_{a}^{j}, &  & \mathcal{P}_{ab}^{0}\mathcal{P}_{c}^{i}=\delta_{i0}\delta_{bc}\mathcal{P}_{ab}^{i},\\
\\
\mathcal{P}_{ab}^{0}\mathcal{P}_{cd}^{0}=\delta_{ad}\delta_{bc}\mathcal{P}_{a}^{0}, &  & \mathcal{P}_{c}^{i}\mathcal{P}_{ab}^{0}=\delta_{i0}\delta_{ac}\mathcal{P}_{ab}^{0},
\end{array}\label{eq:53}
\end{equation}
where $i,j=0,1,2$ and $a,b,c,d$ are either $m,s,w$ or \emph{absent}.\footnote{We shall often suppress the indices for simplicity.} 
\nnn
We can show what the associated spin for each
spin projector operators is in a reminiscent fashion to how this was determined for the projectors $\theta_{\mu \nu}$ and $\omega_{\mu \nu}$ in Footnote \ref{fourvecdecomp}.
Using
\begin{equation}
\mathcal{P}^{2}+\mathcal{P}^{1}+\mathcal{P}_{s}^{0}+\mathcal{P}_{w}^{0}= I\Leftrightarrow\left(\mathcal{P}^{2}+\mathcal{P}^{1}+\mathcal{P}_{s}^{0}+\mathcal{P}_{w}^{0}\right)_{\mu\nu\rho\sigma}=\frac{1}{2}\left(\eta_{\mu\rho}\eta_{\nu\sigma}+\eta_{\nu\rho}\eta_{\mu\sigma}\right).\label{eq:54}
\end{equation}
and the orthogonality relations one yields
\begin{equation}
\begin{array}{lll}
\eta^{\mu\rho}\eta^{\nu\sigma}\mathcal{P}_{\mu\nu\rho\sigma}^{2}&=5=2(2)+1 & (\mathrm{spin\textrm{-}2),}\\
\\
\eta^{\mu\rho}\eta^{\nu\sigma}\mathcal{P}_{\mu\nu\rho\sigma}^{1}&=3=2(1)+1 & (\mathrm{spin\textrm{-}1),}\\
\\
\eta^{\mu\rho}\eta^{\nu\sigma}\mathcal{P}_{s,\,\mu\nu\rho\sigma}^{0}&=1=2(0)+1 & (\mathrm{spin\textrm{-}0),}\\
\\
\eta^{\mu\rho}\eta^{\nu\sigma}\mathcal{P}_{w,\,\mu\nu\rho\sigma}^{0}&=1=2(0)+1 & (\mathrm{spin\textrm{-}0).}
\end{array}\label{eq:55}
\end{equation}
Moreover one can easily show that the following relations hold: 
\begin{equation}
\begin{array}{ll}
\eta^{\mu\nu}\mathcal{P}_{\mu\nu\rho\sigma}^{2}=0=\eta^{\mu\nu}\mathcal{P}_{\mu\nu\rho\sigma}^{2} & (\mathrm{traceless),}\\
\\
k^{\mu}\mathcal{P}_{\mu\nu\rho\sigma}^{2}=0=k^{\mu}\mathcal{P}_{\mu\nu\rho\sigma}^{1} & \mathrm{(transverse)}.
\end{array}\label{eq:56}
\end{equation}
We have obtained six spin projection operators
\begin{equation}
\left\{ \mathcal{P}^{2},\mathcal{P}^{1},\mathcal{P}_{s}^{0},\mathcal{P}_{w}^{0},\mathcal{P}_{sw}^{0},\mathcal{P}_{ws}^{0}\right\} 
\end{equation}
which form a basis in terms of which the symmetric four-rank tensor $\mathcal{O}^{\mu\nu\rho\sigma}$
can be expanded, but it should be kept in mind that the last two operators are no genuine \emph{projection} operators) 
\footnote{By the expression ``a symmetric four-rank tensor'' we mean the operator
	$\mathcal{O}^{\mu\nu\rho\sigma}$ that appear in a given parity-invariant
	Lagrangian, like the operator (\ref{eq:42}). See Appendix $B.2$
	for more details.%
}. The properties (\ref{eq:53})-(\ref{eq:54})
imply that four out of six form a complete set of spin projector
operators, 
\begin{equation}
\left\{ \mathcal{P}^{2},\mathcal{P}^{1},\mathcal{P}_{s}^{0},\mathcal{P}_{w}^{0}\right\} ,
\end{equation}
in terms of which a symmetric two-rank tensor can be decomposed in
one spin-$2,$ one spin-$1$ and two spin-$0$ components. The operators
$\mathcal{P}_{sw}^{0}$ and $\mathcal{P}_{ws}^{0}$ are not projectors
as we can see from the relations in Eq. \ref{eq:53}, but are necessary
to close the algebra and form a basis of symmetric four-rank tensors
in terms of which the operator space of the gravitational field equations
can be spanned,
these can mix the
scalar multiplets $\mathcal{P}_s^0$, $\mathcal{P}_w^0$.  \nnn
This basis of projectors represents six field degrees of
freedom. The other four fields in a symmetric tensor field, as
usual, represent unphysical gauge degrees of freedom. $\mathcal{P}^{2}$
and $\mathcal{P}^{1}$ represent transverse and traceless spin-$2$
and spin-$1$ degrees of freedom, accounting for four degrees of freedom, while
$\mathcal{P}_{s}^{0}$ and $\mathcal{P}_{w}^{0}$ represent the spin-$0$
scalar multiplets, both of which have carry 1 degree of freedom. 
\section{Application: Field equations in general relativity \label{grprop} }
In terms of the spin projector operators $h_{\mu\nu}$ decomposes
as
\begin{equation}
h^{\mu\nu}=\mathcal{P}_{\rho\sigma}^{2\mu\nu}h^{\rho\sigma}+\mathcal{P}_{\rho\sigma}^{1\mu\nu}h^{\rho\sigma}+\mathcal{P}_{s,\,\rho\sigma}^{0\mu\nu}h^{\rho\sigma}+\mathcal{P}_{w,\,\rho\sigma}^{0\mu\nu}h^{\rho\sigma}.
\end{equation}
In momentum space, taking into account the presence of the source,
the field equations from Eq. \ref{fullfieldequations} (repeated here for the reader's convenience)
$$
-\kappa\tau^{\mu\nu}
=
\square h^{\mu\nu}-\partial^{\mu}\partial_{\rho}h^{\nu\rho}-\partial^{\nu}\partial_{\rho}h^{\rho\mu}+\eta^{\mu\nu}\partial_{\sigma}\partial_{\rho}h^{\sigma\rho}+\partial^{\mu}\partial^{\nu}h-\eta^{\mu\nu}\square h
$$
assume the form
\begin{equation}\label{fieldeqmomentum}
\begin{aligned}
\frac{\kappa}{k^{2}}\tau_{\mu\nu} &=  \left(\eta_{\mu\rho}\eta_{\nu\sigma}+\eta_{\rho\sigma}\frac{k_{\mu}k_{\nu}}{k^{2}}-\eta_{\mu\sigma}\frac{k_{\nu}k_{\rho}}{k^{2}}-\eta_{\nu\sigma}\frac{k_{\mu}k_{\rho}}{k^{2}}+\eta_{\mu\nu}\frac{k_{\rho}k_{\sigma}}{k^{2}}-\eta_{\mu\nu}\eta_{\rho\sigma}\right)h^{\rho\sigma} \\
&= \Bigl(\eta_{\mu\rho}\eta_{\nu\sigma}+\eta_{\rho\sigma}\omega_{\mu\nu}-\eta_{\mu\sigma}\omega_{\nu\rho}-\eta_{\nu\sigma}\omega_{\mu\rho}+\eta_{\mu\nu}\omega_{\rho\sigma}-\eta_{\mu\nu}\eta_{\rho\sigma}\Bigr)h^{\rho\sigma}.
\end{aligned}
\end{equation}
This equation can be expressed in terms of the spin projector operators.
An equivalent formulation of Eq. \ref{fieldeqmomentum} is
\begin{equation}\label{kappak}
\frac{\kappa}{k^{2}}\tau_{\mu\nu} = \left[{\displaystyle \frac{1}{2}}\left(\eta_{\mu\rho}\eta_{\nu\sigma}+\eta_{\mu\sigma}\eta_{\nu\rho}\right)+\left(\eta_{\rho\sigma}\omega_{\mu\nu}+\eta_{\mu\nu}\omega_{\sigma\rho}\right)-\left(\eta_{\mu\nu}\eta_{\rho\sigma}\right)-{\displaystyle \frac{1}{2}\left(\eta_{\mu\rho}\omega_{\nu\sigma}+\eta_{\mu\sigma}\omega_{\nu\rho}+\eta_{\nu\sigma}\omega_{\mu\rho}+\eta_{\nu\rho}\omega_{\mu\sigma}\right)}\right]h^{\rho\sigma},
\end{equation}
which allows us to write the RHS in terms of spin projection operators 
\begin{equation}\label{kappak2}
\bal
\frac{\kappa}{k^{2}}\tau_{\mu\nu} &= \Big[\left(\mathcal{P}^{2}+\mathcal{P}^{1}+\mathcal{P}_{s}^{0}+\mathcal{P}_{w}^{0}\right)+\left(\sqrt{3}\left(\mathcal{P}_{sw}^{0}+\mathcal{P}_{ws}^{0}\right)+2\mathcal{P}_{w}^{0}\right) \\
&\ \ \ \ \ -\left(3\mathcal{P}_{s}^{0}+\mathcal{P}_{w}^{0}+\sqrt{3}\left(\mathcal{P}_{sw}^{0}+\mathcal{P}_{ws}^{0}\right)\right)-\left(\mathcal{P}^{1}+2\mathcal{P}_{w}^{0}\right)\Big]_{\mu\nu\rho\sigma}h^{\rho\sigma} \\
&= \left(\mathcal{P}^{2}-2\mathcal{P}_{s}^{0}\right)_{\mu\nu\rho\sigma}h^{\rho\sigma},
\eal
\end{equation}
by using the following relations 
\bse
\begin{align}
	 \frac{1}{2}\left(\eta_{\mu\rho}\eta_{\nu\sigma}+\eta_{\mu\sigma}\eta_{\nu\rho}\right) \label{firsteq} &=\left(\mathcal{P}^{2}+\mathcal{P}^{1}+\mathcal{P}_{s}^{0}+\mathcal{P}_{w}^{0}\right)_{\mu\nu\rho\sigma},\\
\eta_{\mu\nu}\omega_{\sigma\rho}+\eta_{\rho\sigma}\omega_{\mu\nu} 
&=\left(\sqrt{3}\left(\mathcal{P}_{sw}^{0}+\mathcal{P}_{ws}^{0}\right)+2\mathcal{P}_{w}^{0}\right)_{\mu\nu\rho\sigma},\\
 	\frac{1}{2}\left(\eta_{\mu\rho}\omega_{\nu\sigma}+\eta_{\mu\sigma}\omega_{\nu\rho}+\eta_{\nu\sigma}\omega_{\mu\rho}+\eta_{\nu\rho}\omega_{\mu\sigma}\right)
 &=\left(\mathcal{P}^{1}+2\mathcal{P}_{w}^{0}\right)_{\mu\nu\rho\sigma},\\
	\eta_{\mu\nu}\eta_{\rho\sigma}&=\left(3\mathcal{P}_{s}^{0}+\mathcal{P}_{w}^{0}+\sqrt{3}\left(\mathcal{P}_{sw}^{0}+\mathcal{P}_{ws}^{0}\right)\right)_{\mu\nu\rho\sigma},
\end{align}
\ese
where Eq. \ref{firsteq} was proved in appendix \ref{appendixspinproj1} and the remaining equality can be proved straightforwardly by substitution.
Eq. \ref{kappak2} shows that $\mathcal{P}^{2}$ and $\mathcal{P}_{s}^{0}$  are the only spin projector operators occurring in the field equation.
\nnn
Note that to rewrite the field equations in terms of the spin projector
operators, we have also rewritten the operator $\mathcal{O}$ in Eq. \ref{eq:42}
in terms of them. From the Lagrangian Eq. \ref{eq:41}, the
associated field equations in momentum-space can be shown to satisfy
\begin{equation}
\mathcal{O}_{\mu\nu\rho\sigma}h^{\rho\sigma}=\kappa\tau_{\mu\nu}.\label{eq:125 field equation}
\end{equation}
Equations \ref{kappak2} and \ref{eq:125 field equation} therefore
entail that 
\begin{equation}\label{O}
\mathcal{O}_{\mu\nu\rho\sigma}=k^{2}\left(\mathcal{P}^{2}-2\mathcal{P}_{s}^{0}\right)_{\mu\nu\rho\sigma}.
\end{equation}
Let us note that the kinetic operator $\mathcal{O}_{\mu\nu\rho\sigma}$ contains both spin-$2$ and spin-$0$ components.
Thus, in total we have $5+1=6$ degrees of freedom: five spin-$2$ and one
spin-$0$ components. Unfortunately, this kinetic operator cannot be inverted because not all spin projection operators have non-zero coefficient in analogy to  the photon propagator in QED.
It will now be discussed how the GR propagator can be obtained and that all spin projection operators except for
$\mathcal{P}^{2}$ and $\mathcal{P}_{s}^{0}$
 drop out
due to gauge invariance.
\nnn

We know that the symmetric four-rank tensor operator which occurred in the parity-invariant Lagrangian density of General Relativity
\begin{equation}\label{key}
\mathcal{L} = {1 \over 2} h_{\mu \nu} \mathcal{O}^{\mu \nu \rho \sigma} h_{\rho \sigma}
\end{equation}
can be decomposed in the spin projection operators which were introduced in the previous Appendix. This entails, temporarily omitting indices, that
\begin{equation}\label{decompp}
\mathcal{O}= \mathcal{A} \mathcal{P}^{2} + \mathcal{P}^{1} + \mathcal{P}_{s}^{0} + \mathcal{P}_{w}^{0} + \mathcal{P}_{sw}^{0}  +\mathcal{P}_{ws}^{0}.
\end{equation}
Comparing Eq. \ref{decompp} with the kinetic operator that was obtained for general relativity, i.e. Eq. \ref{OO}, shows that
\begin{equation}
\begin{array}{ccc}
A=k^{2}, & B=0, & C=-2k^{2},\\
\\
D=0, & E=0, & F=0,
\end{array}
\end{equation}
i.e. only two coefficients are non-zero. Due to the fact that $B,D,E,F$ are zero, the kinetic operator cannot be inverted - a familiar problem since the photon propagator suffers from the same issue. The remedy for this mathematical obstable is identical; adding a gauge fixing term to the Lagrangian.
This term ensures that we get rid of spurious degrees of freedom. 
\nnn
The details of this procedure are not discussed in this paper, since the main focus of this thesis is thermal field theory, an elaborate discussion can be found in \emph{e.g.} \ref{luca}. The important aspects are as follows
\begin{itemize}
	\item {A de Donder gauge fixing term ($\mathcal{L}_{\text{gf}}$) is added to $\mathcal{L}_{\text{EH}}$, where
		\begin{equation}
		\mathcal{L}_{\text{gf}}  = -\frac{1}{2\alpha}\left(\partial_{\rho}h_{\mu}^{\rho}-\frac{1}{2}\partial_{\mu}h\right)\left(\partial_{\sigma}h^{\mu\sigma}-\frac{1}{2}\partial^{\mu}h\right),
		\end{equation}
		where $\alpha$ is the $\mathit{gauge}$ $\mathit{parameter}.$ Now we define $\tilde{\mathcal{O}} = \mathcal{O} + \mathcal{O}_{\text{gf}}$ and the new Lagrangian $\tilde{\mathcal{L}}_{\text{HE}}$ can be written as
		\begin{equation}\label{key}
		\tilde{\mathcal{L}}_{\text{HE}} = \frac{1}{2}h_{\mu \nu} \tilde{\mathcal{O}}^{\mu \nu \rho \sigma} h_{\rho \sigma}
		\end{equation}
		where $\tilde{\mathcal{O}}$ can be expressed in terms of spin projection operators
		\begin{equation}
		\tilde{\mathcal{O}}=k^{2}\left[\mathcal{P}^{2}+\frac{1}{\alpha}\mathcal{P}^{1}+\left(\frac{3}{2\alpha}-2\right)\mathcal{P}_{s}^{0}+\frac{1}{2\alpha}\mathcal{P}_{w}^{0}-\frac{\sqrt{3}}{2\alpha}\mathcal{P}_{sw}^{0}-\frac{\sqrt{3}}{2\alpha}\mathcal{P}_{ws}^{0}\right].
		\end{equation}
	}
	\item {Since all spin projection operators have non-zero coefficients, the kinetic operator has become invertible. Therefore we have obtained the propagator of General Relativity: 
		\begin{equation}
		\Pi_{GR}\equiv\tilde{\mathcal{O}}^{-1}=\frac{1}{k^{2}}\left[\mathcal{P}^{2}+\alpha\mathcal{P}^{1}-\frac{1}{2}\mathcal{P}_{s}^{0}+\frac{4\alpha-3}{2}\mathcal{P}_{w}^{0}-\frac{\sqrt{3}}{2}\mathcal{P}_{sw}^{0}-\frac{\sqrt{3}}{2}\mathcal{P}_{ws}^{0}\right],
		\end{equation}
		which simplifies greatly in the Feynman gauge, \emph{i.e.} $\alpha = 1$.}
	\item {We are particularly interested in the
		physical part of the propagator, i.e. the gauge-independent part. This is because the gauge-dependent
		part is not physical, whereas the gauge-independent part is physical and is needed to calculate, among other things, scattering
		amplitudes. This physical part of the propagator is often referred to as the \textit{saturated propagator},
		or \textit{sandwiched propagator}, because it corresponds to the sandwich of the propagator
		between two conserved currents.

		Since
		$	\mathcal{P}_{\mu\nu\rho\sigma}^{1}\tau^{\rho\sigma}= 0,
		\mathcal{P}_{w,\,\mu\nu\rho\sigma}^{0}\tau^{\rho\sigma}=  \mathcal{P}_{sw,\,\mu\nu\rho\sigma}^{0}\tau^{\rho\sigma}=  \tau^{\mu\nu}\mathcal{P}_{ws,\,\mu\nu\rho\sigma}^{0}= 0$, 
		whereas $\mathcal{P}^{2}$ and $\mathcal{P}_{s}^{0}$ give non-vanishing contributions when acting on the energy-momentum tensor, one yields
		\begin{equation}
		\bal
		&\ \ \ \   \tau^{\mu\nu}(-k)\frac{1}{k^{2}}\left[\mathcal{P}^{2}+\alpha\mathcal{P}^{1}-\frac{1}{2}\mathcal{P}_{s}^{0}+\frac{4\alpha-3}{2}\mathcal{P}_{w}^{0}-\frac{\sqrt{3}}{2}\mathcal{P}_{sw}^{0}-\frac{\sqrt{3}}{2}\mathcal{P}_{ws}^{0}\right]_{\mu\nu\rho\sigma}\tau^{\rho\sigma}(k)  \\
		&=\tau^{\mu\nu}(-k)\frac{1}{k^{2}}\left(\mathcal{P}^{2}-\frac{1}{2}\mathcal{P}_{s}^{0}\right)_{\mu\nu\rho\sigma}\tau^{\rho\sigma}(k).
		\eal
		\end{equation}
		We therefore conclude that the physical part of the graviton propagator in GR is
		\begin{equation}\label{key}
		\Pi_{\text{GR}} = \frac{1}{k^2} \left(\mathcal{P}^2 - \frac{1}{2} \mathcal{P}_{s}^{0} \right).
		\end{equation}

	}
\end{itemize}

\chapter{Infinite derivative theories of gravity \label{gravterms}}
\section{Derivation of the quadratic action for gravity
\label{appendixquadraticaction}
}
We have discussed in section \ref{sectionIDG} that the most general, torsion-free, parity conserving, quadratic action is given by
\begin{equation}\label{bkgmresult}
S_{Q}^{(2)} = \int d^{4}x \  
\left[	
\mathcal{R}^{(1)} \mathcal{F}_1 (\square) \mathcal{R}^{(1)} 
+ \mathcal{R}_{\mu \nu}^{(1)}  \mathcal{F}_{2} (\square) \mathcal{R}^{(1)\mu \nu}  
+ \mathcal{R}_{\mu \nu \lambda \sigma}^{(1)} \mathcal{F}_{3} (\square) \mathcal{R}^{(1) \mu \nu \lambda \sigma}
\right]. 
\end{equation}
and can be appended to the Einstein-Hilbert action. The Einstein Hilbert action has been written in terms of the perturbed metric tensor $h_{\mu \nu}$ in Eq. \ref{LaggHE} and the result is repeated here for the reader's convenience
\begin{equation}\label{key}
\begin{aligned}
S_{\text{EH}} =
\int d^{4}x
\left[
- \frac{1}{2} h_{\mu \nu} \square h^{\mu \nu} 
+h_{\sigma}^{\mu} \partial^{\sigma} \partial^{\nu} h_{\mu}^{\nu}
+h_{\mu \nu} \square^{2} h^{\mu \nu} - h \partial^{\mu} \partial^{\nu} h_{\mu \nu}
\right].
\end{aligned}
\end{equation}
The quadratic contributions from $S_{Q}^{(2)}$ will now be explicitly computed and we aim to write these contributions in terms of $h^{\mu \nu}$ as well
\begin{equation}
\bal
\mathcal{R} \mathcal{F}_1(\square) \mathcal{R} &=
\left( \partial_{\mu} \partial_{\nu} h^{\mu \nu}  - \square  h \right) 
\mathcal{F}_1 (\square)
\left( \partial_{\alpha} \partial_{\beta} h^{\alpha \beta} - \square h \right) \\
&=
\partial_{\mu} \partial_{\nu} h^{\mu \nu} 
\mathcal{F}_1 (\square)
\partial_{\alpha} \partial_{\beta} h^{\alpha \beta} 
- \partial_{\mu} \partial_{\nu} h^{\mu \nu}  
\mathcal{F}_1 (\square)
\square h  
- \square  h 
\mathcal{F}_1 (\square)
\partial_{\alpha} \partial_{\beta} h^{\alpha \beta}
+ \square  h 
\mathcal{F}_1 (\square)
\square h \\
&=
h^{\mu \nu} 
\mathcal{F}_1 (\square)
\partial_{\mu} \partial_{\nu}
\partial_{\alpha} \partial_{\beta} h^{\alpha \beta} 
-  h^{\mu \nu}  
\mathcal{F}_1 (\square)
\square 
\partial_{\mu} \partial_{\nu} h  
-   h 
\mathcal{F}_1 (\square)
\square \partial_{\alpha} \partial_{\beta} h^{\alpha \beta}
+   h 
\mathcal{F}_1 (\square)
\square^{2} h \\
&= \mathcal{F}_1 (\square) \left[
h^{\mu \nu} 
\partial_{\mu} \partial_{\nu}
\partial_{\alpha} \partial_{\beta} h^{\alpha \beta} 
-   2h 
\square \partial_{\alpha} \partial_{\beta} h^{\alpha \beta}
+   h 
\square^{2} h
\right],
\eal
\end{equation} 
\begin{equation}
\bal
\mathcal{R}_{\mu \nu} \mathcal{F}_2(\square) \mathcal{R}^{\mu \nu} &= \frac{1}{4}  \left(
\partial_{\lambda}\partial_{\mu} h_{\nu}^{\lambda} 
- \square h_{\mu \nu}  
+ \partial_{\lambda}\partial_{\nu} h_{\mu}^{\lambda} 
- \partial_{\mu}\partial_{\nu} h \right)
\mathcal{F}_2 (\square)
\left(
\partial_{\rho}\partial^{\mu} h^{ \rho \nu} 
- \square h^{\mu \nu}  
+ \partial_{\rho}\partial^{\nu} h^{\rho \mu} 
- \partial^{\mu}\partial^{\nu} h \right) \\
&= \frac{1}{4} \mathcal{F}_2 (\square) \Big[
\partial_{\lambda}\partial_{\mu} h_{\nu}^{\lambda} \partial_{\rho}\partial^{\mu} h^{ \rho \nu}  
- \partial_{\lambda}\partial_{\mu} h_{\nu}^{\lambda} \square h^{\mu \nu}  
+ \partial_{\lambda}\partial_{\mu} h_{\nu}^{\lambda} \partial_{\rho}\partial^{\nu} h^{\rho \mu} 
- \partial_{\lambda}\partial_{\mu} h_{\nu}^{\lambda} \partial^{\mu}\partial^{\nu} h \\
& \ \ \ \ \ \ \ \ \ \ \ \ \ \ \ - \square h_{\mu \nu}  \partial_{\rho}\partial^{\mu} h^{ \rho \nu} 
+ \square h_{\mu \nu}  \square h^{\mu \nu}  
-\square h_{\mu \nu}  \partial_{\rho}\partial^{\nu} h^{\rho \mu} 
+ \square h_{\mu \nu}  \partial^{\mu}\partial^{\nu} h \\
& \ \ \ \ \ \ \ \ \ \ \ \ \ \ \ + \partial_{\lambda}\partial_{\nu} h_{\mu}^{\lambda} \partial_{\rho}\partial^{\mu} h^{ \rho \nu} 
- \partial_{\lambda}\partial_{\nu} h_{\mu}^{\lambda} \square h^{\mu \nu} 
+ \partial_{\lambda}\partial_{\nu} h_{\mu}^{\lambda} \partial_{\rho}\partial^{\nu} h^{\rho \mu} 
- \partial_{\lambda}\partial_{\nu} h_{\mu}^{\lambda} \partial^{\mu}\partial^{\nu} h \\
& \ \ \ \ \ \ \ \ \ \ \ \ \ \ \ - \partial_{\mu}\partial_{\nu} h \partial_{\rho}\partial^{\mu} h^{ \rho \nu} 
+ \partial_{\mu}\partial_{\nu} h \square h^{\mu \nu}  
- \partial_{\mu}\partial_{\nu} h  \partial_{\rho}\partial^{\nu} h^{\rho \mu} 
+ \partial_{\mu}\partial_{\nu} h \partial^{\mu}\partial^{\nu} h \Big], \\
&= \frac{1}{2}\mathcal{F}_{2} (\square)
\left[
\frac{1}{2} h \square^2 h  + \frac{1}{2} h_{\mu \nu} \square^2 h^{\mu \nu} - h \square \partial_{\mu } \partial_{\nu} h^{\mu \nu}
- h_{\mu}^{\rho} \square \partial_{\rho} \partial_{\nu} h^{\mu \nu} + h^{\lambda \sigma} 
\partial_{\sigma}
\partial_{\lambda}
\partial_{\mu}
\partial_{\nu} h^{\mu \nu}
\right],
\eal
\end{equation}
\begin{equation}
\bal
\mathcal{R}_{\mu \nu \lambda \sigma} \mathcal{F}_3(\square) \mathcal{R}^{\mu \nu \lambda \sigma} &=
\frac{1}{4}\left[\partial_{\nu}\partial_{\lambda}h_{\mu\sigma}+\mathcal{\partial_{\mu}\partial_{\sigma}}h_{\nu\lambda}-\mathcal{\partial_{\sigma}\partial_{\nu}}h_{\mu\lambda}-\partial_{\mu}\partial_{\lambda}h_{\nu\sigma}\right]\mathcal{F}_{3}(\square) \\ 
&\ \ \times\left[\partial^{\nu}\partial^{\lambda}h^{\mu\sigma}+\mathcal{\partial^{\mu}\partial^{\sigma}}h^{\nu\lambda}-\mathcal{\partial^{\sigma}\partial^{\nu}}h^{\mu\lambda}-\partial^{\mu}\partial^{\lambda}h^{\nu\sigma}\right]\\
&= \mathcal{F}_{3} (\square)
\left[
h_{\mu \nu} \square^{2} h^{\mu \nu} 
+ h^{\lambda \sigma} \partial_{\sigma} \partial_{\lambda} \partial_{\mu} \partial_{\nu} h^{\mu \nu}
- h_{\mu}^{\rho} \square \partial_{\rho} \partial_{\nu} h^{\mu \nu} 
\right].
\eal
\end{equation}
We can now construct the general action which is quadratic in $h_{\mu \nu}$
\begin{equation}\label{key}
\begin{aligned}
S_{Q} = - \int d^{4}x \ \ &  \Bigg[
\frac{1}{2} h_{\mu \nu} \square \Big[
1 - \frac{1}{2} \mathcal{F}_{2} (\square) \square  - 2 \mathcal{F}_{3} (\square) \square 
\Big] h^{\mu \nu} \\
& - \frac{1}{2} h \square \Big[
1 + 2 \mathcal{F}_1 (\square) \square + \frac{1}{2} \mathcal{F}_2 (\square) \square
\Big] h \\
&+ 
h  \Big[
1 + 2\mathcal{F}_{1} (\square) \square + \frac{1}{2} \mathcal{F}_2 (\square)  \square 
\Big]\partial_{\mu} \partial_{\nu} h^{\mu \nu} \\
&+ 
h_{\mu}^{\sigma} \Big[
-1 + \frac{1}{2} \mathcal{F}_{2} (\square ) \square + 2 \mathcal{F}_{3} (\square ) \square 		
\Big]\partial_{\sigma} \partial_{\nu} h^{\mu \nu} \\
&+ \frac{1}{2} h^{\lambda \sigma}  \Big[
-2 \mathcal{F}_{1} (\square)  - \mathcal{F}_{2} (\square)  - 2 \mathcal{F}_{3} (\square) 
\Big] \partial_{\sigma}
\partial_{\lambda}
\partial_{\mu}
\partial_{\nu}
h^{\mu \nu}
\Bigg].
\eal
\end{equation}
Let us use the following abbreviations for the combinations of $\mathcal{F}(\square)$'s:
\begin{equation}\label{constraints}
\begin{aligned}
& a(\square) := 1 - \frac{1}{2} \mathcal{F}_{2} (\square) \square  - 2 \mathcal{F}_{3} (\square) \square, \\& 
b(\square) := -1 + \frac{1}{2} \mathcal{F}_{2} (\square) \square  + 2 \mathcal{F}_{3} (\square) \square, \\
& c(\square) := 1 + 2 \mathcal{F}_{1} (\square) \square  + \frac{1}{2} \mathcal{F}_{2} (\square) \square, \\
& d(\square) := -1 - 2 \mathcal{F}_{1} (\square) \square  - \frac{1}{2} \mathcal{F}_{3} (\square) \square, \\
& e(\square) := -2 \mathcal{F}_{1} (\square)  - \mathcal{F}_{2} (\square)  - 2 \mathcal{F}_{3} (\square) . \\
\end{aligned}
\end{equation}
The function $e(\square)$ only occurs in higher derivative theories as was pointed out in Ref. \ref{singfree}. The term containing $e(\square)$ can be written more conveniently
$$ \frac{1}{2} h^{\lambda \sigma}  e(\square) \partial_{\sigma}
\partial_{\lambda}
\partial_{\mu}
\partial_{\nu}
h^{\mu \nu} 
=
\frac{1}{2} 
\partial_{\sigma}
\partial_{\lambda}
h^{\lambda \sigma}  e(\square) 
\partial_{\mu}
\partial_{\nu}
h^{\mu \nu}.
$$
The quadratic action can now be rewritten as:
\begin{equation}\label{actioninf}
\begin{aligned}
S_{Q} = - \int d^{4}x \ \ &  \Bigg[
\frac{1}{2} h_{\mu \nu} \square a(\square) h^{\mu \nu} 
+ 
h_{\mu}^{\sigma} b(\square) \partial_{\sigma} \partial_{\nu} h^{\mu \nu}
+
h  c(\square) \partial_{\mu} \partial_{\nu} h^{\mu \nu} \\
& \ + 
\frac{1}{2} h \square d(\square) h 
+ \frac{1}{2} h^{\lambda \sigma}  e(\square) \partial_{\sigma}
\partial_{\lambda}
\partial_{\mu}
\partial_{\nu}
h^{\mu \nu}
\Bigg].
\end{aligned}
\end{equation}
It can be deduced from Eq. \ref{constraints} that the following combinations of analytic functions vanish
\begin{equation}\label{equalities}
\begin{cases}
a(\square) + b(\square) = 0, \\
c(\square) + f(\square) = 0, \\
b(\square) + c(\square) + e(\square) = 0. \\
\end{cases}
\end{equation}
The term $e(\square)$ only occurs in higher order theories , so the third equality reduces to $b(\square) + c(\square) = 0$
 for the infinite-derivative theories of gravity which we aim to construct.
This implies that the functions $\mathcal{F}$'s satisfy
\begin{equation}\label{f1f2}
2 \mathcal{F}_{1}(\square) +  \mathcal{F}_{2}(\square) + 2 \mathcal{F}_{3}(\square) = 0.
\end{equation}
It was through explicit evaluation of the respective terms that these identities were found originally. These identities can however
be understood as entailments of the Bianchi identities, the reader is referred to \ref{conroy} for an elaborate discussion and derivation of these constraints. \nnn
We are free to specify the three function $\mathcal{F}_i(\square)$ as long as Eq. 
\ref{f1f2} is satisfied. Note that
\begin{equation}\label{key}
\mathcal{F}_3(\square)
=
- \left[ \mathcal{F}_1(\square) 
+ \frac{1}{2} \mathcal{F}_2(\square)
\right]
\end{equation}
so the action can be written as
\begin{equation}\label{key}
S = \int d^4x \left[ \frac{\mathcal{R}}{2} 
+  \mathcal{R} 
 \mathcal{F}_1(\square) 
\mathcal{R} 
+
\mathcal{R}_{\mu \nu }
\mathcal{F}_2(\square) \mathcal{R}^{\mu \nu} 
-
\mathcal{R}_{\mu \nu \rho \sigma }
\left[ \mathcal{F}_1(\square) 
+ \frac{1}{2} \mathcal{F}_2(\square)
\right] \mathcal{R}^{\mu \nu  \rho \sigma } 
\right].
\end{equation}
We can in fact get rid of the $\mathcal{F}_3(\square)$ when working around Minkowski space.
The covariant derivatives become simple
partial derivatives, so we can move the function
$\mathcal{F}_i(\square)$ on the left of the curvatures by
using integrating by parts. Using this methodology, we can get rid of the product of two Riemann tensors by implementing the the Euler topological invariant, which was introduced in Eq. \ref{eulerinvariant} and reads
\begin{equation}
\mathcal{R}^{\mu\nu\rho\sigma}\mathcal{R}_{\mu\nu\rho\sigma}-4\mathcal{R}^{\mu\nu}\mathcal{R}_{\mu\nu}+\mathcal{R}^{2}=\nabla_{\mu}K^{\mu}
\end{equation}
Without loss of generality, we choose the convenient constraint
$\mathcal{F}_3(\square) = 0$, but this only holds true up to the perturbative $\mathcal{O}(h^2)$ level.
\section{Field equations
\label{appendixidgfield}
}

There are no contributions from single derivatives of the field $h_{\mu \nu}$ to the Euler Lagrange equation, \emph{i.e.}
\begin{subequations}
\begin{equation}\label{key}
\partial_{\alpha} \frac{\partial \mathcal{L}_{q}}{\partial ( \partial_{\alpha} h^{\mu \nu} )} = 0.
\end{equation}
Since the Lagrangian density contains higher order derivatives, we also need to consider the functional derivatives
 with respect to the second derivatives of the field $h_{\mu \nu}$ in the Euler-Lagrange equations.
\begin{equation}\label{key}
 \partial_{\alpha} \frac{\partial \mathcal{L}_{q}}{\partial ( \partial_{\alpha} h^{\mu \nu} )} -  \partial_{\alpha}  \partial_{\beta} \frac{\partial \mathcal{L}_{q}}{\partial ( \partial_{\alpha} \partial_{\beta} h^{\mu \nu} )} 
 =
 \frac{\partial \mathcal{L}_{q}}{\partial h^{\mu \nu}} .
 \end{equation}
 We obtain
 \begin{align}	
\begin{aligned}
	\frac{\partial \mathcal{L}_{q}}{\partial ( \partial_{\alpha} \partial_{\beta} h^{\mu \nu} )}  &= 
	-
	\frac{\partial }{\partial ( \partial_{\alpha} \partial_{\beta} h^{\mu \nu} )} 
	\bigg[
	\frac{1}{2} h_{\mu \nu} \eta^{\alpha \beta} \partial_{\alpha} \partial_{\beta} a(\square) h^{\mu \nu} 
	+ 
	h_{\mu}^{\sigma} b(\square) \partial_{\sigma} \partial_{\nu} h^{\mu \nu}
	+
	h_{\rho \omega} \eta^{\rho \omega}  c(\square) \partial_{\mu} \partial_{\nu} h^{\mu \nu} \\
	& \ \ \ \ \ \ \ \ \ \ \ \ \ \ \ \ \ \ \ \ \ \ \ + 
	\frac{1}{2} h_{\rho \omega} \eta^{\rho \omega} \eta^{\alpha \beta} \partial_{\alpha} \partial_{\beta} d(\square) \eta_{\mu \nu} h^{\mu \nu}
	+ \frac{1}{2} 
	\partial_{\sigma}
	\partial_{\lambda}
	h^{\lambda \sigma} 
	 e(\square) 
	\partial_{\mu}
	\partial_{\nu}
	h^{\mu \nu} \bigg] \\
	&= 
-	\frac{1}{2} h_{\mu \nu} \eta^{\alpha \beta}  a(\square)  
	-
	h_{\mu}^{\alpha} b(\square)  
	\delta_{\nu}^{\beta}
	-
	h_{\rho \omega} \eta^{\rho \omega}  c(\square) \delta_{\mu}^{\alpha} \delta_{\nu}^{\beta}  \\
	& \ \ \ - 
	\frac{1}{2} h_{\rho \omega} \eta^{\rho \omega} \eta^{\alpha \beta}  d(\square) \eta_{\mu \nu} 
\underbrace{	- \frac{1}{2} 
	\partial_{\mu}
	\partial_{\nu}
	h^{\alpha \beta} e(\square)
	- \frac{1}{2} 
	\partial_{\mu}
	\partial_{\nu}
	h^{\alpha \beta} e(\square) }_{= -  
	\partial_{\mu}
	\partial_{\nu}
	h^{\alpha \beta} e(\square) }  \\
	\end{aligned}
\end{align}
and
\begin{equation}\label{Lq}
\begin{aligned}
\partial_{\alpha} \partial_{\beta}
\frac{\partial \mathcal{L}_{q}}{\partial ( \partial_{\alpha} \partial_{\beta} h^{\mu \nu} )}
	&= 
-	\frac{1}{2} a(\square) h_{\mu \nu} \square   
-
b(\square)
\partial_{\alpha} \partial_{\nu}
h_{\mu}^{\alpha}   
-
c (\square) \partial_{\mu} \partial_{\nu}
h_{\rho \omega} \eta^{\rho \omega} - 
\frac{1}{2} d (\square)  \eta_{\mu \nu} \eta^{\rho \omega}
\square h_{\rho \omega} 
 -  e(\square)
 \partial_{\alpha}
 \partial_{\beta}
	\partial_{\mu}
	\partial_{\nu}
	h^{\alpha \beta}.
\end{aligned}
\end{equation}
Lastly, we expect contributions from $\frac{
\partial \mathcal{L}_{q}}{\partial h^{\mu \nu}}$ as well.

\begin{equation}
\frac{
	\partial \mathcal{L}_{q}}{\partial h^{\mu \nu}}
=
-\frac{1}{2} a (\square) \square h_{\mu \nu}
- b(\square) \partial_{\mu} \partial_{\alpha}
h_{\nu}^{\alpha}
- c(\square) \eta_{\mu \nu} \partial_{\alpha} \partial_{\beta} h^{\alpha \beta}
- \frac{1}{2} d(\square) \eta_{\mu \nu}
\eta^{\rho \sigma}
\square h_{\rho \sigma}
\end{equation}
\end{subequations}
From Eqs. [\ref{Lq}-e] we obtain
\begin{equation}\label{fieldequations}
	 a(\square) \square    h_{\mu \nu} 
+
b(\square)
\partial_{  \alpha} \partial_{ ( \nu }
h_{\mu )}^{\alpha}   
+
c (\square) \left( \partial_{\mu} \partial_{\nu}
 \eta_{\rho \omega}  h^{\rho \omega} + \partial_{\mu} \partial_{\nu} h \right)   + 
 d (\square)  \eta_{\mu \nu} 
\square h
+  e(\square)
\partial_{\alpha}
\partial_{\beta}
\partial_{\mu}
\partial_{\nu}
h^{\alpha \beta} = 0.
\end{equation}
If there is also a matter contribution in the action, described by energy-momentum tensor
of matter $\tau_{\mu \nu}$ , we have to add the term $- \kappa \tau_{\rho \sigma} h^{\rho \sigma}$ to the RHS of the Lagrangian density which is derived from Eq. \ref{actioninf}. Now 
\begin{equation}\label{key}
 \frac{\partial \mathcal{L}_{q}}{\partial  h^{\mu \nu} } = -\kappa \tau_{\mu \nu}.
\end{equation}

\subsubsection*{Bianchi identities}

We can prove that the energy-momentum tensor is conserved because of the
generalized Bianchi identity \ref{koivisto} due to diffeomorphism invariance:
\begin{equation}\label{key}
\nabla_{\mu} \tau_{\nu}^{\mu} = \nabla^{\mu} \tau_{\mu \nu} = 0.
\end{equation}
If we let the covariant derivative act on both sides of 
Eq. \ref{fieldequations} we obtain
\begin{equation}\label{key}
\begin{aligned}
- \kappa \nabla_{\mu} \tau_{\nu}^{\mu} &= 0 \\
&= \partial^{\mu} \big(	 a(\square) \square    h_{\mu \nu}   
+
b(\square)
\left(
\partial_{ \alpha} \partial_{\nu }
h_{\mu}^{\alpha}   
+
\partial_{  \alpha} \partial_{\mu }
h_{\nu}^{\alpha}   
\right)
+
c (\square) \left( \partial_{\mu} \partial_{\nu}
\eta_{\rho \omega}  h^{\rho \omega} + \partial_{\mu} \partial_{\nu} h \right)  \\ 
& \ \ \ +
d (\square)  \eta_{\mu \nu} 
\square h
+  e(\square)
\partial_{\alpha}
\partial_{\beta}
\partial_{\mu}
\partial_{\nu}
h^{\alpha \beta} \big) \\
&= 	 a(\square) \square \partial_{\mu}   h_{\nu}^{\mu}   
+
b(\square)
\left(
\partial_{\mu} \partial_{ \alpha} \partial_{\nu }
h^{\alpha \mu}   
+
\partial_{  \alpha} \square
h_{\nu}^{\alpha}   
\right)
+
c (\square) \left( \square \partial_{\nu}
\eta_{\rho \omega}  h^{\rho \omega} + \square \partial_{\nu} h \right)  \\ 
&\ \ \ + 
d (\square)  \partial_{ \nu} 
\square h
+  e(\square)
\partial_{\alpha}
\partial_{\beta}
\square
\partial_{\nu}
h^{\alpha \beta}.  \\
\end{aligned}
\end{equation}
We thus obtain
\begin{equation}\label{key}
\left( a(\square) + b(\square) \right) \square \partial_{\mu}   h_{\nu}^{\mu}   
+
\left(
b(\square)
+
c(\square)
+
e(\square)
\right)
\partial_{\nu} \partial_{ \alpha} \partial_{\mu }
h^{\alpha \mu}     
+
\left(c (\square)
+
d (\square) \right)  \square \partial_{\nu} h    = 0.  \\
\end{equation}
Keeping in mind that $a,b,c,d,e$ are functions of the d'Alembertian, we can write
\begin{equation}\label{key}
\left( a + b \right) \square \partial_{\mu}   h_{\nu}^{\mu}   
+
\left(
b
+
c
+
e
\right)
h^{\alpha \mu}_{, \nu \alpha \mu}   
+
\left(c
+
d  \right)  \square  h_{, \nu}    = 0.  \\
\end{equation}
We now realize that the equalities obtained in Eq. \ref{equalities} were in fact consequences of Bianchi identities.

\subsubsection*{Graviton propagator}

The field equations can be written in the form
\begin{equation}\label{key}
\Pi_{\mu \mu}^{-1 \lambda \sigma} h_{\sigma \lambda} = \kappa \tau_{\mu \nu},
\end{equation}
where $\Pi_{\mu \mu}^{-1 \lambda \sigma}$ is the inverse propagator for the graviton.
Let us express the field equation in Eq. \ref{fieldequations} in momentum space and divide both sides by $\square = -k^2$:
\begin{equation}
\bal
&a( - k^2)     h_{\mu \nu} 
+
\frac{ b(- k^2) }{k^2}
k_{  \alpha} k_{ ( \nu }
h_{\mu )}^{\alpha}   
+
\frac{ c(- k^2) }{k^2} \left( k_{\mu} k_{\nu}
\eta_{\rho \omega}  h^{\rho \omega} + k_{\mu} k_{\nu} h \right)  \\ 
&+ 
d(- k^2)   \eta_{\mu \nu}  h
+  \frac{ e(- k^2) }{k^2}
k_{\alpha}
k_{\beta}
k_{\mu}
k_{\nu}
h^{\alpha \beta} = \frac{\kappa}{k^2}.
\eal
\end{equation}
Now we want to exploit the fact that the field equations (in momentum space) can be written in terms of the spin projection operators as
$$
\begin{aligned}
a( - k^2)     h_{\mu \nu} &= a (-k^2) \left[
\mathcal{P}^{2}
+ \mathcal{P}^{1}
+\mathcal{P}_{s}^{0}
+ \mathcal{P}_{w}^{0}
\right] h, \\
- \frac{b(- k^2)}{k^2} 
k_{  \alpha} k_{ \{ \nu }
h_{\mu \}}^{\alpha} 
&= -b(-k^{2})k^{2}\left(\eta_{\mu\rho}\omega_{\nu\sigma}+\eta_{\rho\nu}\omega_{\mu\sigma}\right)h^{\rho\sigma}\\
&= -b(-k^{2}){\displaystyle k^{2}\frac{1}{2}}\left(\eta_{\mu\rho}\omega_{\nu\sigma}+\eta_{\mu\sigma}\omega_{\nu\rho}+\eta_{\nu\rho}\omega_{\mu\sigma}+\eta_{\nu\sigma}\omega_{\mu\rho}\right)h^{\rho\sigma}\\
&= -b (-k^2)  \left[
\mathcal{P}^{1} + 2 \mathcal{P}_{w}^{0}
\right] h,   \\
- \frac{ c(- k^2) }{k^2} \left( k_{\mu} k_{\nu}
\eta_{\rho \omega}  h^{\rho \omega} + k_{\mu} k_{\nu} h \right)
&=
-c(-k^{2})k^{2}\left(\eta_{\mu\nu}\omega_{\rho\sigma}h^{\rho\sigma}+\omega_{\mu\nu}\eta_{\rho\sigma}h^{\rho\sigma}\right)\\
&= -c(-k^{2})k^{2}\left(\theta_{\mu\nu}\omega_{\rho\sigma}+\omega_{\mu\nu}\omega_{\rho\sigma}+\omega_{\mu\nu}\omega_{\rho\sigma}+\omega_{\mu\nu}\theta_{\rho\sigma}\right)h^{\rho\sigma}\\
&= -c(-k^{2})k^{2}\left(2\mathcal{P}_{w}^{0}+\sqrt{3}\left(\mathcal{P}_{sw}^{0}+\mathcal{P}_{ws}^{0}\right)\right)h \\
d(- k^2)   \eta_{\mu \nu}  h &= d(-k^{2})(\theta_{\mu\nu}+\omega_{\mu\nu})\left(\theta_{\rho\sigma}+\omega_{\rho\sigma}\right)h^{\rho\sigma}\\
= & d(-k^{2})\left(\theta_{\mu\nu}\theta_{\rho\sigma}+\theta_{\mu\nu}\omega_{\rho\sigma}+\omega_{\mu\nu}\theta_{\rho\sigma}+\omega_{\mu\nu}\omega_{\rho\sigma}\right)h^{\rho\sigma}\\
= & d(-k^{2})\left(3\mathcal{P}_{s}^{0}+\mathcal{P}_{w}^{0}+\sqrt{3}\left(\mathcal{P}_{sw}^{0}+\mathcal{P}_{ws}^{0}\right)\right)h \\
  \frac{ e(- k^2) }{k^2}
k_{\alpha}
k_{\beta}
k_{\mu}
k_{\nu}
h^{\alpha \beta} 
&=
 e(- k^2) 
k^2 \left(\mathcal{P}_{w}^{0}
\right)
h  ,
\end{aligned}
$$
where indices have been suppressed.
The expressions for
$\left(\mathcal{P}^{2}+\mathcal{P}^{1}+\mathcal{P}_{s}^{0}+\mathcal{P}_{w}^{0}\right)$, $\left(\sqrt{3}\left(\mathcal{P}_{sw}^{0}+\mathcal{P}_{ws}^{0}\right)+2\mathcal{P}_{w}^{0}\right)$,  $\left(\mathcal{P}^{1}+2\mathcal{P}_{w}^{0}\right)$ and $\left(3\mathcal{P}_{s}^{0}+\mathcal{P}_{w}^{0}+\sqrt{3}\left(\mathcal{P}_{sw}^{0}+\mathcal{P}_{ws}^{0}\right)\right)$ were already given in Eqs. \ref{firsteq}-d and the expression for $e(-k^2)$ does not contribute since this term only occur in local field theories \ref{nieuwenhuizen}. The field equations for IDG thus read
\begin{equation}\label{complicatedfield}
\bal
\bigg[&a(-k^{2})\left(\mathcal{P}^{2}+\mathcal{P}^{1}+\mathcal{P}_{s}^{0}+\mathcal{P}_{w}^{0}\right)+b(-k^{2})\left(\mathcal{P}^{1}+2\mathcal{P}_{w}^{0}\right) \\
&+c(-k^{2})\left(2\mathcal{P}_{w}^{0}+\sqrt{3}\left(\mathcal{P}_{sw}^{0}+\mathcal{P}_{ws}^{0}\right)\right) 
+d(-k^{2})\left(3\mathcal{P}_{s}^{0}+\mathcal{P}_{w}^{0}+\sqrt{3}\left(\mathcal{P}_{sw}^{0}+\mathcal{P}_{ws}^{0}\right)\right)+f(-k^{2})\mathcal{P}_{w}^{0}\bigg]h \\
&=\kappa\frac{\left(\mathcal{P}^{2}+\mathcal{P}^{1}+\mathcal{P}_{s}^{0}+\mathcal{P}_{w}^{0}\right)}{k^{2}}\tau.
\eal
\end{equation}
Using the orthogonality relations of the spin projection operators of Eq. \ref{orthogonality} one yields (by multiplying Eq. \ref{complicatedfield} with $\mathcal{P}^2$)
\begin{equation}\label{key}
\mathcal{P}^{2} a(\square) h = \frac{\kappa}{\tau} \mathcal{P}^2 
\end{equation}
and likewise for $\mathcal{P}^1$
\begin{equation}\label{key}
\mathcal{P}^1 \left[a(\square) + b(\square)\right] h = \frac{\kappa}{\tau} \mathcal{P}^1.
\end{equation}
Repeating this procedure (the calculations of which can be found in \ref{classicalproperties}) one obtains the saturated propagator (keeping in mind that both $a$ and $c$ are functions of the d'Alembertian)
\begin{equation}
\tau(-k)\Pi(k)\tau(k)\equiv\tau(-k)\left(\frac{\mathcal{P}^{2}}{ak^{2}}+\frac{\mathcal{P}_{s}^{0}}{(a-3c)k^{2}}\right)\tau(k),\label{eq:203}
\end{equation}
and the gauge-independent (physical) part of the propagator reads
\begin{equation}
\Pi(k)=\frac{\mathcal{P}^{2}}{ak^{2}}+\frac{\mathcal{P}_{s}^{0}}{(a-3c)k^{2}}.
\end{equation}
Since we want to recover
general relativity in the IR, one may require from viable theories of gravity that
$a(\square)$ and $c(\square)$ are analytic around $k^2 = 0$ (hence $\square = 0$).
\nnn
Now we can derive the propagator for Infinite Derivative Gravity. Following our discussion of chapter \ref{chapterIDG} (that we do not want to introduce ghost-like DOFs) which implies that the only spin component of the graviton should be the spin-2 and spin-0 components from general relativity. We therefore require that 
\begin{itemize}
	\item{} $a(\square)$	contains no zeros.
	\item{} $a(\square) - 3c(\square)$  contains only one zero.
\end{itemize}
 In order to satisfy both conditions one should
be able to express $c(\square)$ in the following general form	
\begin{equation}\label{key}
c(\square)  = \frac{a(\square)}{3} \left[
1 + 2 \left(
1- \frac{\square}{m^2}
\right)
\mathcal{B}(\square)
\right],
\end{equation}
where $\mathcal{B}(\square)$ \emph{contains no zeros and is analytic around  $\square = 0$.} In order to circumvent the introduction of tachyonic modes one should consider theories with $m^2 > 0$. 
\nnn
Different choices for the coefficients $a(\square)$ and
$c(\square)$ can be made, all of which correspond to different theories of gravity. 
The difference between these theories hinges on the number of degrees
of freedom which are included in the physical part of the propagator, and whether they 
 massless or massive. The BGKM action is based on the the choices $m^2 \rightarrow 0$ and $a(\square) = c(\square)$ and the resulting graviton only contains
the  propagating degrees of freedom from general relativity, \emph{i.e.}
\begin{equation}\label{key}
\Pi(k) =  \frac{1}{k^2a(-k^2) } \Bigg( \mathcal{P}^2 - \frac{1}{2} \mathcal{P}_{s}^{0}		\Bigg)
=
\frac{1}{a(-k^2)} \Pi_{\text{GR}}(k	).
\end{equation}
 An interesting feature of this theory is one has only one independent function since $a(\square) = c(\square)$, which governs the modification of gravity in
the UV-regime. This type of theories is a suitable candidate for a ghost-free and renormalizable theory of gravity as is is elaborately discussed in section \ref{sectionbgkm}.
\section{Nonsingular Newtonian potential \label{appendixnewtonpotential} }
The Newtonian potential can be calculated from the field equation given in Eq. \ref{fieldequationsupdated} with a matter source $(-\kappa \tau_{\mu \nu})$ at the RHS. \ref{singfree}
In the Newtonian approximation, the potentials $\Psi$ and $\Phi$ are weak and thus satisfy $\abs{\Psi} \ll 1$ and $\abs{\Phi} \ll 1$.
\nnn
The metric is only allowed to have minute perturbations in the Newtonian potential, furthermore it is stationary and the sources are static (or, alternatively, have negligible velocities). Hence the metric
reduces to
\begin{equation}\label{key}
\bal
ds^2 &= (1 + 2\Phi)dt^2 - (1- 2\Psi) \abs{d\mathbf{x}}^2 \\
 &= \underbrace{dt^2 - \abs{d\mathbf{x}}^2}_{\eta_{\mu \nu}} + \underbrace{ \left[2\Phi dt^2 + 2\Psi \abs{d\mathbf{x}}^2  \right] }_{h_{\mu \nu}},\\	
\eal
\end{equation}
where
\begin{equation}\label{newtonhmunu}
h_{\mu \nu} = 2 \begin{pmatrix}
 \Phi & 0 & 0 & 0 \\
 0 & \Psi &  0 & 0 \\
  0 & 0 & \Psi &  0 \\
   0 & 0 & 0 & \Psi \\
\end{pmatrix}.	
\end{equation}
The assumption of the static potential ($\partial_0 h_{\mu \nu} = 0$) and Eq. \ref{newtonhmunu} entail the following useful equalities
\bse
\begin{align}
h = \eta^{\mu \nu} h_{\mu \nu} &= 2(\Phi - 3 \Psi), \label{constraintsh}\\
\partial^{\mu } \partial^{\nu} h_{\mu \nu} = \partial^{i } \partial^{j} h_{ij} &= 2 \nabla^2 \Psi, \\
\square = 	 - \delta^{ij} \partial_i \partial_j &= -\nabla^2.
\end{align}
\ese
Our aim is to find the exact expression for the metric and hence determine the appropriate form for the potentials $\Psi$ and $\Phi$.
Therefore, we consider the trace and the $00$-component of Eq. \ref{fieldequationsupdated} (with a matter source), while keeping in mind
that $\partial_0 h_{\mu \nu} = 0$ in the Newtonian approximation:
\bse
\begin{align}
a(\square) \left[
\square h - 2\partial_{\alpha} \partial_{\beta}h^{\alpha \beta} 
+ (4\partial_{\alpha} \partial_{  \beta}h^{\alpha \beta} + \square h )
-4\square h  \right] & \nonumber	 \\
= 2 a(\square) \left[
 - \square h + 2\partial_{\alpha} \partial_{\beta}h^{\alpha \beta} 
 )
 \right] &= -\kappa \rho \ \boxed{\text{trace}}	 \label{trace}\\
 a(\square) \left[
\square h_{00} + \partial_{  \alpha} \partial_{  \beta} h^{\alpha \beta} - \square h \right] &= -\kappa \rho
\ \boxed{00\text{-component}}
\end{align}
\ese
Eqs. \ref{trace},b can be further simplified by incorporating conditions \ref{constraintsh}-c:
\begin{align}
 2 a(\nabla^2) \left[
2 \nabla^2 (\Phi - 3 \Psi) + 2\nabla^2 \Psi
\right] &	\nonumber \\
= 4 a(\nabla^2) \left[
2 \nabla^2 \Phi - 3 \Psi
\right] & = -\kappa \rho \ 	\boxed{\text{trace}} \label{newtrace} \\
a(\nabla^2) \left[
-2 \nabla^2 \Phi + 2\nabla^2 \Psi + 2 \nabla^2 (\Phi - 3 \Psi)  \right] & \nonumber  \\
= -4 a(\nabla^2)  \nabla^2 \Psi 
&= -\kappa \rho
\ \boxed{00\text{-component}} \label{00comp}
\end{align}
These equations imply (by inserting Eq. \ref{00comp} into Eq. \ref{newtrace}) that the two potentials $\Psi$ and
$\Phi$ satisfy the same equation
\begin{equation}\label{newtonpot}
4 a(\nabla^2)  \nabla^2 \Psi = 4 a(\nabla^2)  \nabla^2 \Phi = \kappa \rho = \kappa m_g \delta^{(3)}(\mathbf{x}),
\end{equation}
where, let me remind you, $m_g$ is the mass of the object that generates the gravitational potential.
Fourier transforming Eq. \ref{newtonpot} yields straightforwardly
\begin{equation}\label{mg}
-4a(-\mathbf{k}^2)\mathbf{k}^2\phi(\mathbf{k}) = \kappa m_g
\end{equation}
from which $\Phi(r)$ can be obtained by Fourier transforming Eq. \ref{mg}  back to position space
\begin{equation}\label{key}
\bal
\Phi(r) &= \frac{-\kappa m_g}{4} \int \frac{d^3k}{(2\pi)^3} \frac{e^{i \mathbf{k}\cdot\mathbf{r}}}{\mathbf{k}^2 a(-\mathbf{k}^2)} \\
&= \frac{-\kappa m_g}{32 \pi^3} \int_0^\infty k \ dk 
\int_{-1}^{1} d(\cos \theta) \int_0^{2\pi} d\phi
\frac{e^{i \mathbf{k}\cdot\mathbf{r}}}{\mathbf{k}^2 a(-\mathbf{k}^2)} \\
&= \frac{-\kappa m_g}{8 \pi^2} \int_0^\infty  \frac{dk}{k}
\frac{ \sin \left( kr \right)  }{ a(-\mathbf{k}^2)}  
\eal
\end{equation}
plugging in our Gaussian choice for the entire function $a(\square)$ yields
\begin{equation}
\Phi(r) = \frac{-\kappa m_g}{8 \pi^2} \int_0^\infty  \frac{dk}{k} \exp \left(- \mathbf{ k }^2/M^2 \right)  \sin \left( kr \right)  
\end{equation}
and, since $\Phi$ and $\Psi$ satisfy the same equation, $\Psi(r)$ is described by the same equation.
From this equation we can already conclude that the potential is modified, besides the Newtonian $1/r$-contribution an additional factor emerges for the BGKM Newtonian potential of the BGKM action.
This integral corresponds to one of the special functions, the error function

\begin{equation}\label{key}
\int \frac{dk}{k} \sin \left( k r \right) \exp[-\mathbf{k}^2 / M^2] = \frac{\pi}{2} \erf \left(
\frac{r M}{2}
\right)
\end{equation}
Plugging in $\kappa = M_{\text{p}}^{-2}$ we obtain
\begin{equation}\label{key}
\Psi(r) = \Phi(r) 
=
- {m_g  \over 16 \pi M_p^2} \frac{1}{r}  \erf \left(
{r M \over 2}
\right).
\end{equation}
The implications of this Newtonian potential are discussed in section \ref{sectionIDG}.
\section{BKGM action at order $\mathcal{O}(h^3)$ 
\label{appendixoh3}
}
We start with the most general covariant quadratic action, which is given by
\begin{equation}\label{key}
S_{Q} = \int d^{4}x \sqrt{-g} \bigg[ \mathcal{R} \mathcal{F}_1 (\square) \mathcal{R} + \mathcal{R}_{\mu \nu}  \mathcal{F}_{2} (\square) \mathcal{R}^{\mu \nu}  + \mathcal{R}_{\mu \nu \lambda \sigma} \mathcal{F}_{3} (\square) \mathcal{R}^{\mu \nu \lambda \sigma}		\bigg].
\end{equation}
Let us now calculate which terms emerge from our action and subsequently enumerate the scalar-type gravitational terms at order $\mathcal{O}(h^3)$.
We should keep in mind that  $\sqrt{-g} = 1 + \frac{h}{2} + \frac{h^2}{8} - \frac{1}{4} h_{\nu}^{\mu} h_{\mu}^{\nu}
+ \mathcal{O}(h^3)$ and that every appearance of the Riemann tensor (and Ricci tensor and Ricci scalar) contributes an $\mathcal{O}(h)$ term in the action (for this reason the terms involving the $ \frac{1}{4} h_{\nu}^{\mu} h_{\mu}^{\nu}$ term don't  contribute to $\mathcal{O}(h^3)$ but to $\mathcal{O}(n)$ with $n \geq 4$). We can therefore immediately see that the term
$$\int d^{4}x \ \frac{h}{2} 
\left[	
\mathcal{R}^{(1)} \mathcal{F}_1 (\square) \mathcal{R}^{(1)} 
+ \mathcal{R}_{\mu \nu}^{(1)}  \mathcal{F}_{2} (\square) \mathcal{R}^{(1)\mu \nu}  
+ \mathcal{R}_{\mu \nu \lambda \sigma}^{(1)} \mathcal{F}_{3} (\square) \mathcal{R}^{(1) \mu \nu \lambda \sigma}
\right]$$ contributes $\mathcal{O}(h^3)$ to the action.
We also expect contributions from products involving any of the following terms
$$\mathcal{R}^{(2) \mu \nu \lambda \sigma}, \ \ \mathcal{R}^{(2) \mu \nu}, \ \ \mathcal{R}^{(2)}$$
all of which contribute $\mathcal{O}(h^2)$ and should consequently be combined with a similar curvature tensor with superscript $(1)$. This gives rise to 6 terms, but this expression can be simplified since Ricci scalars commute
$$ \mathcal{R}^{(1)} \mathcal{F}_1 (\square) \mathcal{R}^{(2)}  + \mathcal{R}^{(2)} \mathcal{F}_1 (\square) \mathcal{R}^{(1)}  = 2\mathcal{R}^{(2)} \mathcal{F}_1 (\square) \mathcal{R}^{(1)}.$$
Additional contributions are expected from
$$\mathcal{R}_{\mu \nu}^{(1)}  \mathcal{F}_{2} (\square) \mathcal{R}^{(2)\mu \nu}  
, \ \ \mathcal{R}_{\mu \nu}^{(2)}  \mathcal{F}_{2} (\square) \mathcal{R}^{(1)\mu \nu}  
, \ \ \mathcal{R}_{\mu \nu \lambda \sigma}^{(1)} \mathcal{F}_{3} (\square) \mathcal{R}^{(2) \mu \nu \lambda \sigma}
\ \ \text{and} \ \ \ \mathcal{R}_{\mu \nu \lambda \sigma}^{(2)} \mathcal{F}_{3} (\square) \mathcal{R}^{(1) \mu \nu \lambda \sigma} $$
and we should also take into account the contributions from the variation of $\mathcal{F}_i$ and thereby variations of $(\square)^{n}$, \emph{i.e.} $\delta (\square)^{n}$ which contributes $\mathcal{O}(h)$. We shall need the following useful mathematical relation \ref{UV}:
\begin{equation}\label{key}
\begin{aligned}
T \delta (\square^{n})S &= T \square \delta (\square^{n - 1})S + T \delta (\square) \square^{n - 1} S \\
&= T \square \delta (\square^{n - 2})S +  T \delta (\square) \square^{n - 2} S  + T \delta (\square) \square^{n - 1} S \\
&= \cdots \\
&= \sum_{m = 0}^{n - 1} \square^{m} T \delta( \square ) \square^{n - m - 1} S,
\end{aligned}
\end{equation}
where S and T are tensors constructed out of the Riemann
curvatures and the metric and $n \geq 1$. This implies that
\begin{equation}\label{key}
\begin{aligned}
T \delta \mathcal{F}(\square) S 
&= \sum_{n = 1}^{\infty} \frac{f_{i_{n}}}{M^{2n}} \sum_{m = 0}^{n - 1}  T \square^{m} \delta( \square ) \square^{n - m - 1} S.
\end{aligned}
\end{equation}
We have now obtained all $\mathcal{O}(h^3)$ terms:

\begin{equation}\label{bkgmresult}
\begin{aligned}
\delta S_{Q} &= \int d^{4}x \ \delta \sqrt{-g} \bigg[ \mathcal{R} \mathcal{F}_1 (\square) \mathcal{R} + \mathcal{R}_{\mu \nu}  \mathcal{F}_{2} (\square) \mathcal{R}^{\mu \nu}  + \mathcal{R}_{\mu \nu \lambda \sigma} \mathcal{F}_{3} (\square) \mathcal{R}^{\mu \nu \lambda \sigma}		\bigg] \\
&+ \int d^{4}x \sqrt{-g} \bigg[ \text{variations of the tensors}	\bigg] \\
&+ \int d^{4}x \sqrt{-g} \bigg[ \mathcal{R} \ \delta \mathcal{F}_1 (\square) \mathcal{R} + \mathcal{R}_{\mu \nu}  \ \delta  \mathcal{F}_{2} (\square) \mathcal{R}^{\mu \nu}  + \mathcal{R}_{\mu \nu \lambda \sigma}  \ \delta \mathcal{F}_{3} (\square) \mathcal{R}^{\mu \nu \lambda \sigma}		\bigg], \\
\end{aligned}
\end{equation}
where we have used Eq. \ref{variationmetric} which states that
\begin{equation}\label{key}
\delta \sqrt{ -g } = \frac{1}{2} \sqrt{ -g }g_{\mu \nu} h. 
\end{equation}
The third order terms are thus given by 
\begin{equation}
\begin{aligned}
S_{Q}^{(3)} &= \int d^{4}x \ \frac{h}{2} 
\left[	
\mathcal{R}^{(1)} \mathcal{F}_1 (\square) \mathcal{R}^{(1)} 
+ \mathcal{R}_{\mu \nu}^{(1)}  \mathcal{F}_{2} (\square) \mathcal{R}^{(1)\mu \nu}  
+ \mathcal{R}_{\mu \nu \lambda \sigma}^{(1)} \mathcal{F}_{3} (\square) \mathcal{R}^{(1) \mu \nu \lambda \sigma}
\right] \\
&+ \int d^{4}x \ 
\Big[	
2 \mathcal{R}^{(2)} \mathcal{F}_1 (\square) \mathcal{R}^{(1)} 
+ \mathcal{R}_{\mu \nu}^{(1)}  \mathcal{F}_{2} (\square) \mathcal{R}^{(2)\mu \nu}  
+ \mathcal{R}_{\mu \nu}^{(2)}  \mathcal{F}_{2} (\square) \mathcal{R}^{(1)\mu \nu}  \\
& \ \ \ \ \ \ \ \ \ \ \ \ \   + \mathcal{R}_{\mu \nu \lambda \sigma}^{(1)} \mathcal{F}_{3} (\square) \mathcal{R}^{(2) \mu \nu \lambda \sigma}
+ \mathcal{R}_{\mu \nu \lambda \sigma}^{(2)} \mathcal{F}_{3} (\square) \mathcal{R}^{(1) \mu \nu \lambda \sigma}
\Big] \\
&+ \sum_{n= 1}^{\infty} \sum_{m = 0}^{n - 1} \int \ d^{4}x \Big[
f_{1_{n}} \square^{m} \mathcal{R}^{(1)} \delta ( \square) \square^{n - m - 1} \mathcal{R}^{(1)} 
+ f_{2_{n}} \square^{m} \mathcal{R}_{\mu \nu}^{(1)} \delta ( \square) \square^{n - m - 1} \mathcal{R}^{(1)\mu \nu} \\
&  \ \ \ \ \ \ \ \ \ \ \ \ \ \ \  \ \  \ \  \ \ \ \ \ + f_{3_{n}} \square^{m} \mathcal{R}_{\mu \nu \lambda \sigma}^{(1)} \delta ( \square) \square^{n - m - 1} \mathcal{R}^{(1) \mu \nu \lambda \sigma} 
\Big]
\end{aligned}
\end{equation}
It should be kept in mind that these terms should be included if one wants to study graviton interactions.


\chapter{The toy model IDG action \label{toymod}}
In section \ref{infderscalar} we obtained the following action $S = S_{\text{free}} + S_{\text{int}}$ for the toy model IDG theory, consisting of
\begin{equation}\label{key}
S_{\text{free}} = \frac{1}{2} \int \ d^{4}x  \phi \square a(\square) \phi ,
\end{equation}
\begin{equation}\label{key}
S_{\text{int}} = \frac{1}{M_{p}} \int \ d^{4}x \Big[
\alpha_1 \phi \partial_{\mu} \phi \partial^{\mu} \phi 
+
\alpha_2 \phi \square \phi a(\square) \phi 
+
\alpha_3 \phi \partial_{\mu} \phi \partial^{\mu} a(\square) \phi 
\Big].
\end{equation}
The total action ought to be invariant under the following field transformation
\begin{equation}\label{symmetry2}
\phi \rightarrow (1+ \epsilon)\phi + \epsilon,
\end{equation}
since this describes a \emph{combined scaling and shifting symmetry}.
The interaction term of the action can be rewritten as follows (using integration by parts in the last step)
\begin{equation}\label{key}
\begin{aligned}
S_{\text{int}} &= \frac{1}{M_p} \int \ d^{4}x \Big( 
\alpha_1 \phi \partial_{\mu} \phi \partial^{\mu} \phi 
+
\alpha_2 \phi \square \phi a(\square) \phi 
+
\alpha_3 \phi \partial_{\mu} \phi  a(\square) \partial^{\mu} \phi 
\Big) \\
&= \frac{1}{M_p} \int \ d^{4}x \Big( 
\left( \alpha_1 + \alpha_3 \right) \phi \partial_{\mu} \phi \partial^{\mu} \phi 
+
\alpha_2 \phi \square \phi a(\square) \phi 
+
\alpha_3 \phi \partial_{\mu} \phi  \tilde{a}(\square) \partial^{\mu} \phi  
\Big) \\ 
&= \frac{1}{M_p} \int \ d^{4}x \Big( 
\left( \alpha_1 + \alpha_3 \right) \phi \partial_{\mu} \phi \partial^{\mu} \phi 
+
\alpha_2 \phi \square \phi \tilde{a}(\square) \phi 
+
2 \alpha_2 \phi \phi \square   \phi 
+
\alpha_3 \phi \partial_{\mu} \phi  \tilde{a}(\square) \partial^{\mu} \phi  
\Big) \\
&= \frac{1}{M_p} \int \ d^{4}x \Big( 
\left( \alpha_1 -2 \alpha_2 + \alpha_3 \right) \phi \partial_{\mu} \phi \partial^{\mu} \phi 
+
\alpha_2 \phi \square \phi \tilde{a}(\square) \phi
+
\alpha_3 \phi \partial_{\mu} \phi   \tilde{a}(\square) \partial^{\mu} \phi  
\Big)
\end{aligned}
\end{equation}
and the free action is written as
\begin{equation}\label{key}
\begin{aligned}
S_{\text{free}} 
= \frac{1}{2} \int \ d^{4}x \left[ \phi \square  \phi + \phi \square \tilde{a}\left(\square \right)  \phi \right], \\
\end{aligned}
\end{equation}
where 
\begin{equation}\label{key}
\tilde{a}(\square) = a(\square) - 1.
\end{equation}
\nnn
The scalar action can thus be written 
\begin{subequations}
\begin{align}
	S &= 
  \int \ d^{4}x \left[ \frac{1}{2} \phi \square  \phi + \frac{1}{M_p}  \ 
	\left( \alpha_1 -2\alpha_2 + \alpha_3 \right) \phi \partial_{\mu} \phi \partial^{\mu} \phi \right] \label{first} && (*)\\ 
	&+ \int \ d^{4}x \ \left[ \frac{1}{2} \phi \square \tilde{a}\left(\square \right)  \phi  + \frac{1}{M_p} \left(
	\alpha_2 \phi \square \phi \tilde{a}(\square) \phi 
	+
	\alpha_3 \phi \partial_{\mu} \phi \partial^{\mu}  \tilde{a}(\square) \phi \right) \right]. && (\dagger) \label{second} 
\end{align}
\end{subequations}
Since both equations \ref{first} and \ref{second} should  be invariant under the scaling and shifting symmetry of Eq. \ref{symmetry2} we yield
\begin{subequations}
	\begin{equation}\label{key}
	\begin{aligned}
	S_{(*)} &= \int \ d^{4}x \Bigg[ \frac{1}{2}  \left( (1+ \epsilon)\phi + \epsilon \right) \square  \left((1+ \epsilon)\phi + \epsilon \right) \\
	&+ \frac{1}{M_p}  \ \left( \alpha_1 -2\alpha_2 + \alpha_3 \right) \left((1+ \epsilon)\phi + \epsilon \right) \partial_{\mu} \left((1+ \epsilon)\phi + \epsilon \right) \partial^{\mu} \left((1+ \epsilon)\phi + \epsilon \right) \Bigg] \\
	&=  \int \ d^{4}x \Bigg[
	\frac{1}{2} 
	 \Big( 
	\phi \square \phi 
	+ \phi \square \epsilon \phi
	+  \stkout{\phi \square \epsilon}
	+ \epsilon \phi \square \phi
	+ \stkout{\epsilon  \phi \square \epsilon \phi }
	+  \stkout{\epsilon \phi \square \epsilon}
	+\epsilon  \square \phi 
	+ \stkout{\epsilon  \square \epsilon \phi }
	+ \stkout{ \epsilon  \square \epsilon}  \Big) \\
	&+ \frac{1}{M_p}  \ \left( \alpha_1 -2\alpha_2 + \alpha_3 \right) 
	\left(\phi +  \epsilon\phi  + \epsilon \right) \partial_{\mu} \left(
	\phi +  \epsilon\phi  + \epsilon \right)	\partial^{\mu}
	\left(	\phi +  \epsilon\phi  + \epsilon \right) 
	\Bigg].
	\end{aligned}
	\end{equation}	
	Neglecting terms $\epsilon^\omega$ terms with $\omega \geq 2$ the contributions from $\left(	\phi +  \epsilon\phi  + \epsilon \right)^3$ are
	$$
	\phi^3, \ \  
	3\epsilon \phi \phi \phi, \ \
	\epsilon \phi \phi. \ \
	$$
	so we yield
	\begin{equation}
	\begin{aligned}
		S_{(*)} =  \int \ d^{4}x \left[  \frac{1}{2} \left( 
	\phi \square \phi 
	+ 2 \epsilon \phi \square \phi
	+\epsilon  \square \phi \right)
	+ \frac{1}{M_p}  \ \left( \alpha_1 -2\alpha_2 + \alpha_3 \right) \Big( \phi  \partial_{\mu} 
	\phi 	\partial^{\mu}
	\phi 
	+ 3 \epsilon\phi \partial_{\mu} 
	\phi 	\partial^{\mu}
	\phi
	+  \epsilon  \partial_{\mu} 
	\phi 	\partial^{\mu}
	\phi 
	\Big) \right].
	\end{aligned}
	\end{equation}
\end{subequations} 
The variation of $S_{(*)}$ is consequently
\begin{equation}\label{key}
\begin{aligned}
\delta S_{(*)}
&= \frac{1}{2} \int \ d^{4}x \left[ \left( 
2 \epsilon \phi \square \phi
+ \stkout{\epsilon  \square \phi} \right)
+ \frac{2}{M_p}  \ \left( \alpha_1 -2\alpha_2 + \alpha_3 \right) \Big( 3 \epsilon\phi \partial_{\mu} 
\phi 	\partial^{\mu}
\phi
+  \epsilon  \partial_{\mu} 
\phi 	\partial^{\mu}
\phi 
\Big)
\right] \\
&= \int \ d^{4}x \left[
\epsilon \phi \square \phi
+ \frac{3 \epsilon}{M_p}  \ \left( \alpha_1 -2\alpha_2 + \alpha_3 \right) \phi \partial_{\mu} 
\phi 	\partial^{\mu}
\phi
+ \frac{ \epsilon}{M_p}  \ \left( \alpha_1 -2\alpha_2 + \alpha_3 \right)     \partial_{\mu} 
	\phi 	\partial^{\mu}
	\phi 
\right] \\
&= \int \ d^{4}x \left[
\epsilon \phi \square \phi
+ \frac{3 \epsilon}{M_p}  \ \left( \alpha_1 -2\alpha_2 + \alpha_3 \right) \phi \partial_{\mu} 
\phi 	\partial^{\mu}
\phi
- \frac{ \epsilon}{M_p}  \ \left( \alpha_1 -2\alpha_2 + \alpha_3 \right)      
\phi 	\square
\phi 
\right] \\
&= \int \ d^{4}x \left[  
\epsilon \left(1 - \alpha_1 + 2\alpha_2 - \alpha_3 \right) \phi \square \phi
+ \frac{3 \epsilon}{M_p}  \ \left( \alpha_1 -2\alpha_2 + \alpha_3 \right) \phi \partial_{\mu} 
\phi 	\partial^{\mu}
\phi
\right]
. \\
\end{aligned}
\end{equation}


\begin{equation}\label{key}
\begin{aligned}
S_{(\dagger)} =
\int \ d^{4}x \ \Bigg[   & \frac{1}{2}\Big(
\phi \square \tilde{a}(\square) \phi 
+ 2 \epsilon \phi \tilde{a}(\square) \phi
+\epsilon  \tilde{a}(\square) \phi \big)  + \frac{1}{M_p} \Big(
\alpha_2 \phi \square 
\phi 	\tilde{a}(\square)	\phi + 3\alpha_2   \epsilon\phi   \square 
\phi 	\tilde{a}(\square) 	\phi \\
&+  \alpha_2  \epsilon\square 
\phi 	\tilde{a}(\square) 	\phi   
+\alpha_3 \phi \partial_{\mu} \phi \partial^{\mu}	\tilde{a}(\square)	\phi 
+ 3\alpha_3   \epsilon\phi   \partial^{\mu} \phi \partial^{\mu}	\tilde{a}(\square) 	\phi 
+  \alpha_3  \epsilon \partial^{\mu} 
\phi \partial^{\mu}	\tilde{a}(\square) 	\phi   \Big) \Bigg]. \\
\end{aligned}
\end{equation}
\begin{equation}\label{key}
\begin{aligned}
\delta S_{(\dagger)} &=  \int \ d^{4}x \ \Bigg( 
\epsilon \phi \square \tilde{a}(\square) \phi
+ \frac{\epsilon}{2}  \square \tilde{a}(\square) \phi  + \frac{1}{M_p} \Bigg(
3\alpha_2   \epsilon\phi   \square 
\phi 	\tilde{a}(\square) 	\phi \\
&+  \alpha_2  \epsilon\square 
	\phi 	\tilde{a}(\square) 	\phi 
+ 3\alpha_3   \epsilon\phi   \partial^{\mu}
\phi \partial^{\mu}	\tilde{a}(\square) 	\phi 
+  \alpha_3  \epsilon \partial^{\mu}
	\phi \partial^{\mu}	\tilde{a}(\square) 	\phi   \Bigg) \Bigg) \\ 
&= \int \ d^{4}x \  \epsilon \Bigg( 
\phi \square \tilde{a}(\square) \phi
+ \frac{1}{2} \square \tilde{a}(\square) \phi  + \frac{1}{M_p} \Bigg(
3\alpha_2   \phi   \square 
\phi 	\tilde{a}(\square) 	\phi \\
&+  \alpha_2  
\phi \square 	\tilde{a}(\square) 	\phi   
+ 3\alpha_3   \phi  \partial^{\mu}
\phi \partial^{\mu}	\tilde{a}(\square) 	\phi 
-  \alpha_3  
\phi \square  	\tilde{a}(\square) 	\phi   \Bigg) \Bigg) \\ 
&= \int \ d^{4}x \ 
\epsilon \Bigg( \left(1 + \alpha_2 - \alpha_3 \right)
\phi \square \tilde{a}(\square) \phi
+ \stkout{\frac{1}{2} \square \tilde{a}(\square) \phi } + \frac{3}{M_p} \Bigg(
\alpha_2   \phi   \square 
\phi 	\tilde{a}(\square) 	\phi  
+ \alpha_3   \phi  \partial^{\mu}
\phi \partial^{\mu}	\tilde{a}(\square) 	\phi \Bigg) \Bigg).
\\ 
\end{aligned}
\end{equation}
$\delta S_2$ should be proportional to the original action, which implies
\begin{equation}\label{key}
1 + \alpha_2 - \alpha_3 = \frac{3}{2}.
\end{equation}
Likewise, $\delta S_1$ should be proportional to the original action
\begin{equation}\label{key}
1 - \alpha_1 + 2 \alpha_2 - \alpha_3 = \frac{3}{2}.
\end{equation}
This gives rise to the following interaction term in the scalar toy Infinite Derivative model
\begin{equation}\label{key}
S_{\text{int}} = \frac{1}{4M_{p}} \int d^{4}x \Bigg( 	\phi \partial_{\mu} \phi \partial^{\mu} \phi + \phi \square \phi a (\square) \phi -  \phi \partial_{\mu} \phi a(\square)\partial^{\mu} \phi	\Bigg)
\end{equation}
which mimics gravity.

\chapter{The spectral function and frequency sums \label{freqsumappendix} }
\section{The spectral function and propagators}
It is convenient to introduce 
two different kinds of 
two-point correlation functions at finite temperature
\begin{subequations}
	\begin{align}
	D^{+}(x, y) &= \expval{\phi(x) \phi(y)}_{\beta} = \frac{1}{\mathcal{Z}} \sum_{n} 
	e^{-\beta E_n}
	\bra{n}   \phi(x) \phi(y) \ket{n}, \label{D+}\\
	D^{-}(x, y) &= \expval{\phi(y) \phi(x)}_{\beta} = \frac{1}{\mathcal{Z}} \sum_{n} 
	e^{-\beta E_n}
	\bra{n}   \phi(y) \phi(x) \ket{n}
	= D^{+}(y, x), \label{D-},
	\end{align}
\end{subequations}
where the states $n$ form a complete set of eigenstates
of the Hamiltonian and of the momentum operator. The imaginary time propagator is a useful propagator in thermal field theory and this propagator can be constructed from $D^{+}$ and $D^{-}$.
\nnn
Due to translational invariance, the Green's functions can only depend on the differences $(x - y)$ and the matrix elements at $x$ are related to those at $x = 0$ through
\begin{equation}\label{key}
\bra{m} \phi(x) \ket{n} = e^{i (p_n - p_m) \cdot x} \bra{m} \phi(0) \ket{n} 
\end{equation}
Inserting a complete set of states (using \textit{completeness})
\begin{equation}\label{key}
\sum_{m} \ket{m} \bra{m} = \mathbf{1}
\end{equation}
and noting that, due to translational invariance, the Green function can only depend on the differences $(\mathbf{x} - \mathbf{y})$ and $\tau_x - \tau_y$, Eq. \ref{D+} can be rewritten conveniently. The  following equality 
holds true in the Heisenberg representation
\begin{equation}\label{key}
\bra{n}\phi(\mathbf{x}, \tau) \ket{m} = \bra{m} e^{H \tau} \phi(\mathbf{x},0) e^{-H \tau} \ket{n} = e^{E_m \tau} \bra{m}  \phi(\mathbf{x},0)  \ket{n} e^{-E_n \tau}
\end{equation}
and will now be used to rewrite the propagator $D^{+}(x - y)$
\begin{equation}\label{dx-y}
\begin{aligned}
\mathcal{D}^{+} (x - y)
&=
\expval{\phi(x) \phi(y)}_{\beta} = \frac{1}{\mathcal{Z}} \sum_{m, n} 
e^{-\beta E_n}
\bra{n}   \phi(x) \ket{m} \bra{m} \phi(y) \ket{n} \\
&=
\expval{\phi(x) \phi(y)}_{\beta} = \frac{1}{\mathcal{Z}} \sum_{m, n} 
e^{-\beta E_n}
e^{i (p_n - p_m) \cdot x}
\bra{n}   \phi(0) \ket{m}
e^{i (p_m - p_n) \cdot y}
\bra{m} \phi(0) \ket{n} \\
&=
\expval{\phi(x) \phi(y)}_{\beta} = \frac{1}{\mathcal{Z}} \sum_{m, n} 
e^{-\beta E_n}
e^{i (p_n - p_m) \cdot (x - y)}
\abs{ \bra{n}   \phi(0) \ket{m} }^{2}.
\end{aligned}
\end{equation}
A convergence of this sum is required in order to have a meaningful propagator. The \emph{temporal part} of Eq. \ref{dx-y} now reads
\begin{equation}\label{key}
D^{+} (t - t')
=
\frac{1}{\mathcal{Z}} \sum_{m, n} 
e^{-\beta E_n}
e^{i (E_n - E_m)(t - t')}
\abs{ \bra{n}   \phi(0) \ket{m} }^{2}
\end{equation}
and one is led to conclude that this propagator is only defined on the strip $-\beta < \Im(t - t') < 0$ for $t, t' \in \mathrm{C}$. Similarly,
\begin{equation}\label{D-t}
D^{-} (t - t')
=
\frac{1}{\mathcal{Z}} \sum_{m, n} 
e^{-\beta E_n}
e^{i (E_m - E_n)(t - t')}
\abs{ \bra{n}   \phi(0) \ket{m} }^{2},
\end{equation}
is only defined on the strip $0 < \Im(t - t') < \beta$. Eq. \ref{D-t} implies that $D^{+} (t - i \beta) = D^-(t)$, which basically is the KMS condition dressed in a different form. \newline \newline
\noindent
Fourier transforming the propagators into frequency space, where we set $t' = 0$ without loss of generality, relates $D^-(k^0)$ and $D^+(k^0)$
\bse
\begin{align}
D^{+}(k^0) &= \int dt \ e^{ik^0 t} D^{+}(t), \label{D+k0} \\
D^{-}(k^0) &= \int dt \ e^{ik^0 t} D^{-}(t) = \int dt \ D^{+}(t - i\beta). \label{D-k0}
\end{align}
\ese
Equations \ref{D+k0},b entail a useful relation between the two-point correlators
\begin{equation}\label{relation}
D^{-}(k^0) = e^{-\beta k^0} D^{+}(k^{0}).
\end{equation}
The \textit{spectral intensity} $\rho(k^0)$, whose applications are described below, can now be obtained since it is defined in terms of aforementioned propagators
\bse
\begin{equation}\label{key}
\rho(k^0) = D^{+}(k^0) - D^{-}(k^0).
\end{equation}
The spectral intensity is remarkably useful since different kinds of propagators can be deduced from this function, such as the Minkowskian advanced and retarded propagators ($D^{A}, D^{R})$ but also the imaginary time and real time propagators. 
Eq. \ref{relation} enables the spectral intensity to be be expressed as
\begin{equation}\label{key}
\rho(k^0) = (e^{\beta k^0} - 1)D^{-}(k^0) = D^{-}(k^0) / n_{\text{BE}}(k^0),
\end{equation}
\ese
where $n_{\text{BE}}(k^0)$ is the Bose-Einstein (BE) distribution factor. The propagators $D^+(k^0)$ and $D^-(k^0)$ can now be expressed in terms of the spectral intensity function
\bse
\begin{align}
D^{-}(k^0) &=  n(k^0)\rho(k^0), \label{d-k0} \\ 
D^{+}(k^0) &=  \left[1 + n(k^0)\right]\rho(k^0).  \label{d+k0}
\end{align}
\ese 
The spectral density for non-interacting scalar fields can be derived in a straightforward fashion
\begin{equation}\label{spectralfree}
\bal
\rho_F(k^0) &= {1 \over \mathz} \int dt \ e^{i k^0 t} \sum_n e^{\beta E_n} \bra{ n} \phi(t) \phi(0) - \phi(0) \phi(t) \ket{n} \\
&= {2 \pi \over 2E_\mathbf{k}} \left[
\delta(k^0 - E_\mathbf{k}) 
-
\delta(k^0 + E_\mathbf{k})
\right] \\
&= 2 \pi \epsilon(k^0) \delta
\left( 
\left(k^0\right)^2 - E_\mathbf{k}^2
\right),
\eal
\end{equation}
where $\epsilon(k^0)$ is the sign-function of $k^0$. Interactions modify the shape of the spectral function, but the spectral function remains bounded from above, a proof of which can be found in this footnote.\footnote{
	The spectral intensity function satisfies the sum rule
	\begin{equation}\label{sumrule}
	\int \frac{dk^0}{2\pi} k^0 \rho(k^0) = 1
	\end{equation}
	as will be proved now.
	$$D^{+}(t) - D^{-}(t) = \expval{[\phi(t), \phi(0)]}_{\beta} = \int \frac{dk^0}{2\pi} e^{-ik^0 t} \underbrace{[D^+(k^0) - D^-(k^0)]}_{= \rho(k^0)}.$$ We now differentiate both sides to $t$ and note that the following ETCT relation holds true for $\phi$ and $\pi$:
	$$[\phi(t), \pi(t)] = i.$$
	$$\expval{[\pi(t), \phi(0)]} = -i \int \frac{dk^0}{2\pi}  e^{-ik^0 t} k^0 \rho(k^0) \rightarrow -i = -i \int \frac{dk^0}{2\pi} k^0  \rho(k^0),$$
	where the arrow is to be interpreted as $t \rightarrow 0$. The sum rule of Eq. \ref{sumrule} follows directly.
}
\nnn
The most important propagator for this thesis is the imaginary time propagator, or \textit{Matsubara propagator}, which allows us to perform thermal loop corrections in Euclidean space.   The Matsubara propagator is defined as
\begin{equation}\label{key}
\bal
\mathcal{D}(\tau) &= \frac{1}{\mathz} \sum_n \bra{n} e^{\beta H} \phi(\tau) \phi(0) \ket{n}
\mathcal{D}(\tau) &= \mathcal{D}^+(t = -i \tau) \\
&= \int \frac{\dd k^0}{2 \pi} e^{-k^0 \tau} [1 + n(k^0)]\rho(k^0),
\eal
\end{equation}
where the Fourier transform of Eq. \ref{d+k0} has been inserted.
The Matsubara propagator is thus given by
\begin{equation}\label{key}
\bal
\mathcal{D}(i \omega_n) &= \int_0^\beta \ \dd \tau e^{i \omega_n \tau} \int \frac{dk^0}{2 \pi} e^{-k^0 \tau}
[1 + n(k^0) ] \rho(k^0) \\
&= \int \frac{dk^0}{2 \pi} \int_0^\beta \ \dd \tau e^{(i \omega_n - k^0) \tau}  \left[1 + [e^{\beta k^0} - 1]^{-1} \right] \rho(k^0) \\
&= \int \frac{dk^0}{2 \pi} \frac{ e^{(i \omega_n - k^0) \tau} }{i \omega_n - k^0} 
\Big \rvert_{0}^{\beta}
\left[1 + [e^{\beta k^0} - 1]^{-1} ] \right] \rho(k^0) \\
&= \int \frac{dk^0}{2 \pi} 
\frac{ e^{-\beta k^0} - 1 }{i \omega_n - k^0} 
\left[1 + [e^{\beta k^0} - 1]^{-1} ] \right] \rho(k^0) \\
&= \int \frac{dk^0}{2 \pi} 
\frac{ 1 }{i \omega_n - k^0} 
\left[ { 1 - e^{\beta k^0}    \over - 1 + e^{\beta k^0} } \right] \rho(k^0) \\
&= - \int \frac{dk^0}{2 \pi} 
\frac{ \rho(k^0) }{i \omega_n - k^0} 
.
\eal
\end{equation}
The Matsubara propagator for free (real and massive) scalar fields, whose free spectal function is given by Eq. \ref{spectralfree}, is consequently\footnote{The penultimate equality in Eq. \ref{spectralfree} has been used.}
\begin{equation}\label{key}
\bal
\mathcal{D}(i \omega_n, \mathbf{k}) &= - \int \frac{dk^0}{2 \pi} 
\frac{ 1 }{i \omega_n - k^0} \cdot
{2 \pi \over 2E_\mathbf{k}} \left[
\delta(k^0 - E_\mathbf{k}) 
-
\delta(k^0 + E_\mathbf{k})
\right] \\
&=  {1 \over 2E_\mathbf{k}}
\left[
\frac{ 1 }{i \omega_n + E_\mathbf{k}} 
-
\frac{ 1 }{i \omega_n - E_\mathbf{k}} 
\right] \\
&= \frac{1}{\omega_n^2 + E_\mathbf{k}^2} \ \left( = \frac{1}{(2\pi n T)^2 + E_k^2} \right),
\eal
\end{equation}
in accordance with the Euclidean momentum-space propagator which was deduced in section \ref{sectionpropagators} by realizing that the Matsubara propagator ought to solve the (Fourier transform of the) inhomogeneous partial differential equation
\begin{equation}\label{key}
\square \mathcal{D}(x-y) = \left( \partial_{\tau}^{2} + \nabla^{2} 			\right)\mathcal{D}(x-y) = \delta(\tau - \tau') \delta^{(3)}(\mathbf{x} - \mathbf{x}'),
\end{equation}
where $k_0$ takes the discrete and imaginary values $k_0 = i\omega_n$.
Fourier transforming $\mathcal{D}(i\omega_n, \mathbf{k})$ by using
\begin{equation}\label{key}
\mathcal{D}(\tau, \mathbf{x}) = \frac{1}{\beta}
\sum_n
\int \frac{d^3k}{(2\pi)^3}
e^{-i(\omega_n \tau - \mathbf{k} \cdot \mathbf{x})}
\mathcal{D}(i\omega_n, \mathbf{k})
\end{equation}
and performing the Matsubara frequency sum yields
\begin{equation}\label{key}
\mathcal{D}(\tau, \mathbf{x}) = \int \frac{d^3k}{(2\pi)^3} \frac{e^{\mathbf{k} \cdot \mathbf{x}}}{2 \omega_k} \left[(1 + n_B(\omega_k)) e^{-\omega_k \tau} + n_B(\omega_k) e^{\omega_k \tau}
\right]
\end{equation}
as is shown in section \ref{freqsumsection} where frequency sums are discussed. 
\section{Frequency sums
\label{appendixfreqsum}
}
We want to calculate the following frequency sum in which two bosonic propagators appear  
$$
\bal
S(i\omega_m, \map) &= 
- \frac{1}{2} \int_{C_3} \frac{\dd k^0}{2 \pi i } \frac{1}{(k^0)^2 - E_1^2} \frac{1}{(k^0 - i \omega_m)^2 - E_2^2}  \coth \left(
\frac{\beta k^0}{2}
\right) .
\eal
$$
We note that four poles 
\begin{equation}\label{key}
k^0 = \pm E_1, \ k_0 = i\omega_m  \pm E_2
\end{equation}
contribute to the integral with contour $C_3$ in the complex $k^0$ plane.

Let us first calculate the two residuals corresponding to $i\omega_n = E_1$, i.e.
\begin{subequations}
\begin{equation}\label{key}
\bal
\boxed{k^0 = E_1}: \ \ \  &- \frac{1}{2}  \frac{1}{2E_1} \frac{1}{(E_1 - i \omega_m)^2 - E_2^2} \left[
1 + \frac{2}{e^{\beta E_1} - 1 }
\right] \\
&= - \frac{1}{2}  \frac{1}{2E_1} \frac{1}{E_1^2 - E_2^2 -2E_1\omega_m i - \omega_m^2} \left[
1 + 2n(E_1)
\right] \\
&= - \frac{1}{2}  \frac{1 + 2n(E_1)}{2E_1 \cdot 2E_2}
\left[ \frac{1}{ i \omega_m - E_1 - E_2}  - \frac{1}{ i \omega_m - E_1 + E_2}  
\right],\\
\eal
\end{equation}
\begin{equation}\label{key}
\bal
\boxed{k^0 = -E_1}: \ \ \  &- \frac{1}{2}  \frac{1}{-2E_1} \frac{1}{(-E_1 - i \omega_m)^2 - E_2^2} \left[
1 + \frac{2}{e^{\beta E_1} - 1 }
\right] \\
&= - \frac{1}{2}  \frac{1}{2E_1} \frac{1}{E_1^2 - E_2^2 -2E_1\omega_m i - \omega_m^2} \left[
1 + 2n(E_1)
\right] \\
&= - \frac{1}{2}  \frac{1 + 2n(E_1)}{2E_1 \cdot 2E_2}
\left[ \frac{1}{ i \omega_m + E_1 - E_2}  - \frac{1}{ i \omega_m + E_1 + E_2}  
\right],\\
\eal
\end{equation}
where the property has been used that
the expression for $\coth \left( \beta k^0 \over 2 \right)$ for $k^0 >0$ differs from the expression for $k^0 < 0$:
\begin{equation}\label{key}
\bal
\coth \left( \beta k^0 \over 2 \right) &=
\frac{e^{\beta k^0 / 2} + e^{- \beta k^0 / 2} }{e^{\beta k^0 / 2}
	- e^{- \beta k^0 / 2} }  \\
&= 
\begin{cases}
\frac{e^{\beta k^0 } + 1 }{e^{\beta k^0 } - 1} 
= 1 + \frac{2}{e^{\beta k^0 - 1} }
& \text{for } k^0 > 0, \\
\frac{1 + e^{-\beta k^0 } }{1 - e^{- \beta k^0 }} 
= -1 - \frac{2}{e^{- \beta k^0 - 1}}  & \text{for } k^0 < 0. 
\end{cases}
\eal
\end{equation}
The computation of the remaining two residuals exhibits a remarkable similarity to the previous two and the equality $\exp(i \beta \omega_m) = 1$, which holds true for bosons, is used as well in order to obtain the answer
\begin{equation}\label{key}
\bal
\boxed{k^0 = i\omega_m + E_2}: \ \ \  &- \frac{1}{2}  \frac{1}{(i\omega_m + E_2)^2 - E_1^2} \frac{1}{2E_2} \left[
1 + \frac{2}{e^{\beta (i \omega_m + E_2)} - 1 }
\right] \\
&= - \frac{1}{2}  \frac{1 + 2n(E_2)}{2E_1 \cdot 2E_2}
\left[ \frac{1}{ i \omega_m - E_1 + E_2}  - \frac{1}{ i \omega_m + E_1 + E_2}  
\right],\\
\eal
\end{equation}
\begin{equation}\label{key}
\bal
\boxed{k^0 = i\omega_m - E_2}: \ \ \  &- \frac{1}{2}  \frac{1}{(i\omega_m - E_2)^2 - E_1^2} \frac{1}{-2E_2} \left[
-1 - \frac{2}{e^{\beta (-i \omega_m + E_2)} - 1 }
\right] \\
&= - \frac{1}{2}  \frac{1 + 2n(E_2)}{2E_1 \cdot 2E_2}
\left[ \frac{1}{ i \omega_m - E_1 - E_2}  - \frac{1}{ i \omega_m + E_1 - E_2}  
\right].\\
\eal
\end{equation}
\end{subequations}
We conclude that 
\begin{equation}\label{key}
\bal
S(i\omega_m, \map) = - \frac{1}{4E_1 E_2} 
\Bigg[
&\left[1 + n(E_1) + n(E_2)
\right]
\left[ \frac{1}{ i \omega_m + E_1 - E_2}  - \frac{1}{ i \omega_m + E_1 + E_2}  
\right] \\
&+ \left[-n(E_1) + n(E_2)\right]
\left[\frac{1}{ i \omega_m - E_1 + E_2}
-
\frac{1}{ i \omega_m + E_1 - E_2}
\right]
\Bigg].
\eal
\end{equation}
It should be kept in mind that this result cannot be generalized to bosonic loops whose external momenta are provided by fermionic particles. In that case we should use $\exp(i \beta \omega_n) = -1$ and, consequently, several changes of signs
should be included.

\chapter{Local and nonlocal $\phi^4$ theory
\label{appendixlocalnonlocal}
}

\section{Breakdown of the perturbative expansion in scalar $ \phi^4$ theory
\label{appendixbreakdownperturbation}
}
\begin{figure}[h!]
	\centering
	\includegraphics[scale = 0.8]{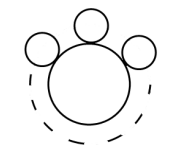}
	\label{ring}
	\caption{\textit{Ring diagrams are the diagrams whose degree of infrared divergence is the highest. The dotted line displays the possibility of more loops being attached to the main, big loop.}}
\end{figure}

Finite temperature field theories cannot contain more UV-divergent diagram than their zero temperature counterparts, because the vacuum and matter contributions decouple and the latter is Boltzmann suppressed at high temperatures/energies \ref{kapustabook}. Infrared divergences on the other hand are rather common at finite temperature and
these, in fact,
typically proliferate in bosonic field theories (note that the $n_{\text{BE}} \rightarrow \infty$ for $k \rightarrow 0$ whereas the Fermi-Dirac distribution function is not IR-divergent!). 
\nnn
The first infrared divergence in the perturbative expansion of the local scalar $\phi^4$ theory is encountered in 
o the partition function in $\lambda \phi^4$ theory, where Eq. \ref{taylor} gives rise to two inequivalent terms.

\begin{equation}\label{B1}
\ln \mathz_2 = - \frac{1}{2}  \left(	\frac{ \int \ \mathcal{D} \phi e^{S_0} S_{\text{I}} }{ \int \mathcal{D} \phi e^{S_0}  }	\right)^{2} + \frac{1}{2} 	\frac{ \int \ \mathcal{D} \phi e^{S_0} S_{\text{I}}^2 }{ \int \mathcal{D} \phi e^{S_0}  }	
\end{equation}
The first term can be written down immediately, since it equals $-\frac{1}{2} (\ln \mathz_1 )^2$:
\begin{equation}\label{B2}
- \frac{1}{2}  \Bigg(
3 \
\begin{tikzpicture}[baseline=(v3), node distance=1cm and 1cm]
\begin{feynman}[inline=(v3)]
\coordinate[] (v3);
\coordinate[vertex, right=of v3] (v4);
\coordinate[right=of v4] (v5);
\semiloop[fermion]{v3}{v4}{0};
\semiloop[fermion]{v4}{v3}{180};
\semiloop[fermion]{v4}{v5}{0};
\semiloop[fermion]{v5}{v4}{180};
\end{feynman}
\end{tikzpicture} 
\
\bigotimes
\ 
3
\
\begin{tikzpicture}[baseline=(v3), node distance=1cm and 1cm]
\begin{feynman}[inline=(v3)]
\coordinate[] (v3);
\coordinate[vertex, right=of v3] (v4);
\coordinate[right=of v4] (v5);
\semiloop[fermion]{v3}{v4}{0};
\semiloop[fermion]{v4}{v3}{180};
\semiloop[fermion]{v4}{v5}{0};
\semiloop[fermion]{v5}{v4}{180};
\end{feynman}
\end{tikzpicture} 
\Bigg).
\end{equation}
The second term in Eq. \ref{B1} will be solved diagramatically, by drawing two crosses corresponding to the factors $\phi^{4}(\mathbf{x}, \tau)$
and
$\phi^{4}(\mathbf{x}', \tau')$ which are contained in $\expval{S_{\text{I}}^{2}}$
\begin{equation}\label{B3}
\begin{tikzpicture}[baseline=(v3), node distance=1cm and 1cm]
\begin{feynman}[inline=(v3)]
\coordinate[] (v1);
\coordinate[right=of v1] (v2);
\coordinate[above=of v2] (v3);
\coordinate[left=of v3] (v4);
\draw (v1) -- (v3);
\draw (v2) -- (v4);
\end{feynman}
\end{tikzpicture} 
\ \ \ \ \
\begin{tikzpicture}[baseline=(v3), node distance=1cm and 1cm]
\begin{feynman}[inline=(v3)]
\coordinate[] (v1);
\coordinate[right=of v1] (v2);
\coordinate[above=of v2] (v3);
\coordinate[left=of v3] (v4);
\draw (v1) -- (v3);
\draw (v2) -- (v4);
\end{feynman}
\end{tikzpicture} 
\end{equation}
and subsequently take into account
all possible Bloch-de Dominicis contractions. We obtain (taking into account combinatorial factors):
\begin{align}\label{key}
\bal
\ln \mathz_2 =
&
- \frac{1}{2} \cdot 3 \
\begin{tikzpicture}[baseline=(v3), node distance=1cm and 1cm]
\begin{feynman}[inline=(v3)]
\coordinate[] (v3);
\coordinate[vertex, right=of v3] (v4);
\coordinate[right=of v4] (v5);
\semiloop[fermion]{v3}{v4}{0};
\semiloop[fermion]{v4}{v3}{180};
\semiloop[fermion]{v4}{v5}{0};
\semiloop[fermion]{v5}{v4}{180};
\end{feynman}
\end{tikzpicture} 
\ \bigotimes 3 \
\begin{tikzpicture}[baseline=(v3), node distance=1cm and 1cm]
\begin{feynman}[inline=(v3)]
\coordinate[] (v3);
\coordinate[vertex, right=of v3] (v4);
\coordinate[right=of v4] (v5);
\semiloop[fermion]{v3}{v4}{0};
\semiloop[fermion]{v4}{v3}{180};
\semiloop[fermion]{v4}{v5}{0};
\semiloop[fermion]{v5}{v4}{180};
\end{feynman}
\end{tikzpicture} \\
&+ 
\frac{1}{2} \cdot 3 \
\begin{tikzpicture}[baseline=(v3), node distance=1cm and 1cm]
\begin{feynman}[inline=(v3)]
\coordinate[] (v3);
\coordinate[vertex, right=of v3] (v4);
\coordinate[right=of v4] (v5);
\semiloop[fermion]{v3}{v4}{0};
\semiloop[fermion]{v4}{v3}{180};
\semiloop[fermion]{v4}{v5}{0};
\semiloop[fermion]{v5}{v4}{180};
\end{feynman}
\end{tikzpicture} 
\ \bigotimes 3 \
\begin{tikzpicture}[baseline=(v3), node distance=1cm and 1cm]
\begin{feynman}[inline=(v3)]
\coordinate[] (v3);
\coordinate[vertex, right=of v3] (v4);
\coordinate[right=of v4] (v5);
\semiloop[fermion]{v3}{v4}{0};
\semiloop[fermion]{v4}{v3}{180};
\semiloop[fermion]{v4}{v5}{0};
\semiloop[fermion]{v5}{v4}{180};
\end{feynman}
\end{tikzpicture} 
+ 36 \
\begin{tikzpicture}[baseline=(v3), node distance=1cm and 1cm]
\begin{feynman}[inline=(v3)]
\coordinate[] (v3);
\coordinate[vertex, right=of v3] (v4);
\coordinate[vertex, right=of v4] (v5);
\coordinate[right=of v5] (v6);
\semiloop[fermion]{v3}{v4}{0};
\semiloop[fermion]{v4}{v3}{180};
\semiloop[fermion]{v4}{v5}{0};
\semiloop[fermion]{v5}{v4}{180};
\semiloop[fermion]{v5}{v6}{0};
\semiloop[fermion]{v6}{v5}{180};
\end{feynman}
\end{tikzpicture} 
+ 12 
\vcenter{
\includegraphics[scale=0.42]{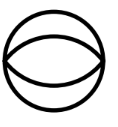}
},
\eal
\end{align}
where the first line represents the contributions from Eq. \ref{B2} and the second line those arising from contractions of Eq. \ref{B3}.
We conclude that the disconnected diagrams 
$$\begin{tikzpicture}[baseline=(v3), node distance=1cm and 1cm]
\begin{feynman}[inline=(v3)]
\coordinate[] (v3);
\coordinate[vertex, right=of v3] (v4);
\coordinate[right=of v4] (v5);
\semiloop[fermion]{v3}{v4}{0};
\semiloop[fermion]{v4}{v3}{180};
\semiloop[fermion]{v4}{v5}{0};
\semiloop[fermion]{v5}{v4}{180};
\end{feynman}
\end{tikzpicture} 
\
\bigotimes
\
\begin{tikzpicture}[baseline=(v3), node distance=1cm and 1cm]
\begin{feynman}[inline = (v3)]
\coordinate[] (v3);
\coordinate[vertex, right=of v3] (v4);
\coordinate[right=of v4] (v5);
\semiloop[fermion]{v3}{v4}{0};
\semiloop[fermion]{v4}{v3}{180};
\semiloop[fermion]{v4}{v5}{0};
\semiloop[fermion]{v5}{v4}{180};
\end{feynman}
\end{tikzpicture} 
$$
vanish - a feature which is not coincidental. Disconnected diagrams can \textit{always} be omitted and consequently do not contribute to the partition function. A proof of this statement can be found in many thermal field theory text books \ref{kapustabook} \ref{bellac} (and, in fact, also in quantum field theory text books \ref{zee} \ref{peskin}), but let me explain it from a more intuitive point of view in this footnote.\footnote{If there exists a contribution from $n$ pieces of disconnected
diagrams, then each of the disconnected pieces will contribute a factor of $V$ and consequently lead to
an overall factor of $V^n$. The logarithm of the partition function, however, is proportional to
the free energy which is an extensive quantity. Since $\ln \mathz \propto V$, the
number of the disconnected pieces must be only one. In other words, all contributing diagrams
must be connected.}
\nnn
Two Feynman diagrams have to be evaluated at second order in our perturbation theory
and the sunset diagram can be computed \ref{kapustabook}. The infrared diverence problem emerges in the \emph{three-loop necklace diagram}
\begin{equation}\label{key}
\begin{aligned}
\begin{tikzpicture}[baseline=(v3), node distance=1cm and 1cm]
\begin{feynman}[inline=(v3)]
\coordinate[] (v3);
\coordinate[vertex, right=of v3] (v4);
\coordinate[vertex, right=of v4] (v5);
\coordinate[right=of v5] (v6);
\semiloop[fermion]{v3}{v4}{0}[$(\mathbf{k}, \omega_{m})$][left];
\semiloop[fermion]{v4}{v3}{180};
\semiloop[fermion]{v4}{v5}{0}[$(\mathbf{p}, \omega_{n})$];
\semiloop[fermion]{v5}{v4}{180}[$(\mathbf{p}, \omega_{n})$][below];
\semiloop[fermion]{v5}{v6}{0}[$(\mathbf{k}, \omega_{m})$][right];
\semiloop[fermion]{v6}{v5}{180};
\end{feynman}
\end{tikzpicture}
\sim (-\lambda)^2  
\left[
\SumInt_{m, \mathbf{k}}
(\omega_m^2 + \mathbf{k}^2)^{-1}
\right]^{2}
\SumInt_{n, \mathbf{p}}
(\omega_n^2 + \mathbf{p}^2)^{-2},
\\
\end{aligned}
\end{equation}
where the short-hand notation $\SumInt$ now contains an additional subscript to indicate which Matsubara frequencies are summed over, \emph{i.e.}
\begin{equation}\label{irdiv}
\SumInt_{m, \mathbf{k}} = T \sum_{m} \int \frac{d^{3} k}{(2\pi)^3}.
\end{equation}
The ``central loop'' is responsible for the infrared divergence since it contains \emph{two identical propagators}. The corresponding momentum integral 
now reads (for the $n=0$ thermal mode)
\begin{equation}\label{key}
\int {d^3 p \over (2\pi)^3} \frac{1}{(\omega_n^2 + \mathbf{p}^2)^{2}} \Big\rvert_{n = 0} 
\sim
\int {dp \over p^2}
\end{equation}
and blows up in the low momentum limit $p \rightarrow 0$. Additional IR-divergent diagrams are encountered at higher orders in perturbation theory as, in fact, 
\emph{ring diagrams} occur at every order of the perturbative expansion and all of these have a similar mathematical structure sa the three-loop necklace diagram. It can be shown that \textit{ring diagrams} (displayed in Fig. \ref{ring}) are the strongest IR-divergent diagrams at every order of perturbation theory, where the divergence becomes more
severe at successively higher orders. 
\nnn
This can be understood as follows.
The finite temperature Feynman rules dictate that an additional term $\frac{1}{\beta} \sum_n \int_{\map} \mathcal{D}(i \omega_n, \map)$ has to be included for every propagator and therefore the central loop becomes more severely divergent. When $N$ loops are attached to the central loop, the central loop therefore has a $\int \frac{dp}{p^{2(N-1)}}$ divergence.
We wish to study this kind of ring diagrams and note that, due to the similarity in structure, it is actually possible to sum over the infinite series of ring diagrams.
The ring diagram at order $N$ should be evaluated as
\begin{equation}\label{key}
\beta V
(-\lambda)^N 
\left[
\SumInt_{m, \mathbf{k}}
(\omega_m^2 + \mathbf{k}^2)^{-1}
\right]^{N}
\left[
\SumInt_{n, \mathbf{p}}
(\omega_n^2 + \mathbf{p}^2)^{-N}
\right]
\end{equation}
and one can yield for their \emph{collective contribution to the self-energy} (see \ref{bellac} \ref{kapustabook} for a derivation)
\begin{equation}\label{key}
\bal
\Pi_{\text{IR}} &= 
\frac{1}{2} \beta V \SumInt_{p} \sum_{N = 2}^{\infty} \frac{1}{N} 
\left[
- \Pi_1(i\omega_n, \map) \mathcal{D}_0(i\omega_n, \map)
\right]^{N} \\
&= -
\frac{V}{2} \sum_{n} \int \frac{\dd^3p}{(2\pi)^3} 
\big[ \ln(1 + \Pi_1(i\omega_n, \map) \mathcal{D}_0(i\omega_n, \map))
- \Pi_1(i\omega_n, \map) \mathcal{D}_0(i\omega_n, \map)
\big]^{N}. \\
\eal
\end{equation}
The leading-order contribution to the self-energy is given by $\lambda T^2$ as was shown in section \ref{lambdaphi4}.  We therefore yield (taking into account that the dominant contribution to the divergence is given by $n = 0$)
\begin{equation}\label{key}
\bal
\Pi_{\text{IR}} &= -
\frac{V}{2} \sum_{n} \int \frac{\dd^3p}{(2\pi)^3} 
\left[ \ln(1 + \frac{ \lambda T^2 }{\omega_n^2 + \mathbf{p}^2})
- \frac{ \lambda T^2 }{\omega_n^2 + \mathbf{p}^2}
\right] \\
&= -
\frac{V}{8 \pi^2} \sum_{n} \int p^2 \ \dd p
\left[ \ln(1 + \frac{ \lambda T^2 }{ \mathbf{p}^2})
- \frac{ \lambda T^2 }{ \mathbf{p}^2}
\right] \\
&= -
\frac{V \lambda^{3/2} T^3}{16 \pi^2} \sum_{n} \int 
\dd x
\sqrt x 
\left[ \ln(1 + \frac{ 1 }{ x} )
- \frac{ 1}{ x}
\right] \\
&= -
\frac{V \lambda^{3/2} }{12 \pi^2} T^3,
\eal
\end{equation}
where we used the following substitution of variables $$x = \map^2 / \lambda T^2 \text{, }
\ \ \ dp  = \frac{T}{2} \sqrt{ \frac{\lambda}{x} } dx \ \ \text{ and } \ \ p^2 \ dp = \frac{\lambda^{3/2} T^3 }{2} \sqrt{x} dx.$$
This proves that the second-order correction gives us a term which is proportional to $\lambda^{3/2}$ rather than $\lambda^{2}$. We can conclude two things regarding our perturbative expansions
\begin{itemize}
	\item{Although all ring diagrams are infrared divergent, it turned out to be possible to sum over \emph{all ring diagrams and compute their collective, finite contribution to the self-energy.}}
	\item {Our naive perturbative methods therefore breaks down as a result of IR divergences - note that this conclusion holds true for the perturbative expansions of both the partition function and the self-energy.} 
\end{itemize}

\section{The $\mathbf{T \rightarrow M/2\sqrt{\pi}}$ limit in nonlocal $\phi^4$ theory
\label{appendixTM}
}

\subsubsection*{Low temperature expansion}
It is interesting to examine the regime where $T \rightarrow M/2\sqrt \pi$ rather than $T \ll M$, which is mathematically tantamount to having a large $M$ rather than infinite $M$, such that $e^{- M^2 / 4T^2} \not\approx 1$ in Eq. \ref{f(T/M)}. 
Since $T$ is no longer the characteristic parameter of the theory (now it is accompanied by $M$) the thermodynamic functions may differ from those of the local field theory. Since $\left[\mathz \right] = E^{3}$ one may obtain terms $T^4/M, T^2M, M^3$ rather than merely $T^3$ as in the case of local field theories.
\nnn
Finding a more compact expression for $
\left[
\sum_{n = 1}^{\infty}
{1 \over n^2}
\exp\left( - \frac{n^2 M^2 }{ 4 T^2  } \right)
\right]$ proves cumbersome, on the other hand, since the low-temperature expansion is only employed for $T< M/2\sqrt \pi$, the exponential
$\exp[- n^2 M^2 / 4 T^2 ]$ satisfies
$$e^{- \frac{n^2 M^2 }{ 4 T^2  }} < e^{-n^2 \pi}, $$
where $\exp\left(-n^2 \pi \right)$ is the expression for the exponential at $T = M/2\sqrt{\pi}$. The subdominant ($n=2$) contribution will be neglected since $\exp[-4\pi] \approx 0$ ($\frac{\exp[-4\pi]}{\exp[-\pi]} \approx 8.07 \cdot 10^{-5}$ implies that the second correction is considerably smaller than the first correction) and only the $m=\pm1$ thermal modes are henceforth considered non-negligible.
Having obtained
\begin{equation}\label{key}
f(T/M)
=
\frac{ M}{ 4 T  \sqrt{\pi}}
+
{\pi \sqrt \pi T \over 3M } 
- \frac{2T}{M \sqrt \pi}
e^{- \frac{M^2 }{4 T^2  }}
\end{equation}
the partition function reads (where the additional terms compared to the local field theory case are shown in magenta and the zero-point energy in red)
\begin{equation}\label{lowtempcorrections}
\bal
\ln \mathz(V, T) &= 
{V \pi^2 \over 90} T^3 
- \frac{3  \lambda V }{16 \pi^3 } M^2 T \left[
\frac{M}{4 \sqrt{\pi} T}	
+
{\pi \sqrt \pi T \over 3M } 
- 
{ \color{magenta}
	\frac{2T}{M \sqrt \pi}
	e^{- \frac{M^2 }{4 T^2  }}
}
\right]^2 \\ 
& =
{V \pi^2 \over 90} T^3 
- \frac{3  \lambda V }{16 \pi^3 } \left[
{\color{red}
	\frac{M^4}{16 \pi T}	}
+
{\pi^3 T^3  \over 9 }
{\color{magenta}
	+  \frac{\pi M^2 T }{6}
	-  \frac{M^2 T  }{\pi  } e^{-\frac{M^2}{4T^2}}
	- \frac{4 \pi T^3}{3  }
	e^{- \frac{M^2 }{4 T^2  }}
	+ \frac{4T^3}{ \pi} e^{-\frac{M^2}{2T^2} }
}
\right] \\
& \overset{!}{=} 
{V \pi^2 \over 90} T^3 
- \frac{3  \lambda V }{16 \pi^3 } \left[
{\pi^3  \over 9 }T^3 
{\color{magenta}
	+  \frac{\pi  }{6} M^2 T
	-  \frac{1  }{\pi  } M^2 T e^{-\frac{M^2}{4T^2}}
	- \frac{4 \pi }{3  } T^3
	e^{- \frac{M^2 }{4 T^2  }}
	+ \frac{4}{ \pi} T^3 e^{-\frac{M^2}{2T^2} }
}
\right], \\
\eal
\end{equation}
where ``$\overset{!}{=}$'' in the last step means that we normalized the vacuum term  to zero like we did before in Eq. \ref{vacuumpartition}.
Approaching the scale of nonlocality, minute modifications to the partition function emerge 
(represented in magenta)
which bring about $\mathcal{O}\left(
\exp\left(
-M^2/4T^2
\right)
\right)$ corrections to thermodynamic functions.\footnote{And higher order corrections such as $\mathcal{O}\left(
	\exp\left(
	-M^2/T^2
	\right)
	\right)$ 
	if one does not want to omit minute corrections.} \nnn 
In the next section, we will compute the partition function in the $T > M$ regime and subsequently check whether identical expressions are obtained for the partition function $T = M/2\sqrt{\pi}$ when approaching it from the UV. 

\subsubsection*{High temperature expansion}
Before comparing our results with results from the string field theory literature, let us examine whether the thermodynamic state function are continuous when approaching $T = M/2\sqrt \pi$ from the IR and approaching it from the UV. A discontinuous jump of the partition function would entail a non-smooth transition for the thermodynamic state functions and indicate a second-order phase transition.
\mathleft
\begin{equation}\label{ZIR}
\bal
\underset{\text{IR}}{\ln \mathz}(T = M/2\sqrt{\pi})
&=
\lim_{T \rightarrow M/2\sqrt \pi}
{V \pi^2 \over 90} T^3 
- \frac{3  \lambda V }{16 \pi^3 } \left[
{\pi^3  \over 9 }T^3 
+  \frac{\pi  }{6} M^2 T
-  \frac{1  }{\pi  } M^2 T e^{-\frac{M^2}{4T^2}}
- \frac{4 \pi }{3  } T^3
e^{- \frac{M^2 }{4 T^2  }}
+ \frac{4}{ \pi} T^3 e^{-\frac{M^2}{2T^2} }
\right]
\\
&=
{V \sqrt{\pi} \over 720} M^3 
- \frac{3  \lambda V }{16 \pi^3 } \left[
{\pi \sqrt{\pi}  \over 72 }M^3 
+  \frac{ \sqrt \pi }{12} M^3
-  \frac{1  }{2 \pi \sqrt \pi  } M^3  e^{- \pi }
- \frac{1 }{6 \sqrt{\pi}  } M^3
e^{- \pi }
+ \frac{1}{2 \pi^2 \sqrt \pi} M^3 e^{-2\pi }
\right]
\\
&= 
\frac{V}{32}
\left[
\frac{ \pi^2 }{20 \pi \sqrt \pi }
- \frac{3  \lambda  }{\pi^3 } \left[
{ \sqrt{\pi} (\pi + 6)  \over 36 }
-  \frac{\pi + 6  }{ 6\pi \sqrt \pi  } e^{-\pi}
+ \frac{1}{ \pi^2 \sqrt \pi} e^{- 2\pi }
\right]
\right] M^3 \\
\eal
\end{equation}
\begin{equation}\label{ZUV}
\bal
\underset{\text{UV}}{\ln \mathz}(T = M/2\sqrt{\pi})
&=
\lim_{T \rightarrow M/2\sqrt \pi}
\left[
\frac{V \pi^2 }{90 } T^3
- \frac{ 3 \lambda  V }{16 \pi^3 } M^2 T
- \frac{ 3 \lambda  V }{64 \pi^5 } \frac{M^4}{T}
\left[
e^{- 4 \pi^2 T^2 / M^2 }
+ \frac{1 }{4 \pi^{2} } \frac{M^2}{T^2} e^{- 8 \pi^2 T^2 / M^2 } 
\right]
\right] \\
&=
\frac{V}{32}
\left[
\frac{ \pi^2 }{20 \pi \sqrt \pi } 
- \frac{ 3 \lambda   }{ \pi^3 \sqrt \pi } 
- \frac{ 3 \lambda   }{ \pi^4 \sqrt \pi } 
\left[
e^{- \pi  }
+ \frac{1 }{ \pi }  e^{- 2 \pi } 
\right]
\right]M^3 \\
&=
\frac{V}{32}
\left[
\frac{ \pi^2 }{20 \pi \sqrt \pi } 
- \frac{ 3 \lambda   }{ \pi^3  } \left[\frac{1}{\sqrt \pi}
+ \frac{1}{\pi \sqrt \pi}
\left[
e^{- \pi  }
+ \frac{1 }{ \pi }  e^{- 2 \pi } 
\right]
\right]
\right]M^3
\eal
\end{equation}
\mathcenter
Although the coefficients match for $e^{-2\pi}$ match, the other coefficients do not match ($ {\sqrt{\pi} (\pi + 6)  \over 36 } \neq \frac{1}{\sqrt \pi}$ and $ \frac{\pi + 6  }{ 6\pi \sqrt \pi  } \neq - \frac{1}{\pi \sqrt \pi}$). Note that this may be due to the fact that both the low-temperature and high-temperature expansion contain a summation over Matsubara modes which can neither be solved analytically nor numerically (but can be approximated very accurately in the very low and high temperature regimes, respectively). In the limit $T \rightarrow M/2\sqrt \pi$ and should also include the subdominant contributions from higher (inverse) thermal modes, which have not been included in the partition function of Eqs. \ref{ZIR}, \ref{ZUV}.
\nnn 
In order to examine whether a phase transition occurs at $s=\sqrt{\pi}$ (corresponding to $T=M/2\sqrt{\pi}$).
we expand the $\zeta$ functions
\begin{equation}\label{key}
\zeta(\sqrt{\pi} \pm \epsilon),
\end{equation}
where $\epsilon$ is an infinitesimal real number. We yield the following elliptic theta function for the high-temperature expansion
\begin{equation}\label{key}
\zeta(\sqrt{\pi}+\epsilon) = \sum_{n=-\infty}^{\infty} e^{-n^2(\sqrt{\pi}+\epsilon)^2} = \vartheta_3\left(0,\exp[-(\sqrt{\pi}+\epsilon)^2]\right)
\end{equation}
and likewise for the low temperature expansion
\begin{equation}\label{key}
\zeta(\sqrt{\pi} - \epsilon) = \frac{\sqrt{\pi}}{\sqrt{\pi} - \epsilon} \sum_{m=-\infty}^{\infty} e^{-\pi^2m^2/(\sqrt{\pi}-\epsilon)^2} =
\frac{\sqrt{\pi}}{\sqrt{\pi} - \epsilon} \vartheta_3\left(0,\exp[-\pi^2/(\sqrt{\pi}-\epsilon)^2]\right)
\end{equation}
We conclude that
\begin{equation}\label{key}
\lim_{\epsilon \rightarrow 0} \zeta(\sqrt{\pi}+\epsilon) = \lim_{\epsilon \rightarrow 0} \zeta(\sqrt{\pi}-\epsilon) = \vartheta_3(0, \exp(-\pi^2)) 
\end{equation}
so there is no discontinuity in the partition function. This entails that 
$\underset{\text{IR}}{\ln \mathz}(T = M/2\sqrt{\pi})  = \underset{\text{UV}}{\ln \mathz}(T = M/2\sqrt{\pi})$ when one does not omit subdominant contributions to the partition function.

\section{Absence of thermal duality in open-string tachyon field theory
\label{appendixthermal}
}
	The appearance of a tachyon in conventional field theories usually implies that we are perturbing around an unstable vacuum. Only if a tachyon evolves to its
true vacuum, with positive mass-squared, a vacuum arises around which perturbative
quantum calculations can be  meaningfully performed. Evidence has arisen that this holds true in string theory as well \ref{sftmodels}, \ref{sen}.
The open string tachyons, described by
\begin{equation}
S = \int \ d^{4}x \ \left[	\frac{1}{2} \phi \left[\square - m^2 \right] e^{-\square / M^2} \phi - \lambda \phi^4	\right]
\end{equation}
and it can be noted that this Lagrangian is identical to the nonlocal scalar field considered in chapter \ref{chapternonlocalfield} in the massless limit $m \rightarrow 0$. This allows us \emph{i)} to check our partition function for the nonlocal scalar field and, moreover, \emph{ii)} check whether thermal duality is absent in this reminiscent theory as well.
\nnn
Open string tachyons are thought
to represent the \emph{instability in the various D-brane systems} -
the rolling of the tachyon
from the unstable potential hill to its stable minimum would describe the dissolution of
unstable D-branes into closed string excitations around the true vacuum which no
longer supports open string excitations. 
This process is referred to as \textit{tachyon
	condensation} in the String Field Theory literature and the quantum theory of tachyon condensation has been studied extensively in Ref \ref{sftmodels}. 
The non-zero mass term for the tachyon $(m)$ in Eq. \ref{tachyoncond} leads to a second-order phase transition which is absent for the massless theory considered in section \ref{nonlocallambda4}. The non-zero mass term implies that the minimum (at tree level) is located at
$\phi_0 = \mu / 2\sqrt{\lambda}$ where $\mu = \sqrt{- m^2} > 0$. At finite temperature the minimum is however shifted to smaller values
of $\phi$, which can be accounted for by expanding around the true minimum 
\begin{equation}\label{key}
\phi = v(T) + \phi_f
\end{equation}
where $v(T)$ is independent of space and time but does depend on the temperature,
while $\phi_f$ is the fluctuation around it whose average value is zero. 
The authors of \ref{tachyoncond} concluded that the
values of the condensate $\nu(T)$ and the effective mass both decrease to zero at a critical
temperature $T_c$ and for $T>T_c$ the tachyon has evolved to its true vacuum
\nnn
Let us now discuss the thermodynamic results which were obtained by Biswas, Kapusta and Reddy.
The results which were obtained for temperatures below the critical temperature, \emph{i.e.} $T < T_c$ are nonphysical when taking the massless limit. The true vacuum is realized for $T\geq 0$ in this limit, consequently
no phase transition occurs in this model and  $T_c = 0$. The results for $T > T_c$ thus constitute all physical results in the massless limit. 
\subsubsection*{$\mathbf{T < M/2\sqrt \pi}$}
For the low temperature regime, we have obtained the usual results for local scalar $\lambda \phi^4$ theory. We should expect this behavior since the particles only ``feel'' two derivatives at low temperatures (those from $\square$), since the higher derivatives (those from $a(\square)$) are heavily suppressed by the nonlocality to some even power. 
We have already obtained this result for temperatures far below $M$ in section \ref{lambdaphi4}, now we have proved that these results remain valid for all temperatures $T < M/2\sqrt{\pi}$.
This corroborates Reddy's, Biswas' and Kapusta's local field theory results in the limit $M \rightarrow \infty$ \ref{sftmodels}. The thermodynamic state functions above the critical temperature $T_c$ are given by
\bse
\begin{align}
\epsilon(T) &= 
\left[
\frac{\pi^2}{90} 
- \frac{\lambda}{48}
\right] T^4 
{\color{red} 
	\	+
	\frac{\mu^2 T^2}{8}
	-
	\frac{\mu^4}{16 \lambda}
},\\
P(T) &= 
\left[
\frac{\pi^2}{30} 
- \frac{\lambda}{16}
\right] T^4 
{\color{red} 
	\ +
	\frac{\mu^2 T^2}{24}
	+
	\frac{\mu^4}{16 \lambda}
}, \\
s(T) &= 
4
\left[
\frac{\pi^2}{90} 
- \frac{\lambda}{48}
\right] T^3 
{\color{red} 
	\	+
	\frac{\mu^2 T}{12}
}, 
\end{align}
\ese
where the terms in red vanish in the massless limit (since $m \rightarrow 0$ implies $\mu^2 \rightarrow 0$) for the nonlocal scalar field.
These are identical to the results which were obtained for the local regime of the nonlocalscalar $\phi^4$ theory in section \ref{subsubsectionlowtemperatureexpansion}, where the low-temperature expansion of the $\zeta$ function was utilized. Corrections to these formulas for large but finite $M$ are suppressed by the factor $\exp\left(-M^2 / 4T^2 \right)$ as can be concluded from Eq. \ref{lowtempcorrections}, corroborating the identical claim made in Ref. \ref{sftmodels}. 
\subsubsection*{$\mathbf{T>M/2\sqrt{\pi}}$}
The results which have been obtained in Ref. \ref{sftmodels} for open-string tachyons at high temperatures ($T > M/2\sqrt{\pi}$) are
\bse
\begin{align}\label{tacyhonicpressure}
\epsilon(T) &= \frac{\pi^2}{30} T^4 + \left[
{\color{red}
	\frac{\mu^2}{24}
}
- \frac{3\lambda M^2}{16 \pi^3}
\right] T^2 \
{
	\color{red}
	+
	\frac{\mu^4}{16 \lambda}
}, \\
P(T) &= \frac{\pi^2}{90} T^4 + \left[
{\color{red}
	\frac{\mu^2}{24}
}
- \frac{3\lambda M^2}{16 \pi^3}
\right] T^2 \
{\color{red}
	- \frac{\mu^4}{16 \lambda}
}, \\
s(T) &= \frac{2\pi^2}{45} T^3 + \left[
{\color{red}
	\frac{\mu^2}{12}
}
- \frac{3\lambda M^2}{8 \pi^3}
\right] T, 
\end{align}
\ese
where all terms in red vanish for our massless scalar theory. Note that these results are identical to the thermodynamic state functions where were derived in chapter \ref{chapternonlocalfield}. 
\subsubsection*{Thermal duality}
Thermal duality entails that
the partition function satisfies the following asymptotic behavior
\begin{equation}
\ln \mathz \rightarrow
\begin{cases}
- \Lambda V / T 	& \text{for } T \ll M \\ 
-4 \pi\Lambda  T V / M^2 	& \text{for } T \gg M \\
\end{cases},
\end{equation}
we thus conclude that thermal duality is violated by open-string tachyons. Given that the theories from the local regime are correct (as we have checked by computing these terms for the local scalar field), thermal duality predicts that the thermodynamic functions in the high temperature regime would satisfy
$\epsilon, p \propto \left( T^4 - \frac{T^6}{M^2} \right) $ and $s \propto \left( T^3 - \frac{T^5}{M^2} \right)$, while in fact these functions increase more slowly with temperature in this regime. As has been argued in section \ref{sectionthermalduality}, this may be due to the fact that the Lagrangian arises from the simplest level of truncation of the full SFT action and consequently cannot capture all stringy phenomena.

\chapter*{Bibliography}
\addcontentsline{toc}{chapter}{Bibliography}
\begin{enumerate}[label={[\arabic*]}, noitemsep]

	\item {A. Das (2006), ``Thermal operator representation of feynman graphs, Braz. J. Phys. vol. \textbf{36} no. 4a S$\tilde{a}$o Paulo Dec. 2006.
\label{das}	
}

\item{D. J. Gross, R. D. Pisarski and L. G. Yaffe, Rev. Mod. Phys. \textbf{53} (1981) 43.
\label{refgross}
}

\item{K. Rajagopal and F. Wilczek, Nuc. Phys. \textbf{B204} (1993) 577.
\label{wilczek}
}

\item{R. Kubo, J. Phys. Soc. Japan, \textbf{12} (1957) 570; P. Martin and J. Schwinger, Phys. Rev. \textbf{115}
	(1959) 1342.
\label{refkubo}
}

\item{C. Bloch, Nuc. Phys. \textbf{7} (1958) \textbf{451}.
\label{refbloch}
}

\item{
	H. Umezawa, H. Matsumoto and M. Tachiki (1982)
	``Thermo field dynamics and condensed states,''
North-Holland Pub. Co., 1982 \label{umezawa}
}

\item{
F. Guerra	(2005), ``Euclidean Field Theory,
 	arXiv:math-ph/0510087.
\label{euclideantimeorder}
}

\item{M. L. Bellac (1996), ``Thermal Field Theory'', Cambridge University Press.
\label{bellac}
}

\item{Y. Yang (2011),
``An Introduction to
Thermal Field Theory'', Imperial College of Science, Technology and Medicine.
Master's thesis.
\label{yang}
}

\item{
B. P. Abbott et al. [LIGO Scientific and Virgo Collaborations],
``Observation of Gravitational Waves from a
Binary Black Hole Merger,'' Phys. Rev. Lett. \textbf{116} (2016)
no.6, 061102.
\label{ligo}
}

\item{
 G.	Aad, \emph{et. al.},
ATLAS collaboration (2010),	
	``Observation of a Centrality-Dependent Dijet Asymmetry in Lead-Lead Collisions at $\sqrt{S_{\text{NN}}} = 2.76$ TeV with the ATLAS Detector at the LHC'', Phys. Rev. Lett. \tb{105}, 252303
	\label{atlas}
}

\item{
S. Chatrchyan, \emph{et. al.}. CMS collaboration (2011),
	``Observation and studies of jet quenching in PbPb collisions at nucleon-nucleon center-of-mass energy = 2.76 TeV.''
Phys. Rev. \tb{C84}, 024906.
\label{cms}
}

\item{
Aamodt, K. \emph{et. al.}, ALICE collaboration (2011),	
	``Suppression of charged particle production at large transverse momentum in central Pb-Pb collisions at $\sqrt{S_{\text{NN}}} = 2.76$ TeV.''
Phys. Lett. \tb{B696} (1-2), 30-39.
\label{alice}
}

	\item {A.D. Sakharov (1967), JETP Lett. \textbf{5}, 24 (1967) [Pisma Zh. Eksp. Teor. Fiz. \textbf{5}, 32].	
\label{sakharov}
}

\item{
G. 't Hooft and M. J. G. Veltman, ``One loop divergences in the theory of
gravitation,'' Annales Inst. Poincar\'{e} Phys. Theor. A \textbf{20} (1974) 69
\label{hooft}
}

\item{
 A. E. M. van de Ven, ``Two loop quantum gravity'', Nucl. Phys. \textbf{B378}, 309 (1992).
\label{ven}
}

\item{ 
K.S. Babu, E. Kearns  (2013),
``Baryon Number Violation'',  	arXiv:1311.5285 [hep-ph].
\label{baryonviolation} 
}

	\item {H. Satz (1985), Annu. Rev. Nucl. Part. Sci. \textbf{35} (1985) 245. \label{qcdplasma} }

\item{T. S. Evans and A. C. Pearson, Phys. Review \bf{D52} (1995) 4652
\label{refrealtime}
}

\item{M.W. Clifford (1986), ``Was Einstein right?: putting general relativity to the test,'' New
	York: Basic Books, 1986.

\label{refeinstein}
}

\item{T. Matsubara (1955), ``A New Approach to Quantum-Statistical Mechanics'',
Progress of Theoretical Physics, Volume 14, Issue 4, 1 October 1955, Pages 351–378, https://doi.org/10.1143/PTP.14.351.
\label{refmatsubara}
}
	
	\item {R. Baier, B. Pire and D. Schiff, Phys. Rev. \textbf{D38} (1988) 2814; T. Altherr, P. Aurenche
		and T. Becherrawy, Nucl. Phys. \textbf{B315} (1989) 436; T. Altherr and P. Aurenche,
		Z. Phys. \textbf{C45} (1989) 99; T. Altherr and T. Becherrawy, Nucl. Phys.
		\textbf{B330} (1990) 174; Y. Gabellini, T. Grandou and D. Poizat, Ann. Phys. (NY)
		\textbf{202} (1990) 436.
		\label{plasmares1}	
}
		\item {		 T. Altherr and P. V. Ruuskanen, Nucl. Phys. \textbf{B380} (1992) 377. 
		\label{plasmares2}	
	}

\item{E. Braaten, R.D. Pisarski (1988), ``Scattering amplitudes in hot gauge theories'' R.D., Nucl. Phys. \textbf{B337}, 569 (1990).
\label{pisarski1}
}
		\item { E. Braaten, R. D. Pisarski (1990),
			``Soft Amplitudes in hot gauge theories: a general analysis.'' Nucl Phys. \tb{B337}, 569. \label{pisarki}
\label{plasmares3}	
}

\item {M.E. Peskin \& D. V. Schroeder (1995), ``An Introduction To Quantum Field Theory'', Avalon Publishing. \label{peskin}
}
	
	\item {} M. Gaul and C. Rovelli, ``Loop Quantum Gravity and the Meaning of Diffeomorphism Invariance,''  	arXiv:gr-qc/9910079 \label{gaul}
	
	\item {\label{pol}} T. Biswas, T. Koivisto and A. Mazumdar,  ``Nonlocal theories of gravity: the flat space propagator,''  	arXiv:1302.0532.
	
	\item {} S. Talaganis, T. Biswas and A. Mazumdar, ``Towards understanding the ultraviolet behavior of quantum loops in infinite-derivative theories of gravity.'' In \textit{Proceedings, Barcelona Postgrad
		Encounters on Fundamental Physics}, 2013.
	\label{UV}
	
	\item{J.F. Donoghue \& B.R. Holstein (1983), ``Renormalization and radiative corrections at finite temperature.''
	\label{renorm}}
	
	\item {} J.A.R. Cembranos, T. Biswas and J.I. Kapusta, ``Thermodynamics and Cosmological Constant of Non-Local Field Theories from p-Adic Strings,'' https://arxiv.org/abs/1005.0430
	\label{p-adic}

\item{J. Hwang and H. Noh (2001) Phys. Lett. B  506, 13 [arXiv:astro-ph/0102423].
\label{f(r)inflation1}
}

\item{C. S. Cardone V. F. and A. Troisi (2005) Phys. Rev. D  71, 043503 [arXiv:astro-ph/0501426] 
\label{f(r)inflation2}
}

\item{N. Ohta; ``Quantum equivalence of $f(\mathcal{R})$ gravity and scalar–tensor theories in the Jordan and Einstein frames''. Progress of Theoretical and Experimental Physics, Volume 2018, Issue 3, 1 March 2018, \textbf{033B02}.
\label{f(r)ref}
}
	
	\item {S. Talaganis (2016), ``Towards understanding the ultraviolet behavior of
		quantum loops in infinite-derivative theories of
		gravity'', \url{https://arxiv.org/pdf/1508.07410.pdf} \label{UVinf}}
	
	\item {} S. Talaganis (2017), ``Towards UV Finiteness of Infinite Derivative Theories of Gravity and Field Theories,'' https://arxiv.org/abs/1704.08674 \label{UVfiniteness}
	\item {\label{singfree} } T. Biswas, E. Gerwick, T. Koivisto, and A. Mazumdar.
	``Towards singularity and ghost free theories of gravity.'' \textit{Phys.Rev.Lett.},
	108:031101, 2012.
	\item {} G. Esposito,	``An introduction to quantum gravity,'' https://arxiv.org/abs/1108.3269
	\item {M. Ostrogradsky, ``Memoires sur les equations differentielles relatives au probleme des
		isoperimetretres'', Mem. Acad. St. Petersburg \textbf{6} (1850) 385. \label{ostrogradsky} }
	\item {
		D. Roest \& R. Klein, ``Exorcising the Ostrogradsky ghost in coupled systems'', \url{https://arxiv.org/abs/1604.01719} \label{roest} }
	\item {S.M. Carroll (2004), ``Spacetime and Geometry: An Introduction to General Relativity'', Pearson. \label{carroll} }
	\item {} K.S. Stelle, ``\textit{Renormalization Of Higher Derivative Quantum Gravity}”, Phys. Rev. \textbf{D16}, 953-969 (1977) \label{stelle}
	
	\item {M. H. Goroff and A. Sagnotti (1986), ``The Ultraviolet Behavior of Einstein
		Gravity.'' Nucl. Phys., \textbf{B266}:709–736, 1986.
\label{refeinsteinuv}	
}
	
	\item {D. Lovelock (1971), ``The Einstein Tensor and its Generalizations.'' Journal of
		Mathematical Physics, \textbf{12}(3), 1971.
\label{lovelock2}	
}
	
	\item{D. Lovelock. ``The uniqueness of the Einstein field equations in a four-dimensional
		Space.'' Archive for Rational Mechanics and Analysis, 33(1):54–
		70, 1969.\label{lovelock} }
	
	\item{J.I. Kapusta, ``Finite-temperature Field Theory,
		Principles and Applications" (2006). \label{kapustabook}}
	
\item {S. Capozziello et al. (2009), ``Accelerating cosmologies from non-local higher-derivative
	gravity'', \textit{Phys.Lett.} \textbf{B671} (2009), 193.
\label{capozziello}
}	

\item {H. Yukawa (1950)} \label{yukawa0}
\begin{enumerate}[label=(\alph*)]
	\item {}
``Quantum Field Theory of Nonlocal Fields. Part I: Free Fields,'' Phys.
	Rev. \tb{77}, 219 (1950). \label{yukawa1}
\item {} ``Quantum Field Theory of Nonlocal Fields. Part II: Irreducible Fields
	and Their Interactions,'' Phys. Rev. \tb{80}, 1047 (1950). \label{yukawa2}
\end{enumerate}

\item{S. Deser, R. Woodard, ``Nonlocal Cosmology'', \textit{Phys.Rev.Lett. }\textbf{99} (2007), 111301.
\label{refdarkenergy1}
}

\item {T. Koivisto, ``Dynamics of Nonlocal Cosmology'', \textit{Phys.Rev. }\textbf{D77} (2008), 123513.
\label{refdarkenergy2}
}

\item {T. S. Koivisto, ``Newtonian limit of nonlocal cosmology'', \textit{Phys.Rev. }\textbf{D78} (2008),
	123505.
\label{refstructureform1}
}

\item {S. Park, S. Dodelson, ``Structure formation in a nonlocally modified gravity
	model'', \textit{Phys.Rev. }\textbf{D87} (2013), 024003.
\label{refstructureform2}
}

\item {N. Arkani-Hamed et al., ``Nonlocal modification of gravity and the cosmological
	constant problem'' (2002), arXiv:hep-th/0209227 [hep-th].
\label{refscreening1}
}

\item {Y.l. Zhang, M. Sasaki, ``Screening of cosmological constant in non-local cosmology'',
	\textit{Int.J.Mod.Phys. }\textbf{D21} (2012), 1250006.
\label{refscreening2}
}

\item {M. Soussa, R. P. Woodard, ``A Nonlocal metric formulation of MOND'', \textit{Class.Quant.Grav.}
	\textbf{20} (2003), 2737.
\label{refdarkmatter1}
}
\item {H.J. Blome \emph{et al.}, ``Nonlocal Modification of Newtonian Gravity'', \textit{Phys.Rev.
}	\textbf{D81} (2010), 065020
\label{refdarkmatter2}
}

\item{
	D.A. Lowe and L. Thorlocius (1994),
``Hot String Soup'', \url{https://arxiv.org/pdf/hep-th/9408134.pdf}.
\label{refstringsoup}
}

\item{T. Biswas, A. Mazumdar, and W. Siegel. ``Bouncing universes
	in string-inspired gravity,'' \textit{JCAP}, 0603:009, 2006.	\label{bounce}}

\item {T. Biswas, R. Brandenberger, A. Mazumdar, W. Siegel, ``Non-perturbative Gravity, Hagedorn Bounce \& CMB'', JCAP 0712, 011 (2007) doi:10.1088/1475-7516/2007/12/011
	[hep-th/0610274]  (\url{https://arxiv.org/abs/hep-th/0610274}).
\label{nonperturbCMB}
}

\item {B. Zwiebach, ``Curvature Squared Terms And String Theories,'' Phys. Lett. B \textbf{156}, 315
	(1985).
\label{gaussbonnet}
}

\item {G. Calcagni, B. de Carlos and A. De Felice, ``Ghost conditions for Gauss-Bonnet
	cosmologies,'' Nucl. Phys. B \textbf{752}, 404 (2006) [arXiv:hep-th/0604201].
\label{gaussbonnetanalysis}
}

\item {T. Biswas, T. Koivisto and A. Mazumdar, ``Towards a resolution of the cosmological singularity in non-local higher
	derivative theories of gravity,'' JCAP \textbf{1011}, 008 (2010).
\label{singularityhigherderivative}
}

	\item {T. Biswas, A. S. Koshelev, and A. Mazumdar. Gravitational
		theories with stable (anti-)de Sitter backgrounds. \textit{Fundam. Theor.
			Phys., 183:97–114, 2016.
			\label{sitter} }}
	
	\item {T. Biswas, A. Conroy, A. S. Koshelev, and A. Mazumdar. Generalized ghost-free quadratic curvature gravity.
		\textit{Class.Quant.Grav.}, 31:015022, 2014. \label{genquad} }
	
	\item {J. Edholm, A. S. Koshelev and A. Mazumdar, “Behavior of the Newtonian potential for ghost-free gravity and singularityfree
		gravity,” Phys. Rev. D \textbf{94}, no. 10, 104033 (2016).
	\label{newtonianpotential}	
}

\item {A. S. Cornell, G. Harmsen, G. Lambiase and A. Mazumdar, ``Rotating metric in Non-Singular Infinite Derivative Theories
	of Gravity,'' arXiv:1710.02162 [gr-qc].
\label{rotating} }
	
	\item {Conroy, A. Infinite Derivative Gravity:
		A Ghost and Singularity-free
		Theory. \url{https://arxiv.org/pdf/1704.07211.pdf}. \label{conroy}
	}

\item {T. Biswas, J. Kapusta and A. Reddy, Thermodynamics of String Field Theory Motivated Nonlocal Models, \url{https://arxiv.org/pdf/1201.1580.pdf}. \label{sftmodels} }

\item {V. A. Kostelecky and S. Samuel, Phys. Lett. \tb{B207}, 169 (1988)
\label{tachyonquantum} }
	
\item {S. P. de Alwis, Strings in background fields: $\beta$ functions and vertex operators. Phys. Rev. D \textbf{34}, 3760 (1986). \label{alwis} }	

\item {J. Z. Simon, Higher-derivative Lagrangians, nonlocality, problems and solutions. Phys.Rev. D\textbf{41} (1990). \label{simon} }	

\item {J. A. Wheeler and R. P. Feynman, Rev. Mod. Phys. \textbf{21}, 245 (1949). \label{feynmanwheeler}}

\item {W. Siegel, ``Introduction to string field theory,” hep-th/0107094. \label{siegelbook} }

\item {J.J. Atick and E. Witten (1988), ``The Hagedorn Transition and the Number of Degrees of Freedom of
	String Theory'', Nucl. Phys. \textbf{B310} (1988) 291.
\label{wittenatick}
}

\item{E. Witten (1986), Nucl. Phys. \textbf{B268}, 253 (1986).
\label{witten1986}
}

\item {
T. Biswas, T. Koivisto and A. Mazumdar (2014),	
``Atick-Witten Hagedorn conjecture, near
	scale-invariant matter and blue-tilted gravity power
	spectrum'', \url{https://www.rug.nl/research/portal/files/55385086/Atick_Witten_Hagedorn_conjecture_near_scale_invariant_matter_and_blue_tilted_gravity_power_spectrum.pdf}.
	\label{wittenatick2}
}

\item {K. Krasnov, ``Renormalizable Non-Metric Quantum Gravity?,” hep-th/0611182
\label{asymptoticsafety}
}

\item {A. Zee (2010), ``Quantum Field Theory in a Nutshell.'' Princeton University Press; 2nd edition. \label{zee} }

\item{M. Maggiore (2005), ``A Modern Introduction to Quantum Field Theory'', 2005, Oxford
	University Press.
\label{maggiore}
}

\item { T. Koivisto, “Covariant conservation of energy momentum in modified gravities,”
	Class. Quant. Grav 23 (2006), 4289. \label{koivisto} }

\item {Ch.G. van Weert`(2012), ``An Introduction to
	Real- and Imaginary-time
	Thermal Field Theory.'' Lecture Notes. \label{weert} }

\item{I. Dimitrijevic, B. Dragovich, J. Grujic, and Z. Rakic. ``New Cosmological
Solutions in Nonlocal Modified Gravity.'' \textit{Rom. J. Phys.}, 58(5-6):550–559,
2013. \label{dimitri}}
\item{ G. Calcagni, L. Modesto, and P. Nicolini. ``Superaccelerating
bouncing cosmology in asymptotically-free non-local gravity.''
\textit{Eur. Phys. J.}, C74(8):2999, 2014.}
\item{ I. Dimitrijevic, B. Dragovich, J. Grujic, and Z. Rakic. ``On
Modified Gravity.'' \textit{Springer Proc. Math. Stat.}, 36:251–259, 2013.}
\item{ A. Conroy, A. S. Koshelev, and Anupam Mazumdar.}
\begin{enumerate}
	\item {(2017) ``Defocusing
		of Null Rays in Infinite Derivative Gravity.'' \textit{JCAP}, 1701(01):017,.}
	\item{ (2014) ``Geodesic
		completeness and homogeneity condition for cosmic inflation.'' \textit{Phys. Rev.},
		D90(12):123525.}
\end{enumerate}
\item{ D. Chialva and A. Mazumdar. Cosmological implications of
quantum corrections and higher-derivative extension. \textit{Mod. Phys. Lett.},
A30(03n04):1540008, 2015.}
\item{ T. Biswas, A. S. Koshelev, A. Mazumdar, and S. Y.
Vernov (2012), ``Stable bounce and inflation in non-local higher derivative cosmology.''
\textit{JCAP}, \textbf{1208}:024, 2012.
\label{refstablebounce}
}
\item{ B. Craps, T. D. Jonckheere, and A. S. Koshelev. ``Cosmological
perturbations in non-local higher-derivative gravity.'' \textit{JCAP}, 1411(11):022,
2014. \label{craps}}

\item {M. H. Goroff and A. Sagnotti, “The Ultraviolet Behavior of Einstein Gravity,” Nucl. Phys. B \textbf{266}, 709
	(1986).
	\label{sagnotti1}
}
\item { M. H. Goroff and A. Sagnotti, “Quantum Gravity At Two Loops,” Phys. Lett. B \textbf{160}, 81 (1985).
	\label{sagnotti2}
}

\item {R. P. Geroch, ``What is a singularity in general relativity?,'' Annals Phys. \textbf{48}, 526 (1968). doi:10.1016/0003-
	4916(68)90144-9 R. P. Geroch, ``Local characterization of singularities in general relativity,'' J. Math. Phys. \textbf{9}, 450 (1968).
	doi:10.1063/1.1664599 R. P. Geroch, C. B. Liang and R. M. Wald, J. Math. Phys. \textbf{23}, 432 (1982). doi:10.1063/1.525365 
\label{geroch}
}

\item { R. M. Wald, “General Relativity,” doi:10.7208/chicago/9780226870373.0
\label{waltbook}
}

\item {T. Biswas and S. Talaganis (2015), ``String-inspired infinite-derivative theories of gravity: A brief overview'', \url{https://arxiv.org/pdf/1412.4256.pd} \label{biswasstring}}

\item {C.D. Carone, ``Unitarity and microscopic acausality in a nonlocal theory'', \url{https://arxiv.org/pdf/1605.02030.pdf} \label{carone}}

\item{S. Coleman, in
	Subnuclear Phenomena, Proceedings of the 7th International
	School of Subnuclear Physics
	, edited by A. Zichichi (Academic Press, New York, 1970) p. 28
	2
\label{coleman}
}
\item{
	L. Buoninfante, A. Mazumdar, G. Lambiase (2018),
	``Ghost-free infinite derivative quantum field theory''.
\label{ghostinfinite}
}

\item{L. Buoninfante,
``Ghost and singularity free theories of gravity'' (2016), \url{https://arxiv.org/pdf/1610.08744.pdf}.
\label{luca}
}

\item {Y. V. Kuz’min (1989), ``Finite nonlocal gravity'', Yad. Fiz. \textbf{50}, 1630-1635. \label{kuzmin} }
\item { E. T. Tomboulis, ``Superrenormalizable gauge and gravitational theories,''
\url{hep-th/9702146}. \label{superrenormalizablegauge } }

\item {J. W. Moffat (2011), ``Ultraviolet Complete Quantum Gravity.'' Eur. Phys. J. Plus,
	\textbf{126}:43, 2011.
\label{moffat}
}

\item {J. W. Moffat,} \label{moffatnonlocal}
\begin{enumerate}
\item {}
``Ultraviolet Complete Electroweak Model Without a Higgs Particle,''
	Eur. Phys. J. Plus \tb{126}, 53 (2011).
\item {}	J. W. Moffat and V. T. Toth, ``Redesigning Electroweak Theory: Does the Higgs
	Particle Exist?,'' arXiv:0908.0780 [hep-ph].
\item {}	J. W. Moffat and V. T. Toth, ``A Finite Electroweak Model without a Higgs Particle,''
	arXiv:0812.1991 [hep-ph].
\item {}	J. W. Moffat, ``Electroweak Model Without A Higgs Particle,'' arXiv:0709.4269 [hep-ph].
\end{enumerate}
\item {
	 T. Biswas and N. Okada,
	``Towards LHC physics with nonlocal Standard Model''Nucl. Phys.
	B \textbf{898}, 113 (2015).
\label{biswasokada}
}

\item {A. Sen} 
\begin{enumerate}
	\item {}
Int. J. Mod. Phys. \textbf{A20}, 5513-5656 (2005).
\item {} A. Sen, JHEP \textbf{0204}, 048 (2002).
\label{sen} 
\end{enumerate}

\item {
L. Modesto and L. Rachwal, “Super-renormalizable \& Finite Gravitational Theories,”
arXiv:1407.8036 [hep-th]. \label{modesto} }

\item {E. Tomboulis, “Renormalizability and Asymptotic Freedom in Quantum Gravity,”
	Phys. Lett. B \textbf{97}, 77 (1980).
\label{tomboulis}
}
\item{ P. Van Nieuwenhuizen, ``On ghost-free tensor lagrangians and linearized
	gravitation,'' Nucl. Phys. \tb{B60} (1973), 478. \label{nieuwenhuizen} }

\item {J. Edholm, A. S. Koshelev, and A. Mazumdar (2016), ``Behavior of
	the Newtonian potential for ghost-free gravity and singularity-free gravity.''
	Phys. Rev., \textbf{D94}(10):104033, 2016.
\label{IDGNewtonianpotential}
	}

\item {T. Biswas, A. S. Koshelev and A. Mazumdar, “Gravitational
	theories with stable (anti-)de Sitter backgrounds,”
	Fundam. Theor. Phys. \textbf{183}, 97 (2016). T. Biswas,
	A. S. Koshelev and A. Mazumdar, “Consistent higher
	derivative gravitational theories with stable de Sitter and
	anti?de Sitter backgrounds,” Phys. Rev. D \textbf{95}, no. 4,
	043533 (2017).
	\label{stablebackground}
}

\item {L. Buoninfante, A. S. Koshelev, G. Lambiase \& A. Mazumdar, Classical properties of non-local, ghost- and singularity-free gravity, \url{https://arxiv.org/pdf/1802.00399.pdf}.
\label{classicalproperties}
}

\item { D. J. Kapner, T. S. Cook, E. G. Adelberger, J. H. Gundlach,
	B. R. Heckel, C. D. Hoyle and H. E. Swanson, Phys.
	Rev. Lett. \textbf{98} (2007) 021101.
\label{newtonpotentialtest}
}

\item {A. S. Koshelev, J. Marto, A. Mazumdar, ``Towards non-singular metric solution in infinite derivative gravity'', \url{https://arxiv.org/pdf/1803.00309.pdf}.
	\label{nonsingularsolution}
}

\item {Alexey S. Koshelev, Jo\~{a}o Marto, Anupam Mazumdar, ``Towards resolution of anisotropic cosmological singularity in infinite derivative gravity'', \url{https://arxiv.org/pdf/1803.07072.pdf}.
\label{resolutionanisotropicsingularity}
 }

\item {R. Penrose (1969), ``Gravitational collapse: The role of general relativity'', Riv. Nuovo Cim. 1 (1969) 252-276.
	\label{penrose}
}

\item {R. M. Wald, ``Gravitational collapse and cosmic censorship,'' Fundam. Theor. Phys. \textbf{100}, 69 (1999) doi:10.1007/978-94-
	017-0934-75 [gr-qc/9710068].
\label{wald}
}

\item {T. Biswas, M. Grisaru and W. Siegel, ``Linear Regge trajectories from worldsheet
	lattice parton field theory,'' \textit{Nucl. Phys. B }\textbf{708}, 317 (2005) [hep-th/0409089] \label{regge} }

\item {T. Biswas, J. A. R. Cembranos and J. I. Kapusta, ``Finite Temperature Solitons
	in Non-Local Field Theories from p-Adic Strings,'' \textit{Phys. Rev. }\textbf{D 82}, 085028
	(2010) [arXiv:1006.4098 [hep-th]] \label{hagedorn1} }

\item {T. Biswas, J. A. R. Cembranos and J. I. Kapusta, ``Thermal Duality and Hagedorn Transition from p-adic Strings,'' \textit{Phys. Rev. Lett. }\textbf{104}, 021601 (2010)
	[arXiv:0910.2274 [hep-th]]. \label{hagedorn2} }

\item {T. Biswas, J. A. R. Cembranos and J. I. Kapusta, ``Thermodynamics and Cosmological
Constant of Non-Local Field Theories from p-Adic Strings,'' \textit{JHEP
}\textbf{1010}, 048 (2010) [arXiv:1005.0430 [hep-th]]. \label{hagedorn3}}

\item {J. Polchinski, Commun. Math. Phys. \tb{104}, 37 (1986).
\label{hagedornpolchinski}
}

\item {B. Sathiapalan and N. Sircar, JHEP \tb{0808}, 019 (2008).
\label{hagedornsathiapalan}
}

\item {K. R. Dienes and M. Lennek (2005)
\begin{enumerate}
	\item {	``Thermal Duality Confronts Entropy:
		A New Approach to String Thermodynamics?'' \url{https://arxiv.org/pdf/hep-th/0312173.pdf}
\label{dienes3}	
 }
	\item {	``Re-Identifying the Hagedorn Transition,'' arXiv:hep-th/0505233. 
\label{dienes}	
}
\end{enumerate}
	arXiv:hep-th/0505233.
\label{hagedorndienes}
}

\item {T. Biswas and S. Alexander (2009), 
	``Emergence of a Cyclic Universe from the Hagedorn Soup'',
	Phys. Rev. D 80, 043511.
\label{hagedornsoup}
}

\item {A. Nayeri, R. H. Brandenberger and C. Vafa, Phys. Rev.
	Lett. 97, 021302 (2006), \\ T. Biswas, R. Brandenberger,
	A. Mazumdar and W. Siegel, JCAP 0712, 011 (2007).
\label{hagedornbrandenberger}
}

\item {R. Bluhm, Phys. Rev. D \tb{43}, 4042 (1991)
\label{bluhm}
}

\item {A. Mazumdar \& S. Talaganis (2016),
	``High-Energy Scatterings in Infinite-Derivative Field
	Theory and Ghost-Free Gravity'', \url{https://arxiv.org/pdf/1603.03440.pdf}. \label{IDGscattering} }

\item {T. Clifton, P. G. Ferreira, A. Padilla, and C.
	Skordis. ``Modified Gravity and Cosmology.'' Phys. Rept., \textbf{513}:1–189, 2012. \label{constantinos} }

\item {M. Schnabl, “Analytic Solution for Tachyon Condensation in Open String Field
	Theory,” Adv. Theor. Math. Phys. 10, 433 (2006) [arXiv:hep-th/0511286].

\label{tachyon1}
}

\item {Y. Okawa, “Comments on Schnabl’s Analytic Solution for Tachyon Condensation in
	Witten’s Open String Field Theory,” JHEP 0604, 055 (2006)
	[arXiv:hep-th/0603159].
\label{tachyon2}
}

\item {M. Kiermaier, Y. Okawa, L. Rastelli and B. Zwiebach, “Analytic Solutions for
	Marginal Deformations in Open String Field Theory,” JHEP 0801, 028 (2008)
	[arXiv:hep-th/0701249].
\label{tachyon3}

}

\item { W. Taylor and B. Zwiebach, “D-branes, Tachyons, and String Field Theory,”
	arXiv:hep-th/0311017.
\label{tachyon4}
}

\item { C. de Rham, “The Effective Field Theory of Codimension-two Branes,” JHEP 0801,
	060 (2008) [arXiv:0707.0884 [hep-th]].
\label{codimension}
}

\item {N. Moeller and B. Zwiebach, “Dynamics with Infinitely Many Time Derivatives and
	Rolling Tachyons,” JHEP 0210, 034 (2002) [arXiv:hep-th/0207107]
\label{p-adicmodel}
}

\item {B. Dragovich, “Zeta Strings,” arXiv:hep-th/0703008.
\label{zetastring}
}

\item {M. R. Douglas and N. A. Nekrasov, “Noncommutative Field Theory,” Rev. Mod.
	Phys. 73, 977 (2001) [arXiv:hep-th/0106048].
\label{noncommutative}
}

\item { S. Hossenfelder, “Self-Consistency in Theories with a Minimal Length,” Class.
	Quant. Grav. 23, 1815 (2006) [arXiv:hep-th/0510245
\label{minimallengthmodel}
}

\item { A. Ludu, R. A. Ionescu and W. Greiner, “Generalized KdV Equation for Fluid
	Dynamics and Quantum Algebras,” Found. Phys. 26, 665 (1996)
	[arXiv:q-alg/9612006].
\label{quantumalgebra}
}

\item { 
N. Barnaby	\& N Kamran. (2008),
	``Dynamics with Infinitely Many Derivatives: The Initial
	Value Problem'', \url{https://arxiv.org/pdf/0709.3968.pdf}
	\label{refinitialvalueinfinite}
}

\item{N. Bodendorfer (2016), ``An elementary introduction to loop quantum gravity \label{reflqc}
	''}
\end{enumerate}

\end{document}